\documentclass[longauth]{aa}  
\pdfoutput=1

\usepackage{blindtext}
\usepackage{graphicx}
\usepackage{multicol}
\usepackage{tabularx}
\usepackage{rotating, booktabs}
\usepackage{txfonts}
\usepackage{filecontents}
\usepackage{threeparttable}
\usepackage{natbib}
\usepackage[toc,page]{appendix}
\usepackage{subfig}
\usepackage{lscape}
\bibpunct{(}{)}{;}{a}{}{,}

\usepackage{gensymb}

\usepackage{xcolor}
\usepackage{graphicx}
\usepackage{txfonts}
\def\aox{$\rm {\alpha_{OX}}$}

\def\ltsima{$\; \buildrel < \over \sim \;$}
\def\simlt{\lower.5ex\hbox{\ltsima}}
\def\gtsima{$\; \buildrel > \over \sim \;$}
\def\simgt{\lower.5ex\hbox{\gtsima}}

\def\ergs{{erg s$^{-1}$}}
\def\cm2{{cm$^{-2}$}}
\def\kms{{km s$^{-1}$}}

\def\lbol{L$_\textup{Bol}$}
\def\edd{$\lambda_\textup{Edd}$}
\def\vciv{v$\rm_\textup{50}^\textup{CIV}$}

\def\mpc{Mpc$^{-1}$}
\def\mbh{M$\rm_{BH}$}

\def\err{$\pm$}

\usepackage{hyperref}

\begin{document} 

 \title{SUPER III. Broad Line Region properties of AGN at z$\sim$2}

 \author{G. Vietri \inst{1,2,3}, V. Mainieri\inst{3}, D. Kakkad\inst{3,4},  H. Netzer\inst{5}, M. Perna\inst{6,7}, C. Circosta\inst{3,8},
    C. M. Harrison\inst{9}, L. Zappacosta\inst{10}, B. Husemann\inst{11}, P. Padovani \inst{3}, M. Bischetti \inst{12}, A. Bongiorno\inst{10}, M. Brusa\inst{13,14}, S. Carniani\inst{15},  C. Cicone\inst{16}, A. Comastri\inst{14}, G. Cresci\inst{7}, C. Feruglio\inst{12}, F. Fiore\inst{12}, G. Lanzuisi\inst{14}, F. Mannucci\inst{7}, A. Marconi\inst{7,17}, E. Piconcelli\inst{10}, A. Puglisi\inst{18,19,20}, M. Salvato\inst{21}, M. Schramm\inst{22}, A. Schulze\inst{23}, J. Scholtz\inst{20,24}, C. Vignali\inst{13,14}, G. Zamorani\inst{14}}
          
   \institute{INAF - Istituto di Astrofisica Spaziale e Fisica Cosmica Milano, Via Alfonso Corti 12, 20133 Milano \\ \email{giustina.vietri@inaf.it}
\and
   Cluster of Excellence, Boltzmann-Str. 2, 85748 Garching bei M\"{u}nchen, Germany
\and
   European Southern Observatory, Karl-Schwarzschild-Strasse 2, Garching bei M\"{u}nchen, Germany
   \and 
   European Southern Observatory, Alonso de Cordova 3107, Vitacura, Casilla 19001, Santiago de Chile, Chile       		
   \and
      School of Physics and Astronomy, Tel-Aviv University, Tel-Aviv 69978, Israel
   \and
   Centro de Astrobiolog\'ia (CAB, CSIC--INTA), Departamento de Astrof\'\i sica, Cra. de Ajalvir Km.~4, 28850 -- Torrej\'on de Ardoz, Madrid, Spain
   \and
   INAF - Osservatorio Astrofisico di Arcetri, Largo E. Fermi 5, I-50125, Firenze, Italy
   \and
      Department of Physics \& Astronomy, University College London, Gower Street, London WC1E 6BT, United Kingdom
   \and
   School of Mathematics, Statistics and Physics, Newcastle University, Newcastle upon Tyne, NE1 7RU, UK
   \and 
   INAF – Osservatorio Astronomico di Roma, Via Frascati 33, 00078 Monte Porzio Catone (Roma), Italy
\and
   Max-Planck-Institut f\"{u}r Astronomie, K\"{o}nigstuhl 17, D-69117 Heidelberg, Germany
   \and 
   INAF – Osservatorio Astronomico di Trieste, via G.B. Tiepolo 11, 34143 Trieste, Italy
   \and
   Dipartimento di Fisica e Astronomia dell’Universit\'{a} degli Studi di Bologna, via P. Gobetti 93/2, 40129 Bologna, Italy	
   \and 
      INAF/OAS, Osservatorio di Astrofisica e Scienza dello Spazio di Bologna, via P. Gobetti 93/3, 40129 Bologna, Italy
   \and
   Scuola Normale Superiore, Piazza dei Cavalieri 7, I-56126 Pisa, Italy
   \and
      Institute of Theoretical Astrophysics, University of Oslo, P.O. Box 1029, Blindern, 0315 Oslo, Norway
      \and
         Dipartimento di Fisica e Astronomia, Universit\'{a} di Firenze, Via G. Sansone 1, I-50019, Sesto Fiorentino (Firenze), Italy
   \and    
   CEA, IRFU, DAp, AIM, Universit\'{e} Paris-Saclay, Universit\'{e} Paris Diderot, Sorbonne Paris Cit\'{e}, CNRS, 91191 Gif-sur-Yvette, France 9
\and
 INAF – Osservatorio Astronomico di Padova, Vicolo dell’Osservatorio 5, 35122 Padova, Italy
\and
   Centre for Extragalactic Astronomy, Department of Physics, Durham University, South Road, Durham DH1 3LE, UK
   \and
   Max-Planck-Institut f\"{u}r extraterrestrische Physik (MPE), Giessenbachstrasse 1, D-85748 Garching bei M\"{u}nchen, Germany
   \and 
  Graduate school of Science and Engineering, Saitama Univ.
255 Shimo-Okubo, Sakura-ku, Saitama City, Saitama 338-8570, Japan
\and
   National Astronomical Observatory of Japan, Mitaka, 181-8588 Tokyo, Japan
  \and 
   Chalmers University of Technology, Department of Earth and Space Sciences, Onsala Space Observatory, 43992, Onsala, Sweden
             }

    \date{Received ?; accepted ?}
     
% \abstract{}{}{}{}{} 
% 5 {} token are mandatory
 
  \abstract
 %{
 {}
{The SINFONI survey for Unveiling the Physics and Effect of Radiative feedback (SUPER) was designed to conduct a blind search for AGN-driven outflows on X-ray selected AGN at redshift z$\sim$2 with high ($\sim$ 2 kpc) spatial resolution, and correlate them to the properties of the host galaxy and central black hole. The main aims of this paper are: a) to derive reliable estimates for the black hole mass and accretion rates for the Type-1 AGN in this survey; b) to characterize the properties of the AGN driven winds in the Broad Line Region (BLR).}
{We analyzed  rest-frame optical and UV spectra of 21 Type-1 AGN. We used H$\alpha$, H$\beta$, and MgII line profiles to estimate the black hole mass. We used the blueshift of the CIV line profile to trace the presence of winds in the BLR.}
{We found that the H$\alpha$ and H$\beta$ line widths are strongly correlated, as well as the line continuum luminosity at 5100 \AA\ with H$\alpha$ line luminosity, resulting in a well defined correlation between black hole mass estimated from H$\alpha$ and H$\beta$. We estimate using these lines that the black hole masses for our objects are in the range Log (M$\rm_{BH}$/M$_{\odot}$)=8.4-10.8 and are accreting at $\rm\lambda_{Edd}$ =0.04-1.3. On the other end, we confirm the well known fact that the CIV line width does not correlate with the Balmer lines and the peak of the line profile is blue-shifted with respect to the [OIII]-based systemic redshift. These findings support the idea that the CIV line is tracing outflowing gas in the BLR for which we estimated velocities up to $\sim$4700 km/s. We confirm the strong dependence of the BLR wind velocity with the UV-to-Xray continuum slope, as well as the bolometric luminosity and Eddington ratio. We inferred BLR mass outflow rates in the range 0.005-3 M$_{\odot}$/yr, showing a correlation with the bolometric luminosity consistent with that observed for ionized winds in the NLR and X-ray winds detected in local AGN, and kinetic power $\sim$10$^{-7} - 10^{-4} \times$ L$\rm_{Bol}$. The coupling efficiency predicted by AGN feedback models are much higher than the values reported for the BLR winds in the SUPER sample, however it should be noted that only a fraction of the energy injected by the AGN in the surrounding medium is expected to become kinetic power in the outflow. Finally, we found an anti-correlation between the equivalent width of the [OIII] line with respect to the CIV velocity shift, and a positive correlation with [OIII] outflow velocity. These findings, for the first time in an unbiased sample of AGN at z$\sim$ 2, support a scenario where BLR winds are connected to galaxy scale detected outflows, and are therefore actually capable of affecting the gas in the NLR located at kpc scale.}
   {}
%}

 \titlerunning{SUPER III. Broad Line Region properties of AGN at z$\sim$2}
\authorrunning{G. Vietri et al.}

   \maketitle

 \section{Introduction}\label{sec:intro}
Supermassive black holes (BH) are thought to be ubiquitous at the center of all massive galaxies (\citealt{Magorrian1998}, \citealt{Gebhardt2000}). 
The black hole mass (\mbh) is known to be correlated with the luminosity, velocity dispersion and stellar masses of the host-galaxy, suggestive of co-evolution between the central engine and its host-galaxy (\citealt{Magorrian1998}, \citealt{Gebhardt2000}, \citealt{Ferrarese2000}). A pre-requisite for studying the interplay between the certral AGN and its host is therefore an accurate measurement of \mbh .

A direct measurement of the BH mass is possible via reverberation mapping (RM), a technique which uses the lag between the variability in the AGN continuum and broad emission lines to measure the broad line region (BLR) size.  

In addition, RM experiments  provide empirical relations between the radius of the BLR (R$\rm_{BLR}$) and the AGN luminosity, i.e. R$\rm_{BLR} \propto (\lambda L_{\lambda})^\alpha$, with $\alpha \sim$ 0.5-0.7 (\citealt{Kaspi2000,Kaspi2005}; \citealt{Bentz2009,Bentz2013}). This BLR radius-luminosity relation provides an indirect way for measuring the BH mass when it is not possible to obtain reverberation data (the so-called single epoch (SE) method, see e.g. \citealt{McLure2002}, \citealt{Shen2013}). Line luminosity, e.g. L$\rm (H_{\beta})$, can be used to replace the continuum luminosity L$_{\lambda}$.

Assuming that the BLR is virialized and the clouds are dominated by gravitational motions, the BH mass can be estimated as follows:

\begin{equation}\label{eq:Mbh}
\rm M_{BH}= {\it{f}}\frac{~R_{BLR}~ V_{BLR}^2}{G} \propto {\it{f}}\frac{~(\lambda L_{\lambda})^\alpha~ V_{BLR}^2}{G} 
\end{equation}

\noindent where G is the gravitational constant and  V$\rm_{BLR}$ is the gas velocity, which can be measured from the width of a specific emission line (full width half maximum, FWHM, or velocity dispersion, $\sigma$). The FWHM is more widely used, being less vulnerable to noise in line wings and continuum placement. The alternative velocity dispersion ($\sigma$) is less sensitive to the narrow line removal but it is ill-defined for Lorentzian profiles and is sensitive to the quality of the data. In this paper we use the FWHM as indicator of the virial velocity of the gas in the BLR.

The factor {\it{f}} in Eq. \ref{eq:Mbh} depends on the geometry and kinematics of the BLR, and can be determined by comparing the BH mass derived from alternative methods (\citealt{Woo2015}, \citealt{Graham2016}).  \cite{Mejia-Restrepo2018} suggested a new way to estimate {\it{f}}. This is based on a strong anti-correlation between the BH mass and the FWHM of the broad emission lines, caused probably by line-of-sight inclination effects.

Continuum luminosity at 5100 \AA\ is usually preferred in Eq. \ref{eq:Mbh} given its tight correlation with the BLR size, based on a large number of sources. The line luminosities are useful in case of contamination by host starlight (\citealt{Greene2005}) or in case of radio-loud objects, where the continuum is contaminated by the non thermal emission of the jet (\citealt{Wu2004}).

Different lines have been used to estimate the BH mass, depending on the redshift, i.e. H$\beta$, H$\alpha$, MgII and CIV with different measures of the line width, i.e. FWHM or line dispersion (\citealt{Vestergaard2002}, \citealt{McLure2002}, \citealt{Wang2009}).
The H$\beta$ line width was used to measure the R-L relation in most RM studies of low redshift AGN (e.g. \citealt{Bentz2009}). As earlier studies confirmed, there is a strong correlation among the widths of H$\alpha$, H$\beta$ and MgII (\citealt{Greene2005}; \citealt{Shen2008}; \citealt{McGill2008}; \citealt{Trakhtenbrot2012}; \citealt{Mejia-Restrepo2016}). Because the majority of RM experiments used low-z AGN, the BH mass estimates based on Balmer lines are considered the most reliable. The use of high ionization lines, as CIV, to measure \mbh~is instead highly debated in the literaure. It is well known that CIV usually exhibits a shift of the peak to the blue, associated with gas in a non-virial motion (\citealt{Gaskell1982}, \citealt{Sulentic2000}, \citealt{Baskin2005}, \citealt{Richards2011}, \citealt{Denney2012}, \citealt{Coatman2017}, \citealt{Mejia-Restrepo2018}, \citealt{Vietri2018}), which leads to a biased estimation of the BH mass.
It is now well-established that the blue-shift of the CIV line peak correlates with AGN properties as its luminosity, Eddington ratio
($\rm\lambda_{edd}$), and quasar spectral energy distribution properties (e.g. \citealt{Richards2011}). While these properties are a limitation in the use of CIV as a black hole mass estimator, they offer the possibility to trace AGN winds on the pc-scale of the BLR.

\begin{figure}[]
 \includegraphics[width=1.1\columnwidth]{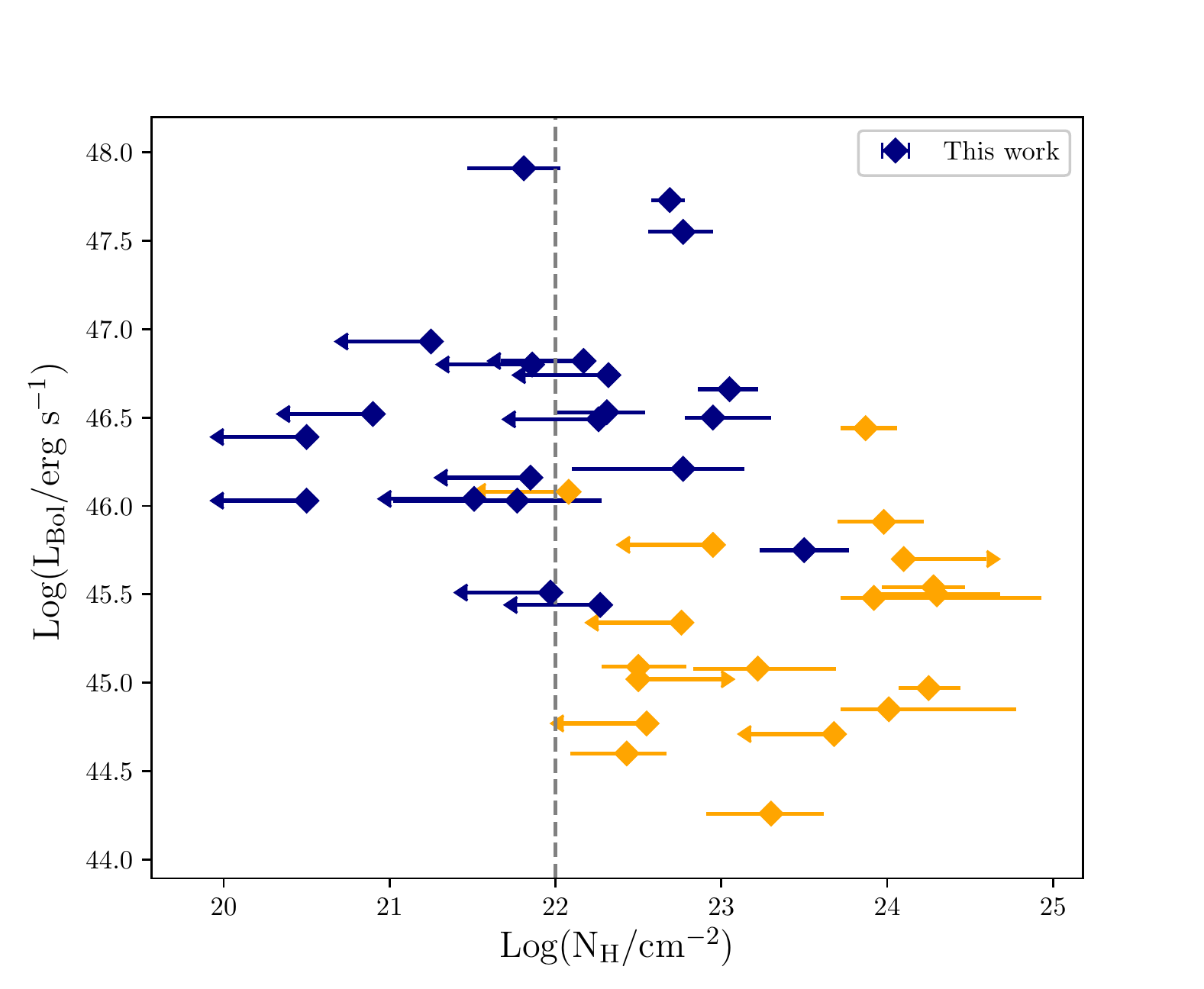}
 \caption{\small{AGN bolometric luminosities versus column densities of the SUPER sample. The dashed line at Log(N$\rm_{H}$/cm$^{-2}$) = 22 marks the assumed separation between X-ray unobscured and obscured AGN. The Type-1 SUPER targets analyzed in the present paper are represented as blue diamonds.}}\label{fig:Lbol_Nh}
\end{figure}

In this paper we analyze the properties of the BLR in the SUPER sample. As described in \cite{Circosta2018}, the SUPER survey consists of SINFONI observations of thirty-nine blindly-selected  X-ray AGN at redshift $\sim$2. The survey provides high-resolution, spatially resolved SINFONI observations in the H and K bands, with the main scientific goal of inferring the impact of outflows on on-going star formation and link outflow properties with AGN and host galaxy parameters (see \citealt{Circosta2018} for further details). The objects are selected from the COSMOS-Legacy (e.g. \citealt{Civano2016}, \citealt{Suh2015}, \citealt{Suh2020}), the wide area XMM-XXL (e.g. 
\citealt{Georgakakis2011}, \citealt{Liu2016}, \citealt{Menzel2016}), Stripe 82 X-ray survey (\citealt{LaMassa2016}) and WISSH surveys (\citealt{Bischetti2017}, \citealt{Martocchia2017}, \citealt{Duras2017}, \citealt{Vietri2018}) and about 58\% are Type 1 AGN.
We focus on the 21 Type-1 AGNs observed with SINFONI (see Fig. \ref{fig:Lbol_Nh}) both in H and K bands, with the aim of measuring the black hole mass M$\rm_{BH}$ and Eddington ratio  from H$\beta$ and H$\alpha$ emission lines, and compare the results with the MgII and CIV-based measurements, thanks to ancillary UV rest-frame data. Furthermore, we use the CIV line profile to trace the winds at pc-scale in order to derive the energetics of the winds and possibly link the BLR wind properties with the winds located at kpc-scales in the NLR.

Throughout this paper we assume H$_{0}$ = 70 \kms \mpc , $\rm\Omega_{\Lambda}$ = 0.7, and $\rm\Omega_{m}$ = 0.3, wavelengths in vacuum and line blueshifts defined as positive values.

\section{Observations and data reduction}

\begin{table*}[t]
        \begin{threeparttable}
        \setlength{\tabcolsep}{4pt}
        \caption{Properties of the SUPER AGN considered in this paper.}\label{tab:data1}
 %{\scriptsize 
         %\centering

                \begin{tabular}{lccccccccccr}
                        \hline
                        \hline
                       ID & RA & Dec  & z & z$\rm_{[OIII]}$&H & K & r$\rm_H$ & r$\rm_K$  & Log (L$\rm_{Bol}/ erg s^{-1}$) & Log (L$\rm_{5100}/ erg\ s^{-1}$)   \\ 
                        (1) & (2) & (3) & (4) & (5) & (6)& (7) & (8) & (9) & (10) & (11) \\
\hline
                              \hline
X\_N\_160\_22 & 02:04:53.81&-06:04:07.82&2.445&2.442 &   19.22&18.79 & 0.5 & 0.4        &  46.74\err0.02 & 45.83\err0.06 \\
X\_N\_81\_44 & 02:17:30.95&-04:18:23.66&2.311&2.317 &18.78&18.43 & 0.45 & 0.4        &  46.80 \err 0.03 & 45.75\err0.06 \\
X\_N\_53\_3 & 02:20:29.84&-02:56:23.41&2.434&2.433&20.60&-& -&0.2                   & 46.21 \err 0.03 &   - \\
X\_N\_66\_23 & 02:22:33.64 & -05:49:02.73 & 2.386 & 2.385&20.56 & 20.33 &0.35 &0.45 &  46.04 \err0.02 & 45.63\err0.06 \\
X\_N\_35\_20 &  02:24:02.71 & 05:11:30.82 & 2.261 & 2.261& 22.07 & 21.70& 0.15 &0.2  &  45.44 \err 0.02 & 43.95\err0.06 \\
X\_N\_12\_26 & 02:25:50.09&-03:06:41.16&2.471& 2.472&19.83&19.53 &0.4 & 0.35         &  46.52 \err 0.02 & 45.53\err0.06 \\
X\_N\_44\_64 & 02:27:01.46& -04:05:06.73 & 2.252 & 2.244&21.31 & 20.77 & - & 0.2     &  45.51 \err 0.07 &   - \\ 
X\_N\_4\_48 & 02:27:44.63&-03:42:05.46&2.317&2.315&19.57&20.43& 0.25 &0.35          & 46.16 \err 0.02 & 45.05\err0.06 \\
X\_N\_102\_35 & 02:29:05.94&  04:02:42.99 &2.190 &2.190&18.76 &18.19 & 0.15 & 0.15  &  46.82 \err 0.02 & 45.52\err0.06 \\
X\_N\_115\_23 & 02:30:05.66&-05:08:14.10&2.342&2.340&19.79&19.26 &0.35 & 0.4        & 46.49 \err 0.02 & 45.55\err0.06 \\
cid\_166& 09:58:58.68 & +02:01:39.22&2.448&2.461&18.55&18.23 & 0.3 & 0.45           & 46.93 \err 0.02 & 45.83\err0.06 \\
cid\_1605 & 09:59:19.82 &  +02:42:38.73& 2.121&2.118 &20.63  & 20.14 & 0.2& 0.2     &  46.03 \err 0.02 & 44.84\err0.06 \\
cid\_346& 09:59:43.41 & +02:07:07.44&2.194& 2.217&19.24&18.95 &0.35 & 0.40           & 46.66 \err 0.02 & 45.61\err0.06  \\
cid\_1205 & 10:00:02.57 & +02:19:58.68& 2.255 & 2.257 & 21.64 & 20.72 & 0.15 & 0.6 & 45.75 \err 0.17 & 44.91 \err 0.06 \\
cid\_467  &10:00:24.48 &+02:06:19.76 &2.288& 2.285&19.34& 18.91& 0.2 &0.2           & 46.53 \err 0.04 & 45.25\err0.06 \\
J1333+1649&13:33:35.79&16:49:03.96&2.089& 2.099&15.72&15.49 & 0.55 & 0.5             &  47.91 \err 0.02 & 47.32\err0.06 \\
J1441+0454&14:41:05.54&+04:54:54.96&2.059& 2.080&17.15&16.53 &0.50 & 0.55                &  47.55 \err 0.02 & 46.66\err0.06 \\
J1549+1245&15:49:38.73&+12:45:09.20&2.365& 2.368&15.92&15.34 & 0.50 & 0.65                &  47.73 \err 0.04 & 47.16\err0.06 \\
S82X1905&23:28:56.35&-00:30:11.74&2.263& 2.273&19.72&19.15 & 0.4 &0.35               &  46.50 \err 0.02 & 45.55\err0.06 \\
S82X1940&23:29:40.28&-00:17:51.68&2.351&2.350&20.80&20.15& 0.3 & 0.3                &  46.03 \err 0.02 & 44.98\err0.06 \\
S82X2058&23:31:58.62&-00:54:10.44&2.308&2.315&19.79&19.29 & 0.3 & 0.35              &  46.39 \err 0.02 & 45.54\err0.06 \\
                                                  
                                                  \hline

                        \hline 
                \end{tabular}
         \begin{tablenotes}[para,flushleft]
         \item \tiny {\bf{Notes}}. Columns give the following information: (1) Target identification,  (2--3) celestial coordinates, (4) redshift from archival optical spectra, (5) redshift from the peak location of the [OIII]$\lambda$5008 in the integrated spectra, (6-7) 2MASS photometric data, (8-9) the radius (in arcsec) of the circular aperture centered on the target used to extract the spectrum, for the H and K bands, (10) Logarithm of the bolometric luminosity derived from SED fitting (\citealt{Circosta2018}) and (11) Logarithm of the extinction-corrected luminosity at 5100\AA\ derived  from the best-fit values of the power-law model representing the AGN continuum (see sect. \ref{sec:analysis}).
                                  \end{tablenotes}
                      %            }
                                       \end{threeparttable}

\end{table*} 

SINFONI observations were carried out as part of the ESO large programme 196.A-0377, with 3"$\times$3" field of view in Adaptive Optics (AO) assisted mode, with a pixel scale of 0.05$\times$0.1 arcsec (final resampled pixel scale 0.05$\times$0.05 arcsec), using H grating with resolution of R=3000, to trace rest-frame optical region H$\beta$-[OIII]$\lambda$5008 and K grating with resolution of R=4000 to trace lines from H$\alpha$ up to [SII]$\lambda$6716,6731 lines. We acquired six observation frames with exposure times of 600 s for each observing block, using the pattern "O-O-O-O-O-O" (O=object) as strategy of observation for a good sky subtraction, and in case of extended sources dedicated sky exposures were acquired (O-S-O-O-S-O). The total on-source exposure time ranges from 1 hr up to 6 hr. While we refer to \cite{Kakkad2020} for a detailed description of the SINFONI data reduction, below we report only the main steps. We used the ESO pipeline (3.1.1) which returns science, PSF and telluric cubes distortion corrected and wavelength calibrated. The background sky emission was removed with the IDL routine SKYSUB (\citealt{Davies2007}). We corrected the telluric absorption in the science frames by removing the telluric lines from the observed telluric standard star, observed close in time and airmass to the object. The response curve was determined by normalizing the telluric free star spectrum divided by a blackbody spectrum and then applied to the science and standard star cubes. The spectrum extracted from the telluric-free star cube was then convolved with a 2MASS transmission filter, H or K according to the SINFONI observation bands, to derive the conversion factor from count to appropriate physical unit, which was then applied to the science cube. The flux calibrated frames were then combined using dedicated SINFONI pipeline recipe. For observations taken during the same night, we stacked the individual observing blocks according to the center position of the image recorded in the header. For observations taken during different nights, we first determined the relative offset of the centroids of the images, performing a two-dimensional Gaussian fit, and then re-aligned and co-added the individual observing blocks. 21 Type-I AGN were observed both in the H and K bands. However, one target was detected neither in continuum nor in emission lines in the H band but was detected in the K band, i.e. X\_N\_53\_3, and one lacks emission lines in the H band but detected in continuum and in lines in the K band, i.e. X\_N\_44\_64.

\section{Spectroscopic analysis}\label{sec:analysis}

We extracted the integrated spectra from a circular region centered at the QSO position, which covers at least 95\% of the total emission. The target center was found using a 2D Gaussian fit on the wavelength-collapsed image from the datacube, over the entire wavelength range of the SINFONI observations. 

The radius of the extraction region is on average $\sim$0.3 ($\sim$0.4) arcsec, with a minimum value of 0.15 (0.15) arcsec and a maximum value of 0.55 (0.65) arcsec for H (K) band (see Table \ref{tab:data1}).
In the following we describe in detail the analysis of the line profile for the H$\beta$-[OIII]$\lambda$5008 complex extracted from the H band datacube, and the H$\alpha$-[NII]-[SII] region extracted from the K band data.

\subsection{Modeling of SINFONI spectra}\label{sec:OIII_analysis}

We performed separately the fit for the H and K band spectra using the python routine {\tt{scipy.optimize.curve\_fit}}.  The sky line residuals were masked from the spectrum during the fitting procedure. We shifted the H and K band spectra to the rest-frame using the spectroscopic redshift obtained from the peak location of the [OIII]$\lambda$5008 (hereafter [OIII]) in the integrated spectrum\footnote{For the two sources undetected in the H-band (X\_N\_53\_3 and X\_N\_44\_64), we used the redshift obtained from the H$\alpha$ narrow component.}.

 {\it{H band}.} We modeled the spectra using a simultaneous fit of the continuum, the [OIII] doublet, the H$\beta$ emission line, and the iron emission lines. A power-law was adopted to model the continuum, while Gaussian components were used to reproduce the emission lines. Specifically, we used (i) up to two Gaussian components to model the emission from the narrow line region (NLR) of the [OIII] doublet and H$\beta$ emission lines, and the addition of the second Gaussian depended on whether it minimizes the reduced chi-square value of the overall model; (ii) a Gaussian component or broken power-law for fitting the H$\beta$ emission from the broad line region (BLR). For the NLR components of [OIII] and H$\beta$, we fixed the wavelength separation and assumed equal broadening for the lines assuming the same gas is responsible for the those emission lines. The flux ratio for the components of the [OIII] doublet was set to 1:3, according to their atomic parameters.

 The iron emission is modeled with three observational templates of \cite{Boroson1992}, \cite{Veron2004}, \cite{Tsuzuki2006}. 
 These empirical templates are anchored to the [OIII] rest-frame and each iron template is convolved with a single Gaussian component with FWHM in the range 1000-5000 \kms. A  $\chi^2$ minimization procedure was used to select the best-fit FeII template \footnote{To test the effect of the three FeII templates on the FWHM measurements, we fitted one spectrum, i.e. X\_N\_81\_44, with all three templates and the BH masses are consistent within the uncertainties, i.e. Log(M$\rm_{BH}/M_{\odot}$)=8.77\err0.30, Log(M$\rm_{BH}/M_{\odot}$)=9.05\err0.30 and Log(M$\rm_{BH}/M_{\odot}$)=8.81\err0.30, for \cite{Veron2004},\cite{Boroson1992} and \cite{Tsuzuki2006}, respectively.}. 

Finally, we also modeled the HeII$\lambda$4686 line with two components: 1) a narrow Gaussian with centroid and velocity dispersion tied to the narrow [OIII] line; 2) a broad Gaussian component with parameters free to vary to reproduce the emission due to the BLR. Only in one case, X\_N\_160\_22, we unambiguously detected the HeII narrow component with FWHM= 920\err 50. We do not detect the HeII BLR component, this is probably due to the difficult deblending of this component from the underlying iron emission. In Fig. \ref{fig:fit_hb} it is shown an example of the H band fit (for the rest of the sample see Appendix \ref{app:app}).

 \noindent We report in Table \ref{tab:beta} the best-fit parameter of the H$\beta$ BLR emission lines component for the 16/21\footnote{The H band datacube of cid\_1205 is contaminated by a bright stripe at the location of H$\beta$ line, preventing us a reliable measurement of the line parameters. We therefore did not estimate BLR H$\beta$ line parameters for this source.} SUPER targets with H$\beta$ BLR detection. 
 
 {\it{K band}}. 
 We  modelled the continuum by adopting a power-law function and emission lines by using the same modeling as for the [OIII] emission profile, for the H$\alpha$, [NII] and [SII] doublet emission lines. In cases, where we used one Gaussian component to reproduce the [OIII] profile, we used one Gaussian model to reproduce the NLR emission of the K band lines and tied together the centroid and velocity dispersion of the K band lines, allowing the centroid to vary in the range of 7 \AA\ and the velocity dispersion to vary up to the values derived for the [OIII] one Gaussian component. In case of two Gaussians for the [OIII] profile, we used two Gaussian components for the NLR emission of the K band, and for the second Gaussian it was adopted the same broadening of the second [OIII] Gaussian component with a fixed wavelength separation between the [OIII] and K band lines centroid.

  We also fitted a BLR Gaussian component for the H$\alpha$ line. This parametrization reduced the possibility of model degeneracy. We note that for cid\_346 we also use a narrow component to reproduce the H$\alpha$-region profile, clearly detected in the spectrum ([OIII] profile was modelled with one Gaussian component).   
 
The flux intensities of the [NII] doublet were set to 1:3, according to their atomic parameters. In Fig. \ref{fig:fit_ha} we show an example of the K band fit (for the rest of the sample see Appendix \ref{app:app}).

\noindent We report in Table \ref{tab:alpha} the best-fit parameter of the H$\alpha$ BLR emission lines component for the all the SUPER targets.

   \begin{figure}[]
   \includegraphics[width=0.5\textwidth]{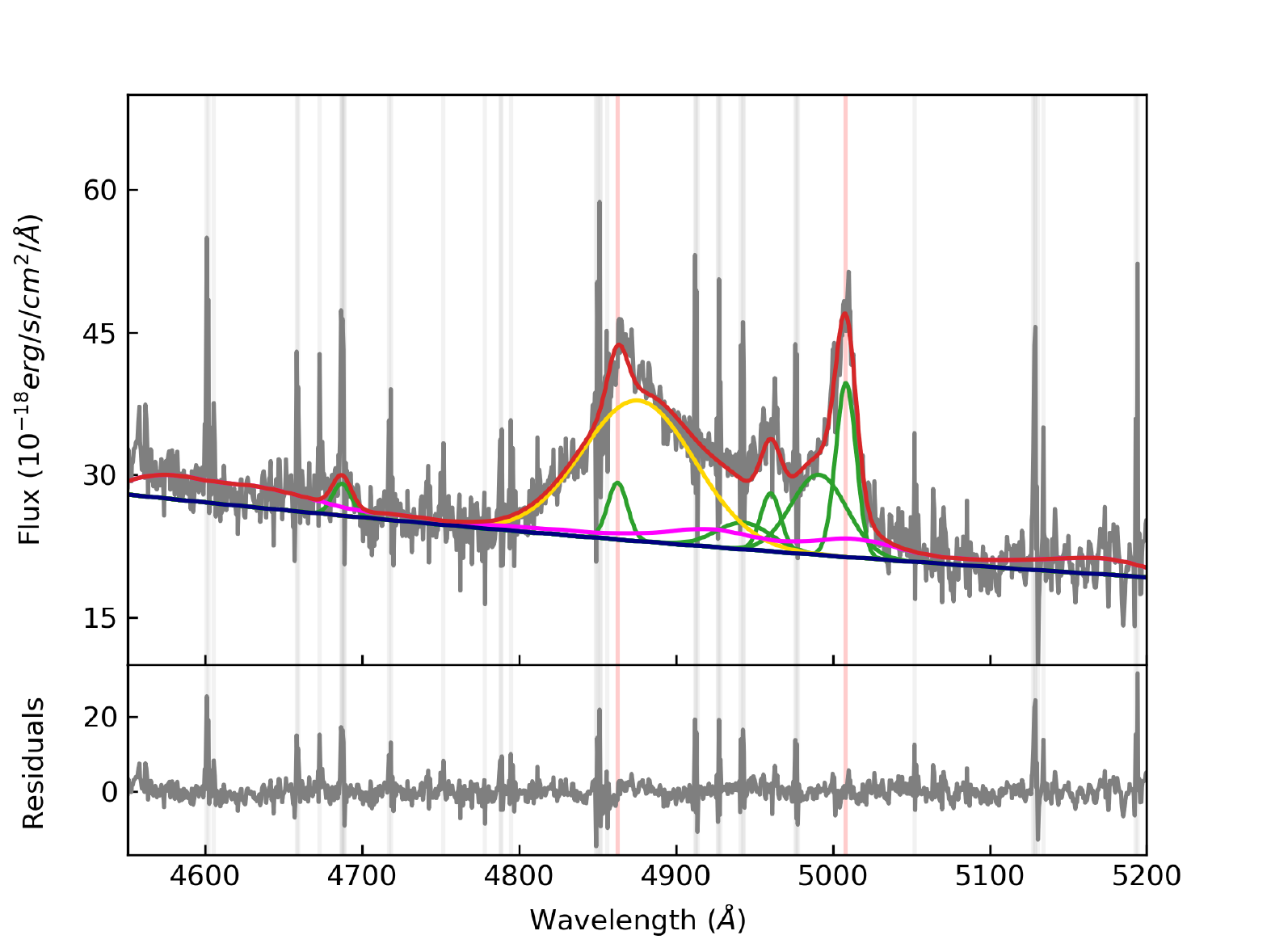}  
 \caption{\small{Parametrization of the H$\beta$-[OIII] region of the SUPER AGN X\_N\_160\_22. The red curve shows the best-fit to the data. Green Gaussian refer to the Gaussian components used to reproduce the line profile of each emission line. Gold Gaussian component indicates the broad component of H$\beta$ associated with BLR emission. FeII emission is marked in magenta. Lower panel shows the fit residuals. Grey bands indicate the sky line residuals masked during the fit procedure. The x- and y- axis show the restframe wavelength and flux (not corrected for extinction), according to the redshift of the target.  }}\label{fig:fit_hb}
\end{figure}

 \begin{table}[]
      %  \center
%             \footnotesize
\begin{threeparttable}
         \small
        \caption{Properties of  BLR components of the H$\beta$ emission line derived from parametric model fits.}\label{tab:beta}% {\bf{add balmer decrement and EBV}}}
        \setlength\tabcolsep{2.3pt}

                        \begin{tabular}{lccccc}
                                \hline
                                \hline  
                               ID & $\rm \lambda^{BLR}_{H\beta}$  & FWHM$\rm ^{BLR}_{H\beta}$ & EW$\rm ^{BLR}_{H\beta}$ & Log (L$\rm ^{BLR}_{H\beta}$/\ erg s$^{-1}$)      \\      
                                                    &\footnotesize{(\AA)} &\footnotesize{(km s$^{-1}$)} & \footnotesize{(\AA)} & \\
                                \hline

X\_N\_160\_22  & 4876  \err   1&  5190  \err   170 & 58 \err 2 & 43.99\err0.04 \\
X\_N\_81\_44   & 4873  \err   1 &  5290  \err   170 & 57  \err 2 & 43.87\err0.04 \\
X\_N\_12\_26   & 4865  \err   1 &  4890  \err   200 & 51  \err 2 & 43.58\err0.04 \\
X\_N\_4\_48    & 4860  \err   4 &  6710  \err   670 & 40  \err 4 & 42.95\err0.06 \\
X\_N\_102\_35  & 4872  \err   1 &  4810  \err   160 & 58  \err 2 & 43.67\err0.04 \\
X\_N\_115\_23  & 4869  \err   2 &  6330  \err   250 & 56  \err 2 & 43.68 \err0.04 \\
cid\_166    & 4867  \err   1 &  6970  \err   130 & 91  \err 2 & 44.17\err0.04 \\
cid\_1605   & 4860  \err   3 &  5040  \err   510 & 156  \err 22 & 43.25\err0.06 \\
cid\_346    & 4870  \err   2 &  6280  \err   340 & 46  \err 2 & 43.63\err0.05 \\
cid\_467    & 4875  \err   5 &  9260  \err   760 & 93  \err 8 & 43.62\err0.05 \\
J1333+1649  & 4841  \err   7 &  6300  \err   250 & 42  \err 2 & 45.34\err0.04 \\
J1441+0454  & 4855  \err   1 &  4030  \err    100 & 47  \err 1 & 44.73\err0.04 \\
J1549+1245  & 4867  \err   2 & 16570  \err   690 & 108  \err 7 & 45.53\err0.05 \\
S82X1905  & 4872  \err   1 &  4960  \err   100 & 71  \err 1 & 43.80\err0.04\\
S82X1940  & 4862  \err   1 &  3710  \err   140 & 75  \err 3 & 43.24\err0.04 \\
S82X2058  & 4860  \err   1 &  6450  \err   150 & 69  \err 1 & 43.78\err0.04 \\

                                                  \hline

                                        \end{tabular}
               
       \begin{tablenotes}[para,flushleft]
         \item {\bf{Notes}}. Columns give the following information for the  BLR component of the H$\beta$ emission line: (1) Target identification,  (2) centroid (\AA), (3) full width at half maximum (km/s), (4)  rest-frame equivalent width (\AA), (5) Logarithm of the extinction-corrected H$\beta$ luminosity.
                                  \end{tablenotes}
                                       \end{threeparttable}
                        \end{table}

 \begin{table}[]
        \center
\begin{threeparttable}
         \small
        \caption{Properties of  BLR components of the H$\alpha$ emission line derived from parametric model fits.}\label{tab:alpha} 
        \setlength\tabcolsep{2pt}

                        \begin{tabular}{lcccc}
                                \hline
                                \hline  
                               ID & $\rm \lambda^{BLR}_{H\alpha}$  & FWHM$\rm ^{BLR}_{H\alpha}$ & EW$\rm ^{BLR}_{H\alpha}$ & Log (L$\rm ^{BLR}_{H\alpha}$/ erg s$^{-1}$)      \\      
                                                    &\footnotesize{(\AA)} &\footnotesize{(km s$^{-1}$)} & \footnotesize{(\AA)} &\\
                                \hline

X\_N\_160\_22  & 6584\err  1 &  5410\err  110 & 289\err7 & 44.46\err0.03 \\
X\_N\_81\_44   & 6580\err  1 &  6320\err  120 & 267\err5 & 44.41\err0.03 \\
X\_N\_53\_3    & 6571\err  1 &  4630\err  180 & 415\err27 & 43.57\err0.03 \\
X\_N\_66\_23   & 6567\err  2 &  6105\err  310 & 254\err15 & 43.86\err0.03 \\
X\_N\_35\_20   & 6564\err  7 &  6440\err 1590 & 317\err92 & 42.80\err0.08\\
X\_N\_12\_26   & 6570\err  1 &  5270\err  120 & 256\err7 & 44.03\err0.03 \\
X\_N\_44\_64   & 6567\err  5 &  7720\err  720 & 187\err21 & 43.15\err0.04 \\
X\_N\_4\_48    & 6573\err  2 &  7700\err  240 & 377\err19 & 44.27\err0.03 \\
X\_N\_102\_35  & 6577\err  1 &  5190\err   100 & 292\err7 & 44.16\err0.03 \\
X\_N\_115\_23  & 6572\err  1 &  6560\err  130 & 335\err8 & 44.28\err0.03 \\
cid\_166    & 6570\err  1 &  6810\err   100 & 437\err8 & 44.76\err0.03 \\
cid\_1605   & 6569\err  2 &  3690\err  230 & 296\err20 & 43.55\err0.04 \\
cid\_346    & 6565\err  2 &  6980\err  260 & 205\err9 & 44.12\err0.03 \\
cid\_1205   & 6564\err  2 &  5100\err  230 & 542\err42 & 43.52\err0.03\\
cid\_467    & 6569\err  2 &  8450\err  230 & 458\err21 & 44.11\err0.03 \\
J1333+1649  & 6572\err  1 &  6190\err   50 & 217\err2 & 45.73\err0.03 \\
J1441+0454  & 6559\err  1 &  4740\err   30 & 169\err2 & 45.09\err0.03 \\
J1549+1245  & 6580\err  1 &  7270\err   50 & 304\err2 & 45.80\err0.03 \\
S82X1905  & 6571\err  1 &  4730\err   50 & 360\err6 & 44.17\err0.03 \\
S82X1940  & 6563\err  1 &  4370\err  160 & 344\err16 & 44.35\err0.03\\
S82X2058  & 6558\err  1 &  6400\err  110 & 322\err6 & 44.17\err0.03 \\

                              \hline

                                        \end{tabular}
               
     \begin{tablenotes}[para,flushleft]
         \item {\bf{Notes}}. Columns give the following information for the BLR component of the H$\alpha$ emission line: (1) Target identification,  (2) centroid (\AA), (3) full width at half maximum (km/s), (4)  rest-frame equivalent width (\AA), (5) Logarithm of the extinction-corrected H$\alpha$ luminosity.
                                  \end{tablenotes}
                                       \end{threeparttable}
                        \end{table}

 \begin{figure}[]
       \includegraphics[width=0.5\textwidth]{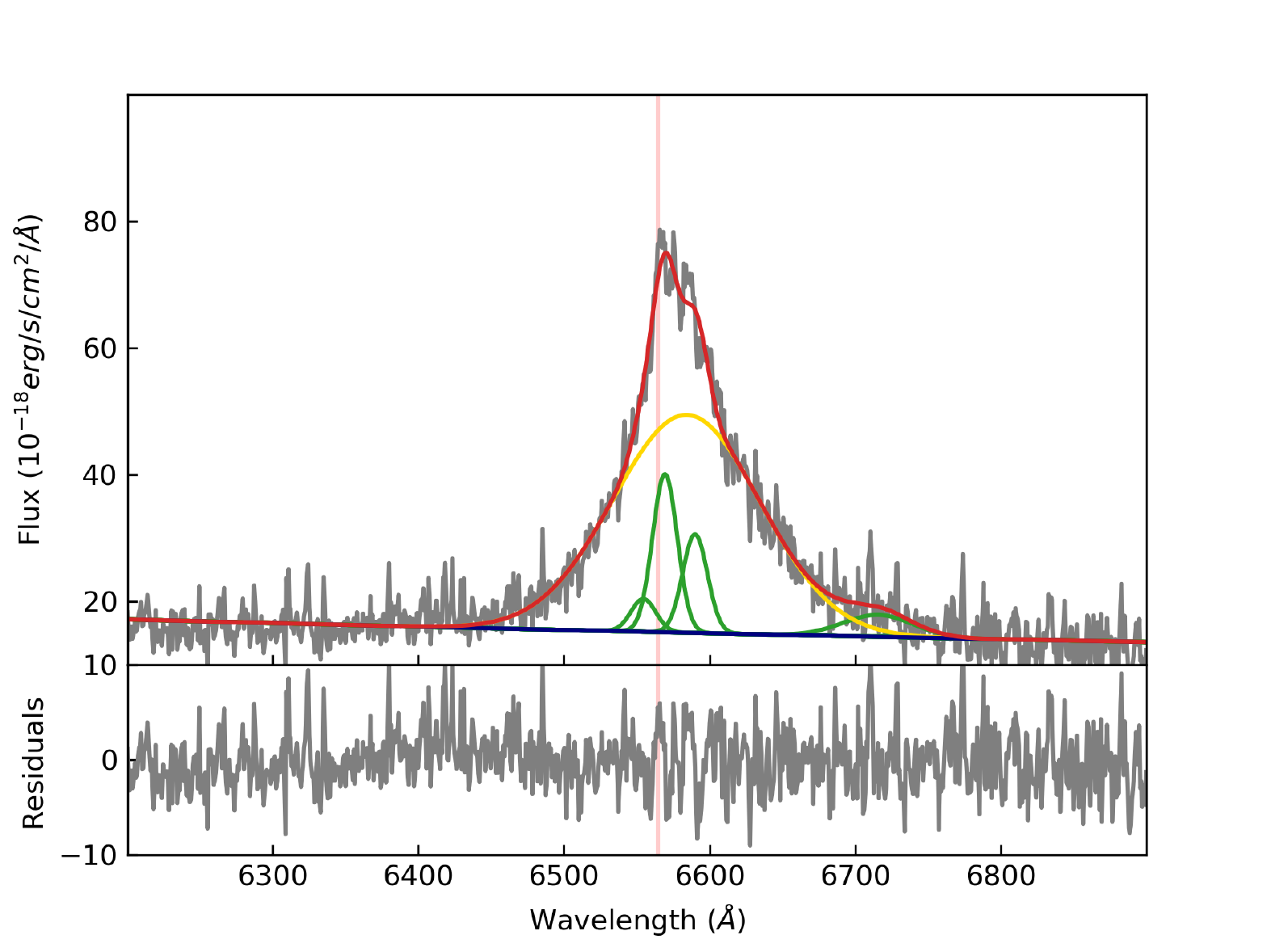}
 \caption{\small{Parametrization of the H$\alpha$ region of the SUPER AGN X\_N\_160\_22. The red curve shows the best-fit to the data. Green Gaussians refer to the narrow and broad components used to reproduce the line profile of  H$\alpha$, [NII] and [SII]. Gold Gaussian component indicates the broad component of H$\alpha$ associated with BLR emission. Lower panel shows the fit residuals. The x- and y- axis show the restframe wavelength and flux (not corrected for extinction), according to the redshift of the target}}\label{fig:fit_ha}
\end{figure}

  \subsection{Modeling of the UV spectra}\label{sec:SDSS}
We retrieved the rest-frame UV spectra for our targets from the SDSS archive  for all but cid\_1205, for which we used archival VIMOS/VLT spectrum. We did not analyze the spectrum of X\_N\_44\_64 because of the very low S/N. 
The observed UV wavelengths were shifted to the systemic redshift (see Sect. \ref{sec:analysis}). 
The presence of strong iron features and Balmer continuum emission in the MgII$\lambda$2800 \AA\ region make the continuum level difficult to estimate. We therefore fit separately the wavelength region from Ly$\alpha$ up to CIII]$\lambda$1909 \AA\ (less affected by the iron emission) and the MgII-FeII region (2600-3000 \AA).

{{\it Ly$\alpha$-CIII] region}}. We fit the continuum in line-free wavelength regions, 1445-1455 $\AA$ and 1973-1983 $\AA$, using a power law function. The UV lines fitted are Ly$\alpha$1216, NV$\lambda$1240, Si II$\lambda$1263, SiV$\lambda$1398 OIV$\lambda$1402, NIV$\lambda$1486, CIV$\lambda$1549, HeII$\lambda$1640, OIII]$\lambda$1663, Al II$\lambda$1671, Al III$\lambda$1857, Si II$\lambda$1887, CIII]$\lambda$1909 simultaneously, using one Gaussian component allowing a shift of the centroid of $\pm$10 \AA\ and FWHM up to 10000 km/s (e.g. \citealt{Matsuoka2011}, \citealt{Mejia-Restrepo2016}), for all but the CIV line for which a shift of $\pm$ 50 \AA\ is used, justified by the usually asymmetric and shifted profile. These lines (except for the CIV line) are not necessary for the purpose of this work except for limiting the continuum placement and would require more accurate modelling to derive line parameters.

Apart from the BAL AGN CID\_346 and SDSS\_J1549 (\citealt{Bruni2019}), which have the CIV line heavily affected by absorption, and for which we only used one Gaussian component to reproduce the CIV line, we performed the line fitting using both a model with a single Gaussian component and one with two Gaussians. The models were compared using the Bayesian information criterion (BIC, see \citealt{Schwarz1978}), defined as BIC= $\chi^2$ + k ln (N), where N is the number of data points and k is the number of free parameters. A BIC difference larger than 10 was the adopted criteria to choose the model with two components. For five of our targets (X\_N\_35\_20, X\_N\_81\_44, cid\_1605, S82X1905, S82X1940) the CIV profile is better reproduced with one Gaussian component. The remaining sources are better reproduced by a two Gaussian model fit.

The model fitting is performed in the spectral range of 1210 \AA\ - 2000 \AA . 
We excluded from the fit the heavy blended spectral regions of OI+SII $\lambda$ 1305, CII$\lambda$1335,  the so-called 1600 \AA\ bump, the undefined feature in 1570-1631 \AA  (see \cite{Nagao2006} for a detailed discussion) and NIV$\lambda$1719, AlII $\lambda$1722, NIII]$\lambda$1750 and  FeII multiplets, i.e. 1286-1357 \AA , 1570-1631 \AA\ and  1687-1833\AA .
 The measured FWHM$\rm_{CIV}$ of the total profile are in the range $\sim$1300-10000 km/s and the velocities shift, defined as \vciv=c$\times$($\rm\lambda_{half}$-1549.48)/1549.48 with $\rm\lambda_{half}$ the wavelength that bisects the cumulative total line flux and c the speed of light, are in the range \vciv $\sim$ -760 up to 4700 km/s. The detailed results from the best-fit model of the CIV emission line for 20/21 SUPER targets are reported in Table \ref{tab:CIV}. 
We note that for the bulk of the SUPER sources, the total CIV profile is blueshifted and therefore it is dominated by gas not at systemic velocity but in an outflowing phase. An example of the fit is shown in the upper panel of Fig. \ref{fig:fit_civ} (for the rest of the sample see Appendix \ref{app:app}).

     \begin{table*}[]
        
\small
	\centering
\begin{threeparttable}
      \footnotesize
        \caption{Properties of CIV$\lambda1549$ emission line derived from parametric model fits (see Sect. \ref{sec:SDSS}).}\label{tab:CIV}
        \setlength\tabcolsep{2pt}
                        \begin{tabular}{lcccccr}
                                \hline
                                \hline  
                               ID & $\rm\lambda_{50}^{CIV}$ & FWHM$\rm_{CIV}$ &  EW$\rm_{CIV}$&\vciv  &Log (L$\rm_{CIV}$/ erg\ s$^{-1}$) &Log (L$\rm_{1350}$/ erg\ s$^{-1}$) \\
& \footnotesize{(\AA)}&\footnotesize{(km s$^{-1})$}& \footnotesize{(\AA)} & \footnotesize{(km s$^{-1})$} & &             \\   
\hline

X\_N\_160\_22  & 1548\err1  & 3180  \err 280  &  38  \err   1  & 250 \err 70&45.03\err0.09  & 46.71\err0.10  \\
X\_N\_81\_44  & 1546\err1  & 6790  \err 250  &  49  \err   2  &700\err110 &44.77\err0.09  & 46.27\err0.10  \\
X\_N\_53\_3  & 1547 \err1  & 4910  \err 480  &  60  \err   2  & 470\err140 & 44.67\err0.09  & 46.09\err0.10  \\
X\_N\_66\_23  & 1548 \err1  & 2350  \err 210  &  63  \err  2  & 240\err30 &44.35\err0.09  & 45.80\err0.10  \\
X\_N\_35\_20  & 1545\err3  & 5550  \err 1240  &  33  \err   7  & 950\err520 & 43.56\err0.13  & 45.32\err0.10  \\
X\_N\_12\_26  & 1545  \err1  & 3860  \err 520  &  30  \err  1  & 920\err140 &44.49\err0.09  & 46.27\err0.10  \\
X\_N\_4\_48  & 1549 \err1  & 5250  \err 620  &  34  \err   2  & 50\err140 &44.41\err0.09  & 46.13\err0.10  \\
X\_N\_102\_35  & 1548  \err 1  & 3250  \err 100  &  53  \err   1  &190\err70 &45.14\err0.09  & 46.68\err0.10  \\
X\_N\_115\_23  & 1549  \err1  & 2280  \err 220  &  61  \err   2  & 140\err70 &44.79\err0.09  & 46.34\err0.10  \\
cid\_166  & 1545  \err1  & 3940  \err173  &  50  \err   1  & 850\err70 &45.24\err0.09  & 46.79\err0.10  \\
cid\_1605  & 1550 \err 1  & 6090  \err 300  &  70  \err   3  & -120\err130 &44.30\err0.09  & 45.76\err0.10  \\
cid\_346  & 1538 \err1 & 7470  \err 190  &  30  \err  1  & 2230\err80 &44.66\err0.09  & 46.49\err0.10  \\
cid\_1205  & 1551  \err1  & 1340  \err 140  &  60  \err   6  & 250\err60 &42.34\err0.10  & 43.72\err0.11  \\
cid\_467  & 1548  \err1  & 3450  \err310  &  60  \err   2  &280\err70 &44.91\err0.09  & 46.51\err0.10  \\
J1333+1649   & 1538  \err1  & 5250  \err 170  &  19\err   1  &2300\err30 &45.65\err0.09  & 47.61\err0.10  \\
J1441+0454  & 1525  \err1  & 9690  \err 200  &  24  \err 1  & 4690\err30 &45.47\err0.09  & 47.33\err0.10  \\
J1549+1245  & 1553  \err1  & 5180  \err 120  &  18  \err  1  &-760\err50 &45.10\err0.09  & 47.01\err0.10  \\
S82X1905  & 1544  \err 1  & 7250  \err 230  &  45  \err   1  &1070\err100 &44.63\err0.09  & 46.24\err0.10  \\
S82X1940  & 1547  \err 1  & 6690  \err 260  &  80  \err  3  &430\err110 & 44.41\err0.09  & 45.83\err0.10  \\
S82X2058  & 1546  \err 1  & 4080  \err 280  &  28  \err   1  & 740\err140 &44.61\err0.09  & 46.44\err0.10  \\

                                \hline
                                
                \end{tabular}
                         \begin{tablenotes}[para,flushleft]
         \item {\bf{Notes}}. Columns give the following information for the BLR component of the CIV emission line: (1) Target identification,  (2) centroid (\AA), (3) full width at half maximum (km/s), (4)  rest-frame equivalent width (\AA), (5) velocity of the CIV at 50\% of the cumulative line flux, (6) Logarithm of the extinction-corrected CIV luminosity and (7) Logarithm of the extinction-corrected luminosity at 1350 \AA\ derived  from the best-fit values of the power-law model representing the AGN continuum (see Sect. \ref{sec:analysis}).
                                  \end{tablenotes}
                                       \end{threeparttable}
                                            \end{table*}

{{\it MgII$\lambda$2800}}. For 17/21 SUPER sources we were able to model the MgII line, the remaining sources have very low S/N on this line and are affected by strong sky-lines residuals. We modelled first the continuum \footnote{We note that no Balmer continuum model is included. This results in an overestimation of the continuum level.} with a power-law plus the UV FeII+FeII templates from \cite{Popovic2019}, convolved with a Gaussian function with a FWHM in the range 1000–5000 \kms. The best fit template was chosen through a $\chi^2$ minimization procedure. A potential velocity shift of the FeII emission lines is not considered in this paper.
The MgII line was then generally modelled with a single Gaussian. For three objects, whose spectra were not affected by strong sky residuals, we performed the line fit using two Gaussian components. We used the BIC criterion to compare the models, and the single Gaussian model fit was favoured in all cases. We derived FWHM$\rm_{MgII}$ in the range $\sim$ 3000-9000 km/s and a velocity shift v$\rm_{50}^{MgII}$ $\sim$ up to 1200 km/s. We note that the MgII emission line and FeII emission surrounding the MgII are affected by sky-line residuals in almost all spectra of the SUPER sample, which can affect the measured line properties. Therefore hereafter the MgII line properties are used with caution.

The result of the emission line fit for the MgII emission line of 17/21 SUPER targets is reported in Table \ref{tab:MgII} and an example of the fit in the bottom panel of Fig. \ref{fig:fit_civ} (for the rest of the sample see Appendix \ref{app:app}). 
 
To estimate the uncertainties on the derived parameters for each of the emission lines discussed above and in Sec. \ref{sec:OIII_analysis}, we created 1000 realizations of each spectrum by adding noise, drawn from a Gaussian distribution with dispersion equal to the rms of the spectrum, to the best-fit model spectrum and repeated the line fitting procedure on these mock spectra. The associated errors are estimated using the 84 and 16 percentiles of the parameter distribution.

 \begin{figure}[]
 \center
   \includegraphics[width=0.5\textwidth]{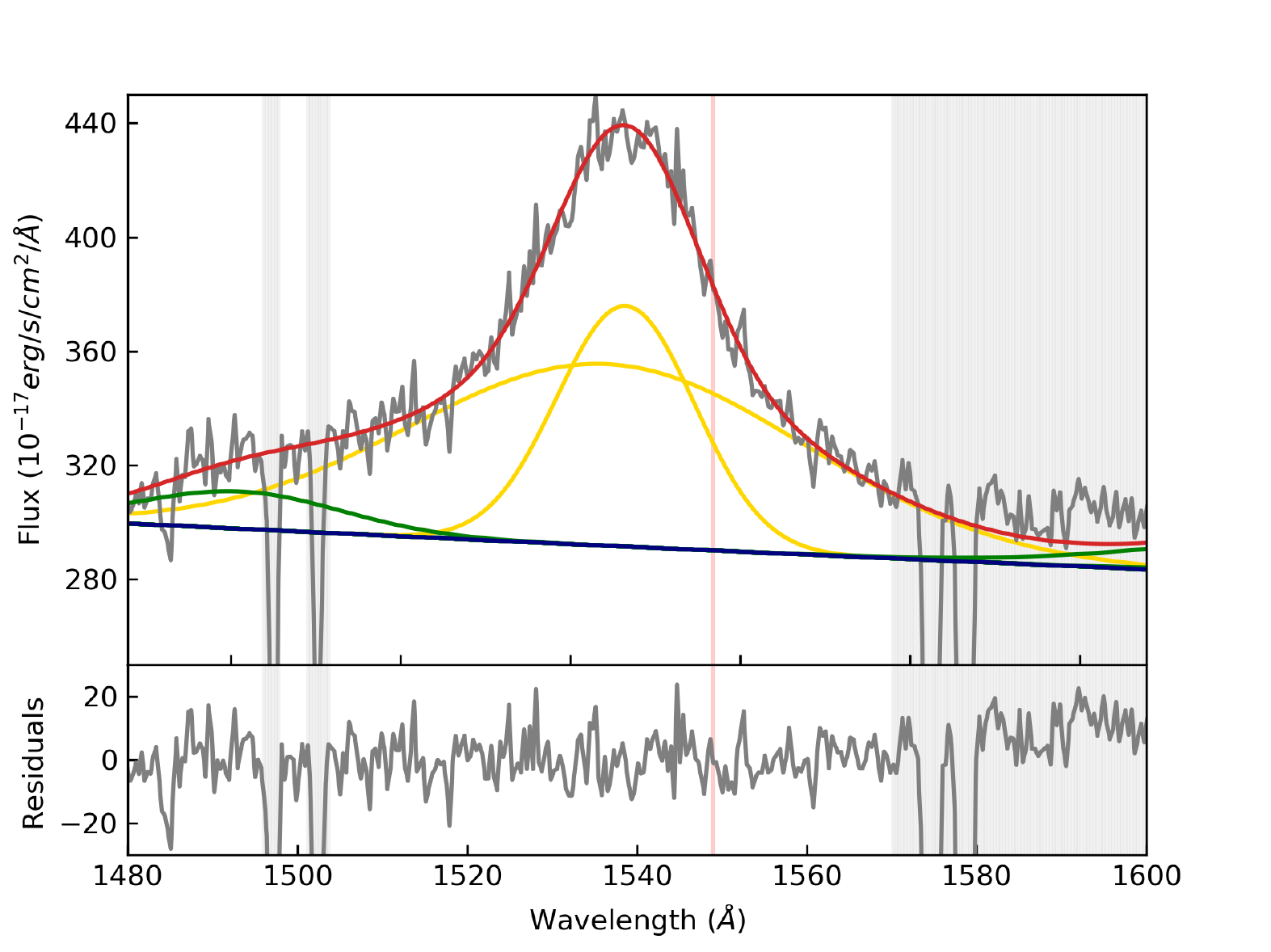}
   \includegraphics[width=0.5\textwidth]{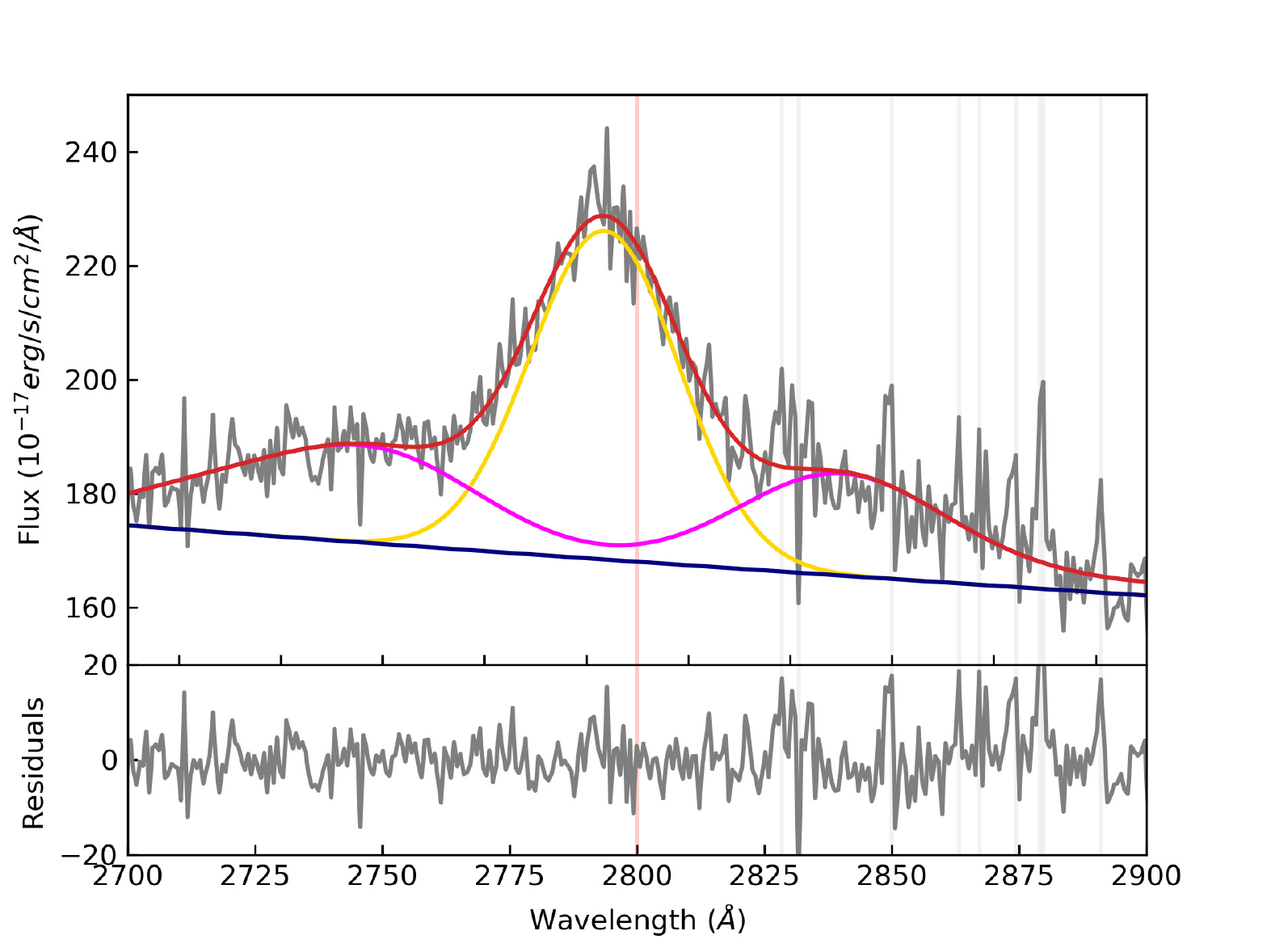}

\caption{\small{(upper) Parametrization of the CIV emission line of the SUPER AGN J1333+1649. The red curve shows the best-fit to the data and the gold curves indicate the multiple Gaussian model used to reproduce the CIV line profile. The red line indicates the expected wavelength of the CIV$\lambda$1549 based on [OIII] systemic redshift. (bottom) Parametrization of the MgII emission line. The red curve shows the best-fit to the data and gold curve the Gaussian component used to reproduce the MgII line profile. The magenta curve represents the best-fit model of the UV FeII emission. The red line indicates the expected wavelength of the MgII$\lambda$2800 based on [OIII] systemic redshift. Grey bands represent the absorption and sky line features masked during the fitting procedure.The x- and y- axis show the restframe wavelength and flux (not corrected for extinction), according to the redshift of the target}}\label{fig:fit_civ}
\end{figure}

      \begin{table*}[]
        
\small
	\centering
\begin{threeparttable}
      \footnotesize
        \caption{Properties of MgII$\lambda$2800 emission line derived from parametric model fits (see Sect. \ref{sec:SDSS}).}\label{tab:MgII}
        \setlength\tabcolsep{1.3pt}
                        \begin{tabular}{lcccccr}
                                \hline
                                \hline  
                               ID & $\rm\lambda_{50}^{MgII}$ & FWHM$\rm_{MgII}$ &  EW$\rm_{MgII}$&v$\rm_{50}^{MgII}$ &Log (L$\rm_{MgII}$/ erg\ s$^{-1}$)& Log (L$_{3000}$/ erg\ s$^{-1}$) \\
                               & \footnotesize{(\AA)}&\footnotesize{(km s$^{-1})$}& \footnotesize{(\AA)} & \footnotesize{(km s$^{-1})$} & &  \\
                                \hline

X\_N\_160\_22    & 2804    \err 1    & 3620    \err 180    &  25    \err   1    &  -390    \err  80    & 44.45  \err 0.07    & 46.46  \err 0.06  \\
X\_N\_81\_44    & 2801    \err 1    & 3910    \err 250    &  36    \err   2    &   -90    \err  100    & 44.37  \err 0.07    & 46.25  \err 0.06  \\
X\_N\_53\_3    & 2795    \err 1    & 2880    \err 290    &  17    \err   2    & 530  \err 120    & 44.18  \err 0.08    & 46.32  \err 0.06  \\
X\_N\_66\_23    & 2795    \err 3    & 3180    \err 880    &  26    \err   8    & 510    \err 340    & 43.70  \err 0.13    & 45.78  \err 0.07  \\
X\_N\_4\_48    & 2790   \err 4    & 8730    \err 1040    &  83    \err  14    & 1030    \err 380    & 44.27  \err 0.09    & 45.70  \err 0.07  \\
X\_N\_102\_35    & 2800    \err 1    & 4110    \err 190    &  23    \err   1    &    -2    \err  80    & 44.33  \err 0.07    & 46.40  \err 0.06  \\
X\_N\_115\_23    & 2803    \err 2    & 4320    \err 550    &  56    \err   8    &  -360    \err 210    & 44.12  \err 0.09    & 45.82  \err 0.07  \\
cid\_166    & 2803    \err 1    & 5910    \err 340    &  35    \err   2    &  -270    \err 140    & 44.64  \err 0.07    & 46.49  \err 0.06  \\
cid\_1605    & 2801    \err 1    & 4210    \err 420    &  42    \err   5    &  -150    \err 160    & 43.77  \err 0.08    & 45.50  \err 0.07  \\
cid\_346    & 2802    \err 2    & 5400    \err 660    &  29    \err   4    &  -220    \err250    & 44.03  \err 0.08    & 45.99  \err 0.06  \\
cid\_467    & 2801    \err 5    & 5830    \err 1340    &  17    \err   4    &   -60   \err 500    & 43.77  \err 0.11    & 45.95  \err 0.06  \\
J1333+1649    & 2794    \err 1    & 3690    \err  80    &  12    \err   0    & 680    \err  30    & 45.15  \err 0.07    & 47.47  \err 0.06  \\
J1441+0454    & 2792    \err 1    & 4010    \err 100    &  20    \err   0    & 880    \err  40    & 44.96  \err 0.07    & 47.05  \err 0.06  \\
J1549+1245    & 2804    \err 1    & 3150    \err  70    &  22    \err  1    &  -410    \err  30    & 45.07  \err 0.07    & 47.15  \err 0.06  \\
S82X1905    & 2805    \err 3    & 4430    \err 770    &  20    \err   4    &  -530    \err 330    & 43.93  \err 0.10    & 45.98  \err 0.06  \\
S82X1940    & 2788    \err 2    & 3410    \err 500    & 370    \err 271    & 1240    \err 190    & 43.98  \err 0.09    & 43.63  \err 0.57  \\
S82X2058    & 2792    \err 3    & 4490    \err 630    &  30    \err   4    & 910    \err 270    & 44.09  \err 0.09    & 46.09  \err 0.06  \\

                                \hline
                                
                \end{tabular}
                        \begin{tablenotes}[para,flushleft]
         \item {\bf{Notes}}. Columns give the following information for the BLR component of the MgII emission line: (1) Target identification,  (2) centroid (\AA), (3) full width at half maximum (km/s), (4)  rest-frame equivalent width (\AA), (5) velocity of the MgII at 50\% of the cumulative line flux, (6) Logarithm of the extinction-corrected MgII luminosity and (7) Logarithm of the extinction-corrected luminosity at 3000 \AA\ derived  from the best-fit values of the power-law model representing the AGN continuum (see Sect. \ref{sec:analysis}).
                                  \end{tablenotes}
                                       \end{threeparttable}
                                            \end{table*}

\section{BLR properties}

 \subsection{Comparison of the broad lines profiles}\label{sec:comp_fwhm}
         
     \begin{figure}[]
 \center   
 \includegraphics[width=1\columnwidth]{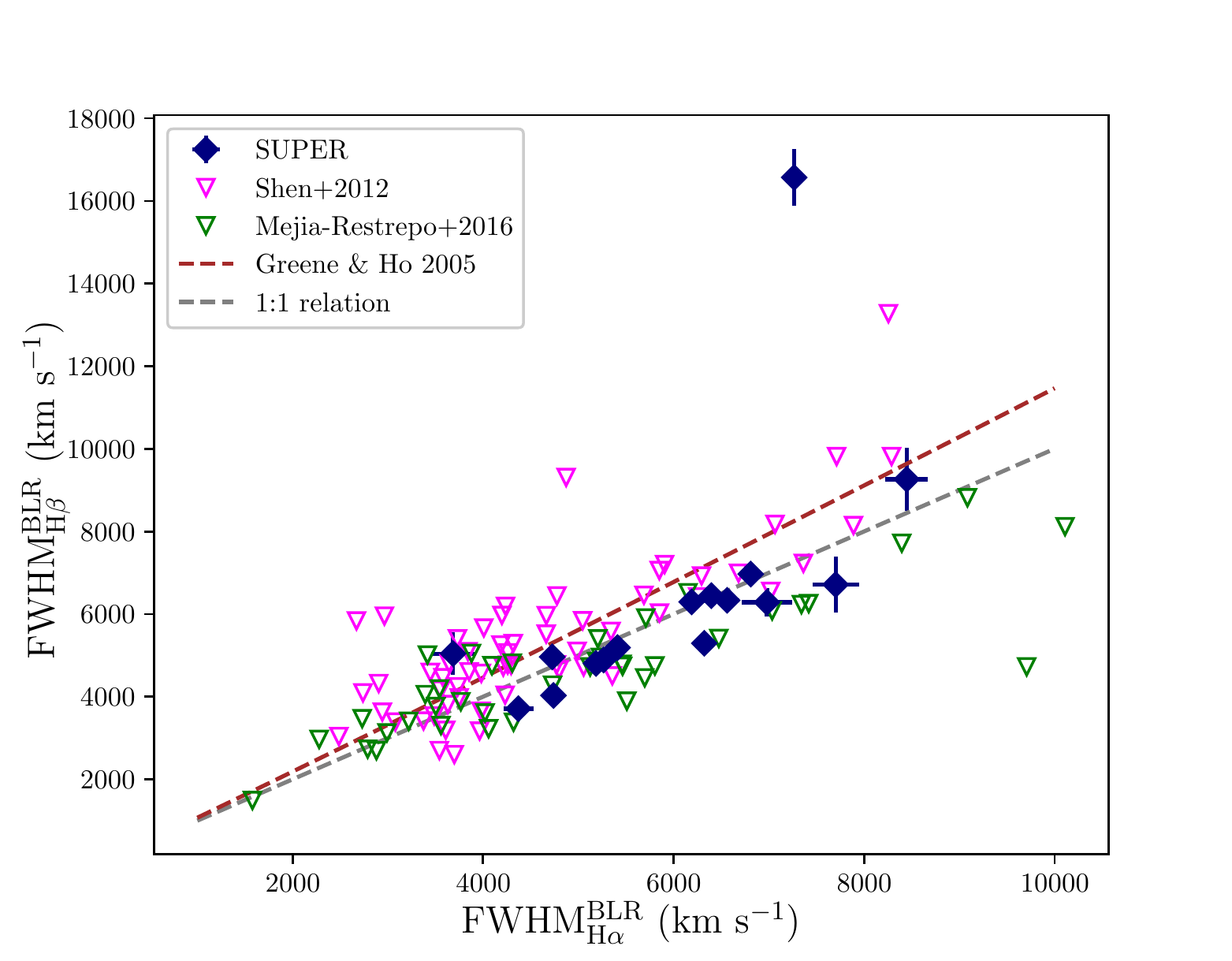}
 
 \caption{Comparison between the FWHM of the BLR component of H$\beta$ and H$\alpha$ emission lines for the SUPER targets (blue diamonds). The SDSS sample from \cite{Shen2012} and the sample from \cite{Mejia-Restrepo2016} are also shown (magenta and green triangles, respectively)}\label{fig:FWHM_ha_hb}

  \end{figure}

 As described in Sec. \ref{sec:analysis}, the line fitting procedure provided luminosities, emission-line centroids and widths for four broad lines in our AGN sample ( H$\alpha$, H$\beta$, CIV and MgII) which we will now compare to derive reliable estimates of the their BH masses. First we compare the best-fit values of the FWHM of the Balmer lines (Fig. \ref{fig:FWHM_ha_hb}).
The FWHM of H$\alpha$ and H$\beta$ for our sample are very similar (slope=1.43\err0.49), consistent with the 1:1 relation. The only exception is the source J1549+1245, for which the H$\beta$ line shows a much broader FWHM with respect to the H$\alpha$ line. We also performed a fit excluding this outlier, resulting in a slope of 0.95\err0.20.

The good agreement between the FWHM of the Balmer lines is consistent with several previous results. In particular  \cite{Greene2005}, analyzing a sample of 229 AGN at z $\sim$ 0.3 and Log (L$\rm_{5100}/erg\ s^{-1}$) $\sim$ 42-45 found such a strong correlation (red-dashed line in Fig. \ref{fig:FWHM_ha_hb}).  More recently, \cite{Mejia-Restrepo2016} also found a correlation consistent with the 1:1 relation for a sample  of 39 Type-1 AGN at z $\sim$ 1.55 with L$\rm_{5100}$ > 10$^{44.3}$ \ergs . This latest work has the further advantage that both lines were observed simultaneously thanks to the wide wavelength coverage of the X-shooter instrument at the VLT, and therefore avoiding any issue related to the time variability of the line profile. The 1:1 correlation between the width of the Balmer lines suggests common kinematics for the H$\beta$ and H$\alpha$ lines, and hence the same emission line region, which is consistent with the similar time lags measured for these lines in reverberation mapping experiments performed to derive the size of the BLR (e.g. \citealt{Kaspi2000}).

\begin{figure}[]
     \center
 \includegraphics[width=1.0\columnwidth]{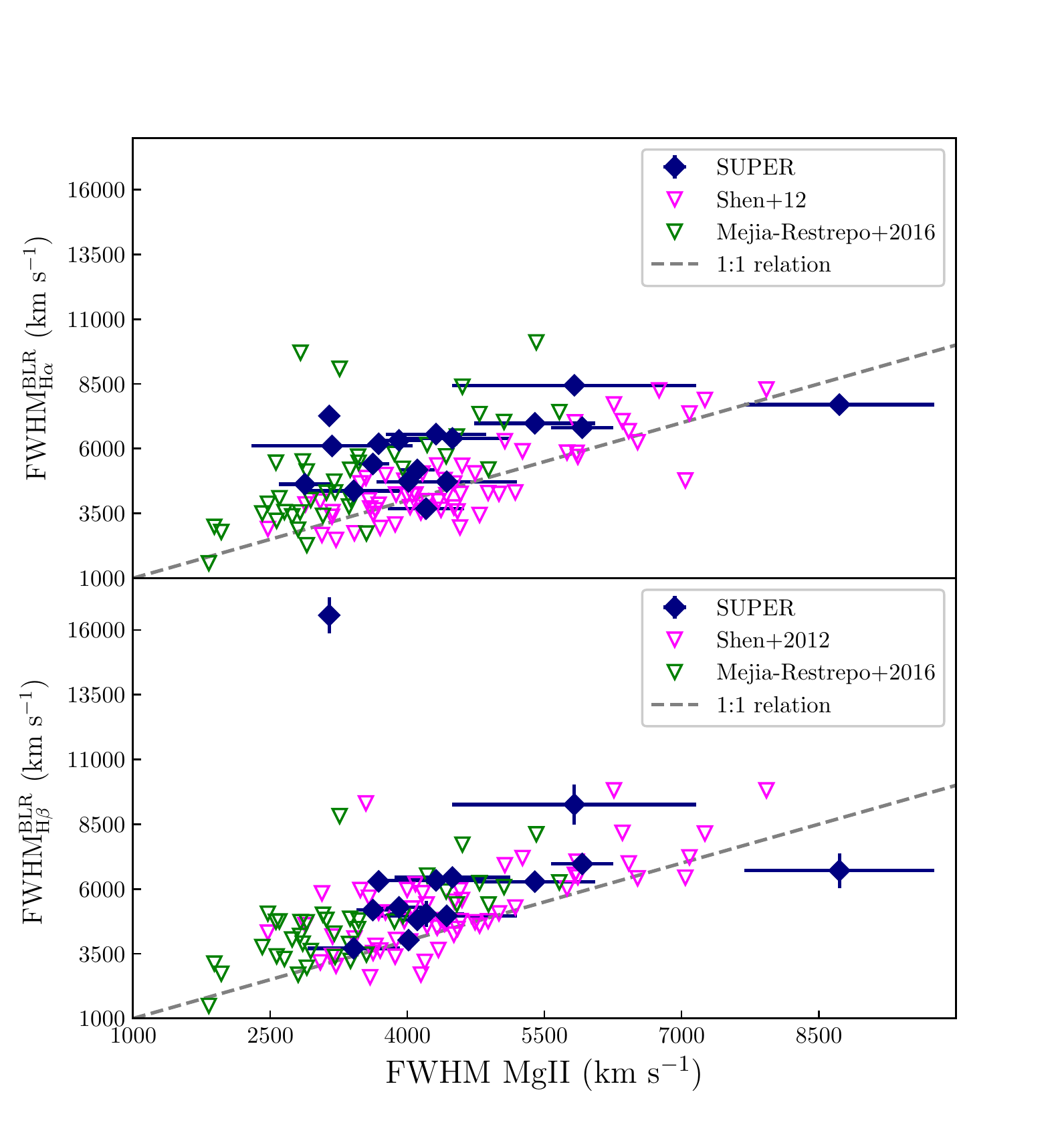}

\caption{Comparison between the FWHM of the BLR component of H$\beta$ and H$\alpha$ emission lines with MgII line for the SUPER targets (blue diamonds). The SDSS sample from \cite{Shen2012} and the sample from \cite{Mejia-Restrepo2016} are also shown (magenta and green triangles, respectively)}\label{fig:FWHM_balmer_MgII}

 \end{figure}

We now compare the FWHM of the MgII with those of the Balmer lines. In this case, excluding the outlier X\_N\_4\_48 showing a very broad FWHM of the MgII, we find a significant positive correlation between the MgII vs H$\alpha$ and H$\beta$ measurements, respectively (Fig. \ref{fig:FWHM_balmer_MgII}), consistent with several previous studies (e.g. \citealt{Shen2012}; \citealt{Mejia-Restrepo2016}). 
As indicated by the slope values reported in Table \ref{tab:corr}, the MgII line is systematically narrower than the corresponding H$\alpha$. We find that on average the MgII lines are narrower than the $H\beta$ lines,  which is consistent with the $30\%$ value reported by \cite{Mejia-Restrepo2016}. 
The most likely explanation for this difference in line width is that MgII is emitted from a region in the BLR further out from the central SMBH than the regions emitting H$\alpha$ and H$\beta$.  Finally, one of our objects, X\_N\_4\_48, shows FWHM(MgII) $>$ FWHM(H$\alpha$,H$\beta$) (however affected by large uncertainties) which may not be surprising given the size of our sample: \cite{Marziani2013} reported that such extreme population, which they named broad-MgII, represents $\approx 10\%$ of bright quasars.     

\begin{figure}[]
 \includegraphics[width=1.\columnwidth]{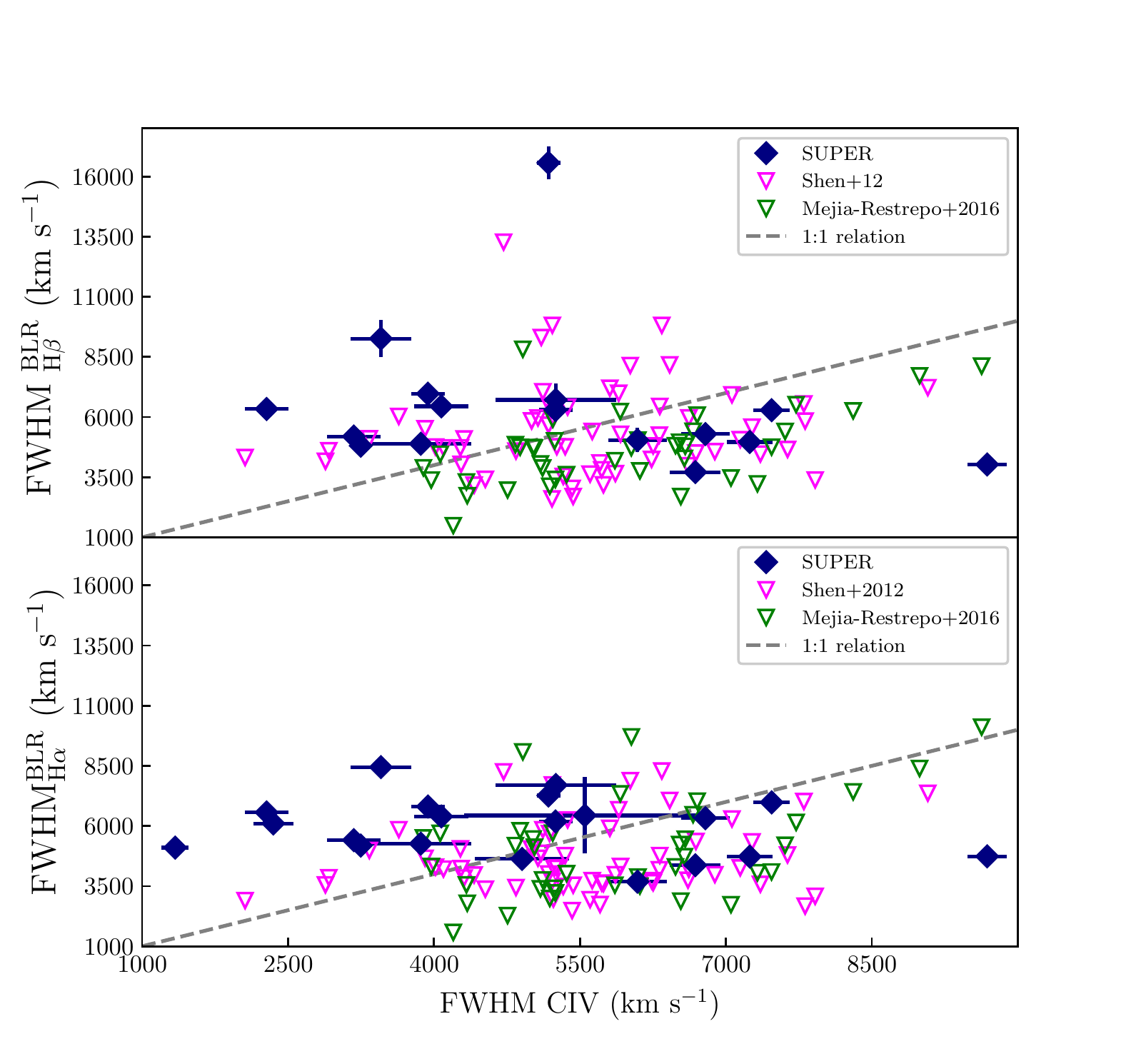}
\caption{Comparison between the FWHM of the BLR component of H$\beta$ and H$\alpha$  emission lines with CIV line for the SUPER targets. Symbols are the same as in Fig. \ref{fig:FWHM_ha_hb}}\label{fig:FWHM_ha_hb_CIV}

 \end{figure}

A completely different story is suggested by the comparison of the FWHM of the CIV with the previous emission lines: H$\beta$, H$\alpha$ and MgII  show a very poor correlation with the CIV FWHM (see Fig. \ref{fig:FWHM_ha_hb_CIV} and Table \ref{tab:corr}). Reverberation mapping experiments, also performed on high redshift quasars (\citealt{Lira2018}), predict that the emission line region of CIV is located closer to the central SMBH than that producing the Balmer lines, therefore we would expect that FWHM(CIV) > FWHM(Balmer lines), instead 10/16 of the SUPER sample, with both H$\beta$ and CIV lines, shows the opposite behavior, i.e. FWHM(CIV) < FWHM(H$\beta$) and 13/20, with both H$\alpha$ and CIV lines, have FWHM(CIV) < FWHM(H$\alpha$). This confirms earlier works based on larger samples of AGN (e.g. \cite{Trakhtenbrot2012}) which suggested that the CIV line is an unreliable probe of the virialized gas in the BLR. In particular,  \cite{Mejia-Restrepo2018} provides a systematic analysis of this behaviour, including a critical analysis of various "correction factors" suggested to cure this issue in the past. They concluded that this is a real effect related, probably, to non-virialized gas motion in the BLR. In addition, the FWHM of the CIV correlates with the velocity shift of the CIV emission line  (\citealt{Gaskell1982}, \citealt{Sulentic2000}, \citealt{Richards2011}, \citealt{Denney2012}, \citealt{Coatman2017}, \citealt{Vietri2018}), leading to a further dispersion in the plane FWHM H$\beta$-H$\alpha$ vs CIV. Based on the above, we will not use therefore CIV as a reliable BH mass estimator, but we will take advantage of the ability to trace non-virialized gas at the scale of the BLR in Section \ref{sec:winds}.

 \begin{table}[]

\caption{Spearman test results for the SUPER sample. We derived linear relations using the BCES (Y|X) regression (\citealt{Akritas1996}).}\label{tab:corr} 
	\begin{threeparttable}
			\begin{tabular}{lccc}
\hline		
				\hline	
			 Correlation & Slope	& $\rho$ & P-value \\
	(1)&(2)&(3)&(4)\\

\hline
	         FWHM H$\beta$ vs.  FWHM H$\alpha$\tnote{1}&	1.43\err0.49&	0.88& 6e-6\\
	         FWHM H$\beta$ vs.  FWHM H$\alpha$\tnote{2}&	0.95\err0.20&	0.87& 2e-5\\

	          FWHM H$\beta$ vs.  FWHM MgII\tnote{1}&	-0.05\err0.70&	0.39& 0.15\\
	          FWHM H$\beta$ vs.  FWHM MgII\tnote{2}&	2.89\err1.89&	0.66& 0.01\\

	          FWHM H$\alpha$ vs.  FWHM MgII\tnote{1}&	1.15\err0.63&	0.41& 0.11\\
	          FWHM H$\alpha$ vs.  FWHM MgII\tnote{2}&	0.63\err0.23&	0.50& 0.04\\

	           FWHM H$\beta$ vs.  FWHM CIV\tnote{1} &	-0.32\err0.14 &	-0.32& 0.24\\
	           FWHM H$\beta$ vs.  FWHM CIV\tnote{2} &	-0.32\err0.14 &	-0.33& 0.23\\

	           FWHM H$\alpha$ vs.  FWHM CIV &	-0.13\err0.11&	-0.15& 0.54\\
	           
				Log L$\rm_{H\alpha}$ vs Log L$\rm_{5100}$  & 0.84\err0.05 & 0.79& 6e-5\\	
				Log L$\rm_{H\beta}$  vs Log L $\rm_{5100}$& 0.97\err0.06 & 0.94& 6e-8\\	
				Log L$\rm_{1350}$ vs Log L$\rm_{5100}$ & 0.78\err0.15 & 0.78& 7e-5\\	
				Log L$\rm_{3000}$ vs Log L$\rm_{5100}$ & 0.97\err0.19& 0.86& 2e-5\\	
				Log L$\rm_{H\beta}$ vs Log L$\rm_{H\alpha}$ & 1.09\err0.08 & 0.78& 4e-4\\	

				\hline
				\end{tabular}
				 \begin{tablenotes}
 \item Outliers objects showing a very broad FWHM of H$\beta$ (J1549+1245, see Fig. \ref{fig:FWHM_ha_hb}) and MgII emission lines (X\_N\_4\_48, see \ref{fig:FWHM_balmer_MgII}) have been included (1) and excluded (2) from the fit.
  \end{tablenotes}	
  
				\end{threeparttable}
				
			\end{table}

  \subsection{Comparison of continuum and line luminosity}\label{sec:comp_lum}

In order to correct the BLR line luminosity for dust extinction we use the ratio of the broad Balmer lines. For a low density environment such as the NLR, the intrinsic Balmer ratio is equal to 2.74-2.86 assuming a case B recombination (\citealt{Osterbrock2006}). The BLR densities are significantly higher, and line optical depths and collisional effects can lead to different Balmer decrements (\citealt{Netzer2013}). While studies of large samples of AGN suggests that this is generally the case, the mean L$\rm_{H\alpha}$/L$\rm_{H\beta}$ of the sources with the bluest continua are surprisingly similar to the Case B prediction with values $\approx 3$ (\citealt{Baron2016}). We also find a similar ratio, with a median value of 3.37 \err 0.09.

The relationship between the Balmer decrement and the color excess is given by:

\begin{equation}
\rm E(B-V)= \frac{E(H\beta - H\alpha)}{k(\lambda_{H\beta})-k(\lambda_{H\alpha})}=\frac{2.5}{k(\lambda_{H\beta})-k(\lambda_{H\alpha})}  log\Biggl[ \frac{(H_{\alpha}/{H_{\beta})}_{obs}}{(H_{\alpha}/{H_{\beta})}_{int}} \Biggr]
\end{equation} 

where k($\rm\lambda_{H\beta}$) and  k($\rm\lambda_{H\alpha}$) are the reddening curves evaluated at H$\beta$ and H$\alpha$ wavelengths respectively, the (H$\rm _{\alpha}/{H_{\beta}})_{obs}$ and (H$\rm _{\alpha}/{H_{\beta}})_{int}$ are the observed and intrinsic Balmer decrement respectively.
Based on the observed Balmer decrement, we derived the color excess E(B-V) assuming a foreground screen, a \cite{Cardelli1989} extinction law, and (H$\rm _{\alpha}/{H_{\beta}})_{int}$=3 (\citealt{Baron2016}). We derived a median E(B-V) of 0.12 mag with a standard deviation of 0.03 mag. We use this median value of E(B-V) to correct the emission lines and continuum luminosities. This means that the luminosities are corrected by a factor of $\sim$1.3, $\sim$1.5, $\sim$1.9 and $\sim$2.4 for the emission lines H$\alpha$, H$\beta$, MgII and CIV, respectively and  $\sim$1.4, $\sim$1.9 and $\sim$2.7 for the continuum at 5100 \AA, at 3000 \AA  and 1350 \AA, respectively. We note that reliable estimates of BH mass for the SUPER sample are based on H$\alpha$ and H$\beta$ lines (see Sec. \ref{sec:mbh}), for which the reddening correction has a low effect on the line luminosity.

We estimated the monochromatic luminosity at 5100 \AA\ using the best-fit values of the power-law model representing the AGN continuum, L$\rm_{5100}$. The contamination from the host-galaxy for the SUPER targets is negligible since L$\rm_{5100}$ $>$ 10$^{45.4}$ erg s$^{-1}$ (\citealt{Shen2011}). The values derived from the fit are consistent with the continuum luminosities at 5100 \AA\ obtained from the SED fitting presented in \cite{Circosta2018}.

  \begin{figure}[]
\center
\includegraphics[width=1.1\columnwidth]{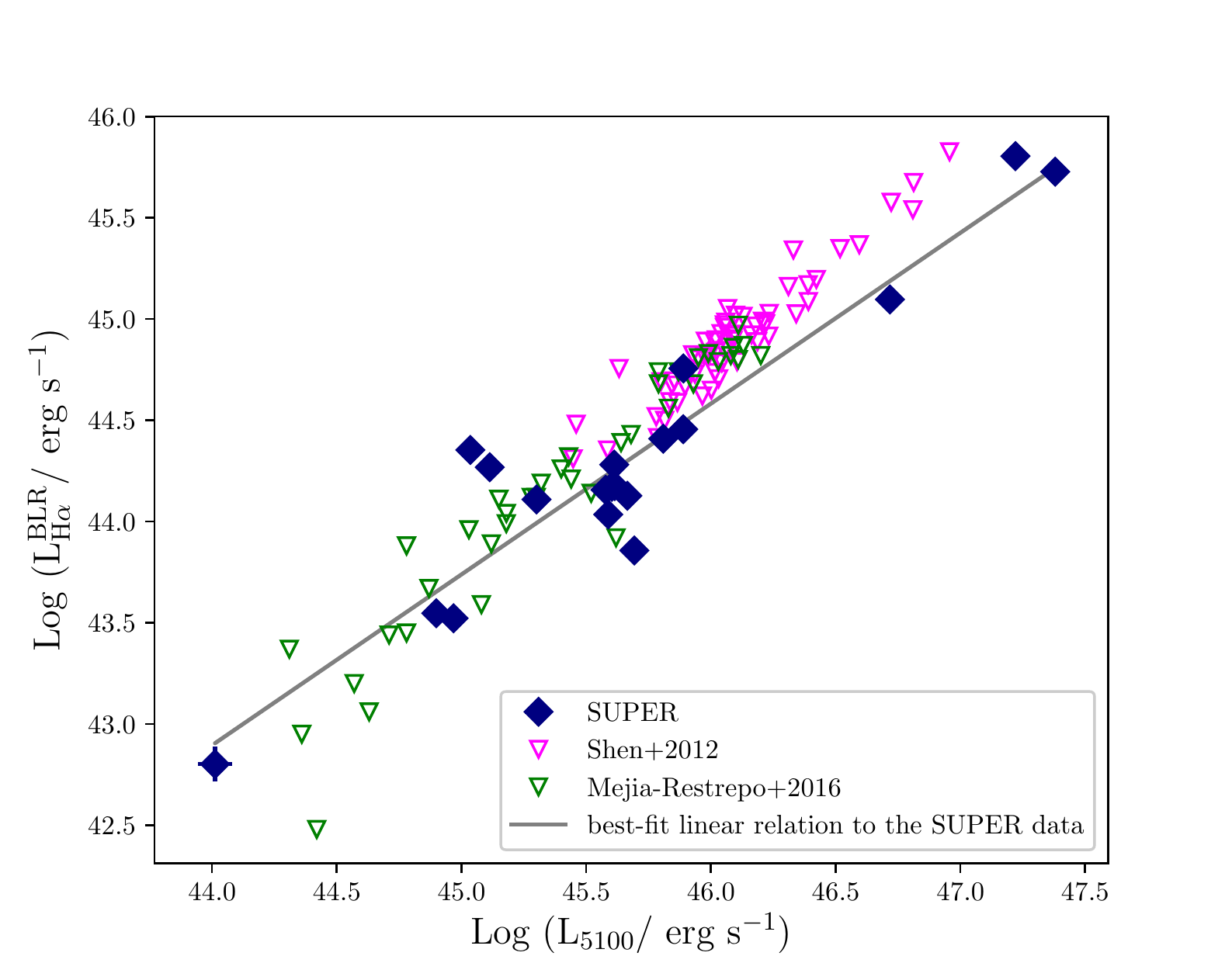}

\caption{Comparison between the continuum luminosity at 5100 \AA\ derived from the best-fit continuum power law model with the luminosity of broad H$\alpha$ emission line, both in logarithmic scale, for the SUPER sample (blue circles). Both continuum and emission line luminosity are corrected for extinction (see Sec \ref{sec:comp_lum}). The SDSS sample from \cite{Shen2012} and the sample from \cite{Mejia-Restrepo2016} are also shown (red and green circles, respectively). The best-fit linear relation (BCES Y|X regression) for the SUPER data is shown in grey solid line.}\label{fig:lum}

 \end{figure}
 
 In Fig. \ref{fig:lum} we show the comparison between the extinction-corrected luminosities at 5100 \AA\ (L$\rm_{5100}$), as estimated from the best-fit continuum power law model, and of the broad H$\alpha$ (L$\rm_{H\alpha}$). We found a tight correlation between the luminosities (see also Table \ref{tab:corr}), which is an important ingredient for the use of H$\alpha$ measurements to estimate the BH mass (see Sec. \ref{sec:mbh}).
 
 We also compared the extinction-corrected continuum luminosities at 5100 \AA\ (L$\rm_{5100}$), 3000 \AA\ (L$\rm_{3000}$ ) and 1350 \AA\ (L$\rm_{1350}$), standard continuum luminosity indicators that were used in earlier works. Both continuum luminosity at 1350 \AA\ and 3000 \AA\ are correlated with L$\rm_{5100}$, with slopes close to unity. We list the slope, Spearman coefficient and p-value of the correlations discussed in Table \ref{tab:corr}. The results from our small SUPER sample are similar, but not identical, to those found in larger samples by \cite{Shen2012} and \cite{Mejia-Restrepo2016}. Note again that unlike the Mejia-Restrepo results, our observations are not simultaneous and variability may be a cause of the different slope and scatter.
 
Finally, we use extinction-corrected L$\rm_{5100}$ to derive the bolometric luminosity assuming a bolometric correction factor, f$\rm_{bol}$= L$\rm_{Bol}$/$\lambda$L$\rm_{\lambda}$ where L$\rm_{\lambda}$ is the monochromatic luminosity. We used the prescription from \cite{Runnoe2012}:
 
 \begin{equation}
\rm Log (L_{Bol})= 4.891+0.912 \times Log (5100\ L_{5100})\label{eq:lbol}
 \end{equation}

The bolometric luminosities obtained from Eq. \ref{eq:lbol} are consistent with those derived from the SED fitting \cite{Circosta2018}. We will use the latest as our fiducial values in the rest of the paper.

\subsection{SMBH masses and Eddington ratios} \label{sec:mbh}

As mentioned in Sect. \ref{sec:intro}, the BH mass can be estimated from single-epoch spectra assuming that the BLR is virialized.
Based on the comparison of the FWHM for the broad lines observed (see Sec. \ref{sec:comp_fwhm}), we will estimate  the virial BH mass from three of those emission lines: H$\beta$, H$\alpha$ and MgII.
The virial BH mass calibrations used for the broad H$\beta$ line, available for 16 SUPER targets, is from \cite{Bongiorno2014}:

\begin{equation}
\resizebox{1 \columnwidth}{!} 
{$\rm Log(M_{BH}/M_{\odot}) = 6.7 + 2\times\rm Log\left(\frac{FWHM_{H\beta}}{10^3 \rm\ km\ s^{-1}}\right) + 0.5\times Log\left(\frac{5100 L_{5100}}{10^{44} \rm\ erg\ s^{-1}}\right)\hspace{0.2cm}
$}\label{eq:mbh_hb}
\end{equation}

where FWHM is the best-fit full width at half maximum of the broad component of H$\beta$ and $\lambda L\rm_{\lambda}$ the best-fit extinction-corrected continuum luminosity at 5100 \AA . This M$\rm_{BH}(H\beta)$ is derived assuming {\it{f}} = 1. 

The \cite{Greene2005} calibration was used to derive the BH mass from the broad H$\alpha$, available for 21 SUPER targets:

\begin{equation}
\resizebox{1 \columnwidth}{!} 
{$\rm Log(M_{BH}/M_{\odot}) = 6.3 + 2.06\times\rm Log\left(\frac{FWHM_{H\alpha}}{10^3 \rm\ km\ s^{-1}}\right) + 0.55\times Log\left(\frac{ L_{H\alpha}}{10^{42} \rm\ erg\ s^{-1}}\right)\hspace{0.2cm}$}\label{eq:mbh_ha}
\end{equation}

with the best-fit value of the FWHM of the profile of H$\alpha$ line and extinction-corrected H$\alpha$ luminosity L$\rm_{H\alpha}$.  

We used the \cite{Bongiorno2014} calibrated formula for the MgII emission line, available for 17 SUPER targets:

\begin{equation}
\resizebox{1 \columnwidth}{!} 
{$\rm Log(M_{BH}/M_{\odot}) = 6.6 + 2\times\rm Log\left(\frac{FWHM_{MgII}}{10^3 \rm\ km\ s^{-1}}\right) + 0.5\times Log\left(\frac{3000 L_{3000}}{10^{44} \rm\ erg\ s^{-1}}\right)
$}\label{eq:mbh_MgII}
\end{equation}

with the best-fit value of the FWHM of the profile of MgII line and extinction-corrected luminosity $\lambda$L$\rm_{\lambda}$  at 3000 \AA  . 

In \cite{Mejia-Restrepo2018} the authors found a relation between the virial factor {\it{f}} and the width of the broad lines:
\begin{equation}
f= \bigg(\rm\frac{FWHM_{obs} (line)}{FWHM_{obs}^0}\bigg)^\beta
\end{equation} 
where FWHM$_{obs} (line)$ is the observed FWHM of a broad line, H$\alpha$, H$\beta$ and MgII in our case, with FWHM$_{obs}^0$=4000\err700 and $\beta$=-1.00\err0.10 for the H$\alpha$ line, FWHM$_{obs}^0$=4500\err1000 and $\beta$=-1.17\err0.11 for the H$\beta$ line and FWHM$_{obs}^0$=3200\err800 and $\beta$=-1.21\err0.24 for the MgII line.

We derived this virial factor {\it{f}} for both H$\alpha$, H$\beta$  and MgII. The mean values are {\it{f}}=0.77\err0.20, {\it{f}}=0.70\err0.12 and {\it{f}}=0.74\err0.23 for H$\beta$, H$\alpha$ and MgII, justifying the assumption of {\it{f}} = 1  in Eq. \ref{eq:mbh_hb}, \ref{eq:mbh_ha} and \ref{eq:mbh_MgII}. 

Regarding CIV-derived masses, this can not be used to estimate BH mass due to the non-virialized CIV emitting gas motion discussed earlier.
 
\begin{figure}[]
 \centering

 \includegraphics[width=1\columnwidth]{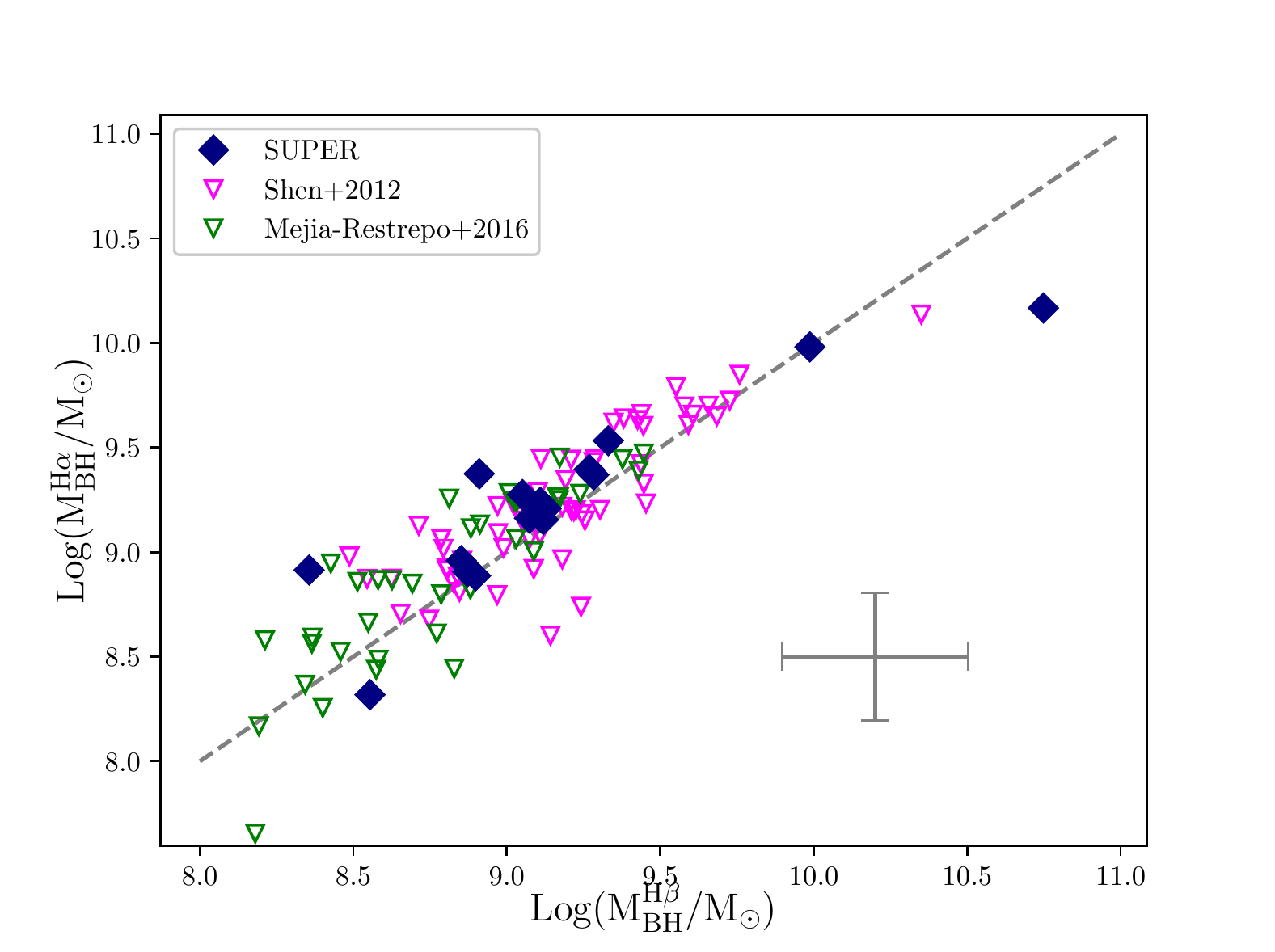}
\caption{H$\alpha$ black hole mass as a function of H$\beta$ BH mass, both in logarithmic scale, for the SUPER targets (blue points) and SDSS targets from \cite{Shen2012} (blue points). The 1:1 relation is shown as dashed line. The typical uncertainty on the H$\alpha$- and H$\beta$-based BH masses for the SUPER sample is also represented in the right-bottom corner}\label{fig:mbhha_mbhhb}

%}
 \end{figure}

\begin{figure}[]
 \centering

 \includegraphics[width=1\columnwidth]{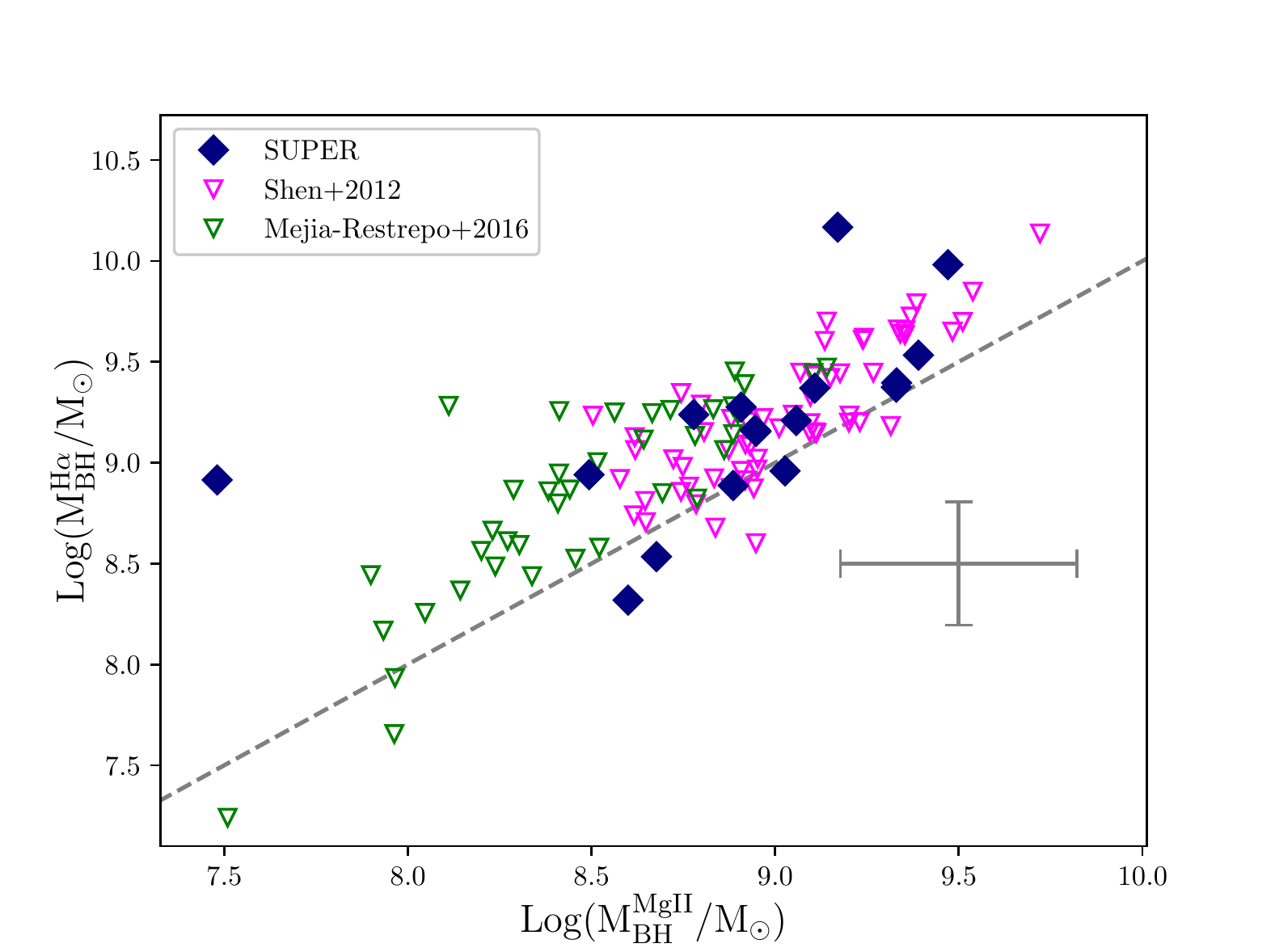}
\caption{H$\alpha$ black hole mass as a function of MgII BH mass, both in logarithmic scale, for the SUPER targets (blue points) and SDSS targets from \cite{Shen2012} (blue points). The 1:1 relation is shown as dashed line. The typical uncertainty on the H$\alpha$- and MgII-based BH masses for the SUPER sample is also represented in the right-bottom corner }\label{fig:mbhMgII_mbhha}

 \end{figure}

The H$\beta$, H$\alpha$-based and MgII BH masses are listed in Table \ref{tab:MBH}. 
The values of the BH mass of the SUPER sample derived from H$\alpha$ and H$\beta$ lines are fairly in agreement as shown in Fig. \ref{fig:mbhha_mbhhb}. The SDSS AGN from the \cite{Shen2012} and \cite{Mejia-Restrepo2016} samples are also plotted\footnote{We derived the BH mass of \cite{Shen2012} and \cite{Mejia-Restrepo2016} samples by using Eq. \ref{eq:mbh_hb}, \ref{eq:mbh_ha}, and \ref{eq:mbh_MgII} and the Eddington ratio from Eq. \ref{eq:edd}}. A correlation is also found between the Balmer-based and MgII-based BH masses (see Fig. \ref{fig:mbhMgII_mbhha}), however the MgII line profile is affected by strong sky-line residuals in almost all spectra of the SUPER AGN, which can affect the measured line properties and hence the derived BH mass.

We used as fiducial virial BH masses the values derived from the H$\beta$ emission lines, which we prefer over the H$\alpha$ line because the latter is blended with the [NII]$\lambda\lambda$ 6548,6583 doublet implying a less reliable measure, for all but five SUPER targets (i.e. X\_N\_53\_3, X\_N\_66\_23, X\_N\_35\_20, X\_N\_44\_64 and cid\_1205), for which we used the BH mass values derived from H$\alpha$ because the H$\beta$ line is not detected in those sources. 

We found that SUPER sample hosts BH with log M$_{BH}$=8.4-10.8 M$_{\odot}$.
From the BH mass and the bolometric luminosity derived from the Spectral Energy Distribution fitting (see \citealt{Circosta2018}), we derived the Eddington ratio for a solar composition gas, defined as:

\begin{equation}
\rm \lambda_{Edd}= \frac{L_{Bol}}{1.5\times 10^{38} M_{BH}}\label{eq:edd}
\end{equation}

We find values in the range \edd =0.04-1.3 (see Table \ref{tab:MBH}).

\begin{figure}[]
 \centering

 \includegraphics[width=1\columnwidth]{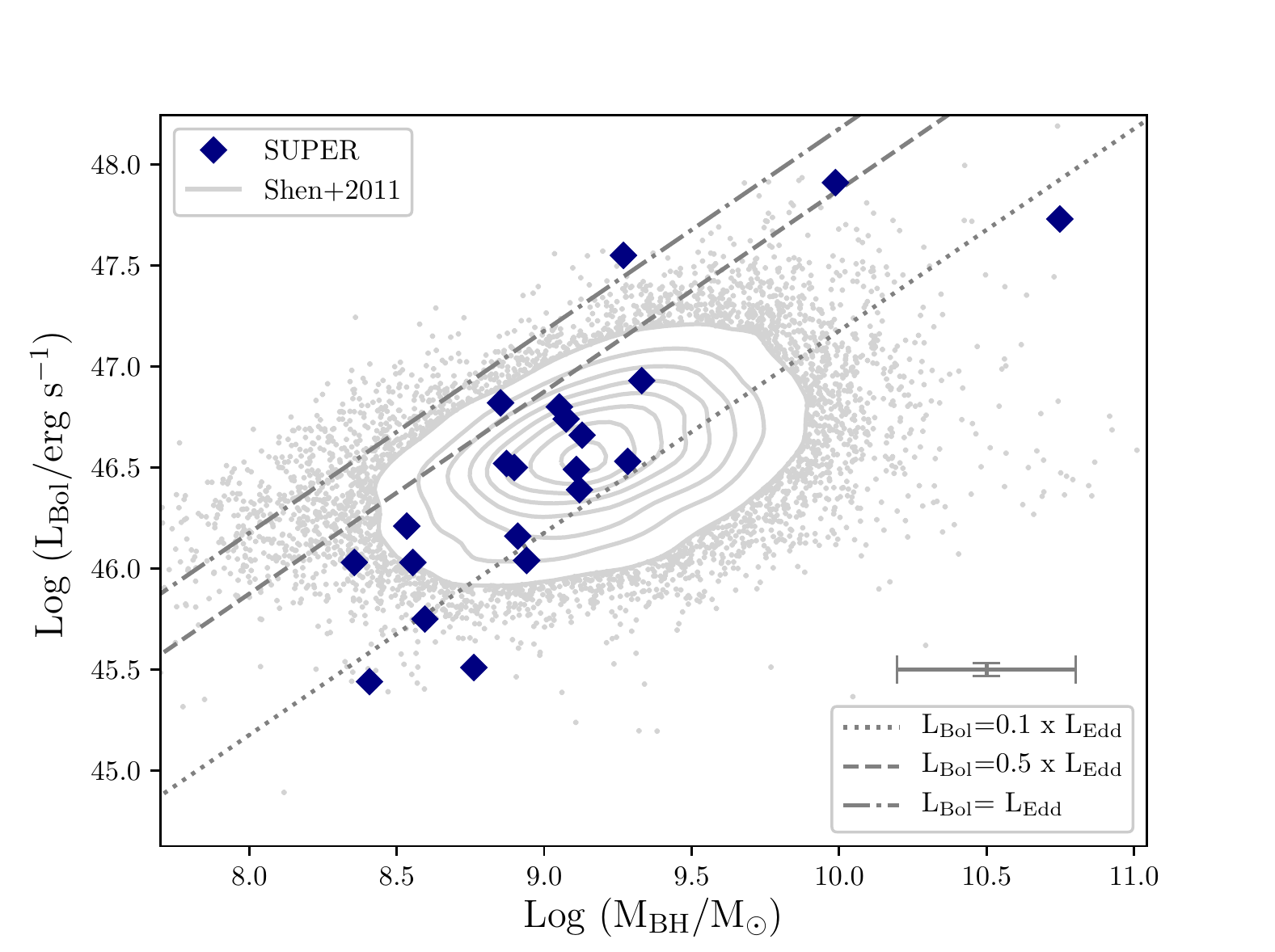}
\caption{Bolometric luminosity as a function of BH mass, both in logarithmic scale, for the SUPER sample. Luminosity in fractions of 0.1, 0.5 and 1 Eddington luminosity are respectively indicated with dot-dashed, dotted and dashed lines. Contours and grey points refer to SDSS DR7 QSOs from \cite{Shen2011}.}\label{fig:mbh_lbol}

 \end{figure}

Fig. \ref{fig:mbh_lbol} shows the comparison of \mbh , \lbol\  and \edd\ measured for the SUPER sample with those derived from a sample of $\sim$23000 SDSS AGN at 1.5$\le$ z $\le$ 2.2 with MgII-based BH mass (\cite{Shen2011}, contour lines). The bolometric luminosity in fraction of 0.1, 0.5 and 1 Eddington luminosity is also reported.
The SUPER sample spans two order of magnitude in bolometric luminosity and in Log (M$\rm_{BH}/M_{\odot})$. 

 \begin{table*}[]
\centering
    % \small
      %\footnotesize
         \begin{threeparttable}
        \caption{H$\beta$-based, H$\alpha$-based and MgII-based SMBH mass, H$\beta$-based and H$\alpha$-based Eddington ratio of the SUPER AGN (see Sect. \ref{sec:mbh} for details)}\label{tab:MBH}
        \setlength\tabcolsep{1.2pt}
        
                        \begin{tabular}{lcccc@{\hspace{0.5cm}}r}
                                \hline
                                \hline  
                                ID  & Log (M$\rm_{BH}^{H\beta}$/ M$_{\odot}$) & Log (M$\rm_{BH}^{H\alpha}$/ M$_{\odot}$)& Log(M$\rm_{BH}^{MgII}$/ M$_{\odot}$)&    $\rm\lambda_{Edd}^{H\beta}$ &    $\rm\lambda_{Edd}^{H\alpha}$ \\[0.5ex]
                                %\footnotesize{(erg s$^{-1})$}
                                %\footnotesize{(erg s$^{-1})$}
                                (1) & (2) & (3) &(4) & (5) & (6) \\
                                \hline
   
 X\_N\_160\_22  & 9.07\err0.30  & 9.16\err0.30  & 8.95\err0.30  & 0.31\err0.21  & 0.25\err0.17 \\
 X\_N\_81\_44  & 9.05\err0.30  & 9.28\err0.30  & 8.91\err0.31  & 0.37\err0.26  & 0.22\err0.15  \\
 X\_N\_53\_3  &   -   & 8.53\err0.30  & 8.68\err0.31  & - & 0.32\err0.22  \\
 X\_N\_66\_23  &   -   & 8.94\err0.30  & 8.49\err0.39  & - & 0.08\err0.06  \\
 X\_N\_35\_20  &   -   & 8.41\err0.38  &   -   & - & 0.07\err0.06  \\
 X\_N\_12\_26  & 8.87\err0.30  & 8.91\err0.30  &   -   & 0.30\err0.21  & 0.27\err0.19 \\ 
 X\_N\_44\_64  &   -   & 8.76\err0.31  &   -   & -  & 0.04\err0.03  \\
 X\_N\_4\_48  & 8.91\err0.31  & 9.37\err0.30  & 9.33\err0.32  & 0.12\err0.09  & 0.04\err0.03  \\
 X\_N\_102\_35  & 8.85\err0.30  & 8.96\err0.30  & 9.03\err0.30  & 0.62\err0.43  & 0.48\err0.34  \\
 X\_N\_115\_23  & 9.11\err0.30  & 9.24\err0.30  & 8.78\err0.32  & 0.16\err0.11  & 0.12\err0.08  \\
 cid\_166  & 9.33\err0.30  & 9.53\err0.30  & 9.39\err0.31  & 0.26\err0.18  & 0.17\err0.12  \\
 cid\_1605  & 8.55\err0.31  & 8.32\err0.31  & 8.60\err0.31  & 0.20\err0.14  & 0.34\err0.24  \\
 cid\_346  & 9.13\err0.30  & 9.21\err0.30  & 9.06\err0.32  & 0.23\err0.16  & 0.19\err0.13  \\
 cid\_1205  &   -   & 8.60\err0.30  &   -   & -  & 0.10\err0.07  \\
 cid\_467  & 9.28\err0.31  & 9.37\err0.30  & 9.11\err0.36  & 0.12\err0.08  & 0.10\err0.07  \\
 J1333+1649  & 9.99\err0.30  & 9.98\err0.30  & 9.47\err0.30  & 0.56\err0.39  & 0.57\err0.39  \\
 J1441+0454  & 9.27\err0.30  & 9.40\err0.30  & 9.33\err0.30  & 1.27\err0.88  & 0.95\err0.66  \\
 J1549+1245  & 10.75\err0.30  & 10.17\err0.30  & 9.17\err0.30  & 0.06\err0.04  & 0.24\err0.17  \\
 S82X1905  & 8.90\err0.30  & 8.89\err0.30  & 8.89\err0.34  & 0.27\err0.18  & 0.27\err0.19  \\
 S82X1940  & 8.36\err0.30  & 8.91\err0.30  & 7.48\err0.43  & 0.31\err0.22  & 0.09\err0.06  \\
 S82X2058  & 9.12\err0.30  & 9.16\err0.30  & 8.95\err0.33  & 0.12\err0.09  & 0.11\err0.08  \\

                                \hline
                                \end{tabular}
                \begin{tablenotes}[para,flushleft]
         \item {\bf{Notes}}. Columns give the following information: (1) Target identification,  (2) Logarithm of H$\beta$-based BH mass, (3)  Logarithm of H$\alpha$-based BH mass, (4) Logarithm of MgII-based BH mass, (5) Eddington ratio derived from H$\beta$-based BH mass and (6) Eddington ratio derived from H$\alpha$-based BH mass. The error associated with the BH masses and hence Eddington ratios includes both the statistical uncertainties affecting the continuum or line luminosities and FWHM values and the systematic uncertainty in the virial relations ($\sim$0.3 dex, e.g. \citealt{Bongiorno2014}).
                                  \end{tablenotes}
                                       \end{threeparttable}         
                               
                        \end{table*}

\section{BLR winds}\label{sec:winds}

\subsection{Connection between CIV velocity shift and AGN properties}
\begin{figure}
 \centering
 \includegraphics[width=0.9\columnwidth]{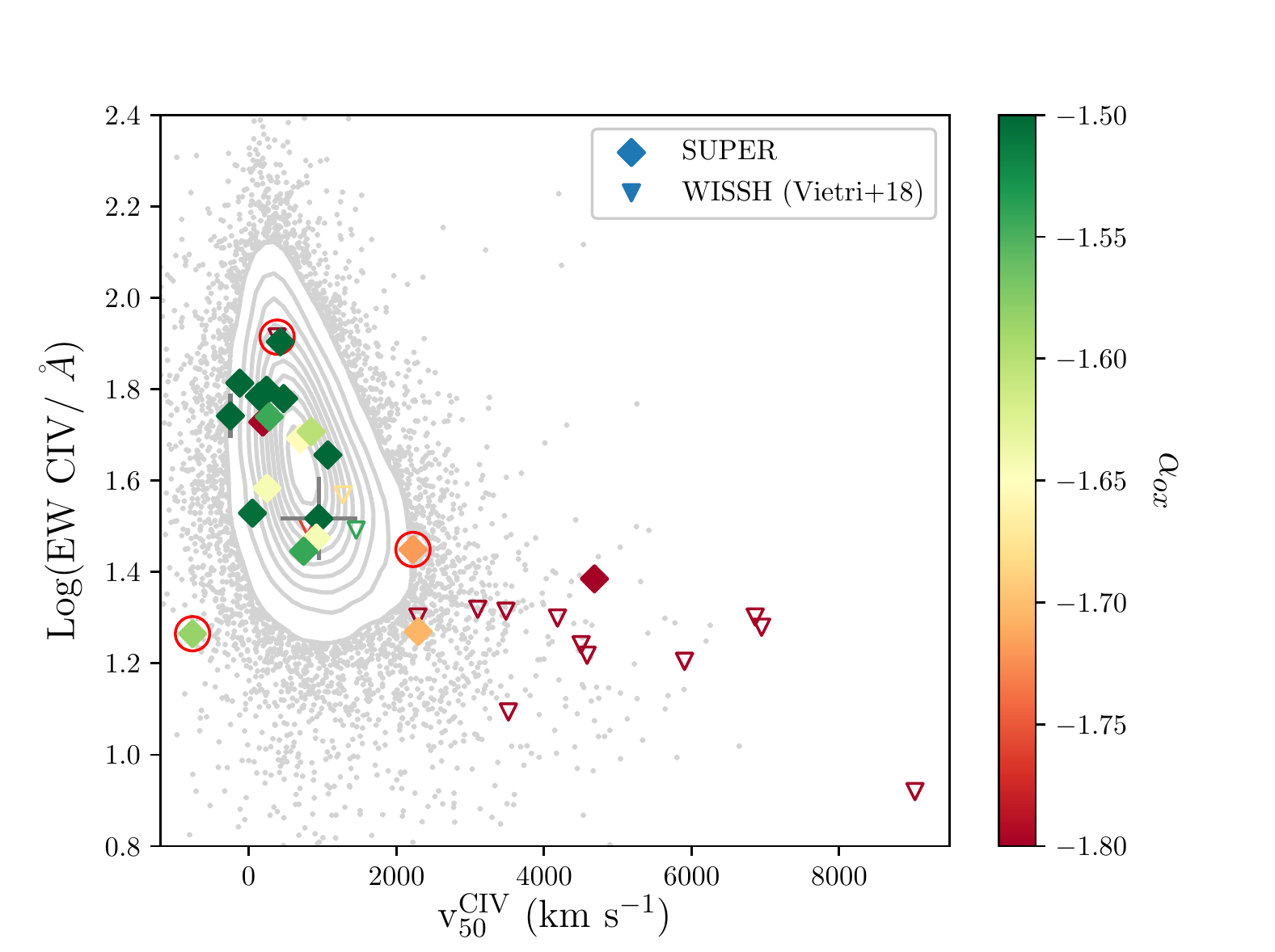}
  \includegraphics[width=0.9\columnwidth]{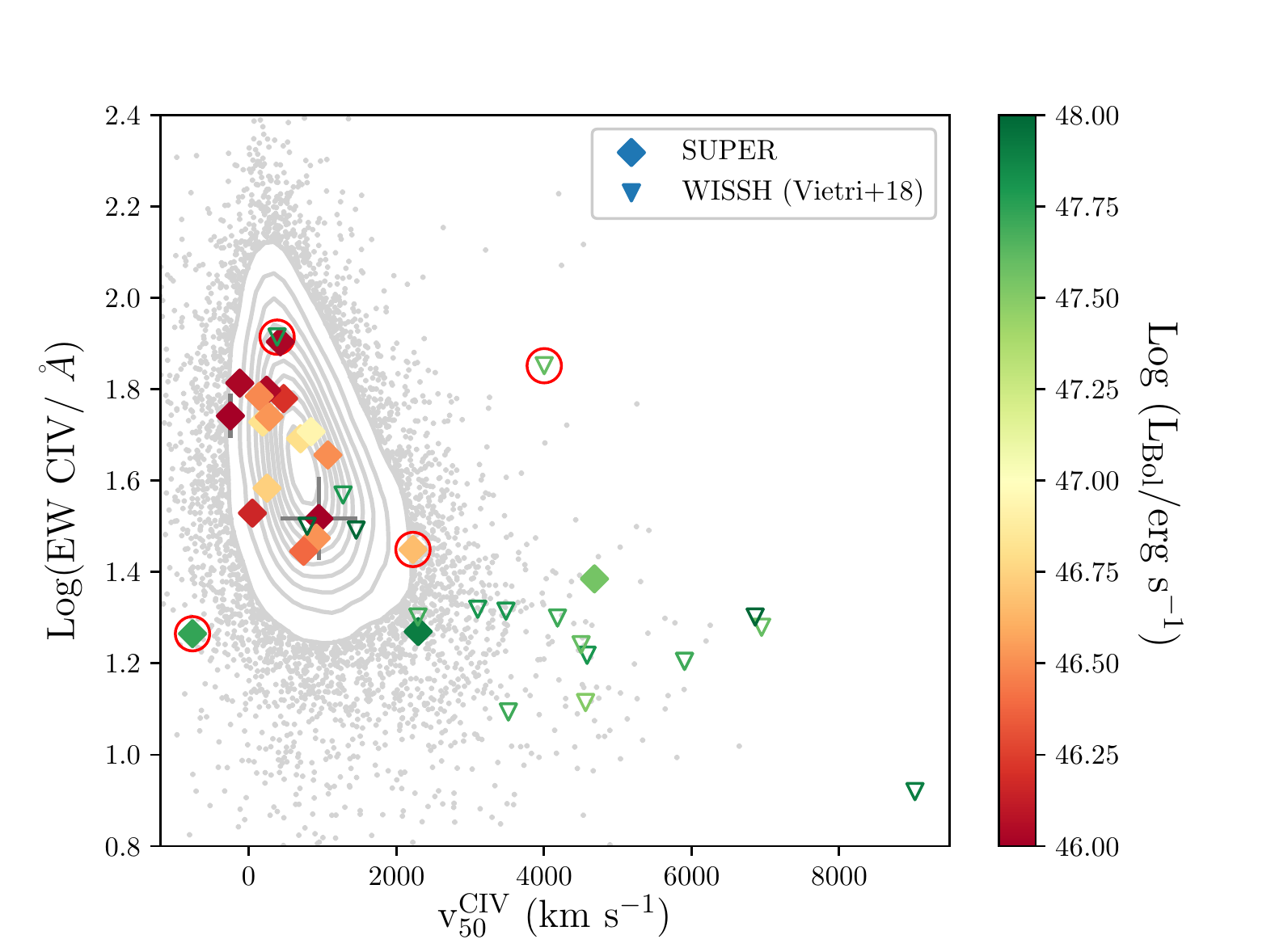}
  \includegraphics[width=0.9\columnwidth]{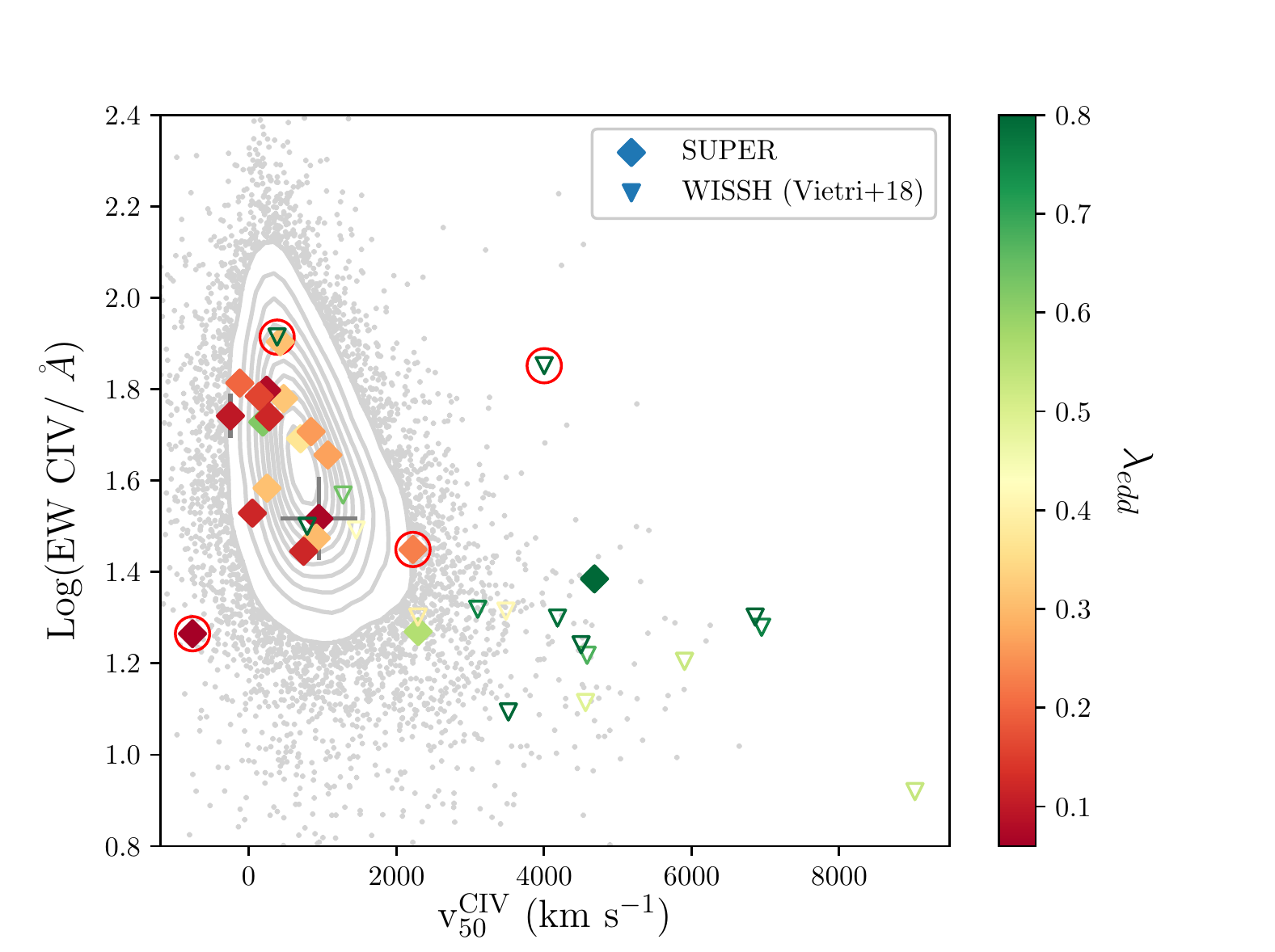}

\caption{Logarithm of equivalent width as a function of the velocity shift of the CIV emission line for the SUPER sample (diamonds), color-coded according to the \aox\ (top), \lbol\ (middle) and \edd\ (bottom). Additionally, the SDSS sample from \cite{Shen2011} (contours and grey points) and the WISSH sample from \cite{Vietri2018} (open triangles) are also reported. BAL AGN are indicated with red circles.} \label{fig:ew_shift}

 \end{figure}

As discussed in Sect. \ref{sec:comp_fwhm}, the FWHM of the CIV line does not correlate with those of the Balmer lines, suggesting a different kinematical state for the gas traced by the CIV emission. Indeed, this line is known to be dominated by non virialized motion components, making the profile asymmetric towards the blue-side of the line (e.g. \citealt{Gaskell1982}, \citealt{Richards2011}). 
While this is a limitation on the use of such line for measuring black hole masses, it offers the opportunity to trace the motion of  the ionized gas in the BLR (e.g. \citealt{Vietri2018}). To achieve this goal, we have measured the velocity shift of the CIV with respect to the expected wavelength based on [OIII] systemic redshift, which is defined as \vciv = c $\times$ ($\rm\lambda_{half}$-1549.48)/1549.48, where $\rm\lambda_{half}$ is the wavelength that bisects the cumulative total line flux, and c the speed of light. In Fig.\ref{fig:ew_shift} we can see that 85\% of the SUPER objects have CIV velocity shift <$\sim$ 2000 km/s and equivalent width $\geq$ 18 \AA~up to $\sim$80 \AA\ and 15\% have velocity larger than 2000 km/s up to $\sim$4700 km/s and EW$\rm_{CIV} \sim$ 20-30 \AA. We have included in the same figure the WISSH targets presented in \cite{Vietri2018} to probe the high-end of luminosity distribution (L$\rm_{Bol}$ > 10$^{47}$-10$^{48}$ erg/s). We overall confirm with our sample the same anti-correlation between the equivalent width and the velocity shift of the CIV line previously reported in the literature (e.g. \citealt{Sulentic2007}; \citealt{Richards2011}: \citealt{Vietri2018}).

In \cite{Vietri2018} we have reported a strong dependence of velocity shift of the CIV emission line on physical parameters as bolometric luminosity and Optical-to-X-ray spectral slope (\aox), as well as  Eddington ratio, despite the non-homogenous parametrization of bolometric luminosity, BH mass and CIV velocity shift of the collected dataset (WISSH and samples from literature). Here we further investigated these correlations, with the advantage of using an unbiased sample as SUPER, covering two orders of magnitude in bolometric luminosity and a wide range of BH mass and Eddington ratio. We will also populate the high-luminosity range by adding to the SUPER sample the Type-1 AGN from the WISSH survey\footnote{The bolometric luminosities of the WISSH QSOs are derived from SED-fitting as reported in \cite{Duras2017}, the BH mass and Eddington ratio are derived by using Eq. \ref{eq:mbh_hb} and \ref{eq:edd}. For the \aox\ we used the values reported in \cite{Martocchia2017} and \cite{Zappacosta2020}} (\citealt{Vietri2018}).  
We computed \aox\ for each of our SUPER quasars following the definition:
\begin{equation}
\rm\alpha_{ox}= \frac{Log(L_{2Kev}/L_{2500\AA})}{Log(\nu_{2Kev}/\nu_{2500})}
\end{equation}
where L$\rm_{2500}$ and L$\rm_{2keV}$ are the restframe monochromatic luminosities at 2500 \AA\ and 2keV. For the calculation of \aox\ we used the rest-frame 2500 \AA\ monochromatic luminosity obtained by SED fitting (\citealt{Circosta2018}) and derived the L$\rm_{2keV}$ assuming L$\rm_{2-10 keV}$ = 1.61 $\times$ L$\rm_{2 keV}$, adopting a power-law X-ray model with $\Gamma$ = 2. As shown in the three panels of Fig.\ref{fig:ew_shift} there is indeed a strong correlation between the \vciv\ with all three of the above mentioned physical quantities: bolometric luminosity, Eddington ratio and  \aox. In particular, in the top panel of Fig.\ref{fig:ew_shift} the data points in the plane EW$\rm_{CIV}$ vs. \vciv\ are color-coded according to the \aox\ values: most of the SUPER sources with \vciv\ < 2000 km/s have \aox\ > -1.6, while the WISSH targets, which are sampling the high luminosity end of the quasar population, have mostly lower values of \aox. This parameter can be used as a measurement of the hardness of the ionizing SED, whereby larger values of \aox\ corresponds to large amount of ionizing radiation for a given optical-UV luminosity.
In the context of a disc-wind scenario, this behaviour suggests that the objects with harder SEDs produce larger amounts of high-ionization gas, i.e. higher values of CIV equivalent width. 
If the ionizing radiation overionizes the gas, the continuum-driving (bound-free absorption) mechanism becomes inefficient, which could explain the smaller velocity shift of the CIV. Instead, the 15\% of the SUPER sample (and the WISSH quasars), which shows softer SEDs,
produces high velocity BLR winds (\vciv\ > 2000 km/s). Therefore, the acceleration is probably correlated with the hardness of the ionizing continuum, the ionization parameter and \aox\ but more specific calculations are required to demonstrate these connections.

In the medium and bottom panels of Fig. \ref{fig:ew_shift}, the plane EW$\rm_{CIV}$ vs. \vciv\ is color-coded according to the bolometric luminosity and Eddington ratio, respectively. SUPER sources with \vciv < 2000 km/s have luminosity range log (L$\rm_{Bol}$/ erg s$^{-1}$)=45.4-46.9, with an exception for a BAL source, J1549, with Log (L$\rm_{Bol}$/ erg s$^{-1}$)=47.7, and \vciv > 2000 km/s in the range log (L$\rm_{Bol}$/ erg s$^{-1}$)=46.7-47.9. The high velocity shift end is also strongly populated by the WISSH sample, with log (L$\rm_{Bol}$/ erg s$^{-1}$) > 47. On the other hand, SUPER sources with \vciv < 2000 km/s have Eddington ratio in the range 0.06-0.62 with a median value of 0.18, while for the high velocity winds the range is 0.23-1.27 with a median value of 0.56. We computed the Spearman's rank correlation coefficient, $\rho$, and associated p-value to quantify the strength and significance of each correlation, considering both SUPER and WISSH samples. We focused on the sources with reliable CIV velocity shift, therefore we excluded the source J1549, whose peak and half of the profile are totally absorbed by the presence of a BAL.
We find that the \vciv\ correlates with \aox\ with a coefficient $\rho$=-0.79 and p-value $<$ 10$^{-5}$ (taking into account upper limits by using the Astronomical Survival Analysis (ASURV) package, available under  IRAF/STSDAS \citealt{Lavalley1992},\citealt{Feigelson1985}, \citealt{Isobe1986}), with Log \lbol\ with a coefficient $\rho$=0.64 and p value=2$\times$10$^{-5}$ and with \edd\ with a coefficient $\rho$ = 0.62 and a p value= 5$\times$10$^{-5}$. These values confirm the visual impression from Fig. \ref{fig:ew_shift} that the properties of the CIV line correlate with all three of these parameters but the highest significant correlation is with \aox. Recently \cite{Zappacosta2020} reported on the WISSH sources a significant correlation between the luminosity in the 2-10 keV band (L$\rm_{2-10 keV}$) and the blueshift of the CIV line, i.e. objects exhibiting high velocity shift have lower L$\rm_{2-10 keV}$. With the SUPER sample we are not able to significantly investigate this relation. Indeed the narrow velocity range spanned by the SUPER sample (compared to the wide range probed by the WISSH sample, see Fig. \ref{fig:ew_shift}) and possibly dependent from L$\rm_{Bol}$ (e.g. \citealt{Fiore2017}) would require a much larger number of sources in order to significantly place constraints on a possible existing correlation at lower luminosities.

\begin{figure}[]
 %\centering
 \includegraphics[width=1\columnwidth]{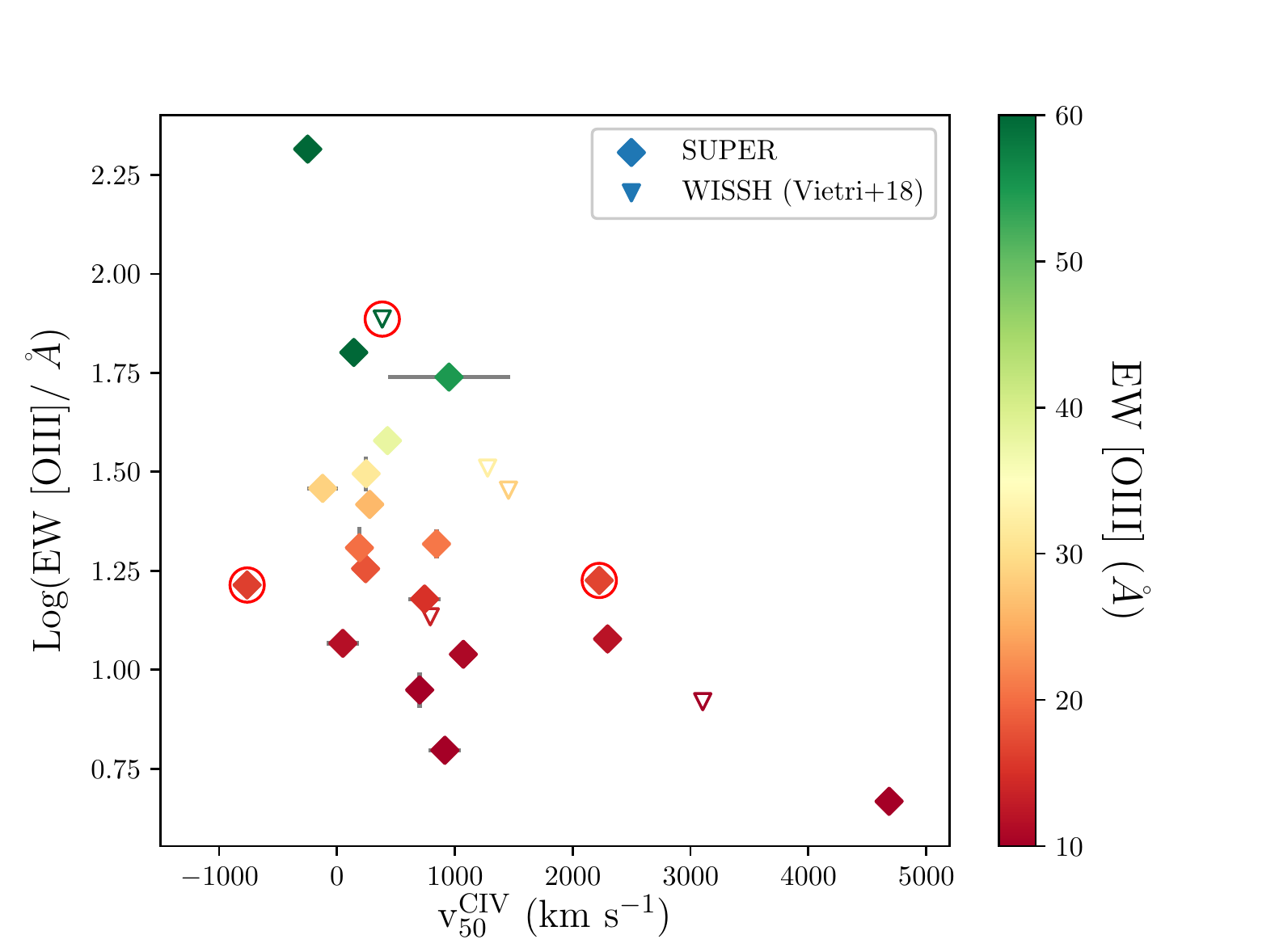}
 \includegraphics[width=0.9\columnwidth]{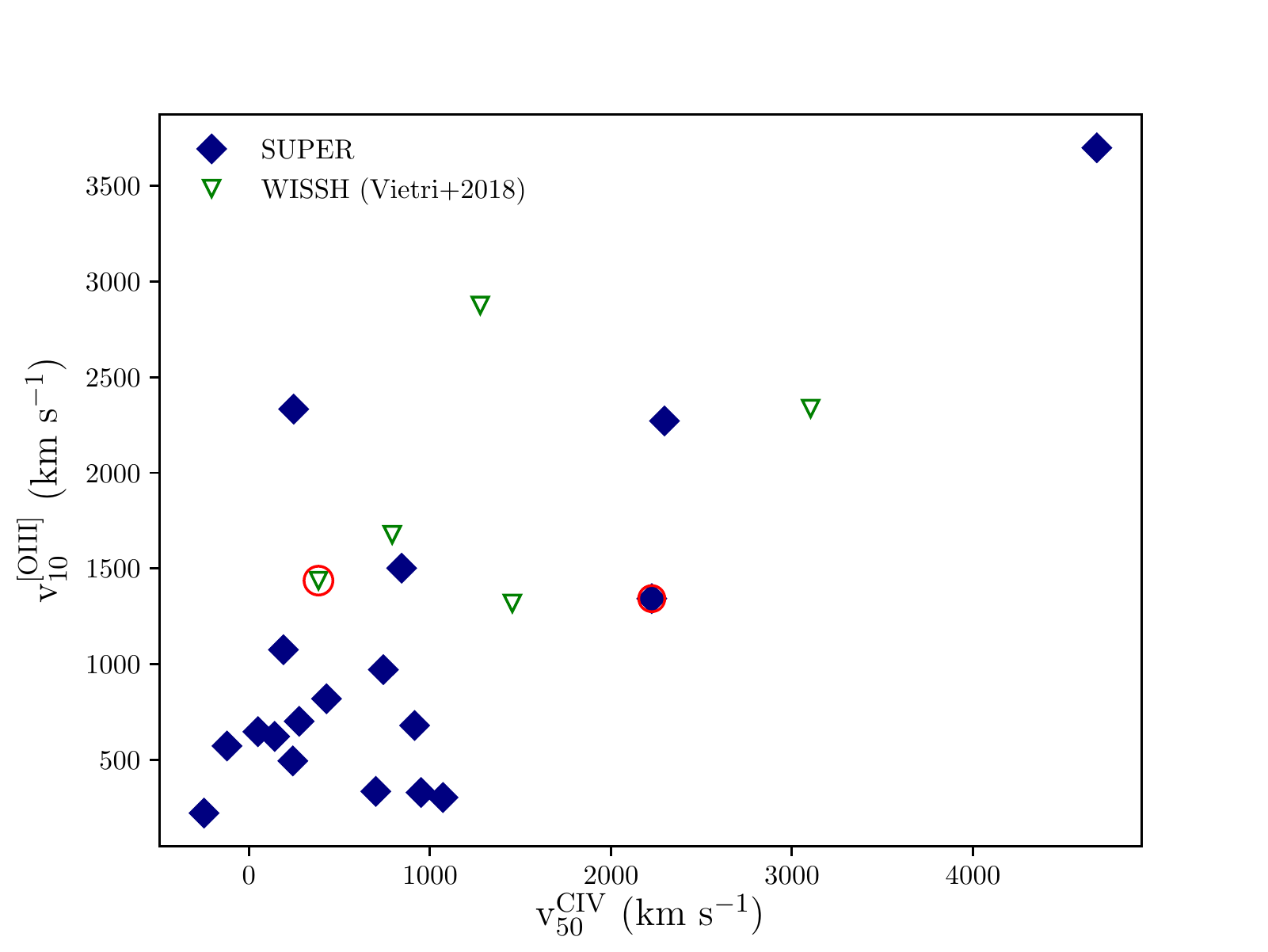}

\caption{(top) \small [OIII] equivalent width in logarithm as a function of the velocity shift of the CIV emission line for the SUPER sample (diamonds), color-coded according to the EW$\rm_{[OIII]}$. Additionally, the WISSH sample with reliable [OIII] measurements is also reported (empty triangles). (bottom) Velocity of [OIII] at 10\% of the line flux as a function of the velocity shift of the CIV emission line for the SUPER sample (blue diamonds). The WISSH sub-sample with reliable [OIII] measurements is also shown as green triangles. BAL AGN are indicated with red circles.}\label{fig:ew_shift_oiii}
 \end{figure}

While we are using CIV to trace the AGN winds in the BLR, the main aim of our survey is to trace the AGN driven outflows in the NLR using the [OIII] line (\citealt{Kakkad2020}). It is therefore interesting to compare the properties of the winds traced at such different physical scales. In Fig. \ref{fig:ew_shift_oiii}, it is shown the [OIII] equivalent width as a function of \vciv. The strength of the [OIII] seems to decrease at increasing CIV velocity shift, i.e. for v < 2000 km/s the EW$\rm_{OIII}$ range probed is 6-207\AA\ with a mean (median) value of 36(21) \AA\ and for v > 2000 km/s the range is 5-17 \AA\ with a mean(median) value of 11(12) \AA. Furthermore, we also compare the kinematic properties of the CIV and [OIII] emission lines. We used the non-parametric definition v$\rm_{10}$, i.e. the velocity of the [OIII] at 10\% of the cumulative line flux, to trace the [OIII] outflow velocity (see \citealt{Kakkad2020}).

We found a significant correlation between v$\rm_{10}^{[OIII]}$ and the CIV blueshift, with a Spearman correlation coefficient $\rho$= 0.54  and  p-value=0.007. \footnote{The probability of the correlation decreases if we exclude from the fit the only object showing v$\rm_{50}^{CIV}$>3500 km/s and v$\rm_{10}^{[OIII]}$>3000 km/s, i.e. $\rho$= 0.48 and p-value=0.02.} 
We also test for the presence of a correlation between \vciv\ and W$\rm_{80}$ i.e. the width containing 80\% of the line flux and v$\rm_{max}$, defined as the shift between the systemic velocity and the broad Gaussian component of [OIII] plus twice the velocity dispersion of the broad Gaussian. In our analysis of the NLR winds we use a cut of W$\rm_{80}$ larger than 600 km/s to distinguish between targets with and without AGN outflows (\citealt{Kakkad2020}). In this case we report a marginally significant correlation with W$\rm_{80}$, i.e. $\rho$=0.42 and p-value=0.05  and a weaker not significant correlation with v$\rm_{max}$, i.e. $\rho$=0.32 and p-value=0.13.

Our findings are in agreement with those presented by  \cite{Coatman2019}, despite the lower statistical significance of our correlations which is likely due to the smaller sample. They analyzed the integrated spectra of 213 quasars in the luminosity range of Log (L$\rm_{Bol}/ erg\ s^{-1}$)=46-49 with redshift z$\sim$2-4 and found a significant correlation (p-value= 6$\times$ 10$^{-7}$) between v$_{10}$ and \vciv\ with $\rho$=0.46. They demonstrated that the correlation is independent of the bolometric luminosity, however we note that both v$\rm_{10}^{[OIII]}$ and v$\rm_{50}^{CIV}$ correlate with bolometric luminosities for the SUPER and WISSH samples, although with a large scatter.  While their analysis was limited by the lack of spatially resolved data to locate the scale at which the gas traced by the [OIII] emission was located, we inferred from the SINFONI IFU data that the [OIII] emission is extended on  kpc scales (\citealt{Kakkad2020}), therefore supporting the idea that BLR outflows may affect the galaxy-scale wide gas in the NLR.

\subsection{Outflow energetics}\label{sec:energetics}

Our estimate of the ionized gas mass, M$\rm_{out}$, is somewhat different than the one used in \citealt{Vietri2018} which followed the expression given in \cite{Marziani2016}.
 That approximation was based on a spherical gas distribution and did not take into account the dependence on the gas temperature. Here we assume a thin-shell geometry for the outflowing gas and normalize the results using photoionization calculations typical of BLR conditions. 

The expression we use is similar to the one given in \cite{Baron2019}, and in several earlier publications,
 for the [OIII] line. It assumes a line emission coefficient, $\rm\gamma_{CIV}$ (L$\rm_{line} \propto \gamma n_e^2$) given by:
\begin{equation}
\rm\gamma_{CIV} = C_{CIV} \times h \nu_{CIV}  \frac {n(C^{+3})}{n(C)} \frac {n(C)}{n(H)} \,
\end{equation}
where  $\rm C_{CIV} \propto \exp[{ \frac {-h \nu}{ kT }}]/\sqrt T $,
and $\rm n(C^{+3})/n(C)$ is the fractional ionization of $\rm C^{+3}$. The line luminosity is obtained by integrating the emissivity over volume and the associated mass is M$\rm_{out} \propto L(CIV)/n_e  \gamma_{CIV}$. 

Using known atomic rates, and calibrating the above expressions against photoionization calculations (see e.g. the specific calculations in \citealt{Netzer2020}) of a thin-shell of gas with constant density, we get:
\begin{equation}
\rm M_{out} \approx 100 \frac {L_{45}(CIV)}{n_9 \times n(C^{+3})/n(C)} \,\,  M_{\odot} \, .\label{eq:mion}
\end{equation}
where $\rm n_9$ is the electron density in units of $10^9$ cm$^{-3}$ and $\rm L_{45}(CIV)$ is the CIV luminosity in units of $10^{45}$ erg/s. 
This estimate is appropriate for the highly ionized part of the cloud, for $\rm 1.5\times 10^4\ K < T < 2\times 10^4\ K$ and for metallicity in the range 1-5 solar.
To derive the mass of the wind we used the CIV luminosity of the total profile, based on the fact that the bulk of our sources show blue-asymmetry according to the CIV shifts.

It is important to note that higher metallicity ($\rm n(C))/n(H)$) tends to cool the gas which compensates for much of the influence of the temperature on the excitation rate. Thus, the main dependence is on the fractional ionization which is determined by the ionization parameter and not on the carbon abundance.  This is why the carbon abundance term is not part of the mass equation.

The fractional ionization of carbon scales with the hydrogen ionization parameter, $U$, and depends also on the metallicity through the gas temperature. For $U=0.05$, $\rm n(C^{+3})/n(C)\simeq 0.5$.
The mass obtained here is about an order of magnitude smaller than the one obtained by using the \cite{Marziani2016} expression. We derived the following range of outflowing mass: M$\rm_{out}$ = 0.1-290 M$\rm_{\odot}$, assuming n = 10$^{9.5}$ cm$^{-3}$.

For the mass outflow rate we use the expression for a thin outflowing shell,
\begin{equation}
\rm \dot{M}_{out} = M_{out} \frac {v}{r_{CIV}} \approx 0.005 M_{out} \frac{v_{5000}}{r_{pc}} \,\, M_{\odot}/yr \,\, , \label{eq:mdot}
\end{equation}
where $\rm v_{5000}$ is the outflow velocity\footnote{The calculations of the maximum velocity of the outflow require numerical integration across the BLR and a point-by-point calculation of the force multiplier (see \cite{Netzer2013} Chapter 5.9.2). This is beyond the scope of the present work.} \vciv\ 
in 5000 km/s and $\rm r_{CIV}$ is the outflow radius estimated from the CIV radius-luminosity relation from \citealt{Lira2018}. 

The inferred outflow radius R$\sim$ 0.002-0.2 pc is listed in Table \ref{tab:energetics}, along with the mass outflow rates: $\rm \dot{M}_{out}$=0.005-3 M$\rm_{\odot}$ yr$^{-1}$ with a mean value 0.4 M$\rm_{\odot}$ yr$^{-1}$.  This is a factor 3 smaller than the one used  in \cite{Vietri2018} for a spherical geometry.

The two main uncertainties in the above mass and mass outflow rate estimates are the unknown gas density and level of ionization. 
A lower limit on the density is imposed from the very different shape of the C III$]$1909 line profile which does not show a blueshifted wing and similarly from other semi-forbidden lines like O III$]1664$ (e.g. \citealt{Richards2011}, \citealt{Netzer2013}). This limit is about $3\times 10^9$ cm$^{-3}$. The upper limit is more difficult to establish. Photoionization calculations (\citealt{Netzer2013}) suggest several other strong broad lines, like NV$\lambda 1240$, that will show blueshifted wings under such conditions. However, we lack high quality spectra of these lines. In eq \ref{eq:mion} we used n$_e$=10$^{9.5}$ cm$^{-3}$. 
As for the level of ionization of the outflowing gas, this can be high and results in little $\rm C^{+3}$. This will increase both $\rm M_{out}$ and $\rm \dot{M}_{out}$. Finally, without proper modeling, we lack information about the amount of neutral gas that can be part of the outflow. Such gas will increase both M$\rm _{out}$ and $\rm \dot{M}_{out}$.

\begin{figure}[]
 \centering 

 \includegraphics[width=1.15\columnwidth]{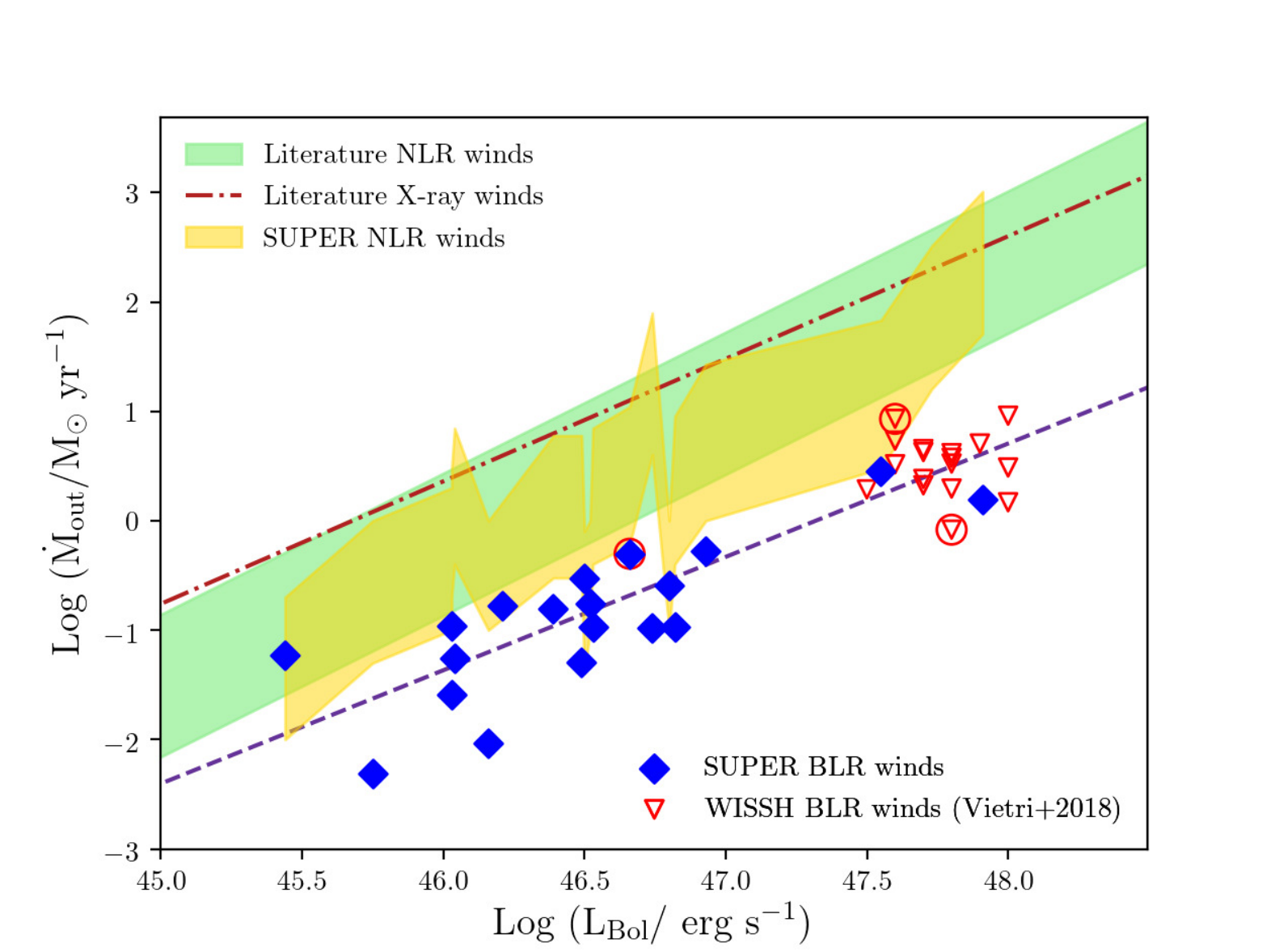}

\caption{Outflow mass rate as a function of the logarithm of the bolometric luminosity for the BLR winds in the SUPER sample. Additionally, the values from the WISSH sample (\citealt{Vietri2018}) (triangles) are also reported. BAL AGNs are denoted with red circles. The blue dashed line is the best-fit correlation of the CIV winds (SUPER+WISSH samples).  Mass outflow rates of NLR winds from the SUPER sample are also reported (Gold shaded area). The green shaded area shows the range probed by the outflow ionized phase in the NLR as found by \cite{Fiore2017}, after rescaling the relation with the same assumption as the Type-1 SUPER targets (see \citealt{Kakkad2020} for further details). The dot-dashed line shows the log linear best fit correlation of the X-ray winds.}.\label{fig:mdot}
 \end{figure}

In Fig. \ref{fig:mdot} we plot the derived mass outflow rate of the CIV winds as a function of the bolometric luminosity for the SUPER sample. We include also the WISSH sample to populate the high-luminosity part and added the correlation of mass outflow rate vs. bolometric luminosities from a compilation of ionized and X-ray winds (see \citealt{Fiore2017}). For the BLR $\rm \dot{M}_{out}$ and L$\rm _{\rm bol}$ we derive a Spearman correlation coefficient $\rho$ = 0.8 and null hypothesis P = 1$\times$ 10$^{-8}$, with a log linear slope Log ($\rm \dot{M}_{out}/ M_{\odot} yr^{-1}$)  $\propto$ 1.01\err 0.09 Log (L$\rm_{Bol}$ /erg s$^{-1}$). Interestingly, the mass outflows rates of the BLR winds seem to have a correlation with bolometric luminosity of the central AGN as steep as that observed for the ionized winds in the NLR and X-ray traced winds close to the accretion disk. The different normalization could be probably due to the efficiency of the coupling with the ISM, which changes in terms of density and composition, but a similar slope suggests that the winds at different scales are linked, namely they have the same functional dependency with a basic physical property of the SMBH, its bolometric luminosity. 

The $\rm \dot{M}_{out}$ values inferred for the BLR winds are lower than the one derived for X-ray winds. 
Further, the range probed by the BLR winds in terms of mass outflow rate is lower than  that of the NLR winds measured for the 18 objects for which we can sample both regions (see also \citealt{Kakkad2020} for the [OIII] analysis). In Fig. \ref{fig:mdot} we plot the mass outflow rates for the SUPER NLR winds assuming a bi-conical outflow model and an electron density from 500 cm$^{-3}$ - 10000 cm$^{-3}$, as reported in \cite{Kakkad2020}. For 6 out of these 18 objects, the mass outflow rate of the BLR winds are consistent with the lower limit found from the NLR analysis.

We also estimated the kinetic power of the outflow defined as:

\begin{equation}
\rm\dot{E}_{kin} = \frac{1}{2} \dot{M}_{out} v_{out}^2 %1.2\ 10^{44} \ L_{45} v_{5000}^3 r_1^{-1}\ (erg\ s^{-1})
\end{equation}\label{eq:ekin}

with v$\rm_{out}$=v$\rm_{50}^{CIV}$ and $\rm\dot{M}_{out}$ as derived in Eq. \ref{eq:mdot}. We derived $\rm\dot{E}_{kin}$ from 7.4$\times$10$^{36}$ \ergs\ up to 2$\times$10$^{43}$ \ergs, for the SUPER object with the most blue-shifted CIV line profile, i.e. J1441+0454.

\begin{figure}[]
 \centering

 \includegraphics[width=1.1\columnwidth]{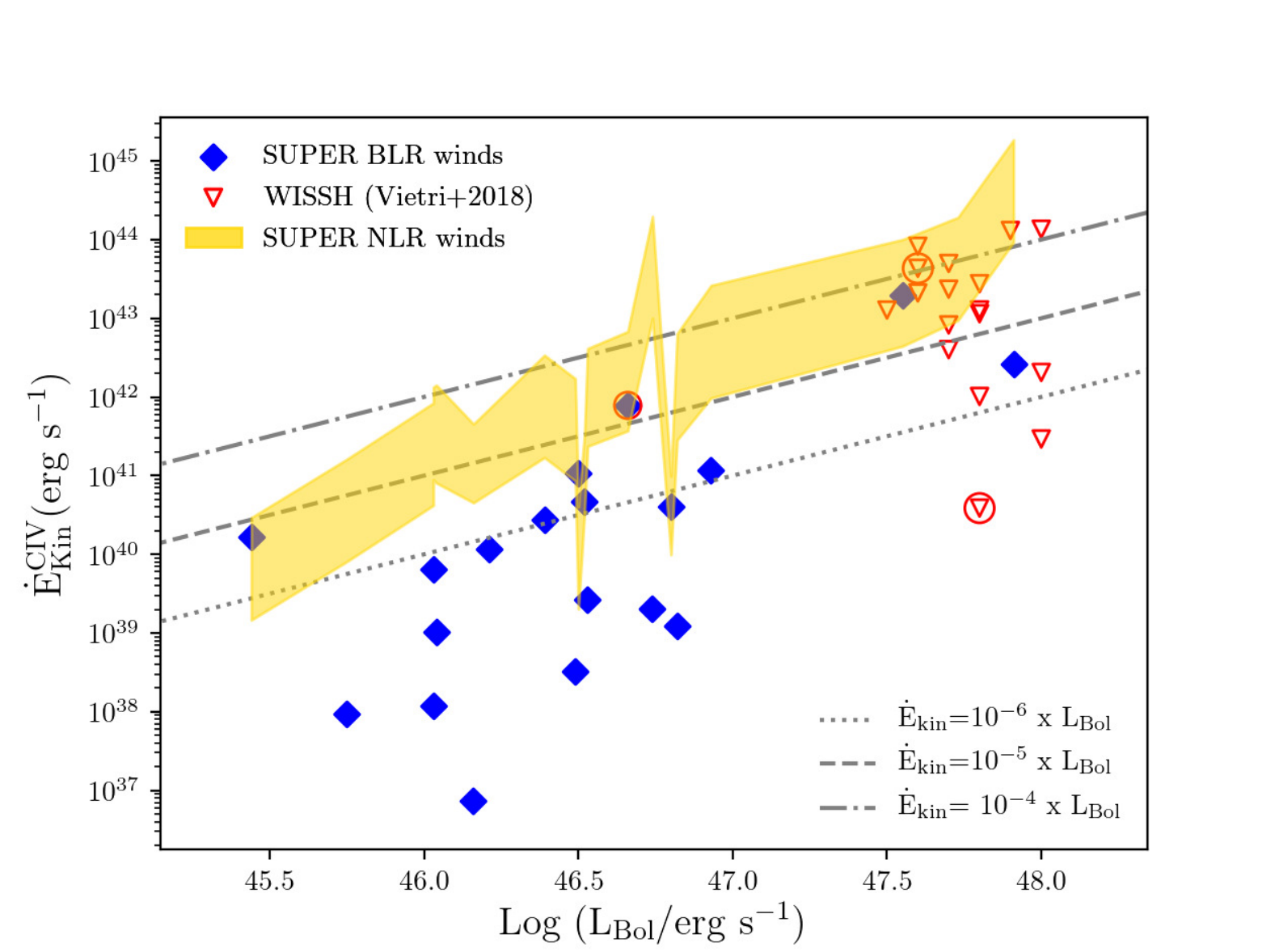}

\caption{Kinetic power of the BLR winds as a function of the logarithm of the bolometric luminosities for the SUPER sample (diamonds). Additionally, the values of the BLR winds from the WISSH sample (\citealt{Vietri2018}) (triangles) are also reported. Kinetic power of NLR winds from the SUPER sample are reported as gold shaded area. Kinetic power in fractions of 10$^{-4}$, 10$^{-5}$, and 10$^{-6}$ bolometric luminosity are respectively indicated with dot-dashed, dashed and dotted lines. BAL sources are denoted with red circles.}\label{fig:ekin}

%}
 \end{figure}

We show in Fig. \ref{fig:ekin} the kinetic power derived using the CIV total profile as a function of bolometric luminosity for the SUPER sample (diamonds).
We also populated the high-luminosity part of the plot, adding the values estimated in a consistent way for the WISSH sample.
Performing a Spearman test on $\rm\dot{E}_{kin}$ and L$\rm_{Bol}$ for both samples, the quantities are strongly correlated with a log linear slope of 2.14\err0.25 and a Spearman correlation coefficient $\rho \sim$ 0.74 and two-sided null hypothesis of p = 3$\times$ 10$^{-7}$. The bulk of the kinetic power for the BLR winds in the SUPER sample is in the range $\rm\dot{E}_{kin} \sim$ 10$^{-7}$ $\times$ L$\rm_{Bol}$ up to 10$^{-4}$ $\times$ L$\rm_{Bol}$ at high bolometric luminosity. It is often reported that the coupling efficiency predicted by feedback models is significantly higher, e.g. 5\%. On the other end, only a fraction of the injected energy will become kinetic power in the outflow, while the rest will be used in doing work for example against the ambient pressure and the gravitational potential (see \citealt{Harrison2018}). From the comparison with the NLR outflows in the SUPER sample, we find that in the majority of the cases (12 out of 18) the NLR kinetic power is larger, in five objects they are comparable, and in one case it is the BLR kinetic power to be larger. Anyway, we do not further speculate on such comparison given the large uncertainties affecting these measurements.

\begin{table*}[]
\centering
    % \small
    %  \footnotesize      
    \begin{threeparttable}
        \caption{Properties of the CIV outflows derived from the total profile of the CIV emission line }\label{tab:energetics}
                \setlength\tabcolsep{2pt}

                        \begin{tabular}{lccccccc}
                                \hline
                                \hline  
                                ID & v$\rm^{CIV}_{peak}$ & R & \vciv & M$\rm_{out}$ & $\rm\dot{M}_{out}$ & $\rm Log (\dot{E}_{kin}/ erg\ s^{-1}$) & $\rm Log (\dot{P}_{out}/ g\ cm\ s^{-2})$ \\[0.5ex]
                                (1) & (2) & (3) & (4) & (5) & (6) & (7) & (8)\\
                                                               &\footnotesize{km s$^{-1}$}& pc & \footnotesize{km s$^{-1}$} & \footnotesize{ M$_{\odot}$} & \footnotesize{ M$_{\odot}$ yr$^{-1}$} & &  \\

                                \hline
X\_N\_160\_22 &   -30 & 0.06 & 250 & 68.46 & 0.10 & 39.31 & 32.21 \\
X\_N\_81\_44 & 700 & 0.04 & 700 & 37.37 & 0.26 & 40.60 & 33.06 \\
X\_N\_53\_3 & 540 & 0.03 & 470 & 29.40 & 0.17 & 40.07 & 32.69 \\
X\_N\_66\_23 & 110 & 0.02 & 240 & 14.14 & 0.06 & 39.02 & 31.93 \\
X\_N\_35\_20 & 950 & 0.01 & 950 & 2.31 & 0.06 & 40.23 & 32.55 \\
X\_N\_12\_26 & 430 & 0.04 & 920 & 19.43 & 0.17 & 40.66 & 33.00 \\
X\_N\_4\_48 & 260 & 0.03 &  50 & 16.28 & 0.01 & 36.87 & 30.47 \\
X\_N\_102\_35 & 400 & 0.06 & 190 & 87.51 & 0.11 & 39.09 & 32.11 \\
X\_N\_115\_23 &  70 & 0.04 & 140 & 38.84 & 0.05 & 38.51 & 31.66 \\
cid\_166 & 780 & 0.06 & 850 & 109.30 & 0.52 & 41.07 & 33.45 \\
cid\_1605 &  -120 & 0.02 &  -120 & 12.64 & 0.03 & 38.07 & 31.29 \\
cid\_346 & 2230 & 0.05 & 2230 & 28.72 & 0.50 & 41.89 & 33.85 \\
cid\_1205 &  -250 & 0.002 &  -250 & 0.14 & 0.005 & 37.98 & 30.88 \\
cid\_467 & 210 & 0.05 & 280 & 51.23 & 0.11 & 39.42 & 32.28 \\
J1333+1649 & 2160 & 0.15 & 2300 & 285.46 & 1.56 & 42.42 & 34.36 \\
J1441+0454 & 3320 & 0.11 & 4690 & 189.15 & 2.85 & 43.29 & 34.93 \\
S82X1905 & 1070 & 0.04 & 1070 & 27.36 & 0.30 & 41.03 & 33.30 \\
S82X1940 & 430 & 0.02 & 430 & 16.40 & 0.11 & 39.81 & 32.47 \\
S82X2058 & 680 & 0.05 & 740 & 25.86 & 0.16 & 40.44 & 32.87 \\

      \hline
                                \end{tabular}
                          \begin{tablenotes}[para,flushleft]
         \item {\bf{Notes}}. Columns give the following information: (1) Target identification, (2) the velocity  at the peak of the CIV total profile, (3) CIV BLR radius derived from the CIV radius-luminosity relation from \cite{Lira2018}, (4) velocity of the CIV at 50\% of the cumulative line flux, (5) mass of the outflow (M$\odot$), (6) mass outflow rate assuming a thin-shell geometry as described in sect. \ref{sec:energetics}, (7) kinetic power of the outflow and (8) outflow momentum load. 
                                  \end{tablenotes}
                                       \end{threeparttable}
                               
                        \end{table*}

\begin{figure}[]

 \includegraphics[width=1.1\columnwidth]{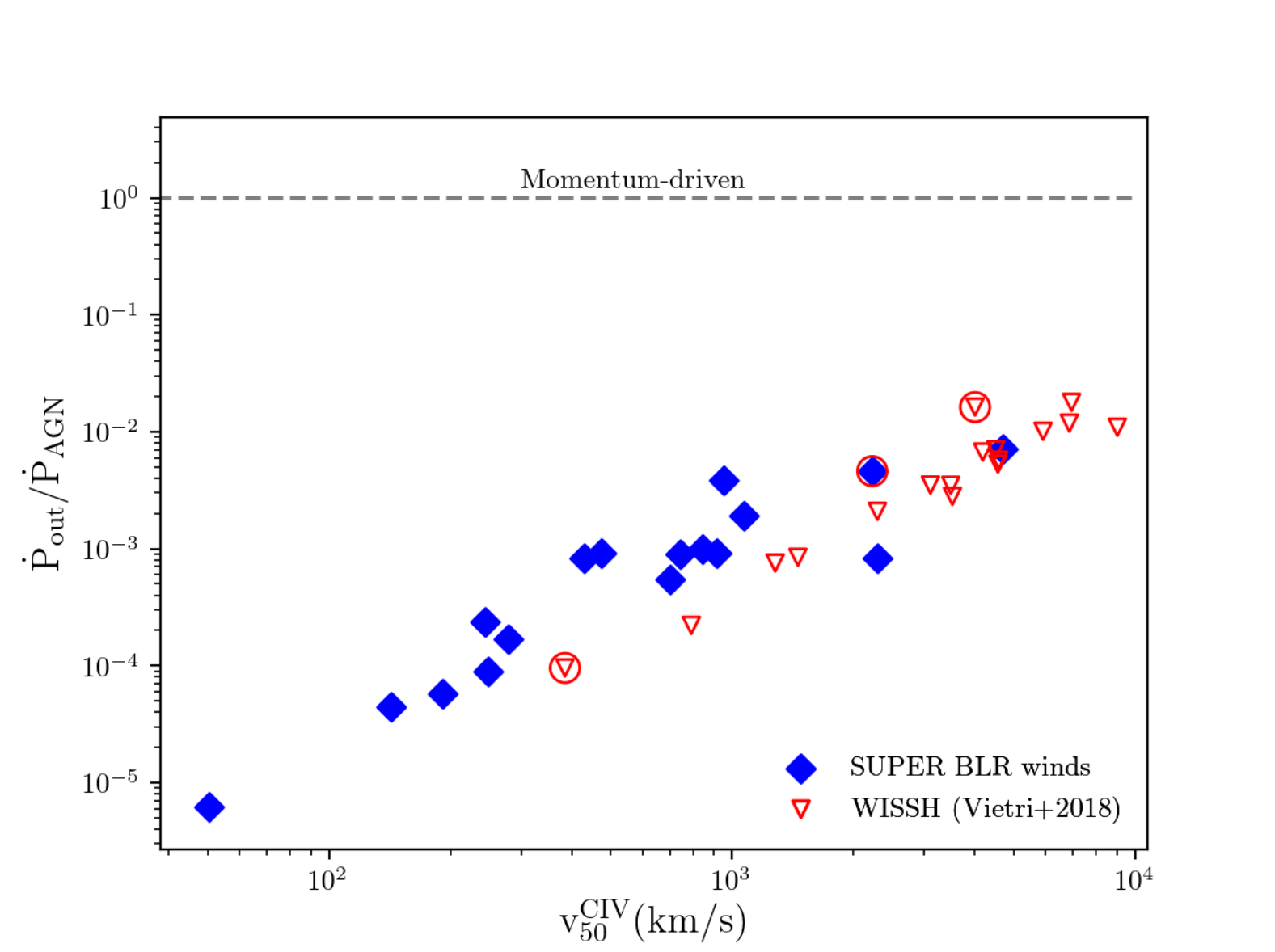}

\caption{Outflow momentum load as a function of the velocity shift of the CIV for the SUPER sample (blue diamonds). The WISSH sample is also reported as red triangles. The dashed line indicates the theoretical expectation for a momentum driven outflow. BAL sources are denoted with red circles.}\label{fig:pdot}

 \end{figure}

Another fundamental parameter of the outflow is the momentum rate defined as $\rm\dot{P}_{out}=\dot{M}_{out}$v$\rm_{50}^{CIV}$. 
In Fig. \ref{fig:pdot} we plot the outflow momentum load, defined as the momentum rate divided by the AGN radiation momentum rate ($\rm\dot{P}_{AGN}$ = L$\rm{_{\rm Bol}}$/c), as a function of the outflow velocity. The values estimated for the BLR winds of the SUPER sample are very low $\sim$[$10^{-5}:10^{-2}$]. For the X-ray winds the theoretical momentum flux is expected to be comparable to L$\rm_{AGN}$/c, i.e a momentum load close to unity (i.e. a momentum conserving outflow). Fig. \ref{fig:pdot} shows the expected momentum load for a momentum-conserving wind model (dashed line). The bulk of the winds in the BLR show lower values than momentum driven winds (with the larger values for the sources with the larger BLR winds velocities) suggesting that a different form of driving mechanism may be acting at these scales.

\section{Conclusions}

We present the results of the analysis of 21 Type-1 X-ray selected AGN from the SUPER survey (\citealt{Circosta2018}) with near-infrared SINFONI IFU and UV spectroscopy. The analysis presented in this paper had two main goals: 1) derive BH masses and Eddington ratios by using virial BH mass estimators based on H$\alpha$, H$\beta$, MgII; 2) trace AGN-driven winds in the broad line region (BLR) using the blueshift of the CIV line profile.

Our main finding can be summarized as follows:

(i) We find that the H$\alpha$ and H$\beta$ line width correlate with each other as does the line continuum luminosity at 5100 \AA\ with the H$\alpha$ line luminosity, resulting in a well defined correlation between BH mass estimated from H$\alpha$ and H$\beta$. The SUPER AGN exhibit SMBHs with mass in the range Log (M$\rm_{BH}$/M$\rm_{\odot}$) = 8.5-10.8 and  Eddington ratios in the range 0.04 < $\lambda$ < 1.3.

(ii) We confirm that the CIV line width does not correlate with the Balmer lines and its peak is blueshifted with respect to the [OIII]-based systemic redshift. We interpreted this findings as the presence of outflows in the BLR with derived velocities up to $\sim$4700 km/s.

(iii) As found previously in \cite{Vietri2018}, we confirm the strong correlation between \vciv\  and the UV-to-Xray continuum slope,  bolometric luminosity, and Eddington ratio, by analysing an unbiased sample of sources. We interpret this in the context of the disc-wind scenario where a high UV luminosity is necessary to launch the wind and softer SEDs can prevent overionization, producing high velocity BLR winds.

(iv) We compare the properties of the BLR and NLR winds, as traced by CIV and [OIII] respectively. We find an anti-correlation between the [OIII] equivalent width and \vciv\ and a significant correlation between v$\rm_{10}$ and \vciv. From SINFONI IFU data, we know that the gas emitting [OIII] is located at kpc scales. We therefore interpret the correlation found as supporting a scenario where BLR winds are capable of affecting the gas in the NLR emission located at kpc scales, likely partially blocking the ionizing photons of the NLR or sweeping up the gas in the NLR.

(v) We derive mass  outflows rates in the range 0.005-3 M$\rm_{\odot}$/yr for the BLR winds. The mass outflow rate of the BLR winds shows a correlation with bolometric luminosity as steep as that observed for winds at sub-parsec scales and in the NLR.
The kinetic power of the BLR winds inferred is $\rm\dot{E}_{kin}$ $\sim$ 10$^{[-7: -4]}$ $\times$ L$\rm_{bol}$, and in the 28\% of the SUPER sample, the kinetic power of the BLR and NLR winds are comparable. Despite the fact that these values are below the coupling efficiency predicted by AGN feedback models, we have to bear in mind that only a fraction of the injected energy will become kinetic power in the outflow.

(vi) The momentum fluxes of the BLR winds normalized by the AGN radiation momentum rate inferred for the SUPER sample are well below the theoretical expectation of a momentum, indicating for these winds that a different driving mechanism may be acting at these scales.

As discussed in this paper we are now able, for the first time in an unbiased sample of AGN at z$\sim 2$, to trace the presence of AGN-driven outflows at z$\sim 2$ from pc-  up to kpc-scales. Moving to even larger scales,  the advent of state-of-art IFU facilities as VLT/MUSE has allowed to uncover the elusive material of the circum-galactic medium (e.g. \citealt{Borisova2016}). 
A clearcut evidence of an outflow propagating over dozens of kpc through a metal-enriched circum-galactic medium was recently reported by \cite{Travascio2020} by analyzing the VLT/MUSE observations of a WISSH QSO  at z$\sim$3.6, highlighting the importance in the future to use such facilities to map the AGN activity up to tens of kpc scales, to study the interplay between nuclear activity and gas content over a very large range in radial distance and trace the impact that the AGN feedback may have on the circum-galactic medium at the peak epoch of the cosmic star-formation and AGN activity.

\vspace{0.5cm}
\noindent

Acknowledgments: This research project was supported by the DFG Cluster of Excellence "Origin and Structure of the Universe" (universe-cluster.de). GV acknowledges financial support from Premiale 2015 MITic (PI B. Garilli). GC and MBi acknowledge support from PRIN MIUR project "Black Hole winds and the Baryon Life Cycle of Galaxies: the stone-guest at the galaxy evolution supper", contract \#2017PH3WAT." MP is supported by the Programa Atracci\'on de Talento de la Comunidad de Madrid via grant 2018-T2/TIC-11715.

\bibliographystyle{aa} % style aa.bst
\bibliography{bib} % your references Yourfile.bib

\begin{thebibliography}{73}
\expandafter\ifx\csname natexlab\endcsname\relax\def\natexlab#1{#1}\fi

\bibitem[{{Akritas} \& {Bershady}(1996)}]{Akritas1996}
{Akritas}, M.~G. \& {Bershady}, M.~A. 1996, \apj, 470, 706

\bibitem[{{Baron} \& {Netzer}(2019)}]{Baron2019}
{Baron}, D. \& {Netzer}, H. 2019, \mnras, 486, 4290

\bibitem[{{Baron} {et~al.}(2016){Baron}, {Stern}, {Poznanski}, \&
  {Netzer}}]{Baron2016}
{Baron}, D., {Stern}, J., {Poznanski}, D., \& {Netzer}, H. 2016, \apj, 832, 8

\bibitem[{{Baskin} \& {Laor}(2005)}]{Baskin2005}
{Baskin}, A. \& {Laor}, A. 2005, \mnras, 356, 1029

\bibitem[{{Bentz} {et~al.}(2013){Bentz}, {Denney}, {Grier}, {Barth},
  {Peterson}, {Vestergaard}, {Bennert}, {Canalizo}, {De Rosa}, {Filippenko},
  {Gates}, {Greene}, {Li}, {Malkan}, {Pogge}, {Stern}, {Treu}, \&
  {Woo}}]{Bentz2013}
{Bentz}, M.~C., {Denney}, K.~D., {Grier}, C.~J., {et~al.} 2013, \apj, 767, 149

\bibitem[{{Bentz} {et~al.}(2009){Bentz}, {Peterson}, {Netzer}, {Pogge}, \&
  {Vestergaard}}]{Bentz2009}
{Bentz}, M.~C., {Peterson}, B.~M., {Netzer}, H., {Pogge}, R.~W., \&
  {Vestergaard}, M. 2009, \apj, 697, 160

\bibitem[{{Bischetti} {et~al.}(2017){Bischetti}, {Piconcelli}, {Vietri},
  {Bongiorno}, {Fiore}, {Sani}, {Marconi}, {Duras}, {Zappacosta}, {Brusa},
  {Comastri}, {Cresci}, {Feruglio}, {Giallongo}, {La Franca}, {Mainieri},
  {Mannucci}, {Martocchia}, {Ricci}, {Schneider}, {Testa}, \&
  {Vignali}}]{Bischetti2017}
{Bischetti}, M., {Piconcelli}, E., {Vietri}, G., {et~al.} 2017, \aap, 598, A122

\bibitem[{{Bongiorno} {et~al.}(2014){Bongiorno}, {Maiolino}, {Brusa},
  {Marconi}, {Piconcelli}, {Lamastra}, {Cano-D{\'{\i}}az}, {Schulze},
  {Magnelli}, {Vignali}, {Fiore}, {Menci}, {Cresci}, {La Franca}, \&
  {Merloni}}]{Bongiorno2014}
{Bongiorno}, A., {Maiolino}, R., {Brusa}, M., {et~al.} 2014, \mnras, 443, 2077

\bibitem[{{Borisova} {et~al.}(2016){Borisova}, {Cantalupo}, {Lilly}, {Marino},
  {Gallego}, {Bacon}, {Blaizot}, {Bouch{\'e}}, {Brinchmann}, {Carollo},
  {Caruana}, {Finley}, {Herenz}, {Richard}, {Schaye}, {Straka}, {Turner},
  {Urrutia}, {Verhamme}, \& {Wisotzki}}]{Borisova2016}
{Borisova}, E., {Cantalupo}, S., {Lilly}, S.~J., {et~al.} 2016, \apj, 831, 39

\bibitem[{{Boroson} \& {Green}(1992)}]{Boroson1992}
{Boroson}, T.~A. \& {Green}, R.~F. 1992, \apjs, 80, 109

\bibitem[{{Bruni} {et~al.}(2019){Bruni}, {Piconcelli}, {Misawa}, {Zappacosta},
  {Saturni}, {Vietri}, {Vignali}, {Bongiorno}, {Duras}, {Feruglio}, {Tombesi},
  \& {Fiore}}]{Bruni2019}
{Bruni}, G., {Piconcelli}, E., {Misawa}, T., {et~al.} 2019, \aap, 630, A111

\bibitem[{{Cardelli} {et~al.}(1989){Cardelli}, {Clayton}, \&
  {Mathis}}]{Cardelli1989}
{Cardelli}, J.~A., {Clayton}, G.~C., \& {Mathis}, J.~S. 1989, \apj, 345, 245

\bibitem[{{Carniani, S.} {et~al.}(2016){Carniani, S.}, {Marconi, A.},
  {Maiolino, R.}, {Balmaverde, B.}, {Brusa, M.}, {Cano-D\'{\i}az, M.}, {Cicone,
  C.}, {Comastri, A.}, {Cresci, G.}, {Fiore, F.}, {Feruglio, C.}, {La Franca,
  F.}, {Mainieri, V.}, {Mannucci, F.}, {Nagao, T.}, {Netzer, H.}, {Piconcelli,
  E.}, {Risaliti, G.}, {Schneider, R.}, \& {Shemmer, O.}}]{Carniani2016}
{Carniani, S.}, {Marconi, A.}, {Maiolino, R.}, {et~al.} 2016, A\&A, 591, A28

\bibitem[{{Circosta} {et~al.}(2018){Circosta}, {Mainieri}, {Padovani},
  {Lanzuisi}, {Salvato}, {Harrison}, {Kakkad}, {Puglisi}, {Vietri}, {Zamorani},
  {Cicone}, {Husemann}, {Vignali}, {Balmaverde}, {Bischetti}, {Bongiorno},
  {Brusa}, {Carniani}, {Civano}, {Comastri}, {Cresci}, {Feruglio}, {Fiore},
  {Fotopoulou}, {Karim}, {Lamastra}, {Magnelli}, {Mannucci}, {Marconi},
  {Merloni}, {Netzer}, {Perna}, {Piconcelli}, {Rodighiero}, {Schinnerer},
  {Schramm}, {Schulze}, {Silverman}, \& {Zappacosta}}]{Circosta2018}
{Circosta}, C., {Mainieri}, V., {Padovani}, P., {et~al.} 2018, \aap, 620, A82

\bibitem[{{Civano} {et~al.}(2016){Civano}, {Marchesi}, {Comastri}, {Urry},
  {Elvis}, {Cappelluti}, {Puccetti}, {Brusa}, {Zamorani}, {Hasinger},
  {Aldcroft}, {Alexand er}, {Allevato}, {Brunner}, {Capak}, {Finoguenov},
  {Fiore}, {Fruscione}, {Gilli}, {Glotfelty}, {Griffiths}, {Hao}, {Harrison},
  {Jahnke}, {Kartaltepe}, {Karim}, {LaMassa}, {Lanzuisi}, {Miyaji}, {Ranalli},
  {Salvato}, {Sargent}, {Scoville}, {Schawinski}, {Schinnerer}, {Silverman},
  {Smolcic}, {Stern}, {Toft}, {Trakhtenbrot}, {Treister}, \&
  {Vignali}}]{Civano2016}
{Civano}, F., {Marchesi}, S., {Comastri}, A., {et~al.} 2016, \apj, 819, 62

\bibitem[{{Coatman} {et~al.}(2017){Coatman}, {Hewett}, {Banerji}, {Richards},
  {Hennawi}, \& {Prochaska}}]{Coatman2017}
{Coatman}, L., {Hewett}, P.~C., {Banerji}, M., {et~al.} 2017, \mnras, 465, 2120

\bibitem[{Coatman {et~al.}(2019)Coatman, Hewett, Banerji, Richards, Hennawi, \&
  Prochaska}]{Coatman2019}
Coatman, L., Hewett, P.~C., Banerji, M., {et~al.} 2019, Monthly Notices of the
  Royal Astronomical Society, 486, 5335

\bibitem[{{Davies}(2007)}]{Davies2007}
{Davies}, R.~I. 2007, \mnras, 375, 1099

\bibitem[{{Denney}(2012)}]{Denney2012}
{Denney}, K.~D. 2012, \apj, 759, 44

\bibitem[{{Duras} {et~al.}(2017){Duras}, {Bongiorno}, {Piconcelli}, {Bianchi},
  {Pappalardo}, {Valiante}, {Bischetti}, {Feruglio}, {Martocchia}, {Schneider},
  {Vietri}, {Vignali}, {Zappacosta}, {La Franca}, \& {Fiore}}]{Duras2017}
{Duras}, F., {Bongiorno}, A., {Piconcelli}, E., {et~al.} 2017, \aap, 604, A67

\bibitem[{{Feigelson} \& {Nelson}(1985)}]{Feigelson1985}
{Feigelson}, E.~D. \& {Nelson}, P.~I. 1985, \apj, 293, 192

\bibitem[{{Ferrarese} \& {Merritt}(2000)}]{Ferrarese2000}
{Ferrarese}, L. \& {Merritt}, D. 2000, \apjl, 539, L9

\bibitem[{{Fiore} {et~al.}(2017){Fiore}, {Feruglio}, {Shankar}, {Bischetti},
  {Bongiorno}, {Brusa}, {Carniani}, {Cicone}, {Duras}, {Lamastra}, {Mainieri},
  {Marconi}, {Menci}, {Maiolino}, {Piconcelli}, {Vietri}, \&
  {Zappacosta}}]{Fiore2017}
{Fiore}, F., {Feruglio}, C., {Shankar}, F., {et~al.} 2017, \aap, 601, A143

\bibitem[{{Gaskell}(1982)}]{Gaskell1982}
{Gaskell}, C.~M. 1982, \apj, 263, 79

\bibitem[{{Gebhardt} {et~al.}(2000){Gebhardt}, {Bender}, {Bower}, {Dressler},
  {Faber}, {Filippenko}, {Green}, {Grillmair}, {Ho}, {Kormendy}, {Lauer},
  {Magorrian}, {Pinkney}, {Richstone}, \& {Tremaine}}]{Gebhardt2000}
{Gebhardt}, K., {Bender}, R., {Bower}, G., {et~al.} 2000, \apjl, 539, L13

\bibitem[{{Georgakakis} \& {Nandra}(2011)}]{Georgakakis2011}
{Georgakakis}, A. \& {Nandra}, K. 2011, \mnras, 414, 992

\bibitem[{{Graham} {et~al.}(2016){Graham}, {Ciambur}, \& {Soria}}]{Graham2016}
{Graham}, A.~W., {Ciambur}, B.~C., \& {Soria}, R. 2016, \apj, 818, 172

\bibitem[{{Greene} \& {Ho}(2005)}]{Greene2005}
{Greene}, J.~E. \& {Ho}, L.~C. 2005, \apj, 630, 122

\bibitem[{{Harrison} {et~al.}(2018){Harrison}, {Costa}, {Tadhunter},
  {Fl{\"u}tsch}, {Kakkad}, {Perna}, \& {Vietri}}]{Harrison2018}
{Harrison}, C.~M., {Costa}, T., {Tadhunter}, C.~N., {et~al.} 2018, Nature
  Astronomy, 2, 198

\bibitem[{{Isobe} {et~al.}(1986){Isobe}, {Feigelson}, \& {Nelson}}]{Isobe1986}
{Isobe}, T., {Feigelson}, E.~D., \& {Nelson}, P.~I. 1986, \apj, 306, 490

\bibitem[{{Kakkad} {et~al.}(2020){Kakkad}, {Mainieri}, {Vietri}, {Carniani},
  {Harrison}, {Perna}, {Scholtz}, {Circosta}, {Cresci}, {Husemann},
  {Bischetti}, {Feruglio}, {Fiore}, {Marconi}, {Padovani}, {Brusa}, {Cicone},
  {Comastri}, {Lanzuisi}, {Mannucci}, {Menci}, {Netzer}, {Piconcelli},
  {Puglisi}, {Salvato}, {Schramm}, {Silverman}, {Vignali}, {Zamorani}, \&
  {Zappacosta}}]{Kakkad2020}
{Kakkad}, D., {Mainieri}, V., {Vietri}, G., {et~al.} 2020, arXiv e-prints,
  arXiv:2008.01728

\bibitem[{{Kaspi} {et~al.}(2005){Kaspi}, {Maoz}, {Netzer}, {Peterson},
  {Vestergaard}, \& {Jannuzi}}]{Kaspi2005}
{Kaspi}, S., {Maoz}, D., {Netzer}, H., {et~al.} 2005, \apj, 629, 61

\bibitem[{{Kaspi} {et~al.}(2000){Kaspi}, {Smith}, {Netzer}, {Maoz}, {Jannuzi},
  \& {Giveon}}]{Kaspi2000}
{Kaspi}, S., {Smith}, P.~S., {Netzer}, H., {et~al.} 2000, \apj, 533, 631

\bibitem[{{LaMassa} {et~al.}(2016){LaMassa}, {Urry}, {Cappelluti},
  {B{\"o}hringer}, {Comastri}, {Glikman}, {Richards}, {Ananna}, {Brusa},
  {Cardamone}, {Chon}, {Civano}, {Farrah}, {Gilfanov}, {Green}, {Komossa},
  {Lira}, {Makler}, {Marchesi}, {Pecoraro}, {Ranalli}, {Salvato}, {Schawinski},
  {Stern}, {Treister}, \& {Viero}}]{LaMassa2016}
{LaMassa}, S.~M., {Urry}, C.~M., {Cappelluti}, N., {et~al.} 2016, \apj, 817,
  172

\bibitem[{{Lavalley} {et~al.}(1992){Lavalley}, {Isobe}, \&
  {Feigelson}}]{Lavalley1992}
{Lavalley}, M., {Isobe}, T., \& {Feigelson}, E. 1992, in Astronomical Society
  of the Pacific Conference Series, Vol.~25, Astronomical Data Analysis
  Software and Systems I, ed. D.~M. {Worrall}, C.~{Biemesderfer}, \&
  J.~{Barnes}, 245

\bibitem[{{Lira} {et~al.}(2018){Lira}, {Kaspi}, {Netzer}, {Botti}, {Morrell},
  {Mej{\'\i}a-Restrepo}, {S{\'a}nchez-S{\'a}ez}, {Mart{\'\i}nez-Palomera}, \&
  {L{\'o}pez}}]{Lira2018}
{Lira}, P., {Kaspi}, S., {Netzer}, H., {et~al.} 2018, \apj, 865, 56

\bibitem[{{Liu} {et~al.}(2016){Liu}, {Merloni}, {Georgakakis}, {Menzel},
  {Buchner}, {Nandra}, {Salvato}, {Shen}, {Brusa}, \& {Streblyanska}}]{Liu2016}
{Liu}, Z., {Merloni}, A., {Georgakakis}, A., {et~al.} 2016, \mnras, 459, 1602

\bibitem[{{Magorrian} {et~al.}(1998){Magorrian}, {Tremaine}, {Richstone},
  {Bender}, {Bower}, {Dressler}, {Faber}, {Gebhardt}, {Green}, {Grillmair},
  {Kormendy}, \& {Lauer}}]{Magorrian1998}
{Magorrian}, J., {Tremaine}, S., {Richstone}, D., {et~al.} 1998, \aj, 115, 2285

\bibitem[{{Martocchia} {et~al.}(2017){Martocchia}, {Piconcelli}, {Zappacosta},
  {Duras}, {Vietri}, {Vignali}, {Bianchi}, {Bischetti}, {Bongiorno}, {Brusa},
  {Lanzuisi}, {Marconi}, {Mathur}, {Miniutti}, {Nicastro}, {Bruni}, \&
  {Fiore}}]{Martocchia2017}
{Martocchia}, S., {Piconcelli}, E., {Zappacosta}, L., {et~al.} 2017, ArXiv
  e-prints

\bibitem[{{Marziani} {et~al.}(2016){Marziani}, {Mart{\'{\i}}nez Carballo},
  {Sulentic}, {Del Olmo}, {Stirpe}, \& {Dultzin}}]{Marziani2016}
{Marziani}, P., {Mart{\'{\i}}nez Carballo}, M.~A., {Sulentic}, J.~W., {et~al.}
  2016, \apss, 361, 29

\bibitem[{{Marziani} {et~al.}(2013){Marziani}, {Sulentic}, {Plauchu-Frayn}, \&
  {del Olmo}}]{Marziani2013}
{Marziani}, P., {Sulentic}, J.~W., {Plauchu-Frayn}, I., \& {del Olmo}, A. 2013,
  \aap, 555, A89

\bibitem[{{Matsuoka, K.} {et~al.}(2011){Matsuoka, K.}, {Nagao, T.}, {Marconi,
  A.}, {Maiolino, R.}, \& {Taniguchi, Y.}}]{Matsuoka2011}
{Matsuoka, K.}, {Nagao, T.}, {Marconi, A.}, {Maiolino, R.}, \& {Taniguchi, Y.}
  2011, A\&A, 527, A100

\bibitem[{{McGill} {et~al.}(2008){McGill}, {Woo}, {Treu}, \&
  {Malkan}}]{McGill2008}
{McGill}, K.~L., {Woo}, J.-H., {Treu}, T., \& {Malkan}, M.~A. 2008, \apj, 673,
  703

\bibitem[{{McLure} \& {Jarvis}(2002)}]{McLure2002}
{McLure}, R.~J. \& {Jarvis}, M.~J. 2002, \mnras, 337, 109

\bibitem[{{Mej{\'\i}a-Restrepo} {et~al.}(2018){Mej{\'\i}a-Restrepo}, {Lira},
  {Netzer}, {Trakhtenbrot}, \& {Capellupo}}]{Mejia-Restrepo2018}
{Mej{\'\i}a-Restrepo}, J.~E., {Lira}, P., {Netzer}, H., {Trakhtenbrot}, B., \&
  {Capellupo}, D.~M. 2018, Nature Astronomy, 2, 63

\bibitem[{{Mej{\'{\i}}a-Restrepo} {et~al.}(2016){Mej{\'{\i}}a-Restrepo},
  {Trakhtenbrot}, {Lira}, {Netzer}, \& {Capellupo}}]{Mejia-Restrepo2016}
{Mej{\'{\i}}a-Restrepo}, J.~E., {Trakhtenbrot}, B., {Lira}, P., {Netzer}, H.,
  \& {Capellupo}, D.~M. 2016, \mnras, 460, 187

\bibitem[{{Menzel} {et~al.}(2016){Menzel}, {Merloni}, {Georgakakis}, {Salvato},
  {Aubourg}, {Brandt}, {Brusa}, {Buchner}, {Dwelly}, {Nandra}, {P{\^a}ris},
  {Petitjean}, \& {Schwope}}]{Menzel2016}
{Menzel}, M.~L., {Merloni}, A., {Georgakakis}, A., {et~al.} 2016, \mnras, 457,
  110

\bibitem[{{Nagao} {et~al.}(2006){Nagao}, {Marconi}, \& {Maiolino}}]{Nagao2006}
{Nagao}, T., {Marconi}, A., \& {Maiolino}, R. 2006, \aap, 447, 157

\bibitem[{{Netzer}(2013)}]{Netzer2013}
{Netzer}, H. 2013, {The Physics and Evolution of Active Galactic Nuclei}

\bibitem[{Netzer(2020)}]{Netzer2020}
Netzer, H. 2020, Monthly Notices of the Royal Astronomical Society, 494, 1611

\bibitem[{{Osterbrock} \& {Ferland}(2006)}]{Osterbrock2006}
{Osterbrock}, D.~E. \& {Ferland}, G.~J. 2006, {Astrophysics of gaseous nebulae
  and active galactic nuclei}

\bibitem[{{Popovi{\'c}} {et~al.}(2019){Popovi{\'c}},
  {Kova{\v{c}}evi{\'c}-Doj{\v{c}}inovi{\'c}}, \& {Mar{\v{c}}eta-Mand
  i{\'c}}}]{Popovic2019}
{Popovi{\'c}}, L.~{\v{C}}., {Kova{\v{c}}evi{\'c}-Doj{\v{c}}inovi{\'c}}, J., \&
  {Mar{\v{c}}eta-Mand i{\'c}}, S. 2019, \mnras, 484, 3180

\bibitem[{{Richards} {et~al.}(2011){Richards}, {Kruczek}, {Gallagher}, {Hall},
  {Hewett}, {Leighly}, {Deo}, {Kratzer}, \& {Shen}}]{Richards2011}
{Richards}, G.~T., {Kruczek}, N.~E., {Gallagher}, S.~C., {et~al.} 2011, \aj,
  141, 167

\bibitem[{{Runnoe} {et~al.}(2012){Runnoe}, {Brotherton}, \&
  {Shang}}]{Runnoe2012}
{Runnoe}, J.~C., {Brotherton}, M.~S., \& {Shang}, Z. 2012, \mnras, 427, 1800

\bibitem[{{Schwarz}(1978)}]{Schwarz1978}
{Schwarz}, G. 1978, Annals of Statistics, 6, 461

\bibitem[{{Shen}(2013)}]{Shen2013}
{Shen}, Y. 2013, Bulletin of the Astronomical Society of India, 41, 61

\bibitem[{{Shen} {et~al.}(2008){Shen}, {Greene}, {Strauss}, {Richards}, \&
  {Schneider}}]{Shen2008}
{Shen}, Y., {Greene}, J.~E., {Strauss}, M.~A., {Richards}, G.~T., \&
  {Schneider}, D.~P. 2008, \apj, 680, 169

\bibitem[{{Shen} \& {Liu}(2012)}]{Shen2012}
{Shen}, Y. \& {Liu}, X. 2012, \apj, 753, 125

\bibitem[{{Shen} {et~al.}(2011){Shen}, {Richards}, {Strauss}, {Hall},
  {Schneider}, {Snedden}, {Bizyaev}, {Brewington}, {Malanushenko},
  {Malanushenko}, {Oravetz}, {Pan}, \& {Simmons}}]{Shen2011}
{Shen}, Y., {Richards}, G.~T., {Strauss}, M.~A., {et~al.} 2011, \apjs, 194, 45

\bibitem[{{Suh} {et~al.}(2020){Suh}, {Civano}, {Trakhtenbrot}, {Shankar},
  {Hasinger}, {Sand ers}, \& {Allevato}}]{Suh2020}
{Suh}, H., {Civano}, F., {Trakhtenbrot}, B., {et~al.} 2020, \apj, 889, 32

\bibitem[{{Suh} {et~al.}(2015){Suh}, {Hasinger}, {Steinhardt}, {Silverman}, \&
  {Schramm}}]{Suh2015}
{Suh}, H., {Hasinger}, G., {Steinhardt}, C., {Silverman}, J.~D., \& {Schramm},
  M. 2015, \apj, 815, 129

\bibitem[{{Sulentic} {et~al.}(2007){Sulentic}, {Bachev}, {Marziani}, {Negrete},
  \& {Dultzin}}]{Sulentic2007}
{Sulentic}, J.~W., {Bachev}, R., {Marziani}, P., {Negrete}, C.~A., \&
  {Dultzin}, D. 2007, \apj, 666, 757

\bibitem[{{Sulentic} {et~al.}(2000){Sulentic}, {Marziani}, \&
  {Dultzin-Hacyan}}]{Sulentic2000}
{Sulentic}, J.~W., {Marziani}, P., \& {Dultzin-Hacyan}, D. 2000, \araa, 38, 521

\bibitem[{{Trakhtenbrot} \& {Netzer}(2012)}]{Trakhtenbrot2012}
{Trakhtenbrot}, B. \& {Netzer}, H. 2012, \mnras, 427, 3081

\bibitem[{{Travascio} {et~al.}(2020){Travascio}, {Zappacosta}, {Cantalupo},
  {Piconcelli}, {Arrigoni Battaia}, {Ginolfi}, {Bischetti}, {Vietri},
  {Bongiorno}, {D'Odorico}, {Duras}, {Feruglio}, {Vignali}, \&
  {Fiore}}]{Travascio2020}
{Travascio}, A., {Zappacosta}, L., {Cantalupo}, S., {et~al.} 2020, \aap, 635,
  A157

\bibitem[{{Tsuzuki} {et~al.}(2006){Tsuzuki}, {Kawara}, {Yoshii}, {Oyabu},
  {Tanab{\'e}}, \& {Matsuoka}}]{Tsuzuki2006}
{Tsuzuki}, Y., {Kawara}, K., {Yoshii}, Y., {et~al.} 2006, \apj, 650, 57

\bibitem[{{V{\'e}ron-Cetty} {et~al.}(2004){V{\'e}ron-Cetty}, {Joly}, \&
  {V{\'e}ron}}]{Veron2004}
{V{\'e}ron-Cetty}, M.-P., {Joly}, M., \& {V{\'e}ron}, P. 2004, \aap, 417, 515

\bibitem[{{Vestergaard}(2002)}]{Vestergaard2002}
{Vestergaard}, M. 2002, \apj, 571, 733

\bibitem[{{Vietri} {et~al.}(2018){Vietri}, {Piconcelli}, {Bischetti}, {Duras},
  {Martocchia}, {Bongiorno}, {Marconi}, {Zappacosta}, {Bisogni}, {Bruni},
  {Brusa}, {Comastri}, {Cresci}, {Feruglio}, {Giallongo}, {La Franca},
  {Mainieri}, {Mannucci}, {Ricci}, {Sani}, {Testa}, {Tombesi}, {Vignali}, \&
  {Fiore}}]{Vietri2018}
{Vietri}, G., {Piconcelli}, E., {Bischetti}, M., {et~al.} 2018, \aap, 617, A81

\bibitem[{{Wang} {et~al.}(2009){Wang}, {Dong}, {Wang}, {Ho}, {Yuan}, {Wang},
  {Zhang}, {Zhang}, \& {Zhou}}]{Wang2009}
{Wang}, J.-G., {Dong}, X.-B., {Wang}, T.-G., {et~al.} 2009, \apj, 707, 1334

\bibitem[{{Woo} {et~al.}(2015){Woo}, {Yoon}, {Park}, {Park}, \&
  {Kim}}]{Woo2015}
{Woo}, J.-H., {Yoon}, Y., {Park}, S., {Park}, D., \& {Kim}, S.~C. 2015, \apj,
  801, 38

\bibitem[{{Wu} {et~al.}(2004){Wu}, {Wang}, {Kong}, {Liu}, \& {Han}}]{Wu2004}
{Wu}, X.~B., {Wang}, R., {Kong}, M.~Z., {Liu}, F.~K., \& {Han}, J.~L. 2004,
  \aap, 424, 793

\bibitem[{{Zappacosta} {et~al.}(2020){Zappacosta}, {Piconcelli}, {Giustini},
  {Vietri}, {Duras}, {Miniutti}, {Bischetti}, {Bongiorno}, {Brusa},
  {Chiaberge}, {Comastri}, {Feruglio}, {Luminari}, {Marconi}, {Ricci},
  {Vignali}, \& {Fiore}}]{Zappacosta2020}
{Zappacosta}, L., {Piconcelli}, E., {Giustini}, M., {et~al.} 2020, \aap, 635,
  L5

\end{thebibliography}

\begin{appendix}
\onecolumn
\section{Integrated spectra of Type-1 SUPER sample}\label{app:app}
 \begin{figure}[h]
 \centering
   \includegraphics[width=0.4\textwidth]{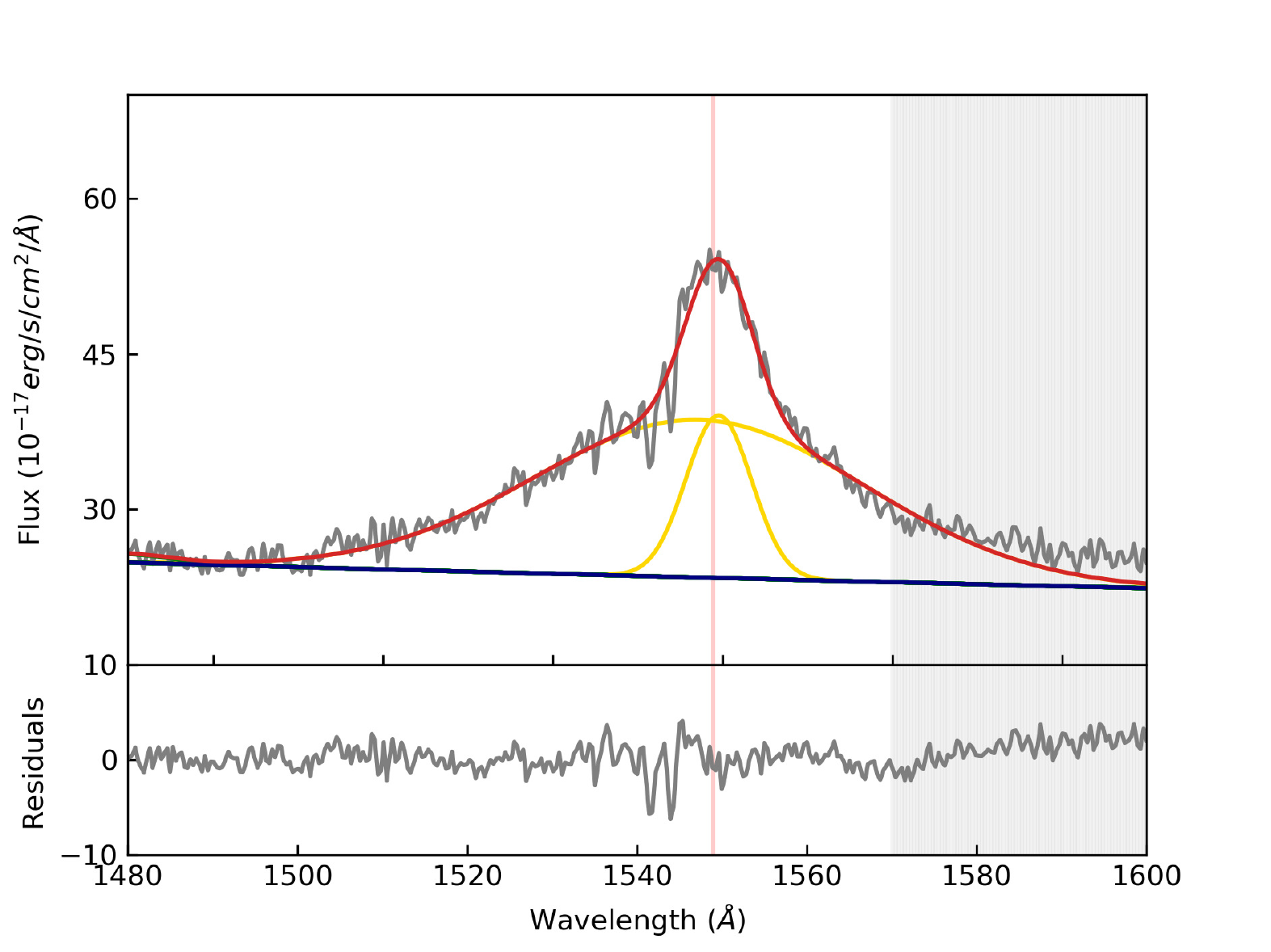}
   \includegraphics[width=0.4\textwidth]{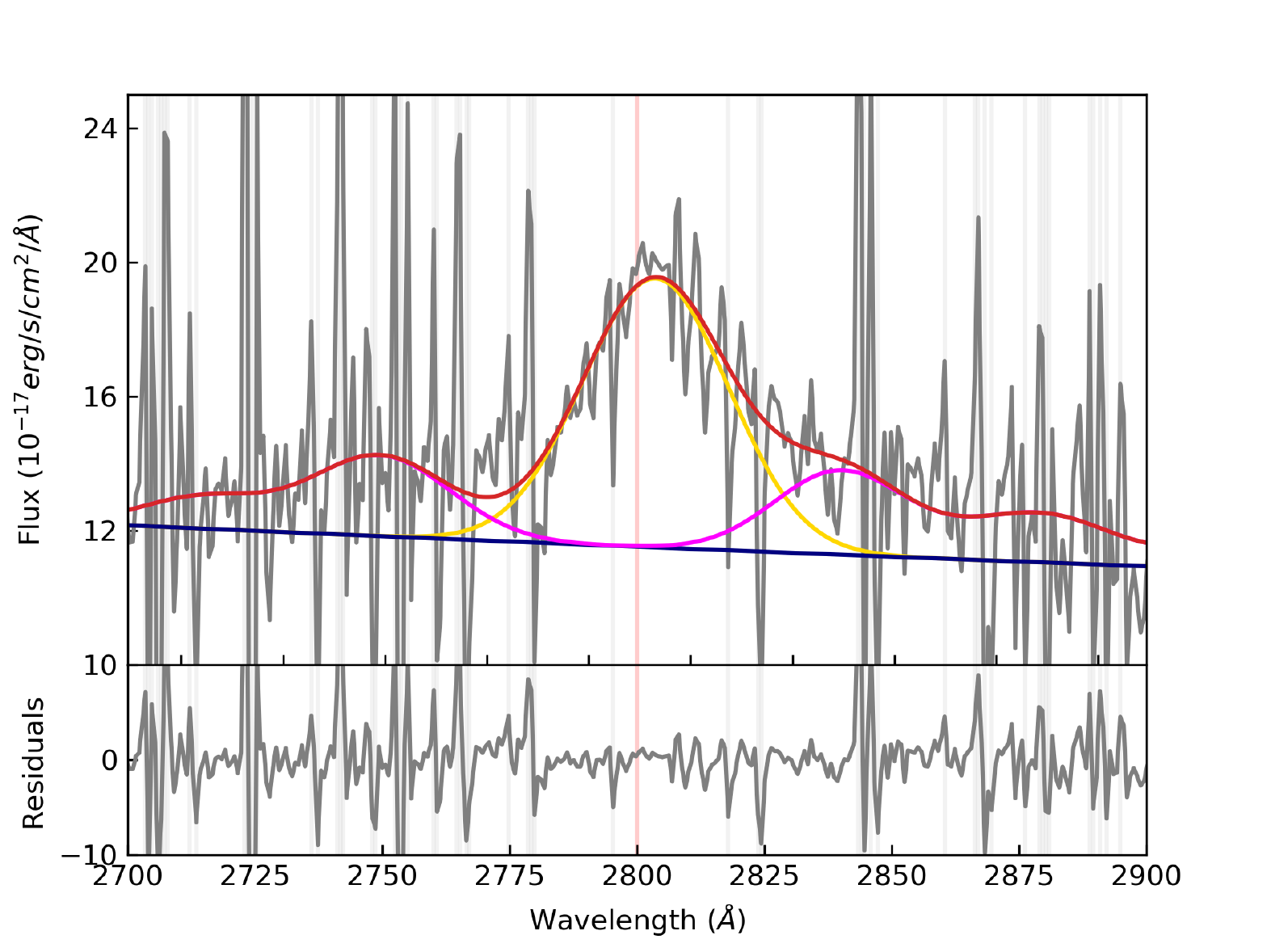}
  \includegraphics[width=0.4\textwidth]{./figure/plot_H_band/Spectrum_XN_160_22_H_finale_1_0_paper-eps-converted-to.pdf}
 \includegraphics[width=0.4\textwidth]{./figure/plot_K_band/Spectrum_XN_160_22_K_finale_0_8_paper-eps-converted-to.pdf}
\caption{Integrated spectra from CIV-MgII region (upper panels) to  H$\beta$-H$\alpha$ region (lower panels) of the SUPER targets X\_N\_160\_22. The grey curve shows the observed spectrum, the red curve shows the reproduced overall emission line model, the magenta curve shows the iron emission, the blue navy curves show the continuum model and the green curves show the individual Gaussian components used to reproduce the profiles of all the emission lines. The BLR component is shown as gold Gaussian profile. The red vertical lines indicate the location of each line at systemic velocity. The vertical grey regions mark the channels with strong skylines which were masked during the fitting procedure. The x- and y-axis show the rest frame wavelength and flux (not corrected for extinction), according to the redshift of each target.}\label{fig:app}
   \end{figure}

      \begin{figure*}
 \center
    \includegraphics[width=0.4\textwidth]{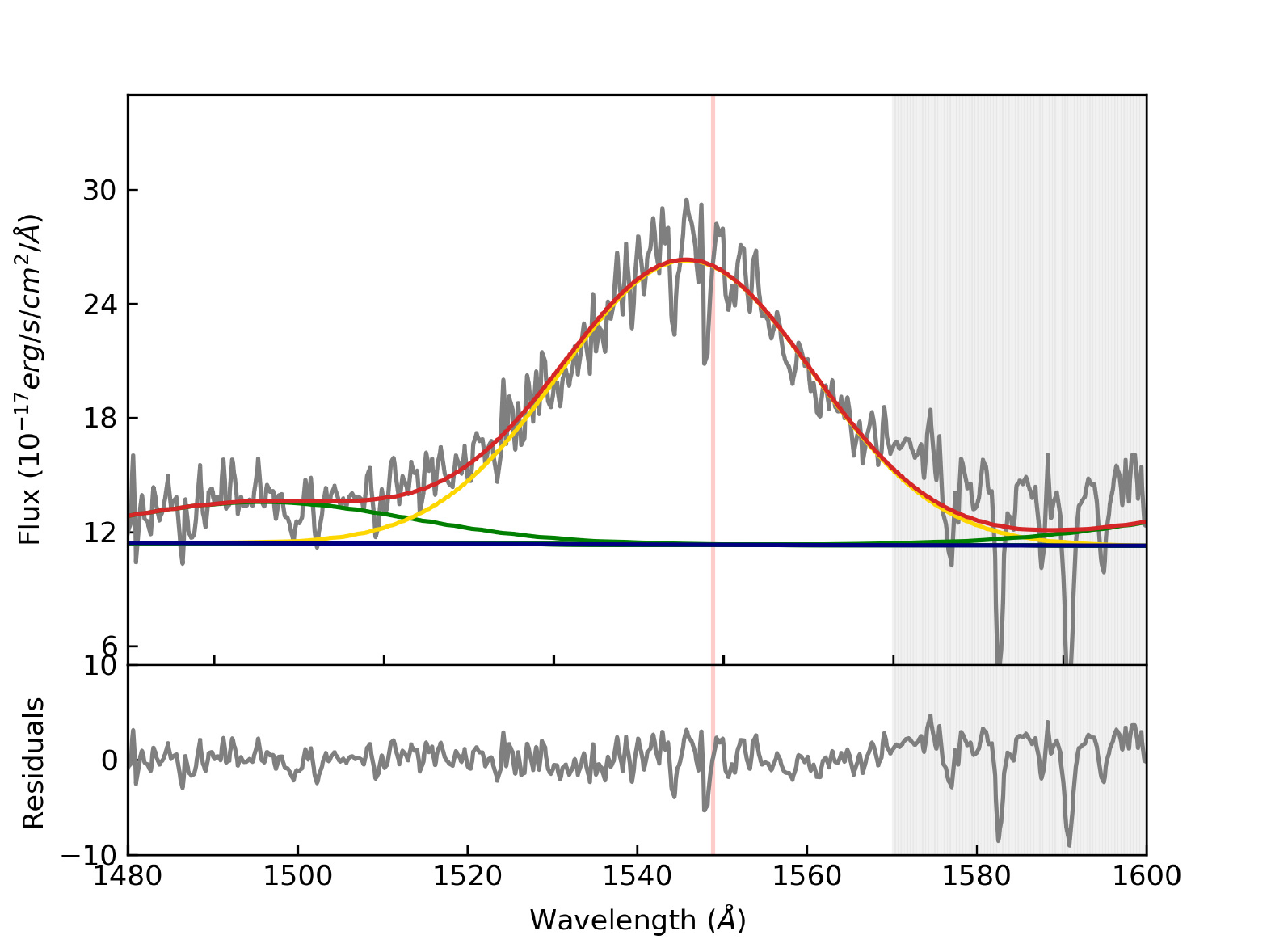}
   \includegraphics[width=0.4\textwidth]{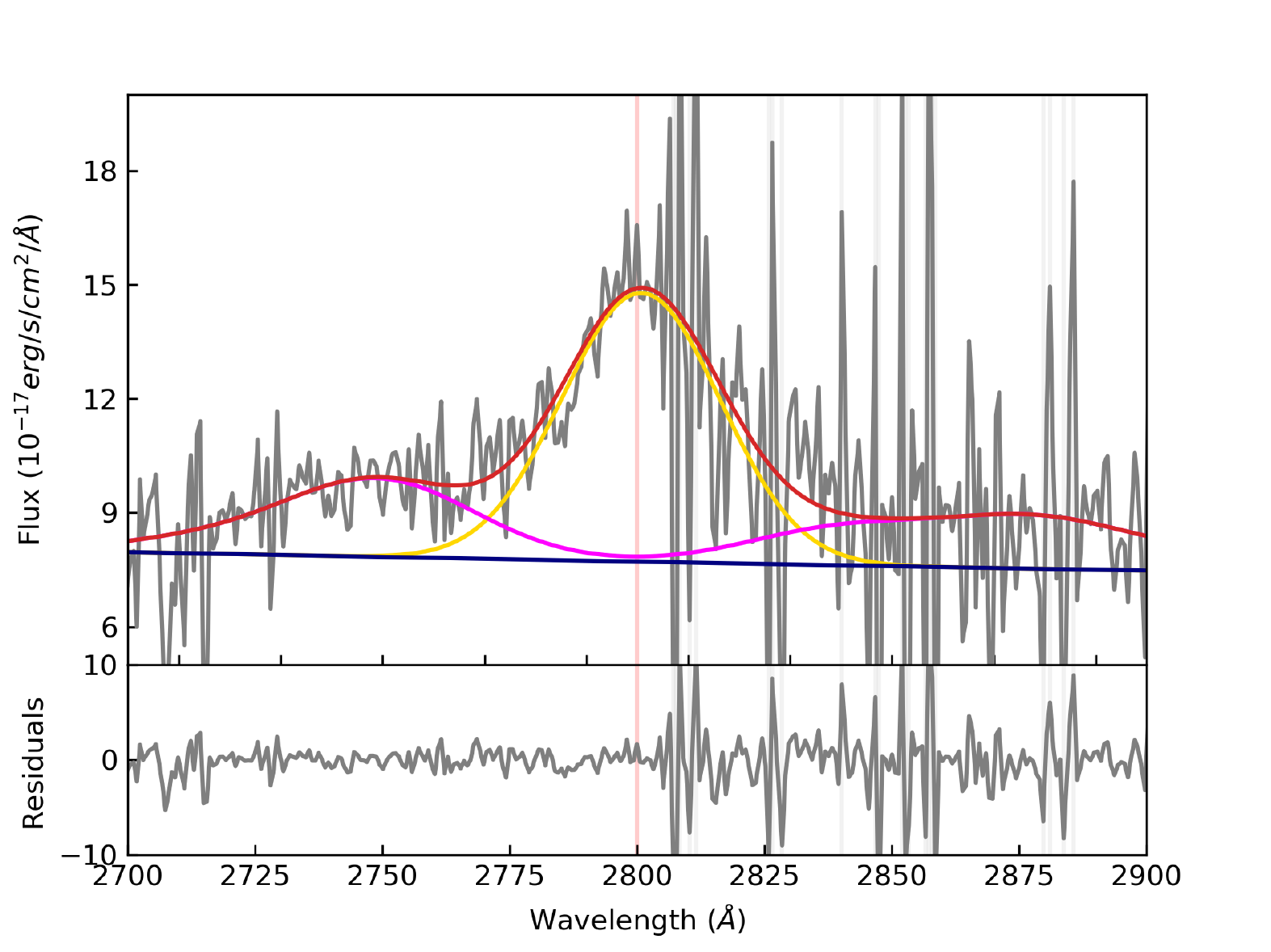}
   \includegraphics[width=0.4\textwidth]{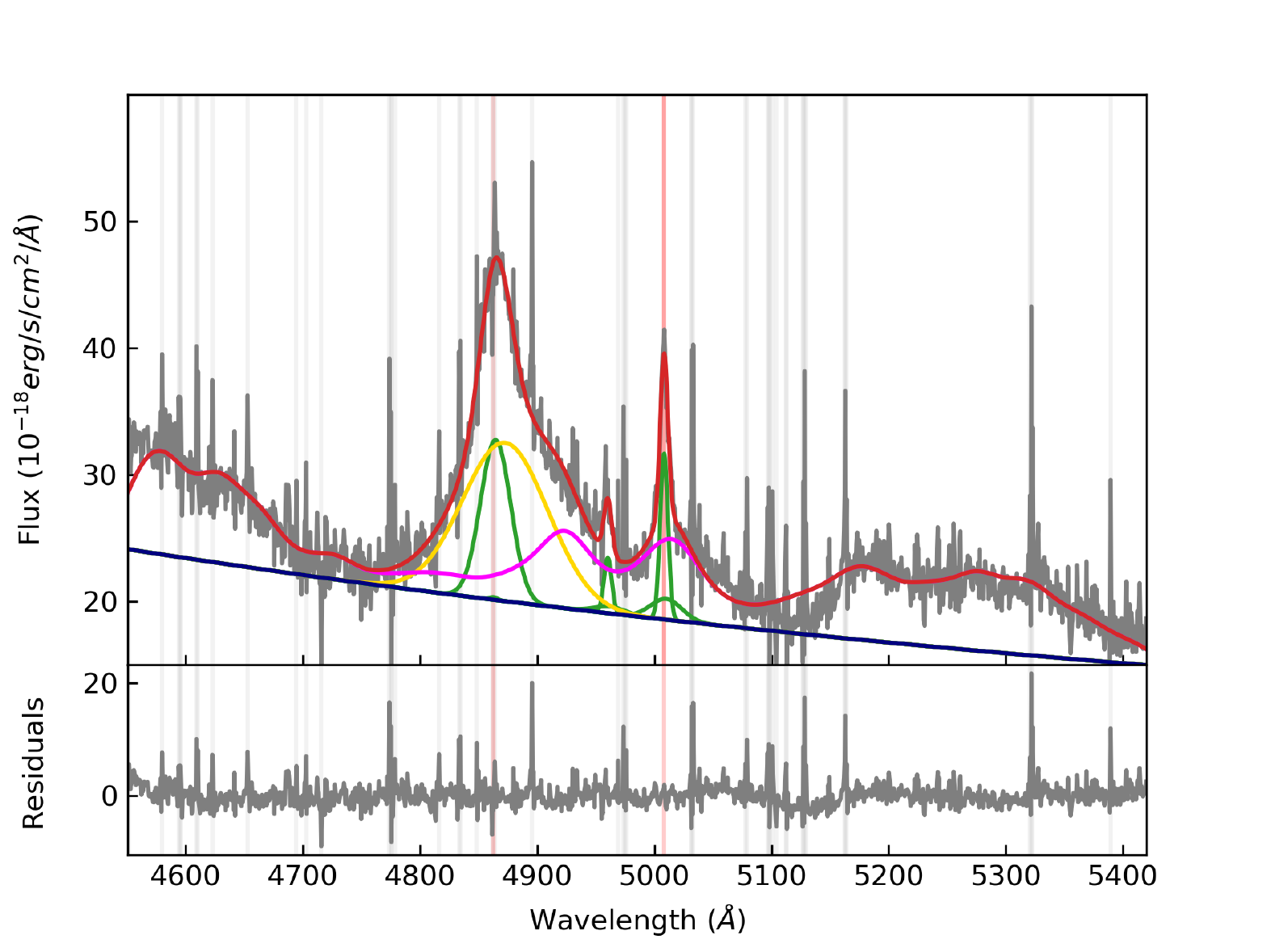}      \includegraphics[width=0.4\textwidth]{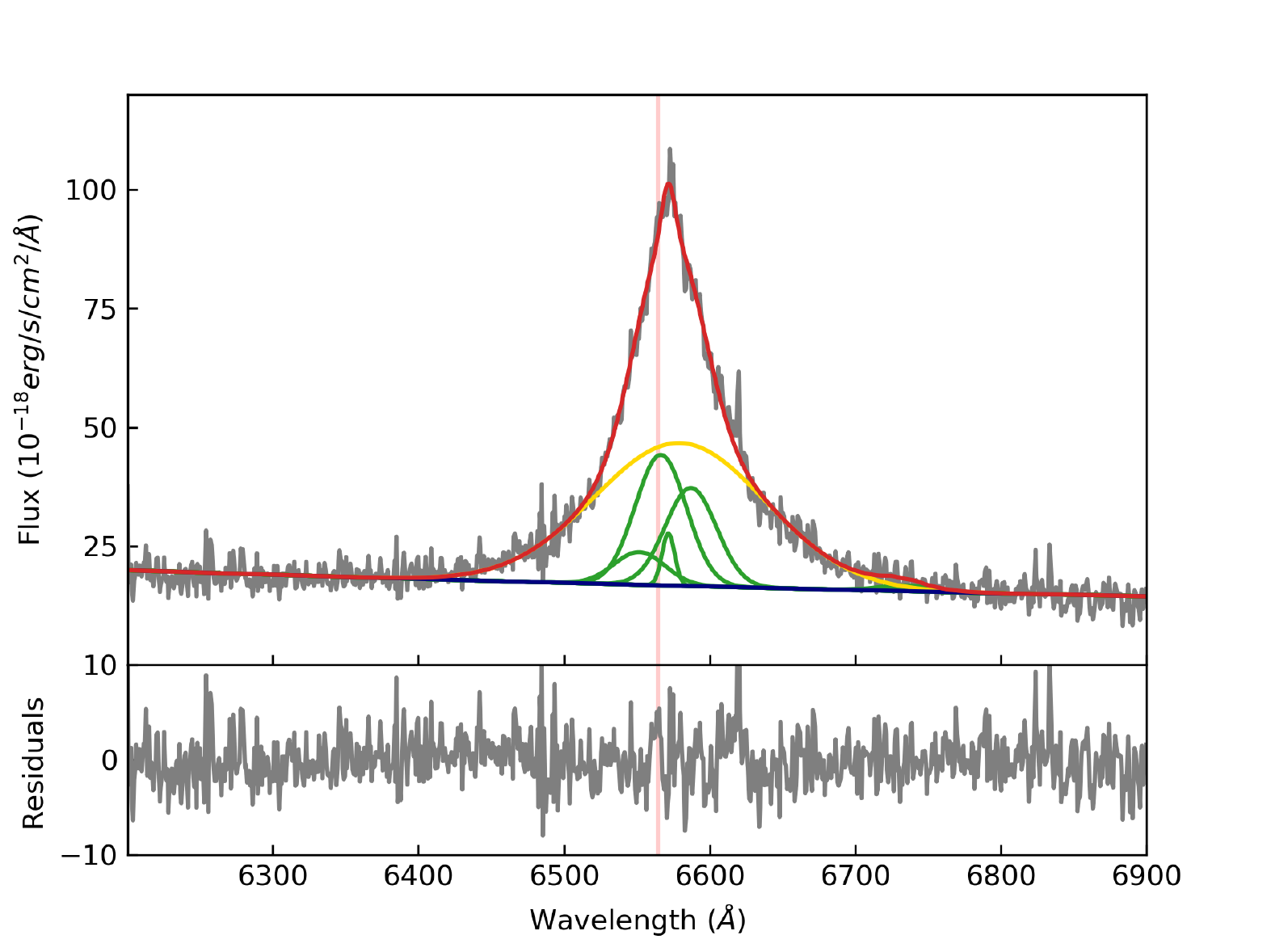}
\caption{X\_N\_81\_44. The modeling is the same as in Fig. \ref{fig:app}}
   \end{figure*}

 \begin{figure*}
 \center
    \includegraphics[width=0.4\textwidth]{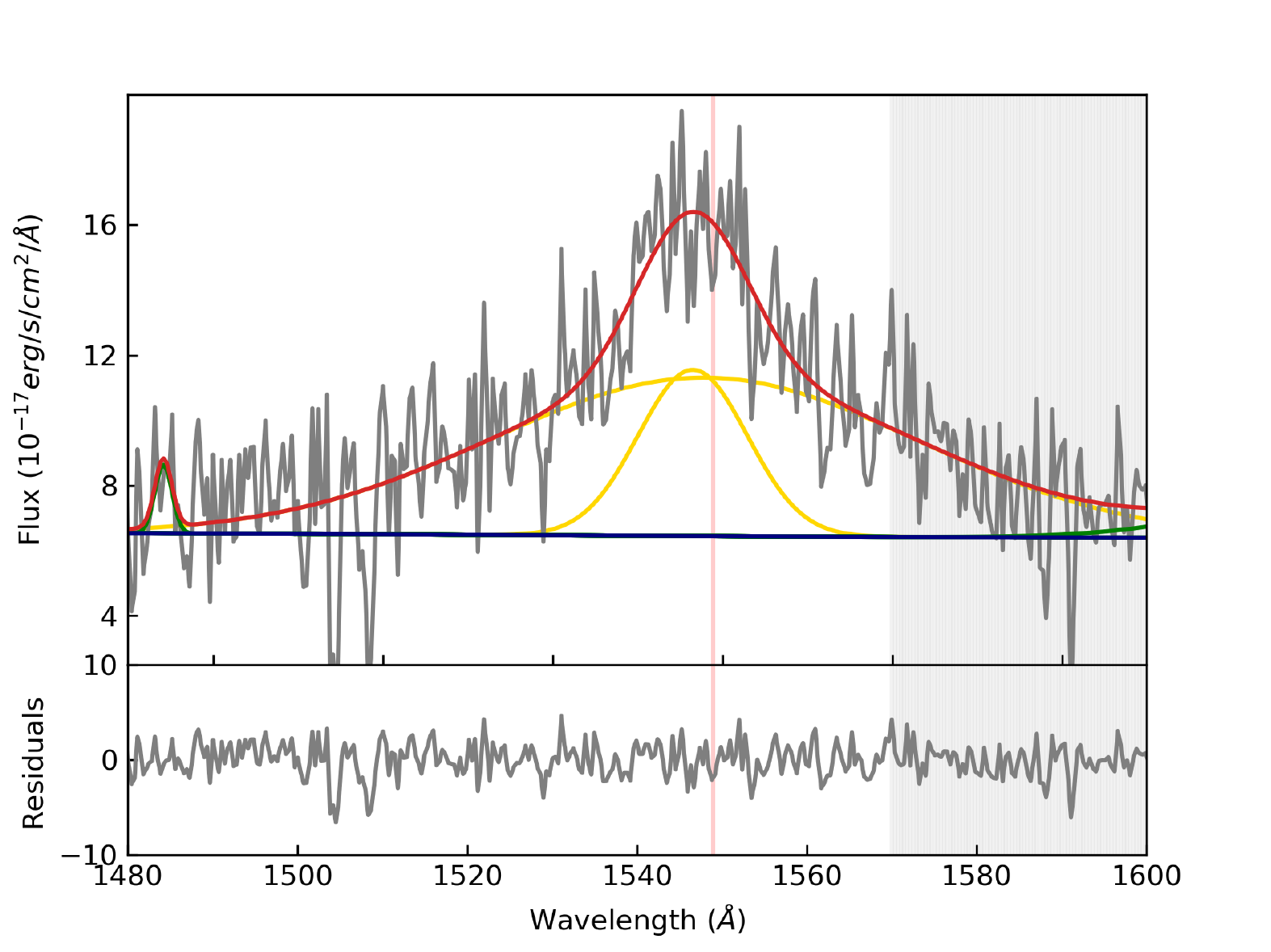}
   \includegraphics[width=0.4\textwidth]{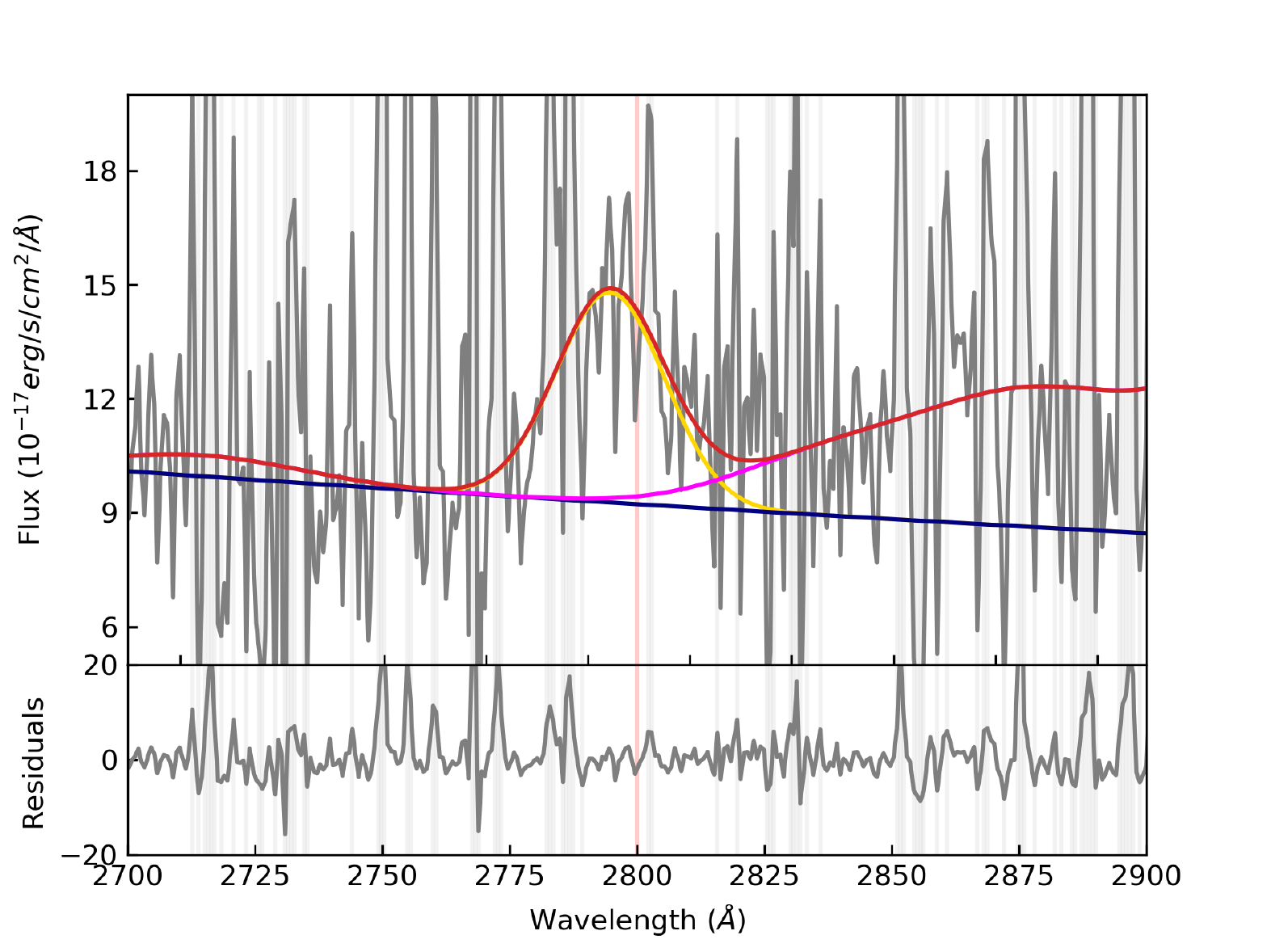}
    \includegraphics[width=0.4\textwidth]{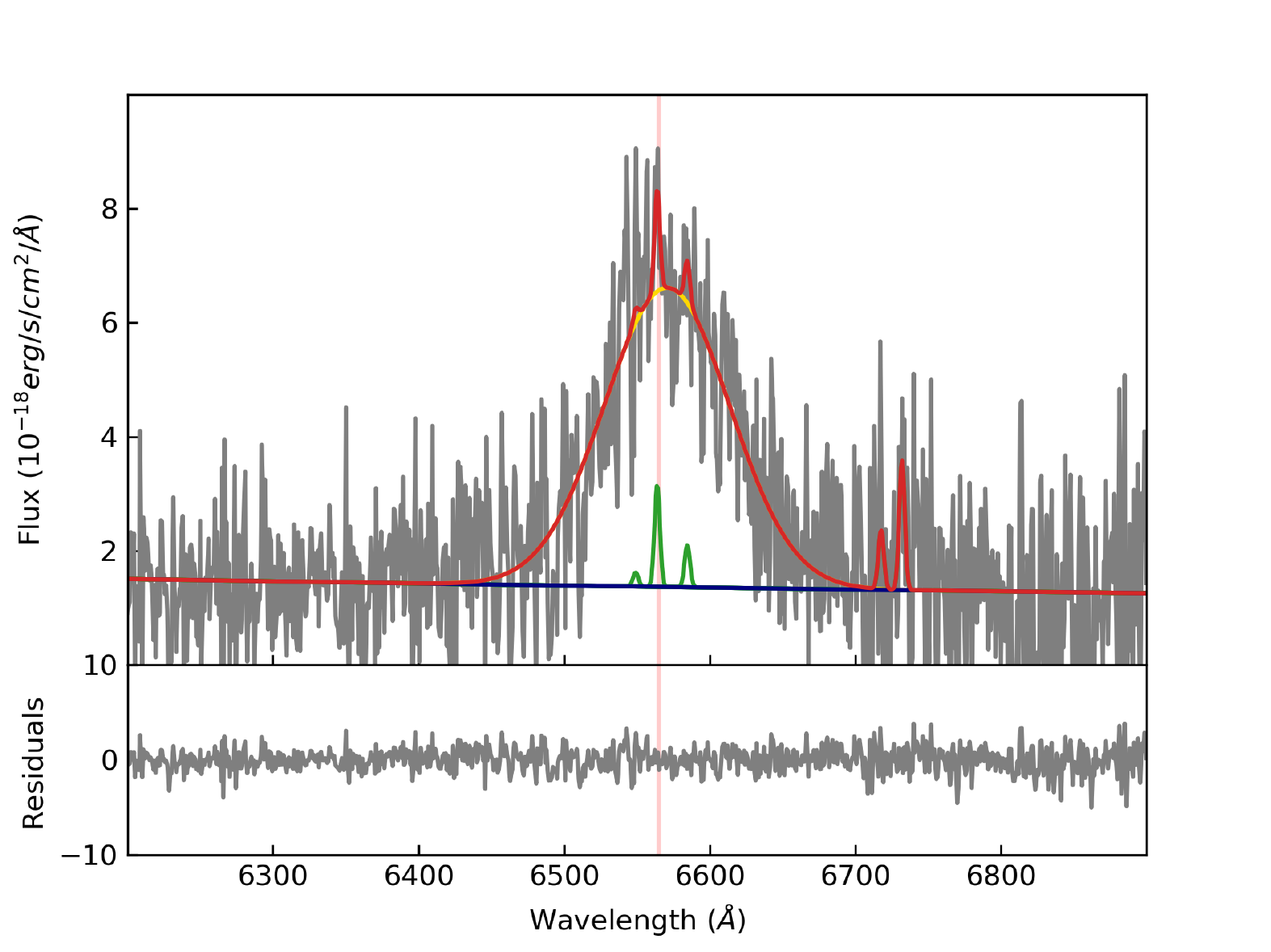}
\caption{X\_N\_53\_3. The modeling is the same as in Fig. \ref{fig:app}}
   \end{figure*}

 \begin{figure*}
 \center
    \includegraphics[width=0.4\textwidth]{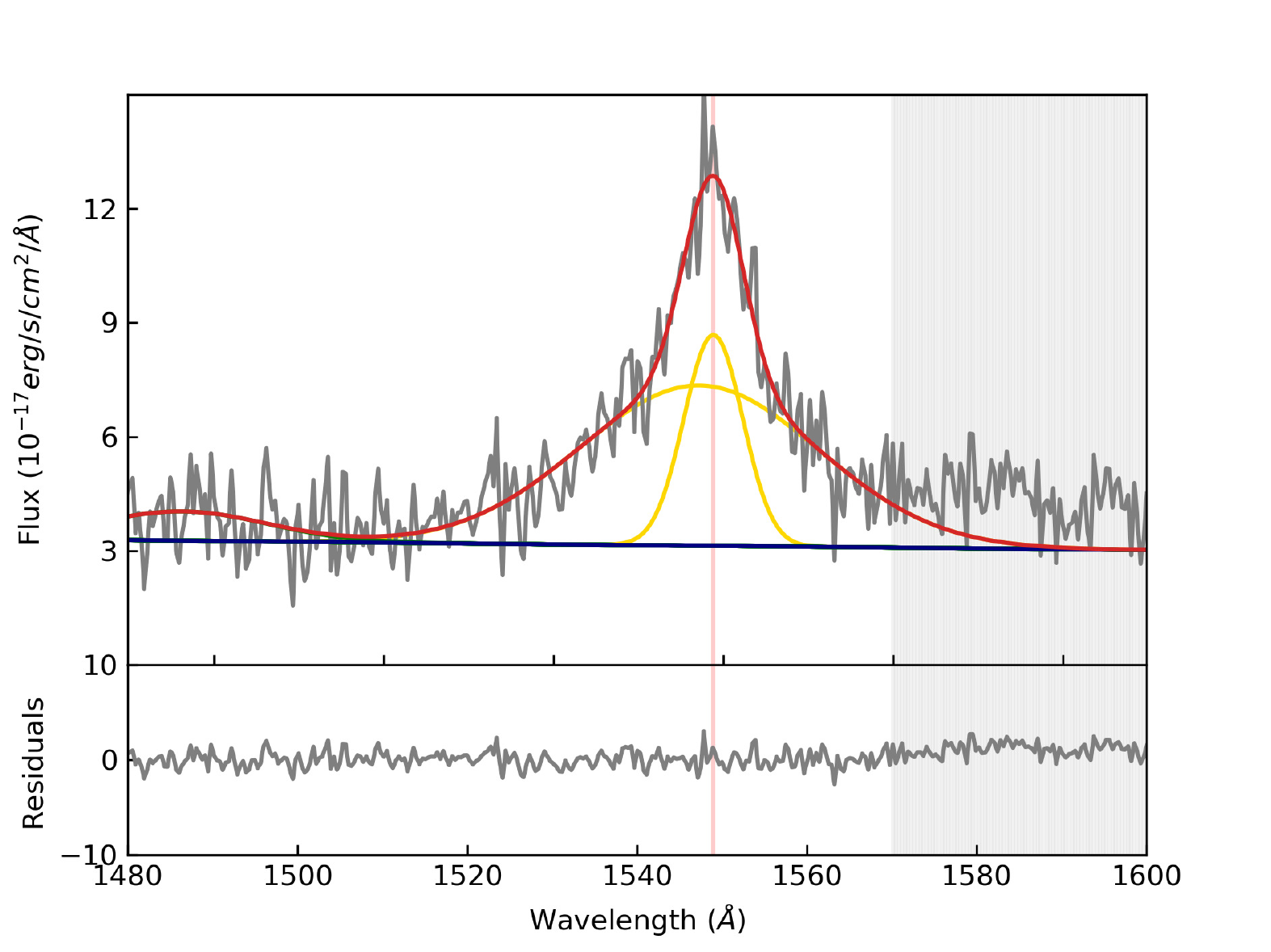}
   \includegraphics[width=0.4\textwidth]{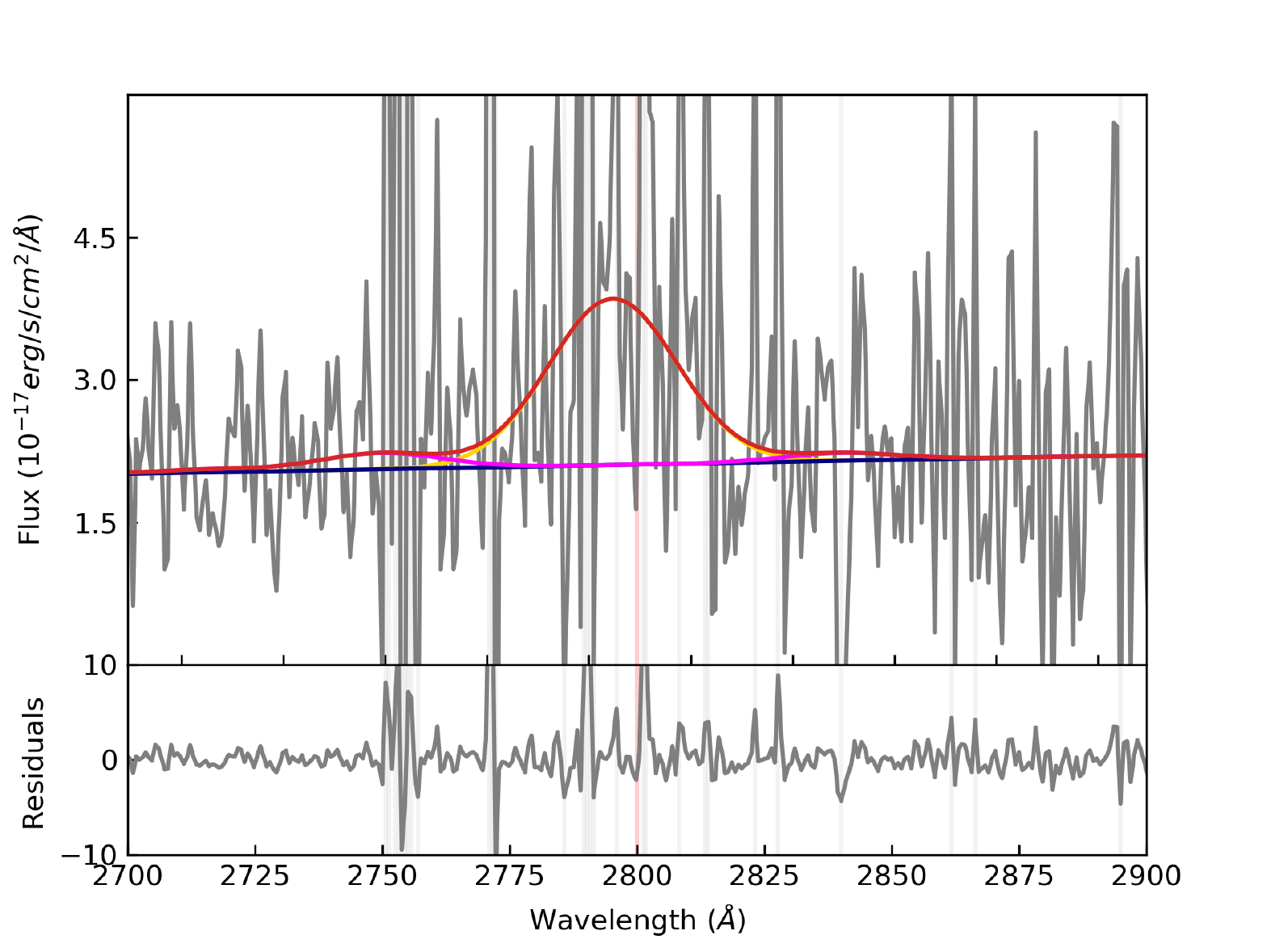}
   \includegraphics[width=0.4\textwidth]{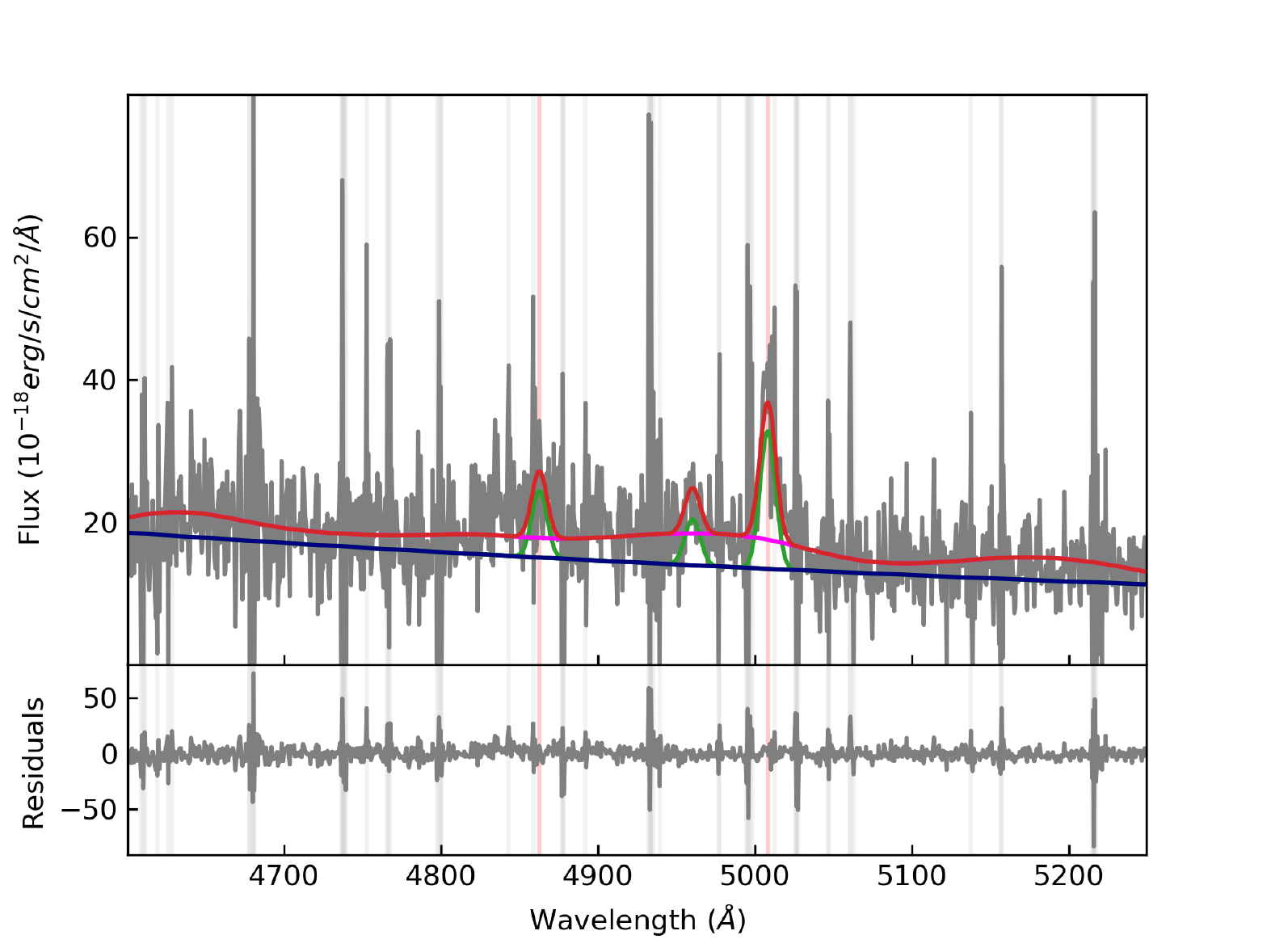}      \includegraphics[width=0.4\textwidth]{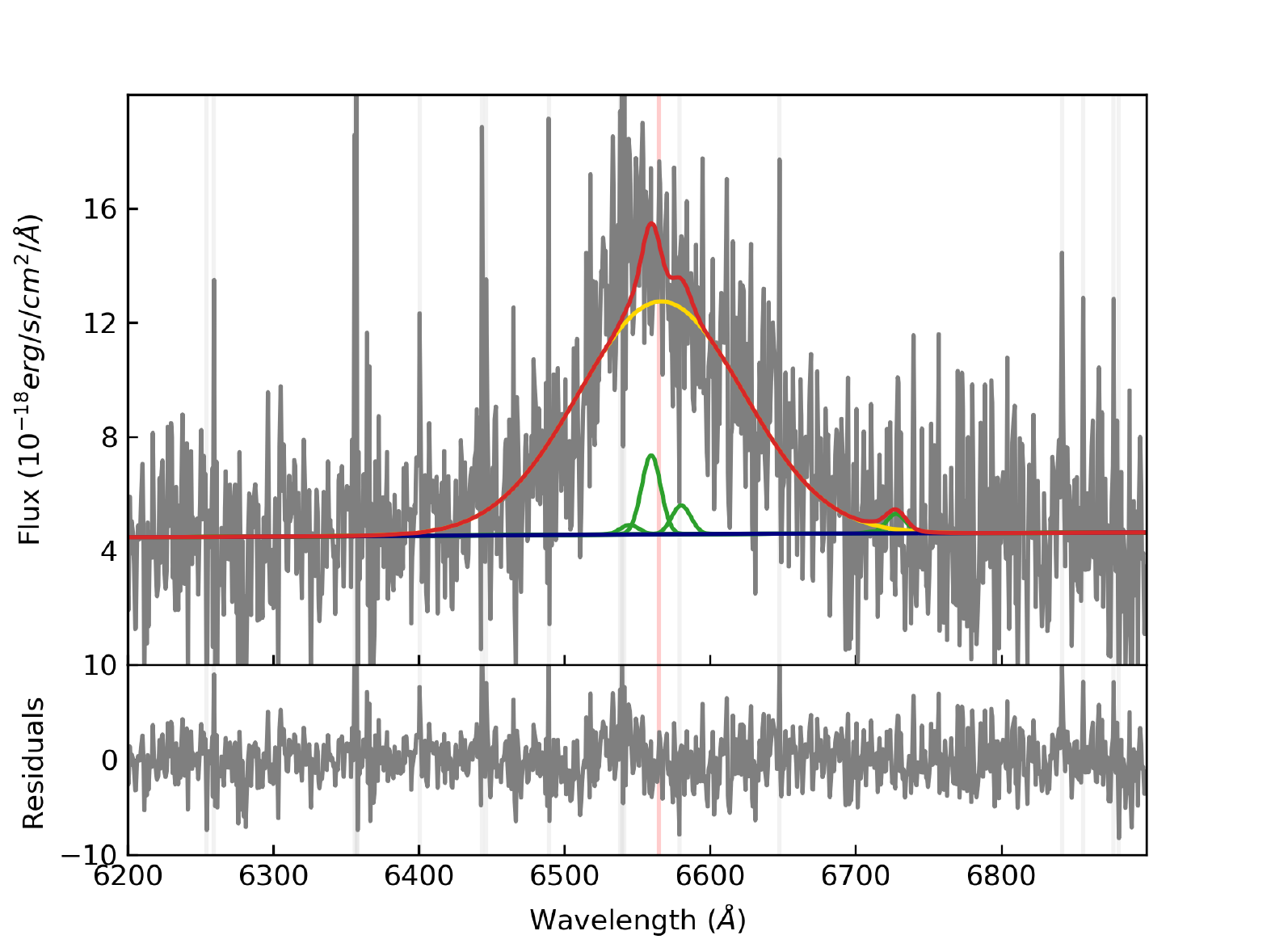}
\caption{X\_N\_66\_23. The modeling is the same as in Fig. \ref{fig:app}}
   \end{figure*}

 \begin{figure*}
 \center
    \includegraphics[width=0.4\textwidth]{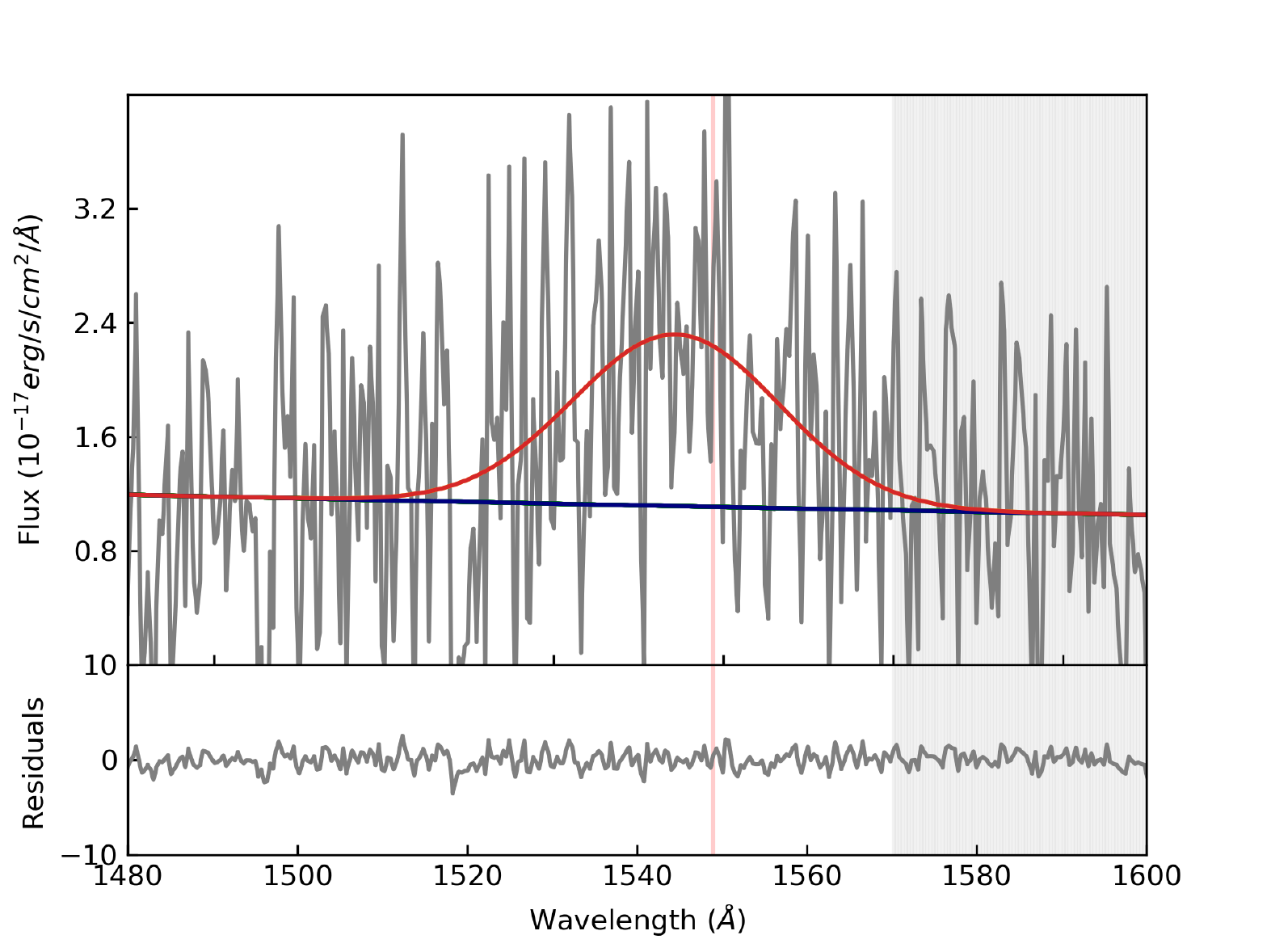}
   \includegraphics[width=0.4\textwidth]{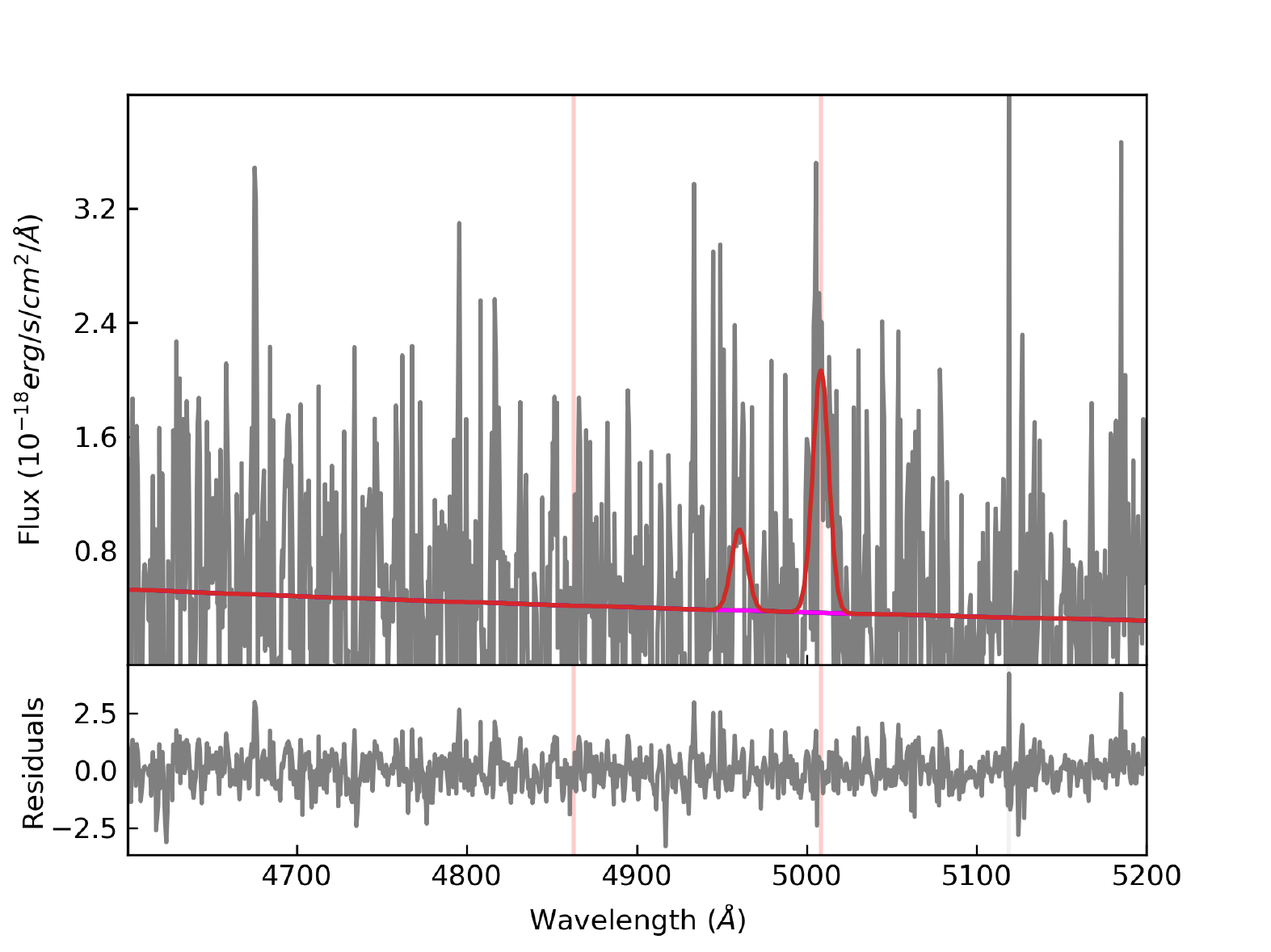}      \includegraphics[width=0.4\textwidth]{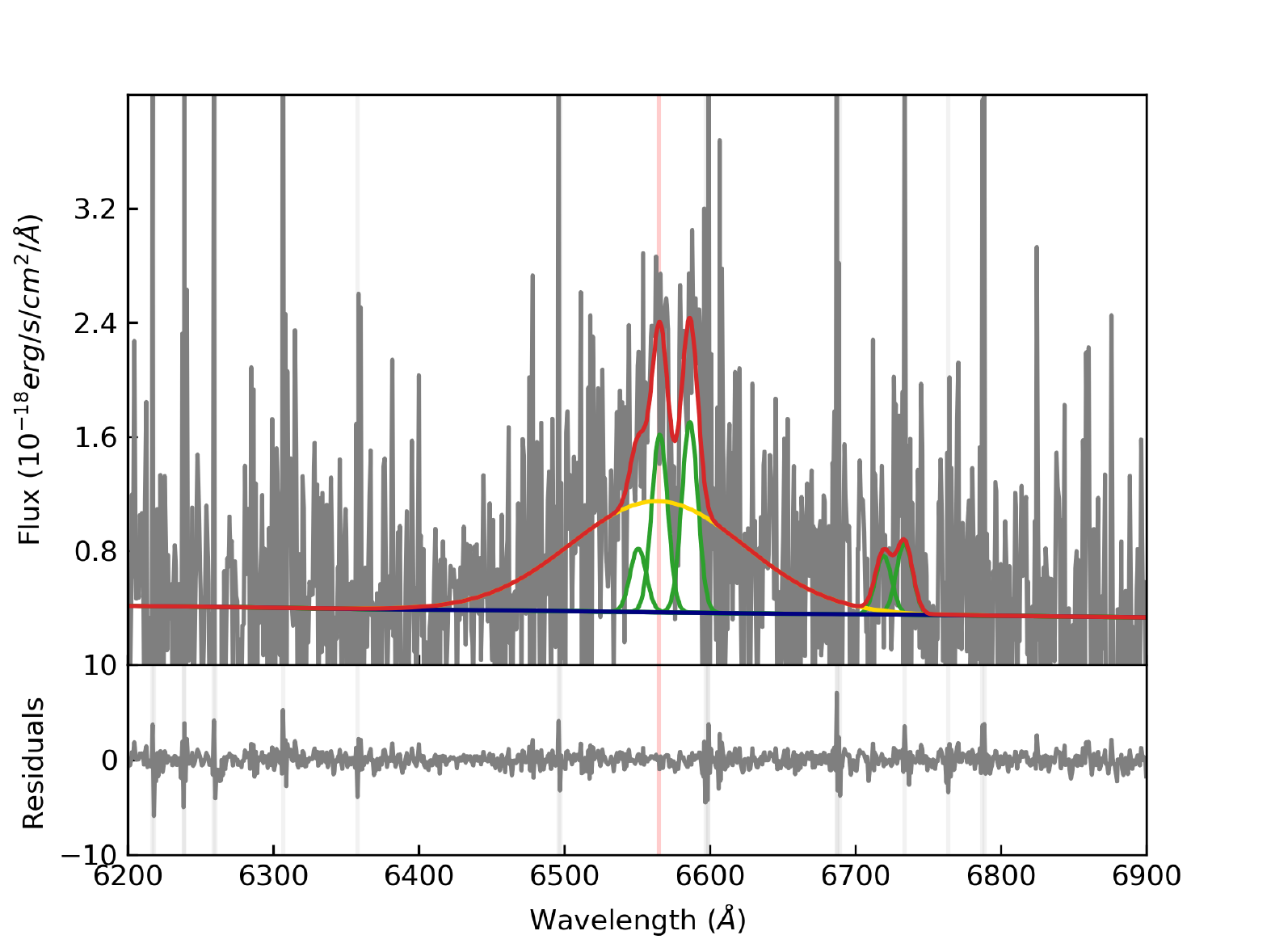}
\caption{X\_N\_35\_20. The modeling is the same as in Fig. \ref{fig:app}}
   \end{figure*}

 \begin{figure*}
 \center
    \includegraphics[width=0.4\textwidth]{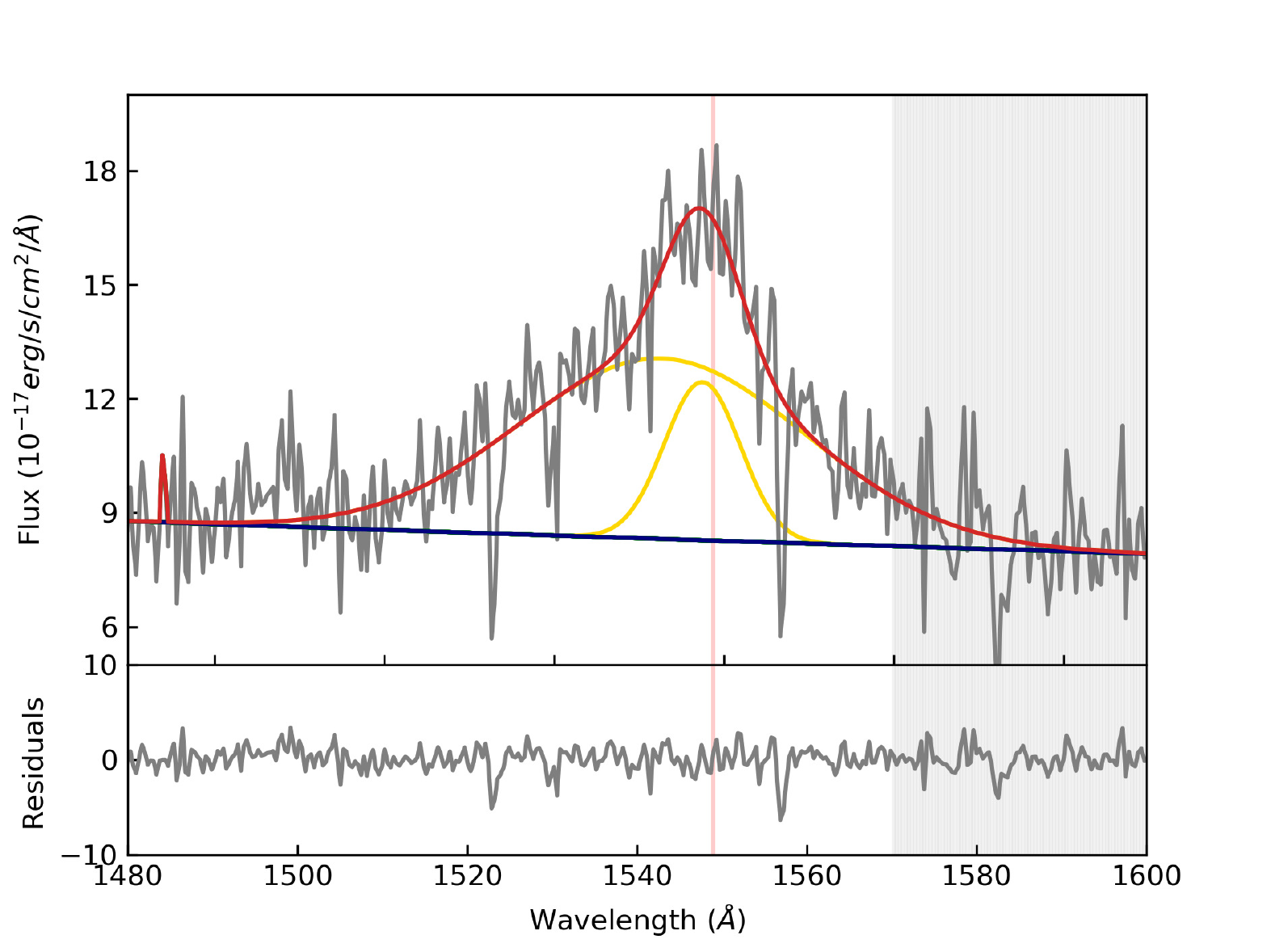}
   \includegraphics[width=0.4\textwidth]{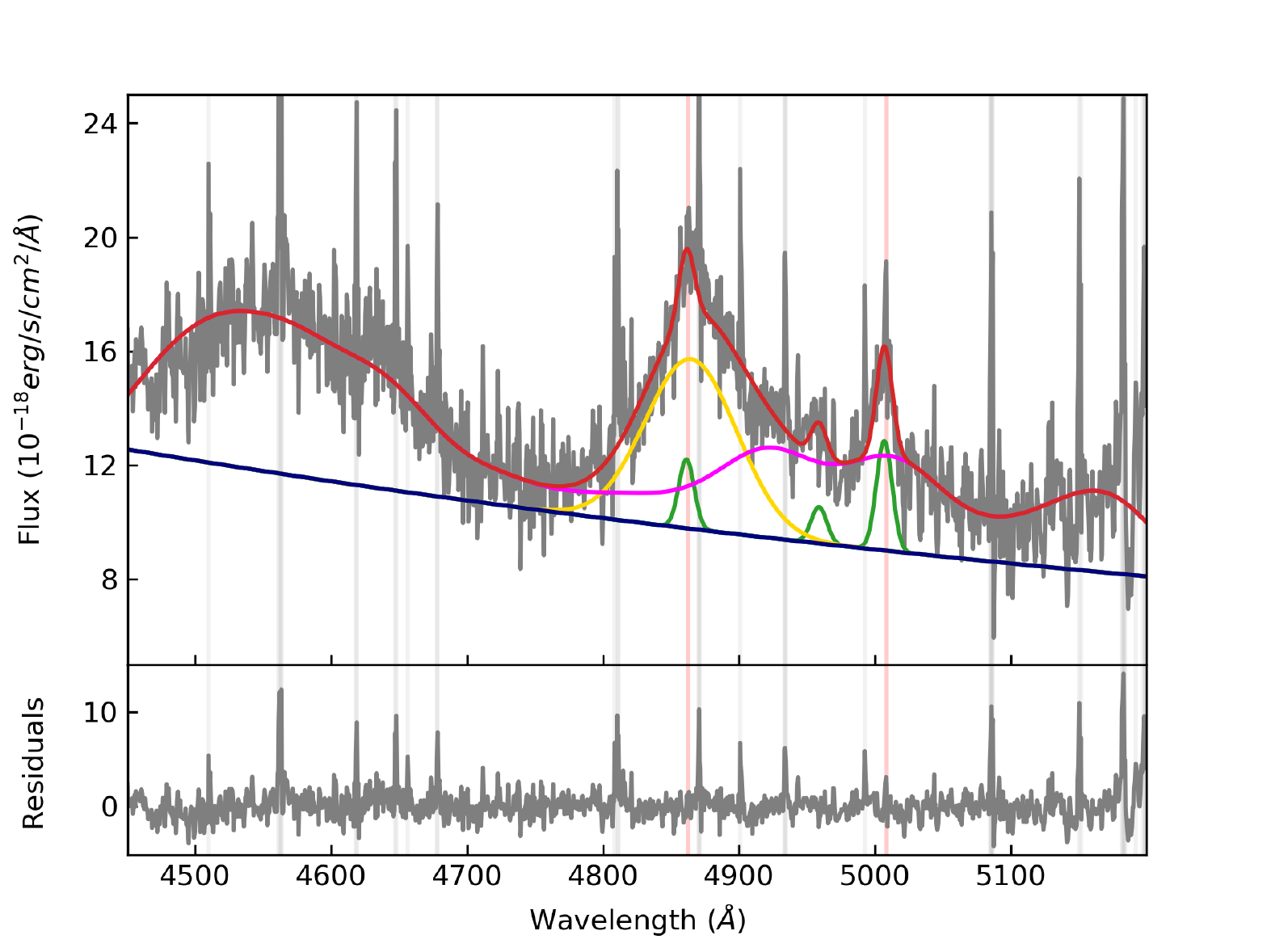}      \includegraphics[width=0.4\textwidth]{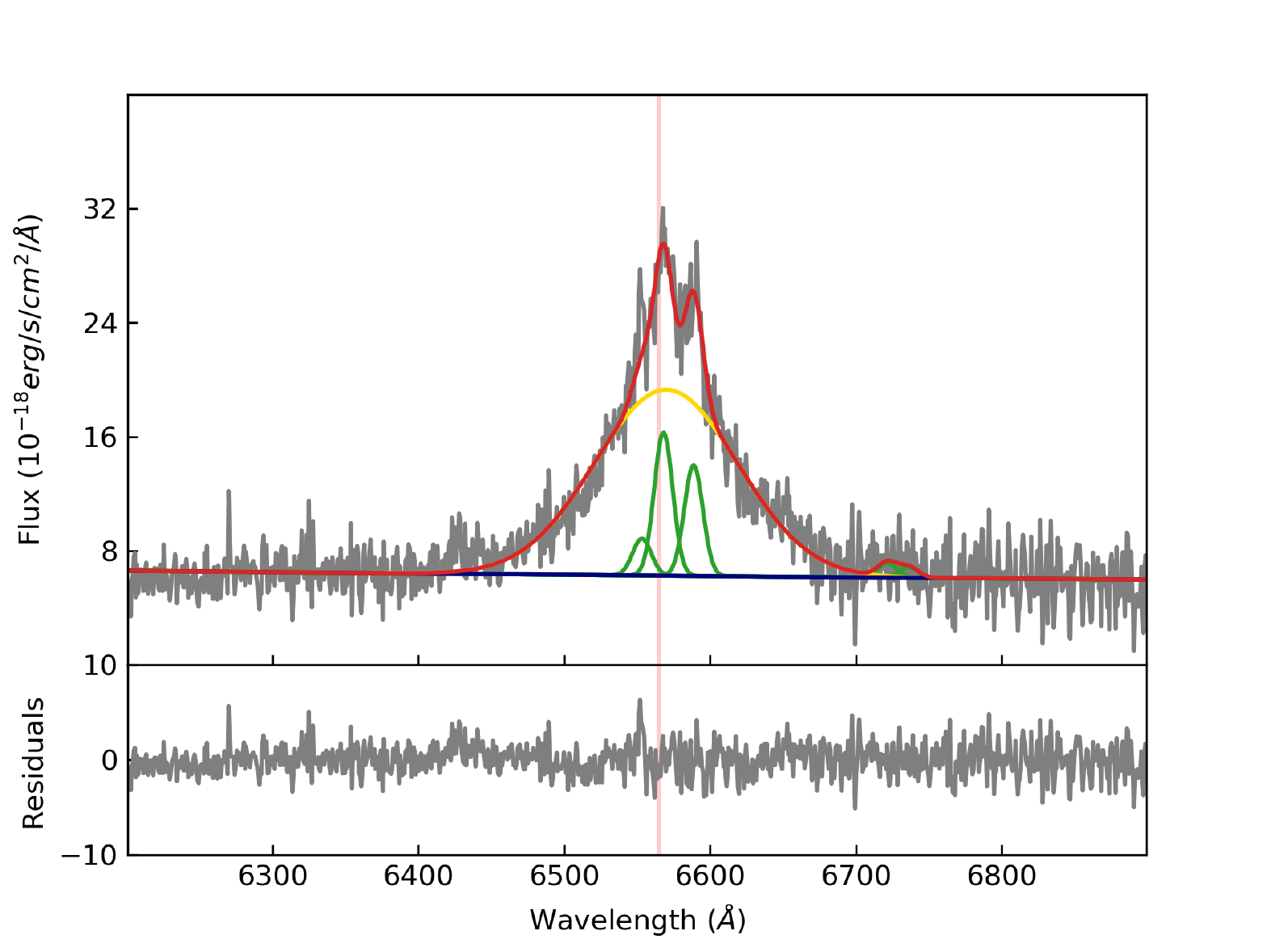}
\caption{X\_N\_12\_26. The modeling is the same as in Fig. \ref{fig:app}}
   \end{figure*}

 \begin{figure*}
 \center
    \includegraphics[width=0.4\textwidth]{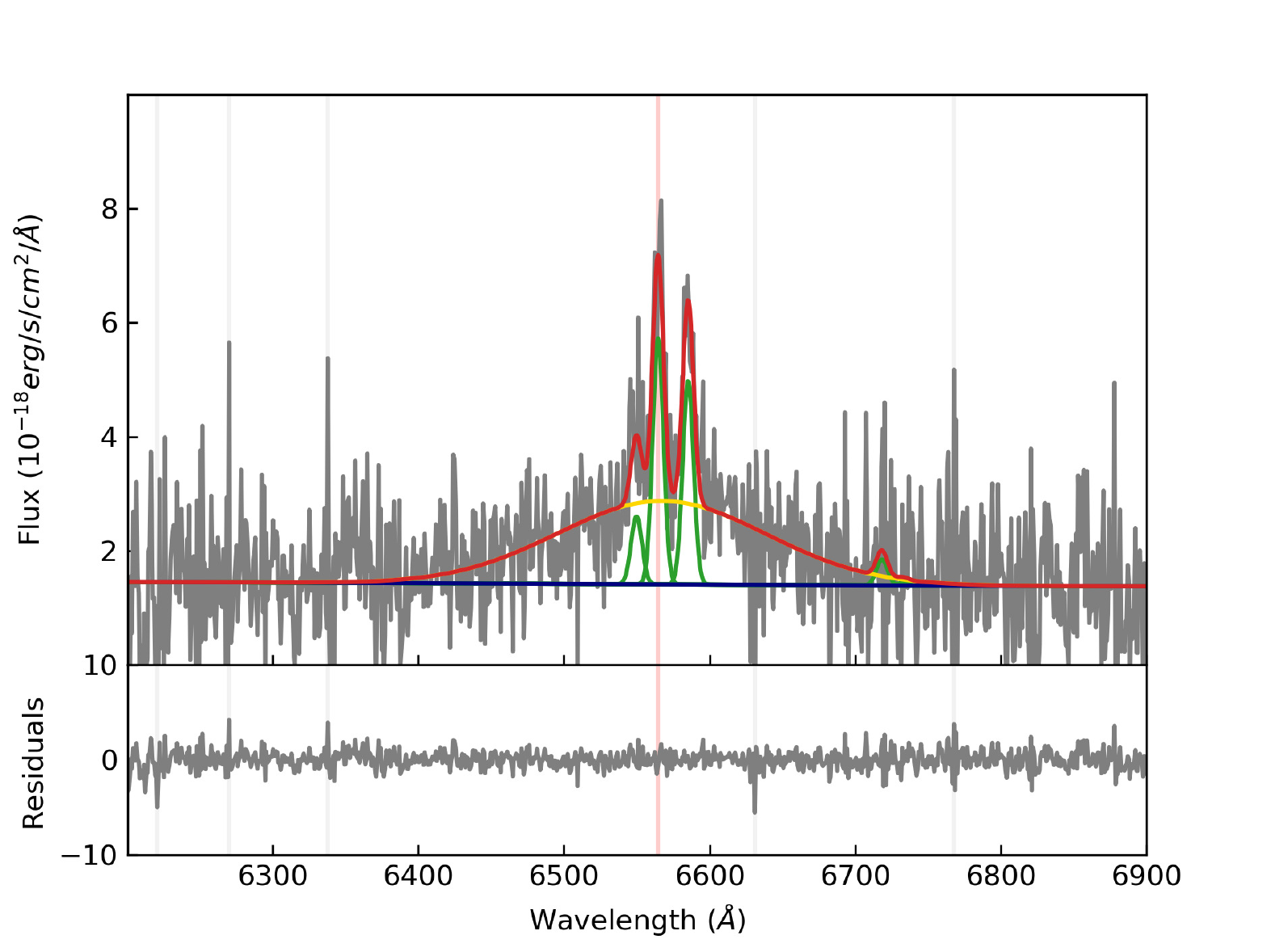}
\caption{X\_N\_44\_64. The modeling is the same as in Fig. \ref{fig:app}}
   \end{figure*}

 \begin{figure*}
 \center
    \includegraphics[width=0.4\textwidth]{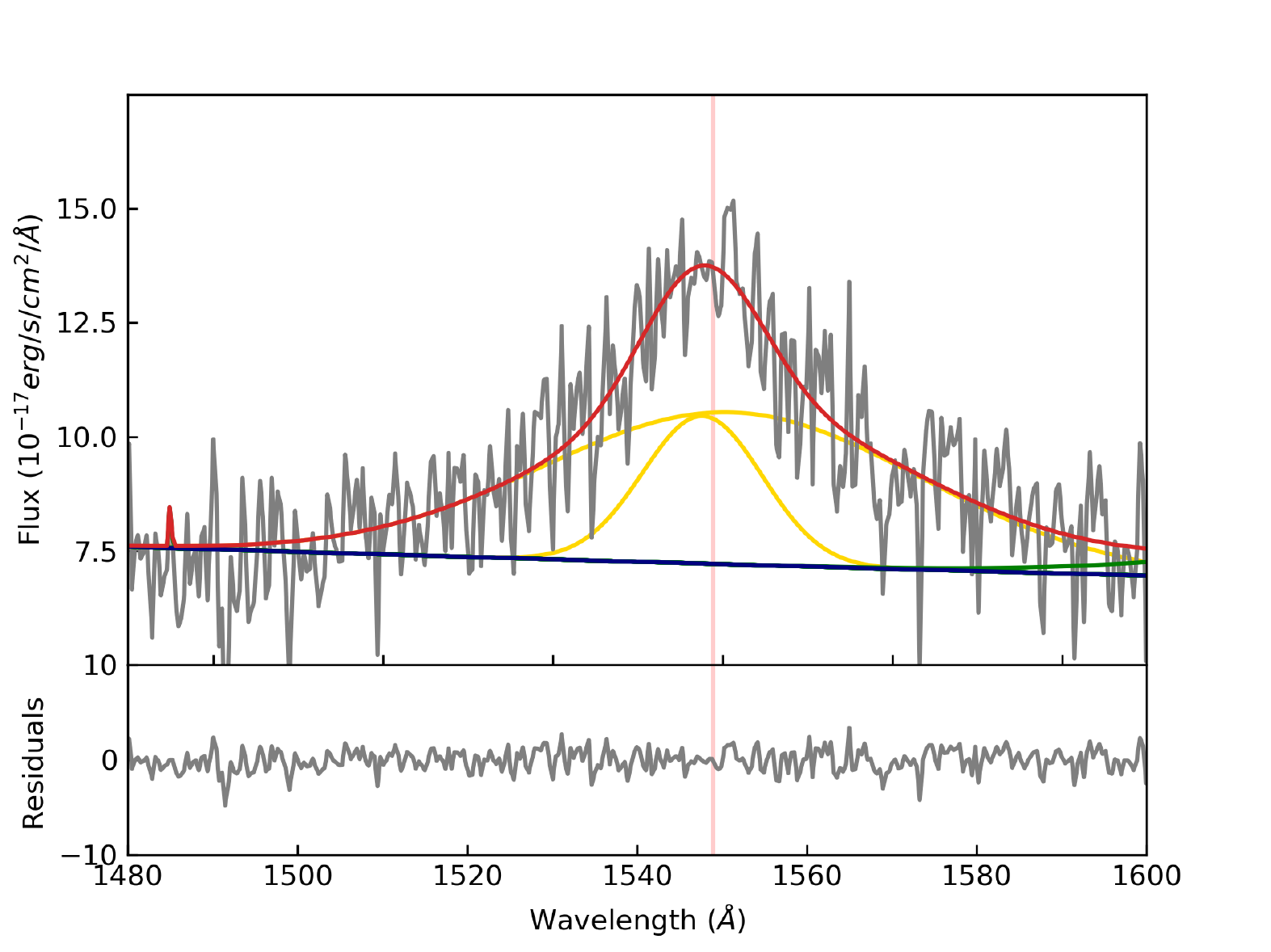}
   \includegraphics[width=0.4\textwidth]{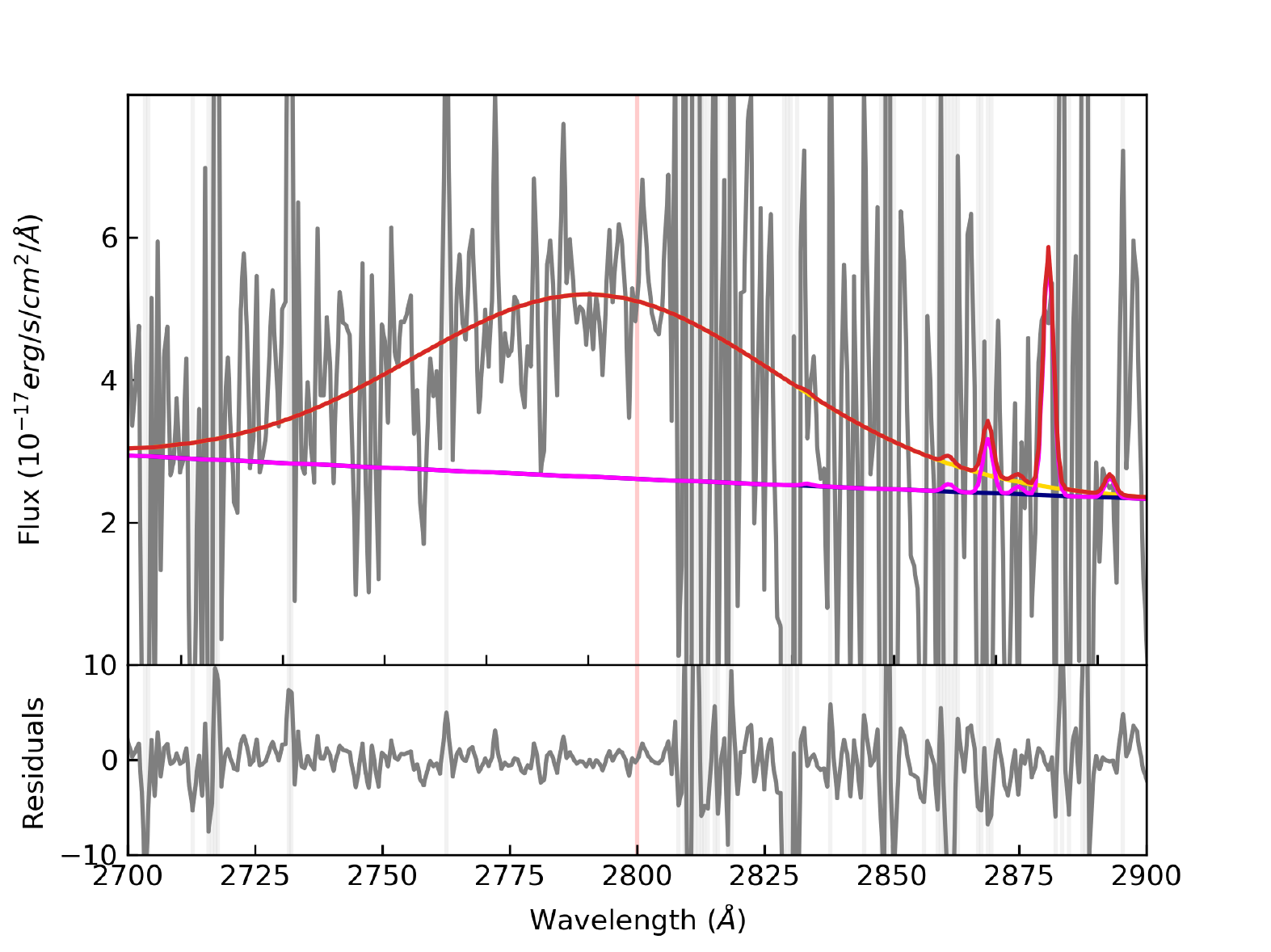}
   \includegraphics[width=0.4\textwidth]{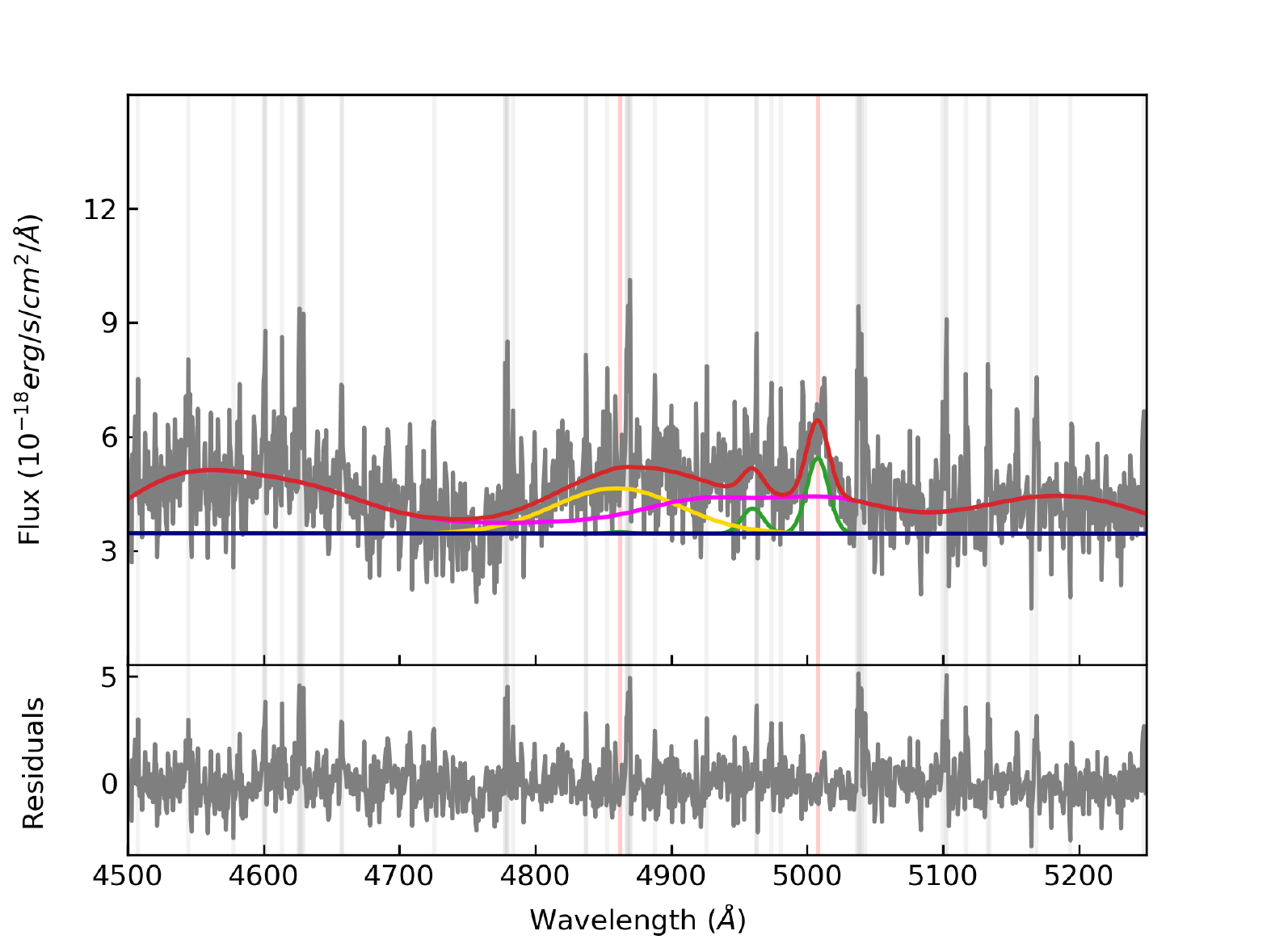}   
      \includegraphics[width=0.4\textwidth]{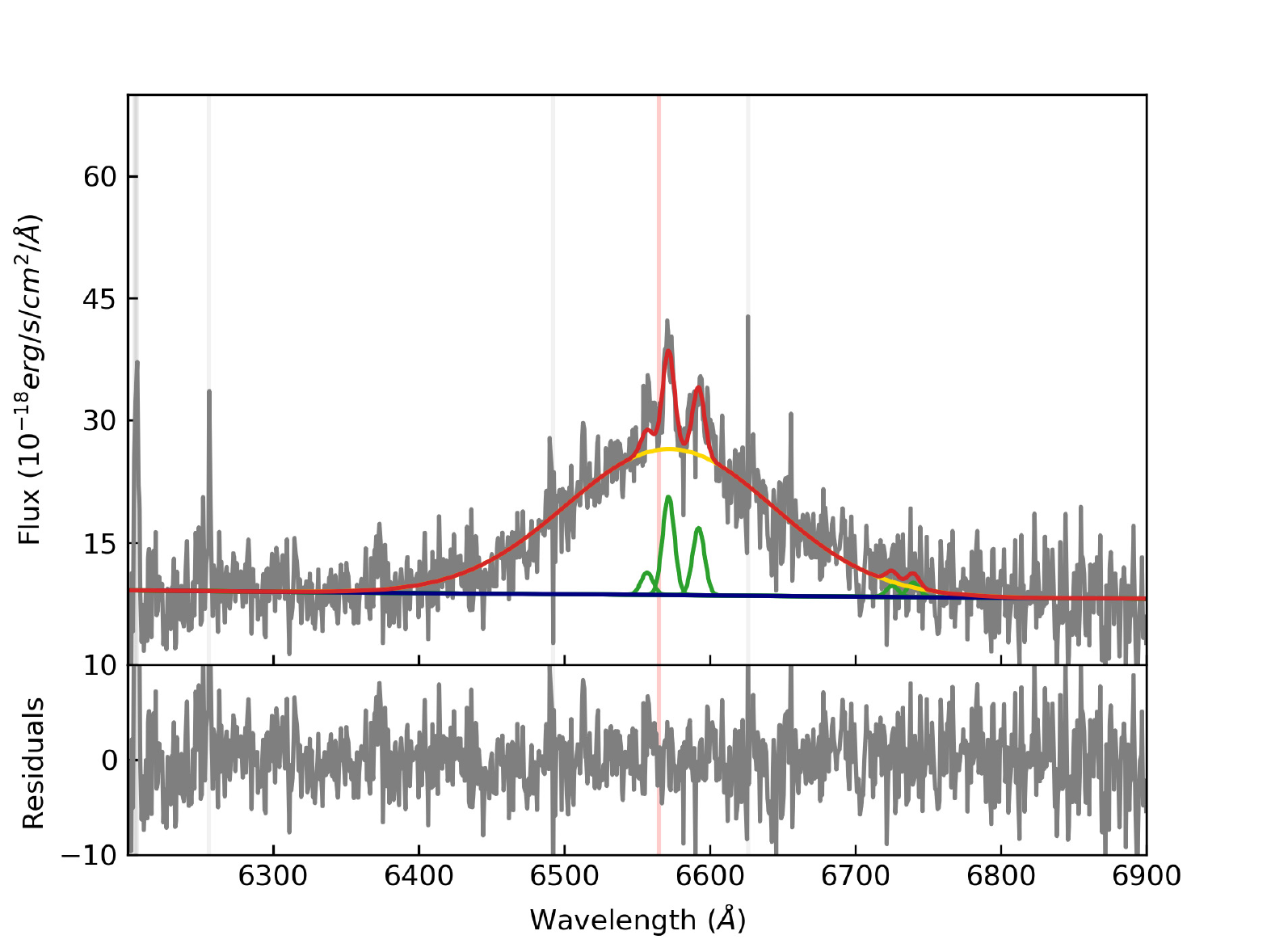}
\caption{X\_N\_4\_48. The modeling is the same as in Fig. \ref{fig:app}}
   \end{figure*}

 \begin{figure*}
 \center
    \includegraphics[width=0.4\textwidth]{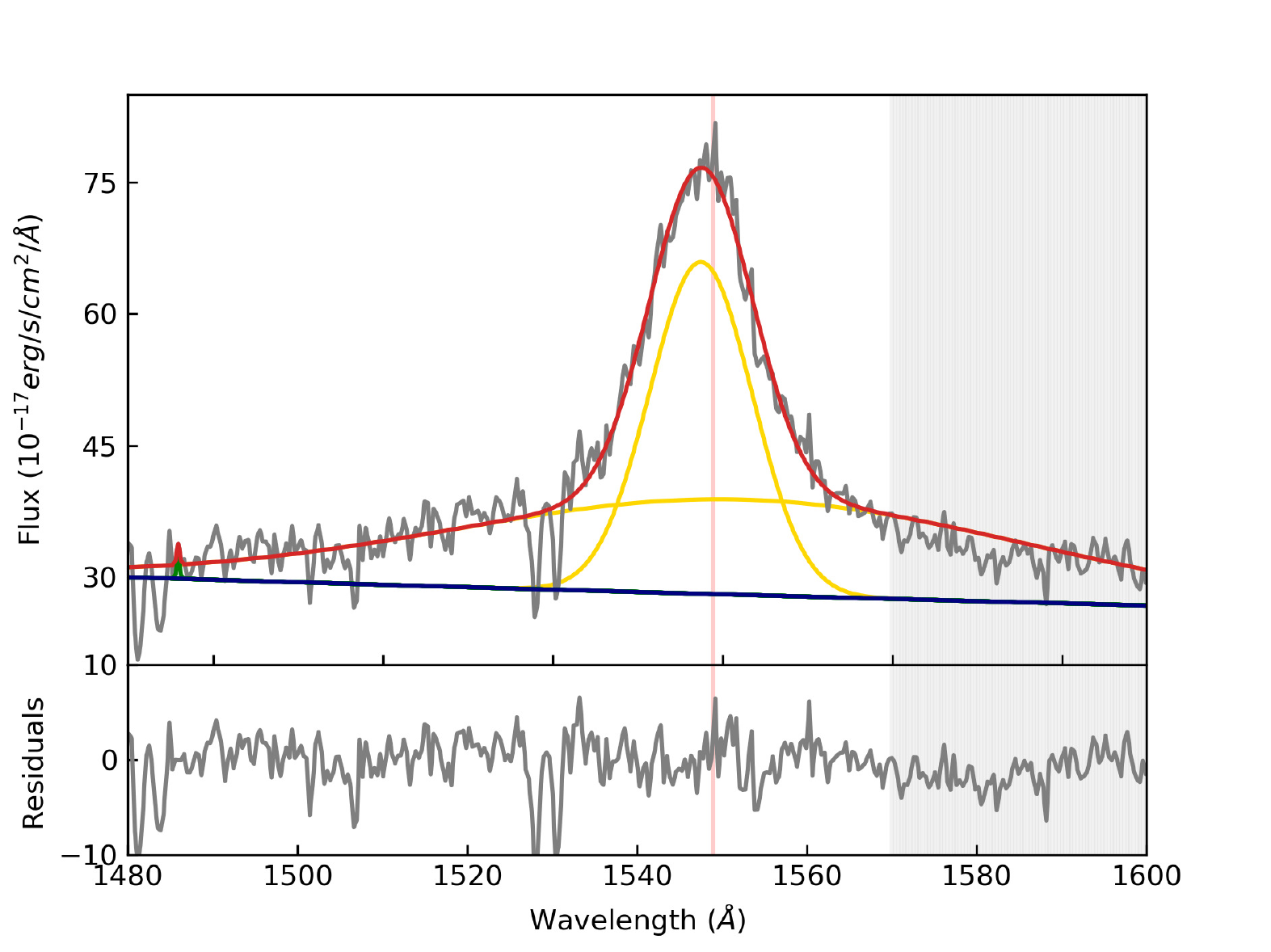}
   \includegraphics[width=0.4\textwidth]{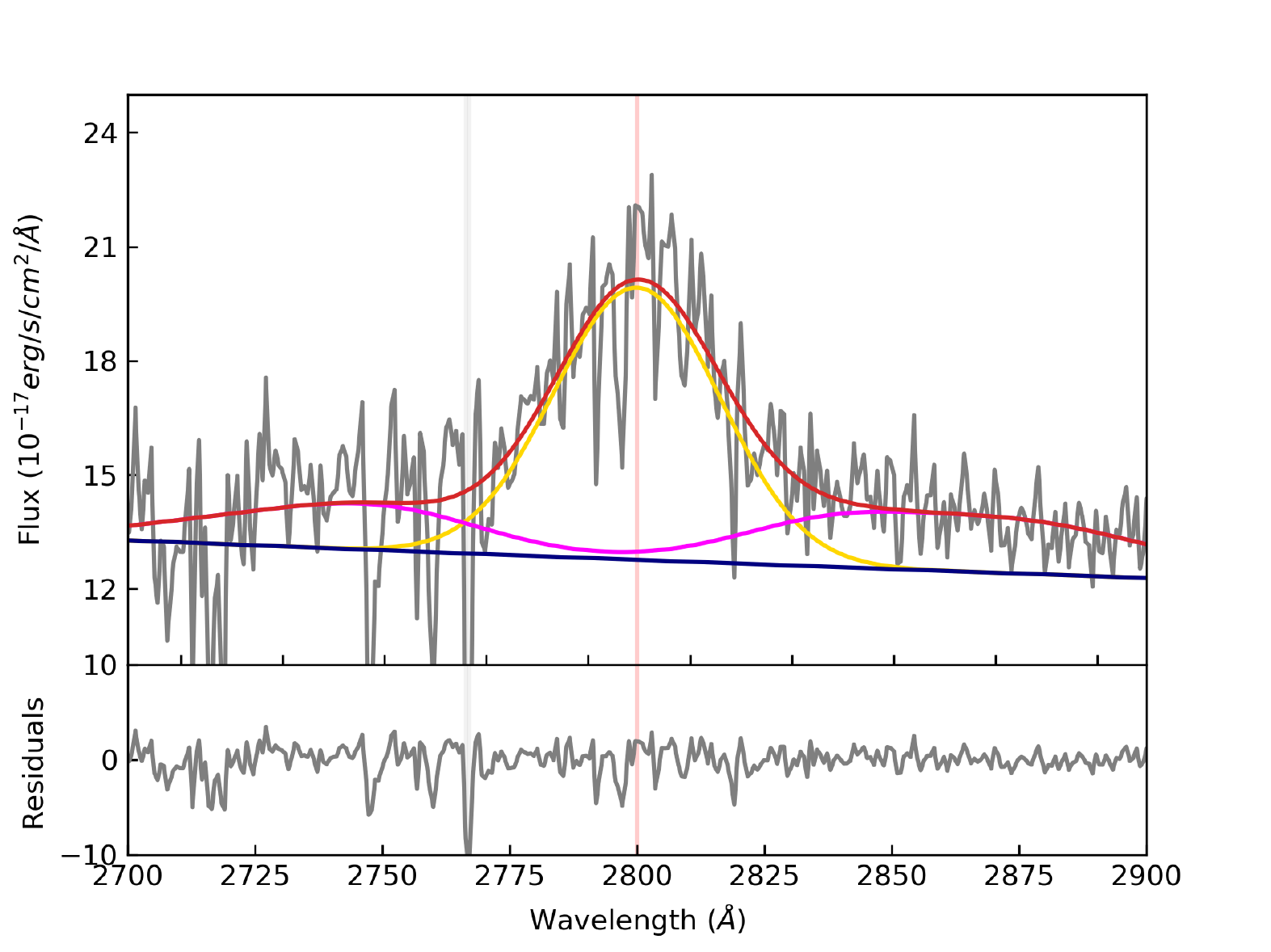}
   \includegraphics[width=0.4\textwidth]{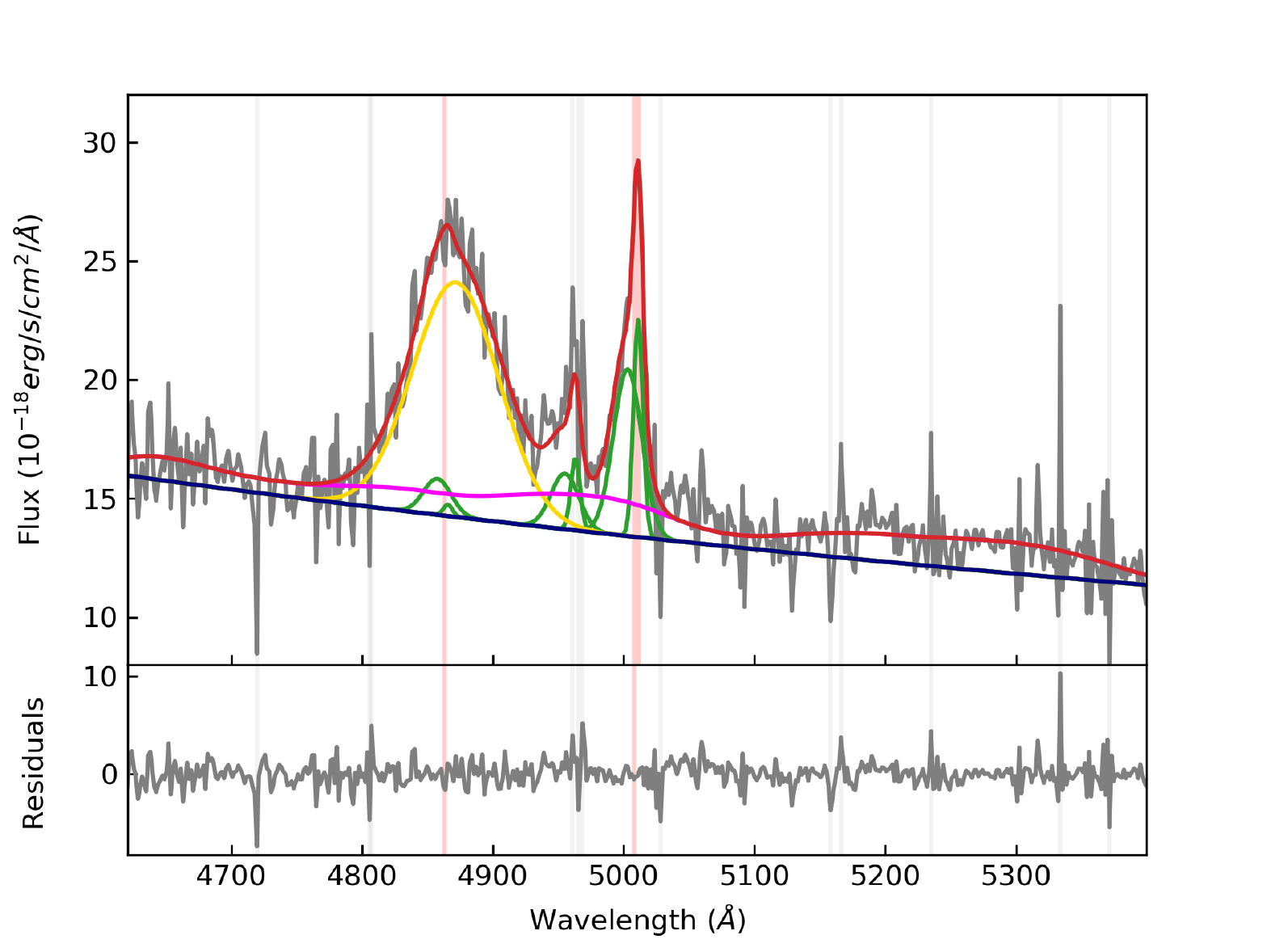}      \includegraphics[width=0.4\textwidth]{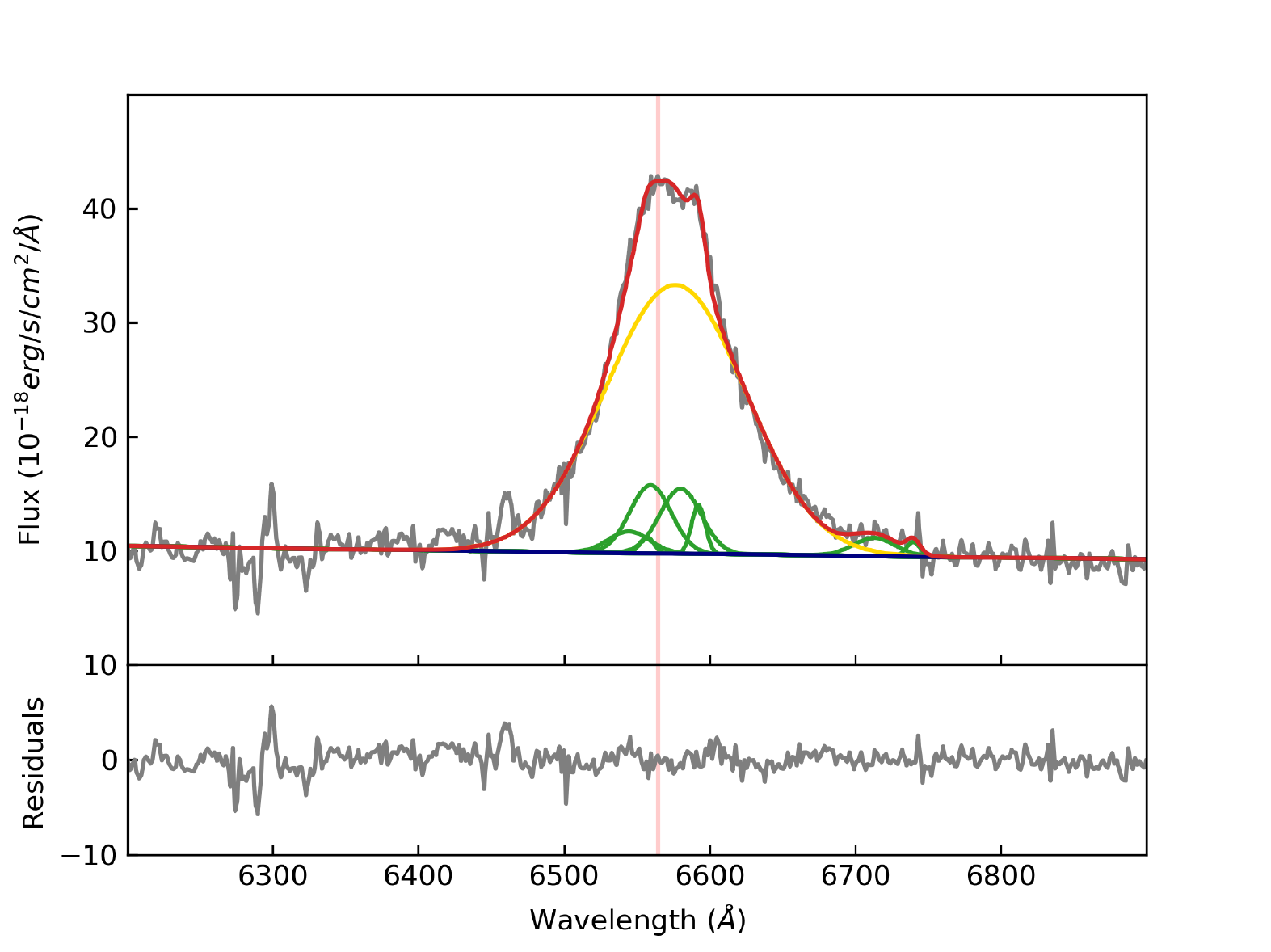}
\caption{X\_N\_102\_35. The modeling is the same as in Fig. \ref{fig:app}}
   \end{figure*}

 \begin{figure*}
 \center
    \includegraphics[width=0.4\textwidth]{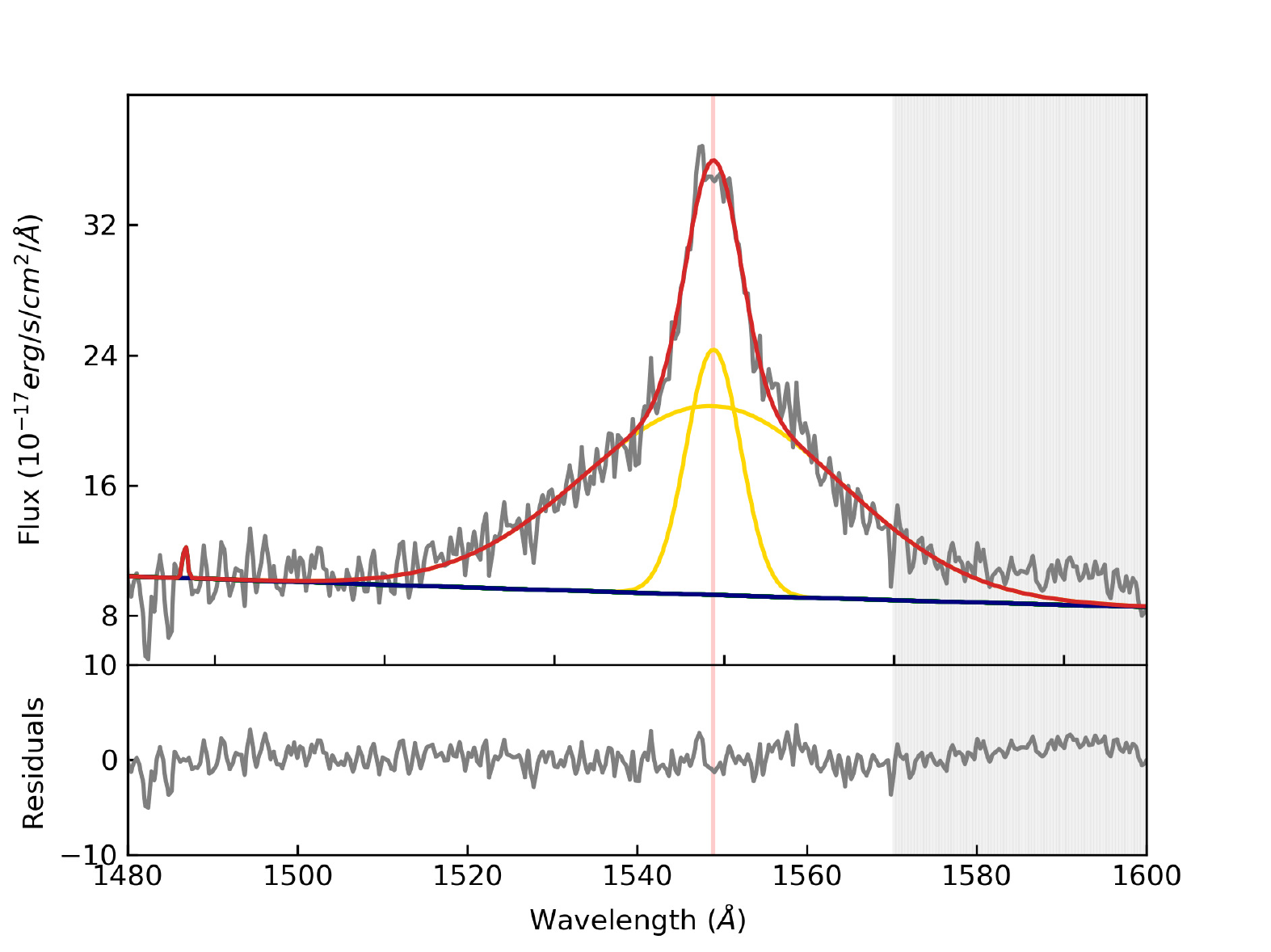}
   \includegraphics[width=0.4\textwidth]{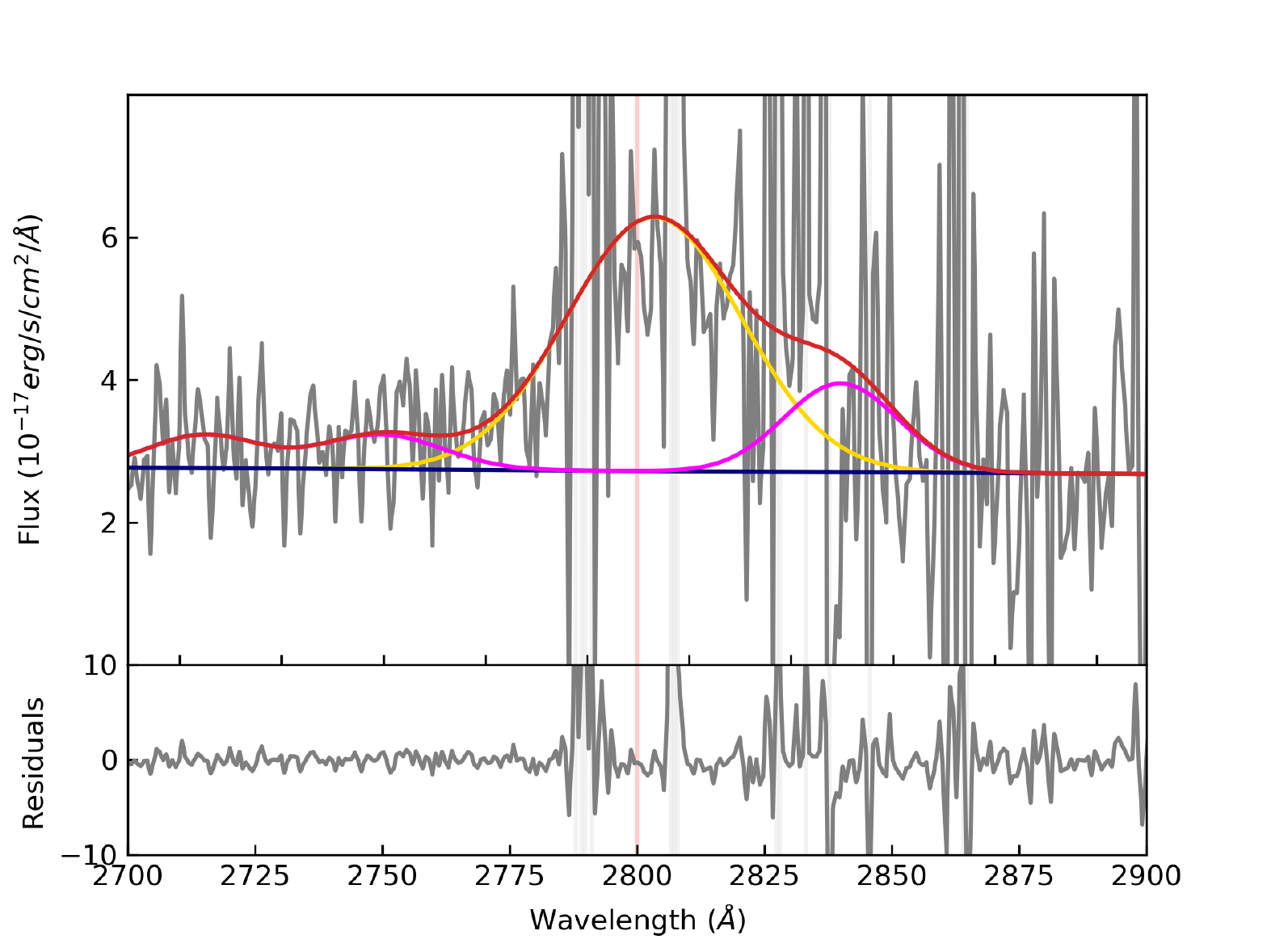}
   \includegraphics[width=0.4\textwidth]{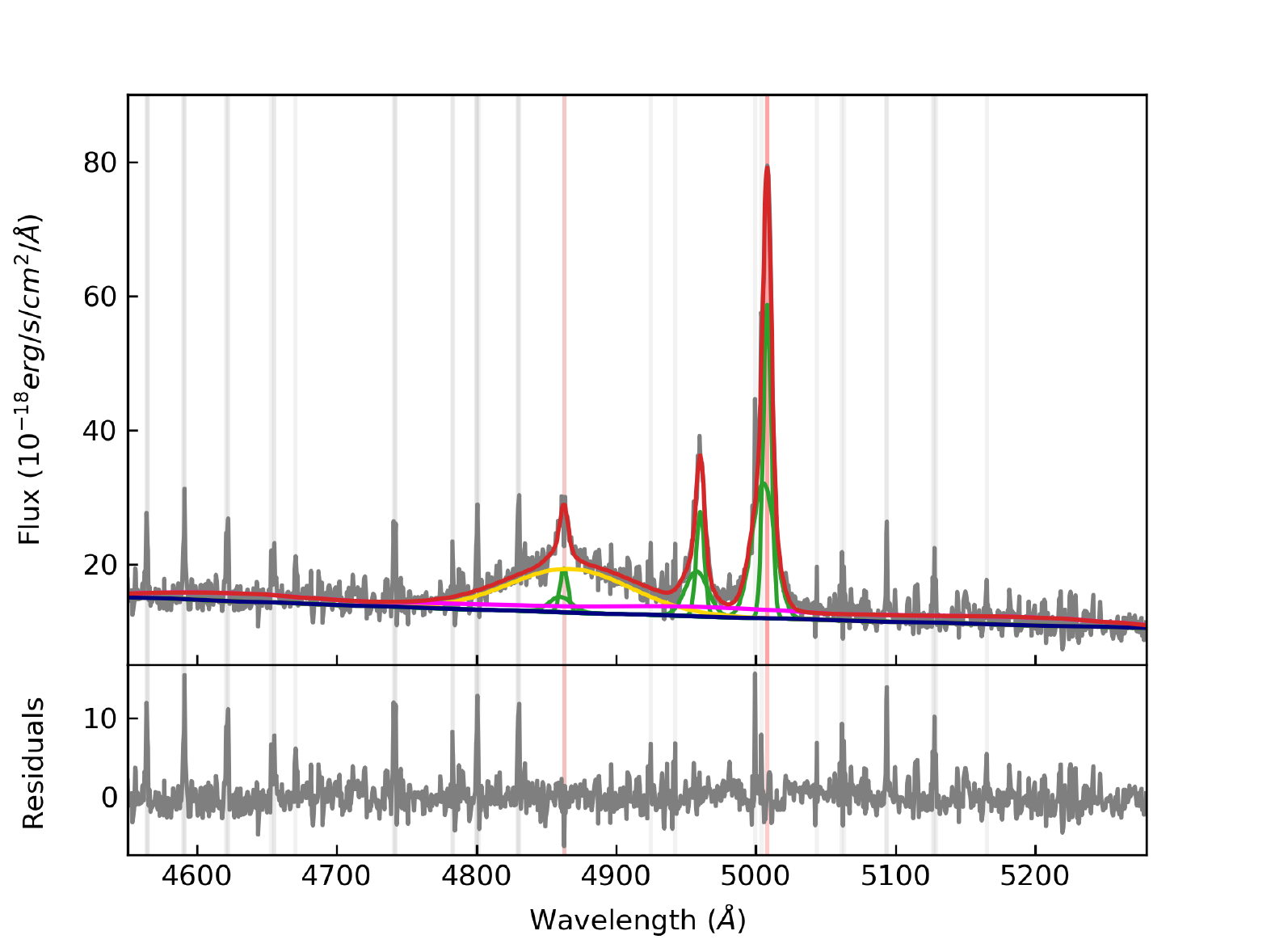}      \includegraphics[width=0.4\textwidth]{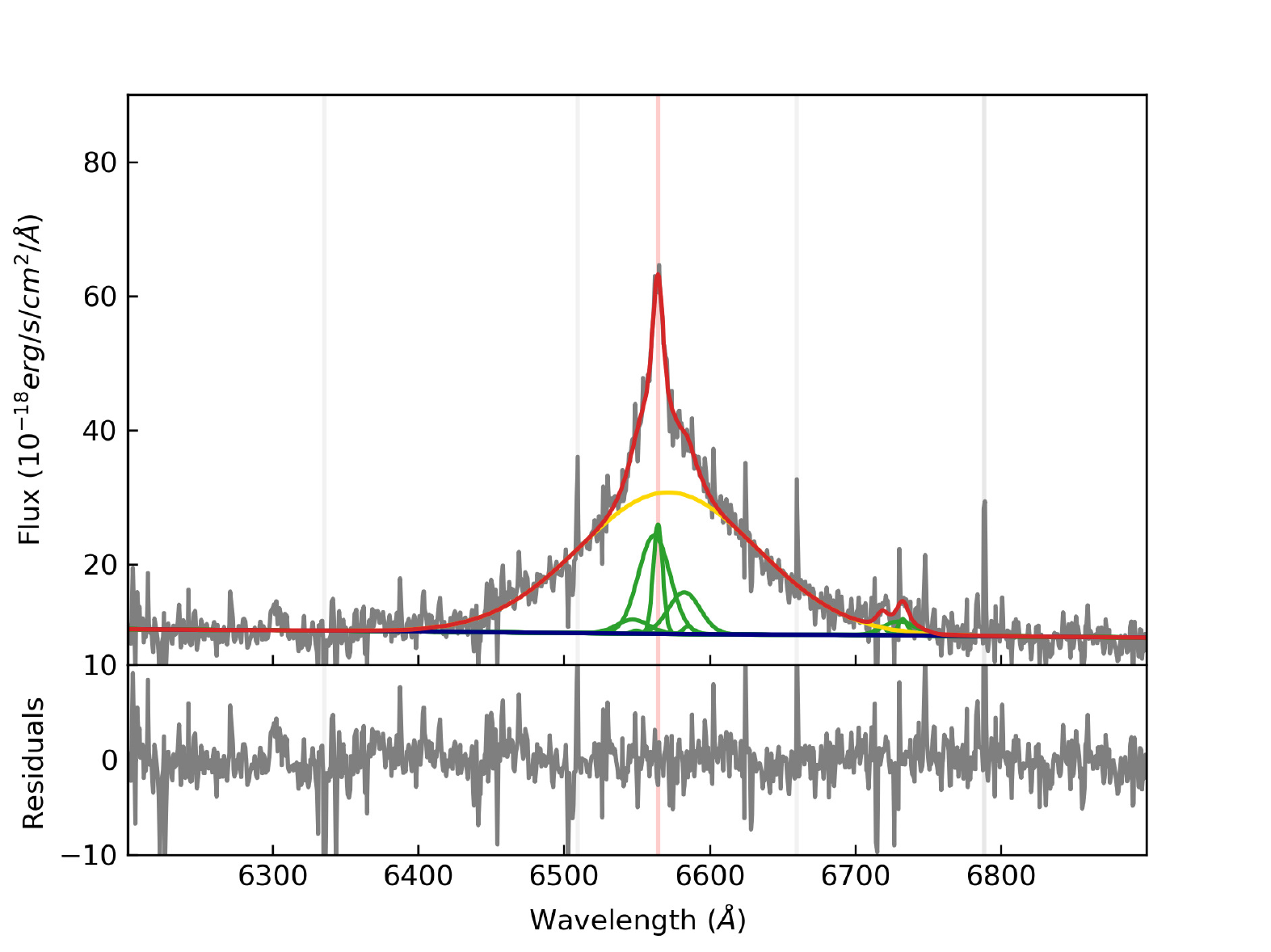}
\caption{X\_N\_115\_23. The modeling is the same as in Fig. \ref{fig:app}}
   \end{figure*}

 \begin{figure*}
 \center
    \includegraphics[width=0.4\textwidth]{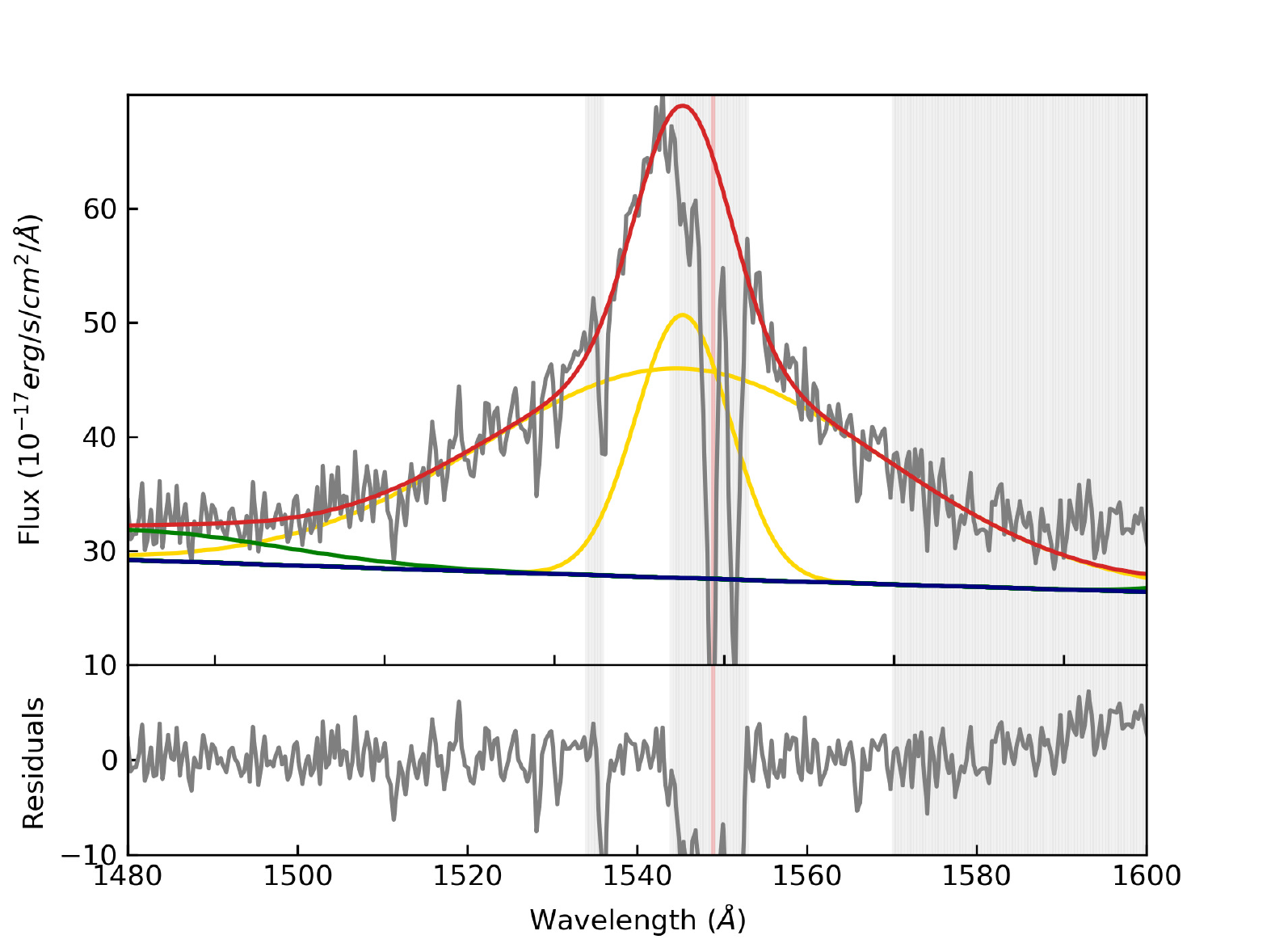}
   \includegraphics[width=0.4\textwidth]{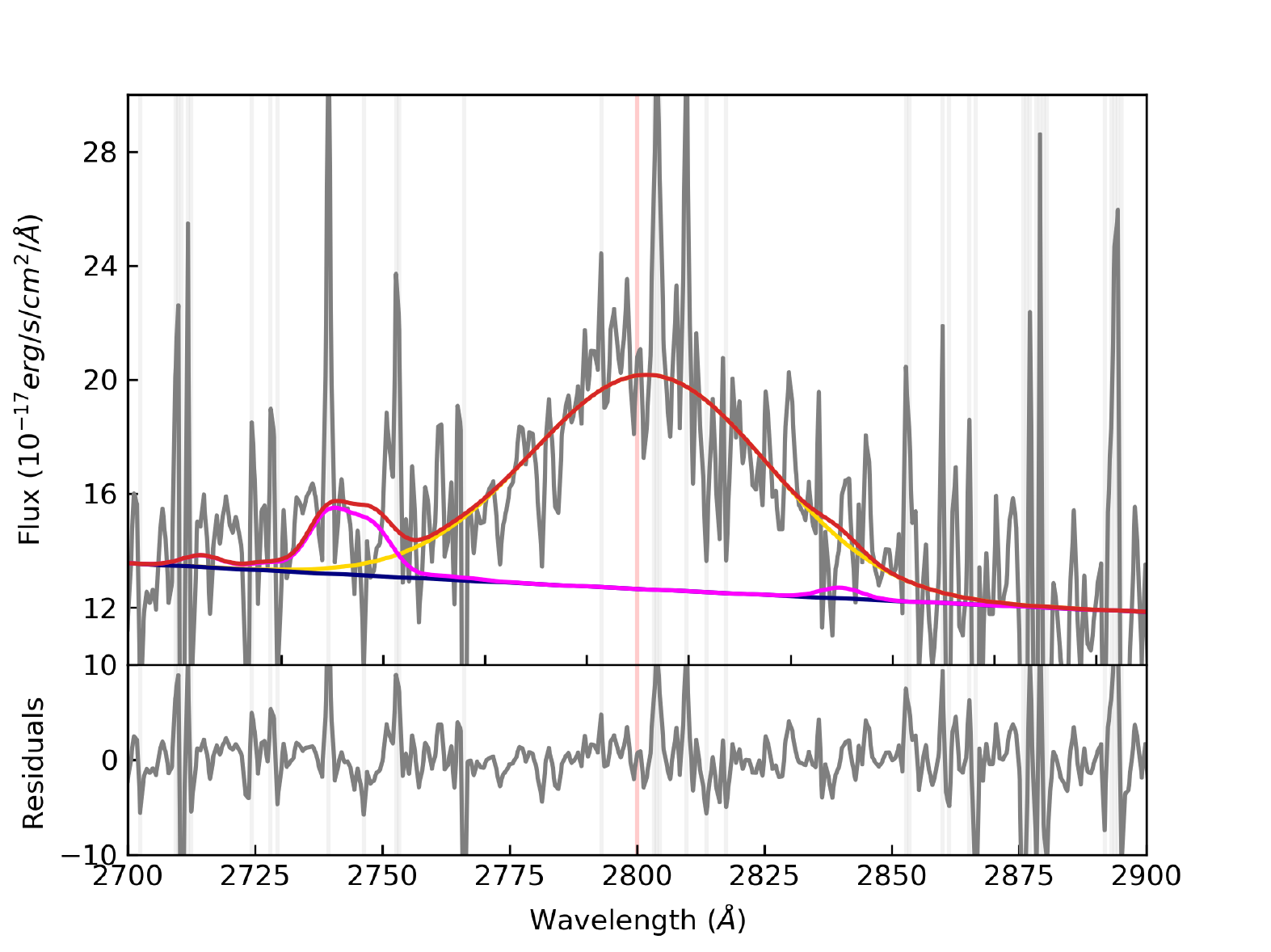}
   \includegraphics[width=0.4\textwidth]{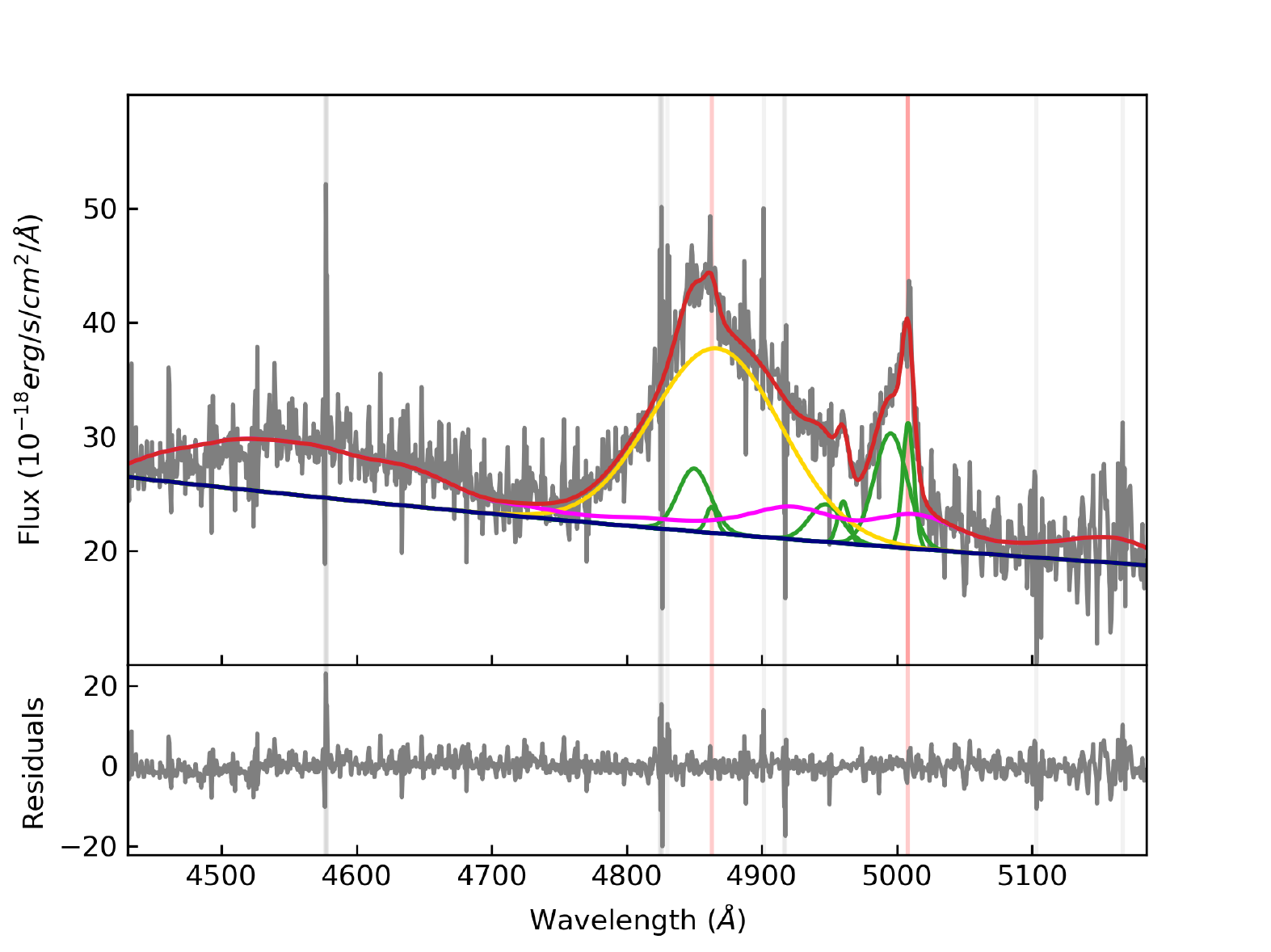}  
       \includegraphics[width=0.4\textwidth]{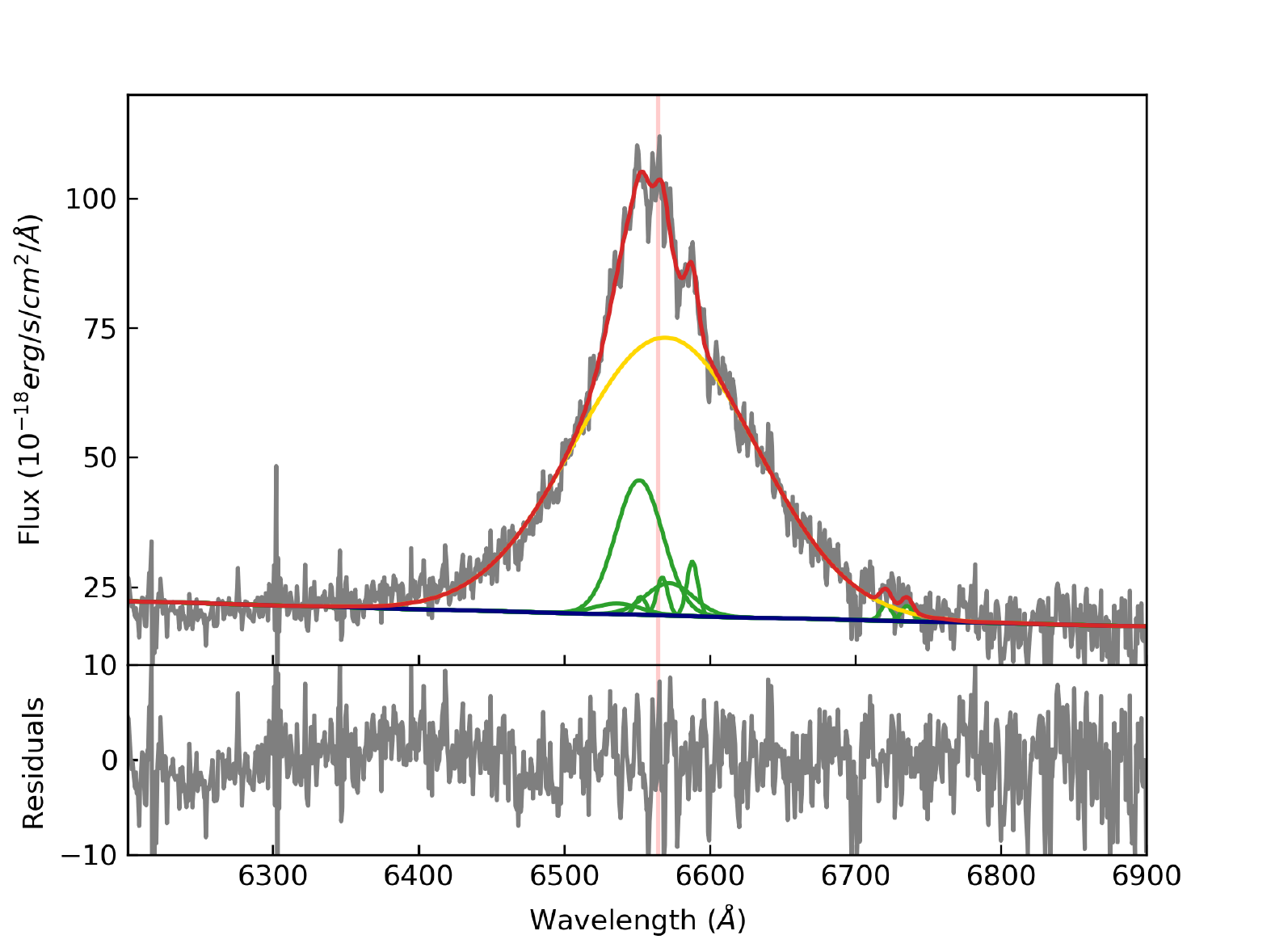}
\caption{cid\_166. The modeling is the same as in Fig. \ref{fig:app}}
   \end{figure*}

 \begin{figure*}
 \center
    \includegraphics[width=0.4\textwidth]{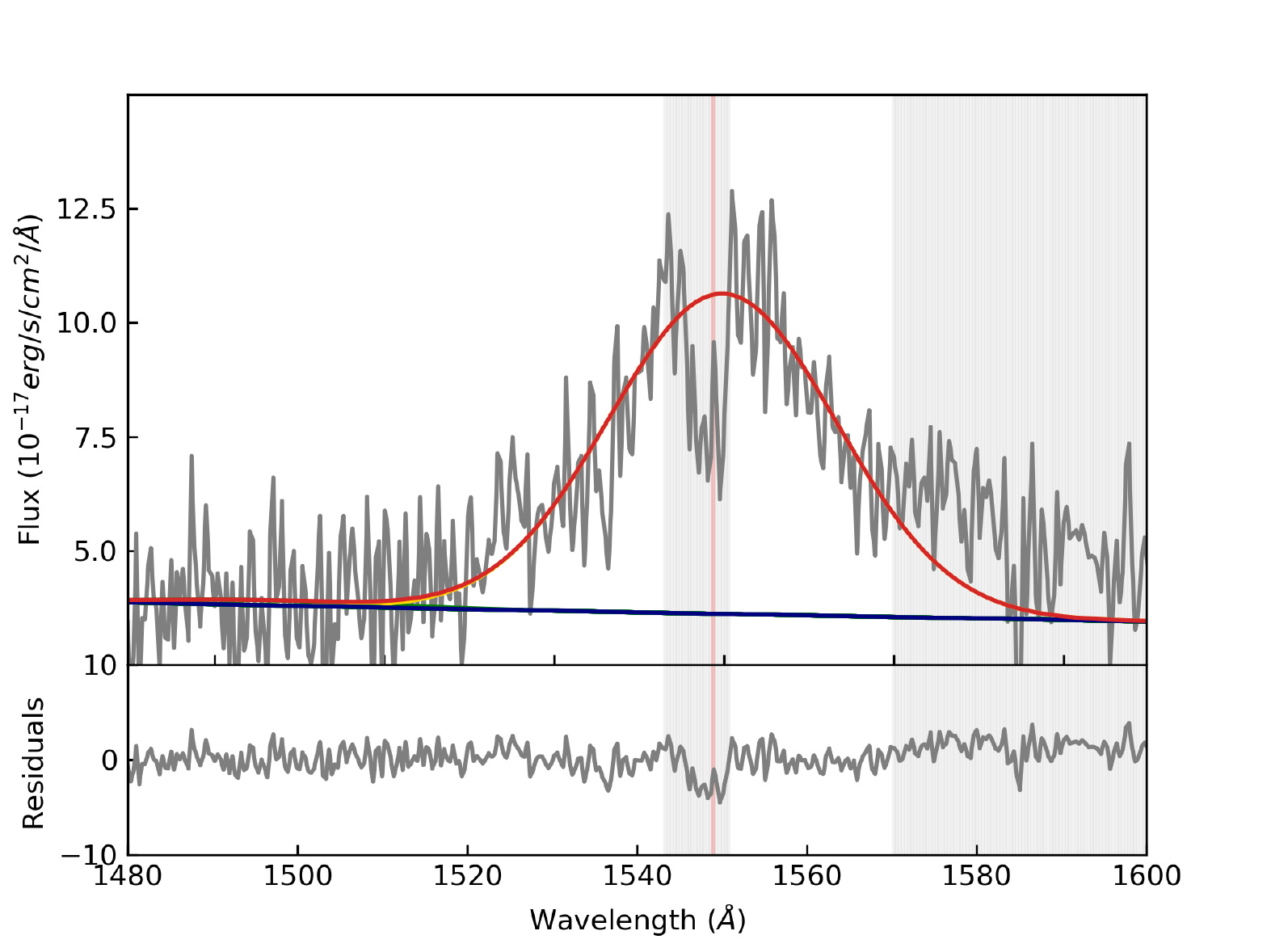}
   \includegraphics[width=0.4\textwidth]{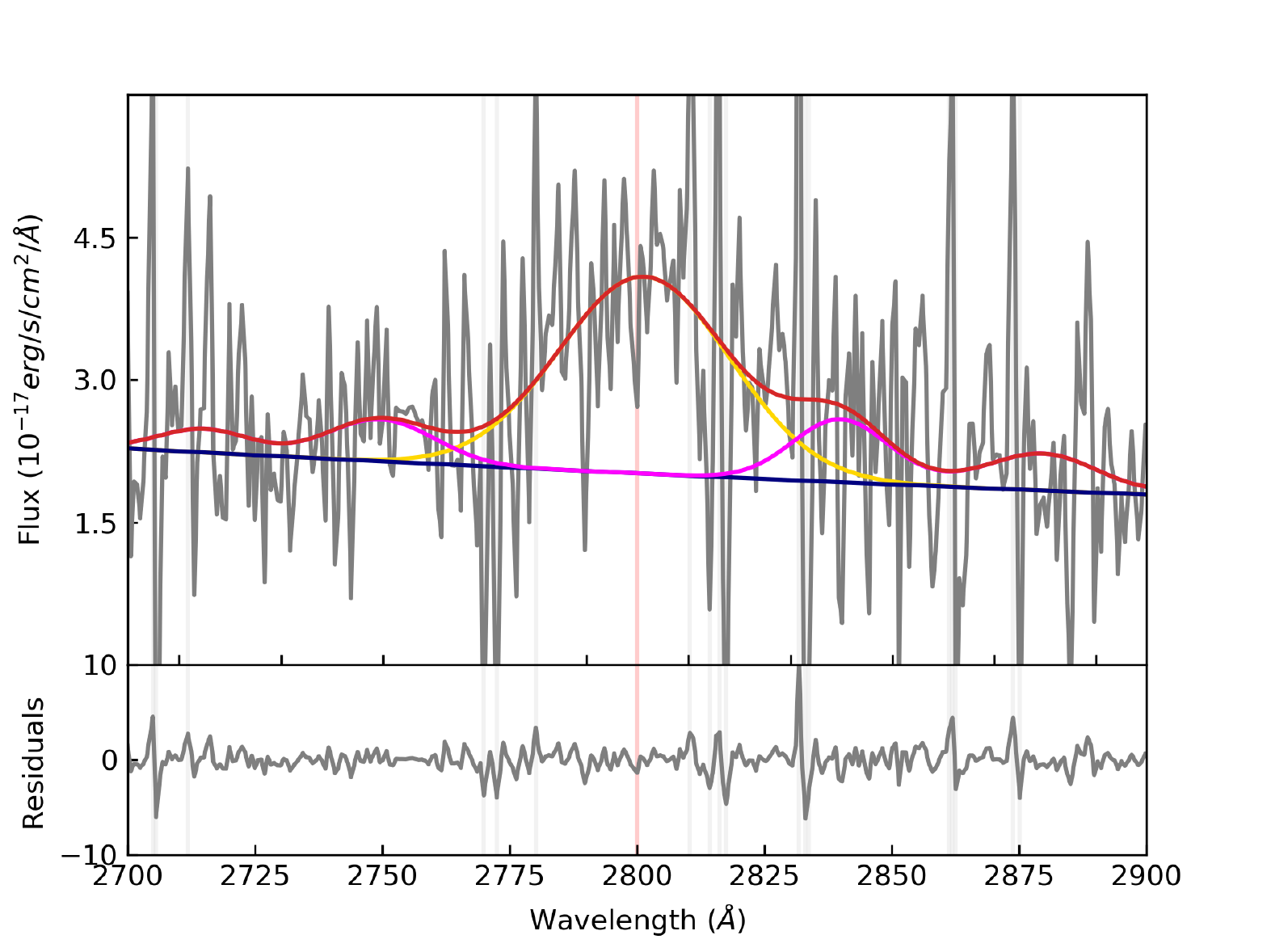}
   \includegraphics[width=0.4\textwidth]{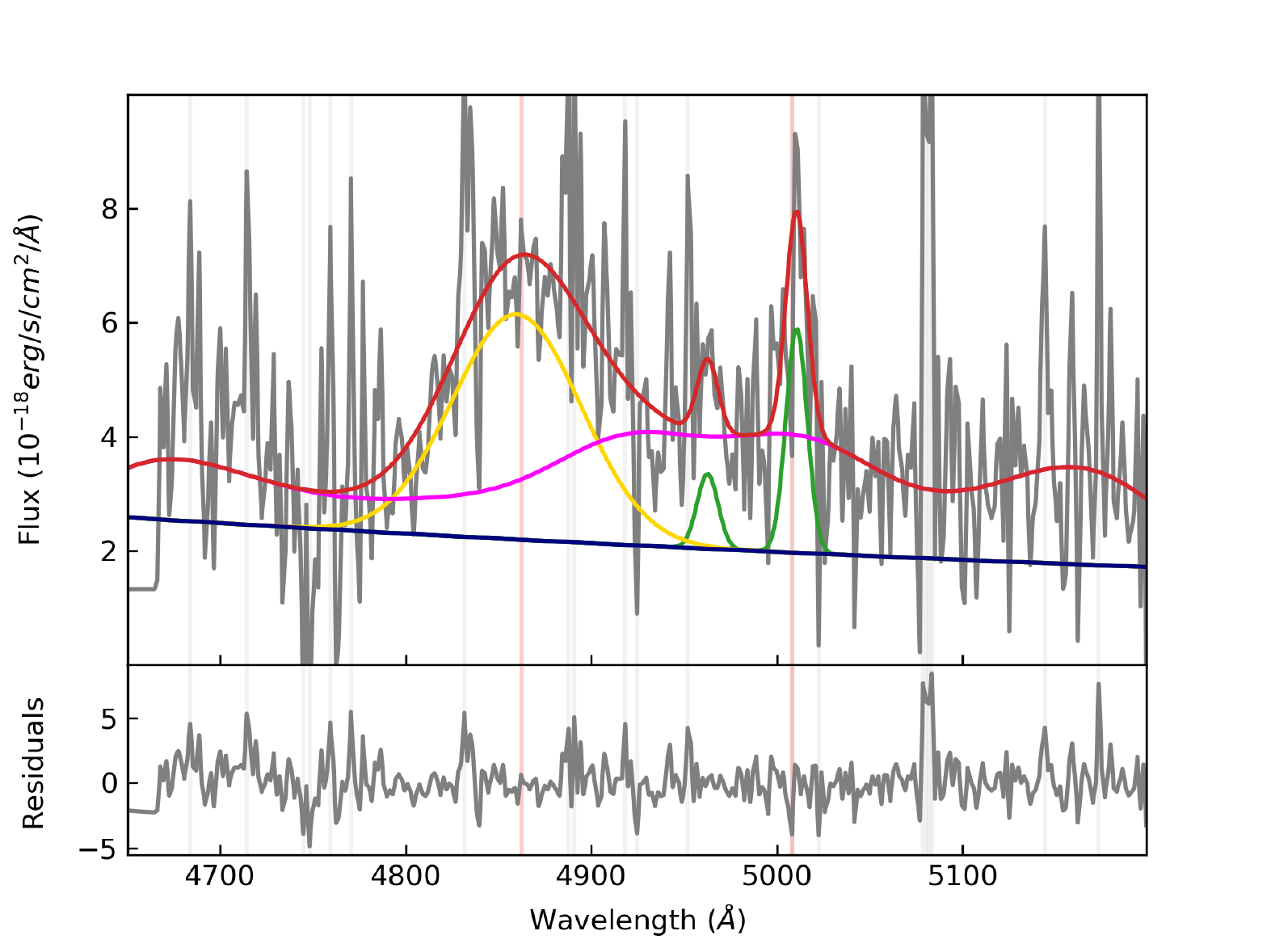}      \includegraphics[width=0.4\textwidth]{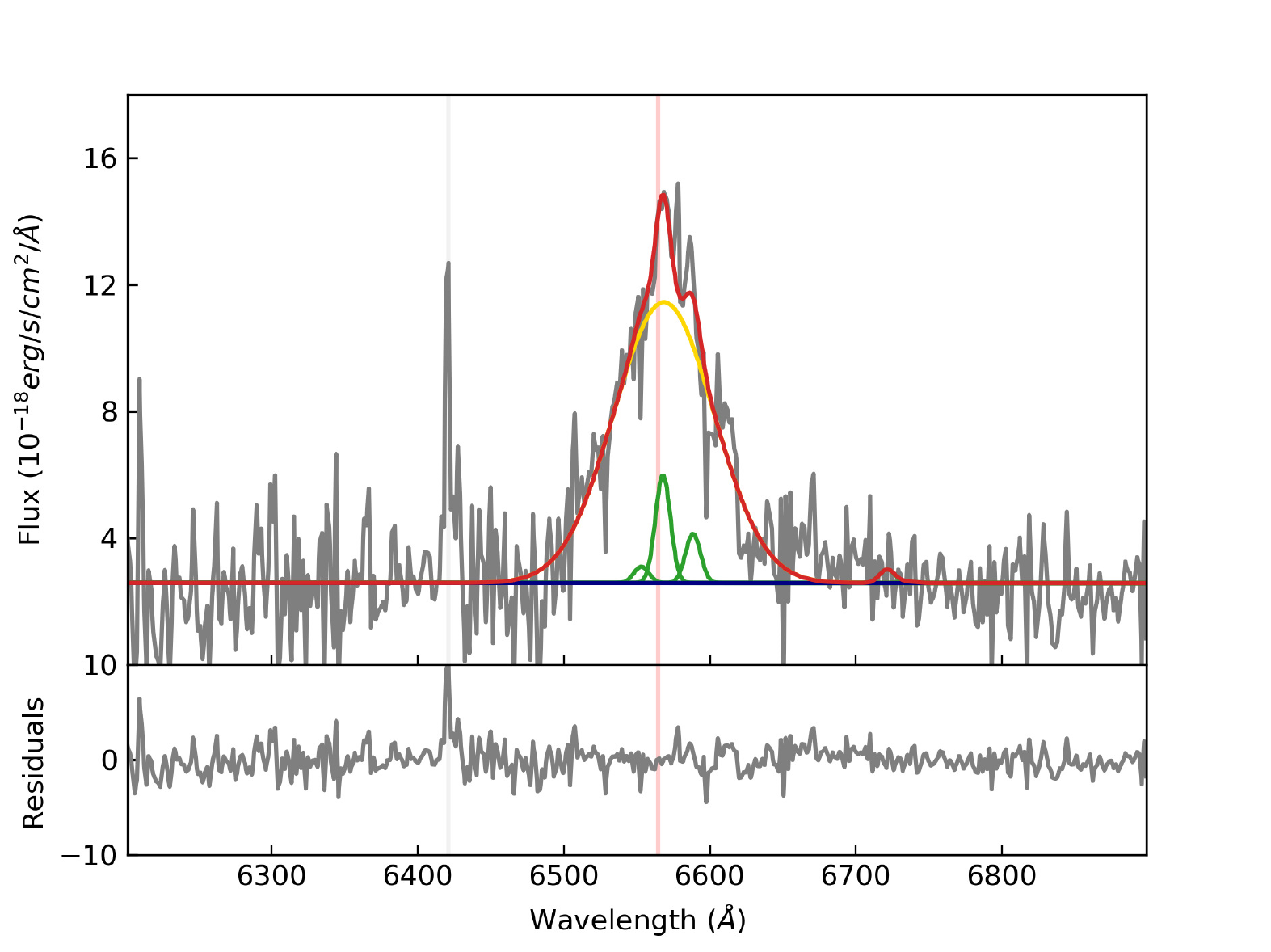}
\caption{cid\_1605. The modeling is the same as in Fig. \ref{fig:app}}
   \end{figure*}   
   
    \begin{figure*}
 \center
    \includegraphics[width=0.4\textwidth]{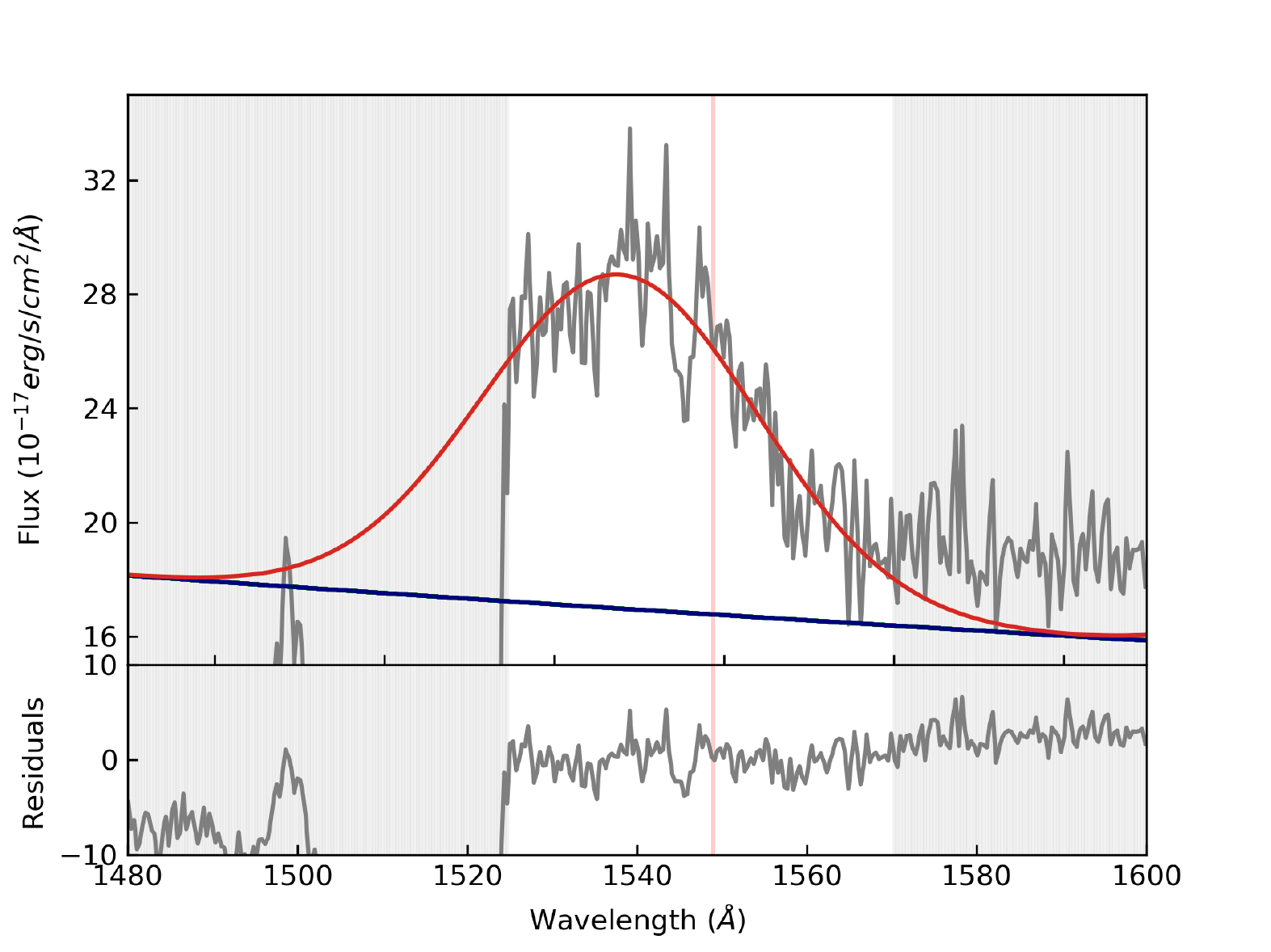}
   \includegraphics[width=0.4\textwidth]{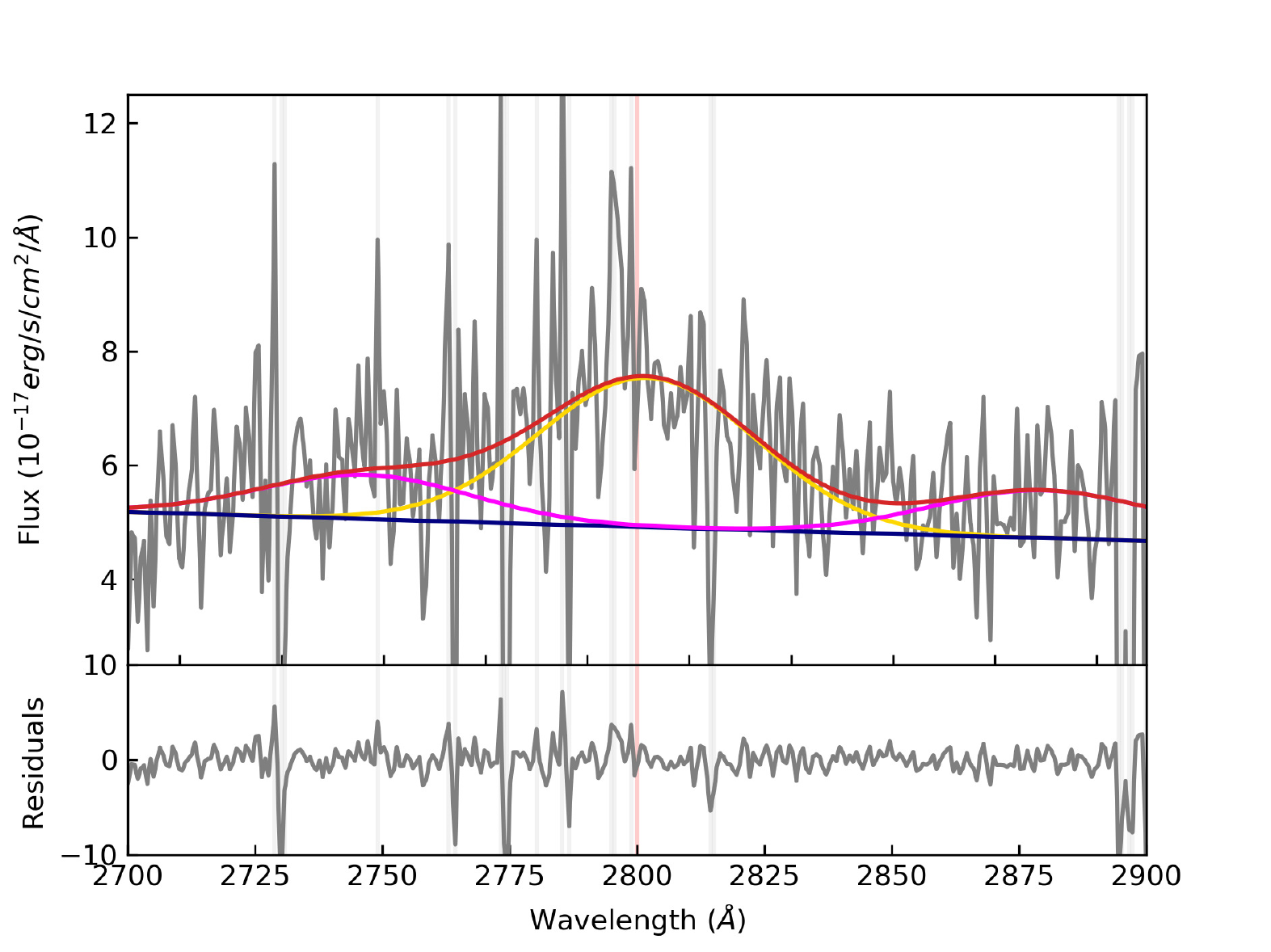}
   \includegraphics[width=0.4\textwidth]{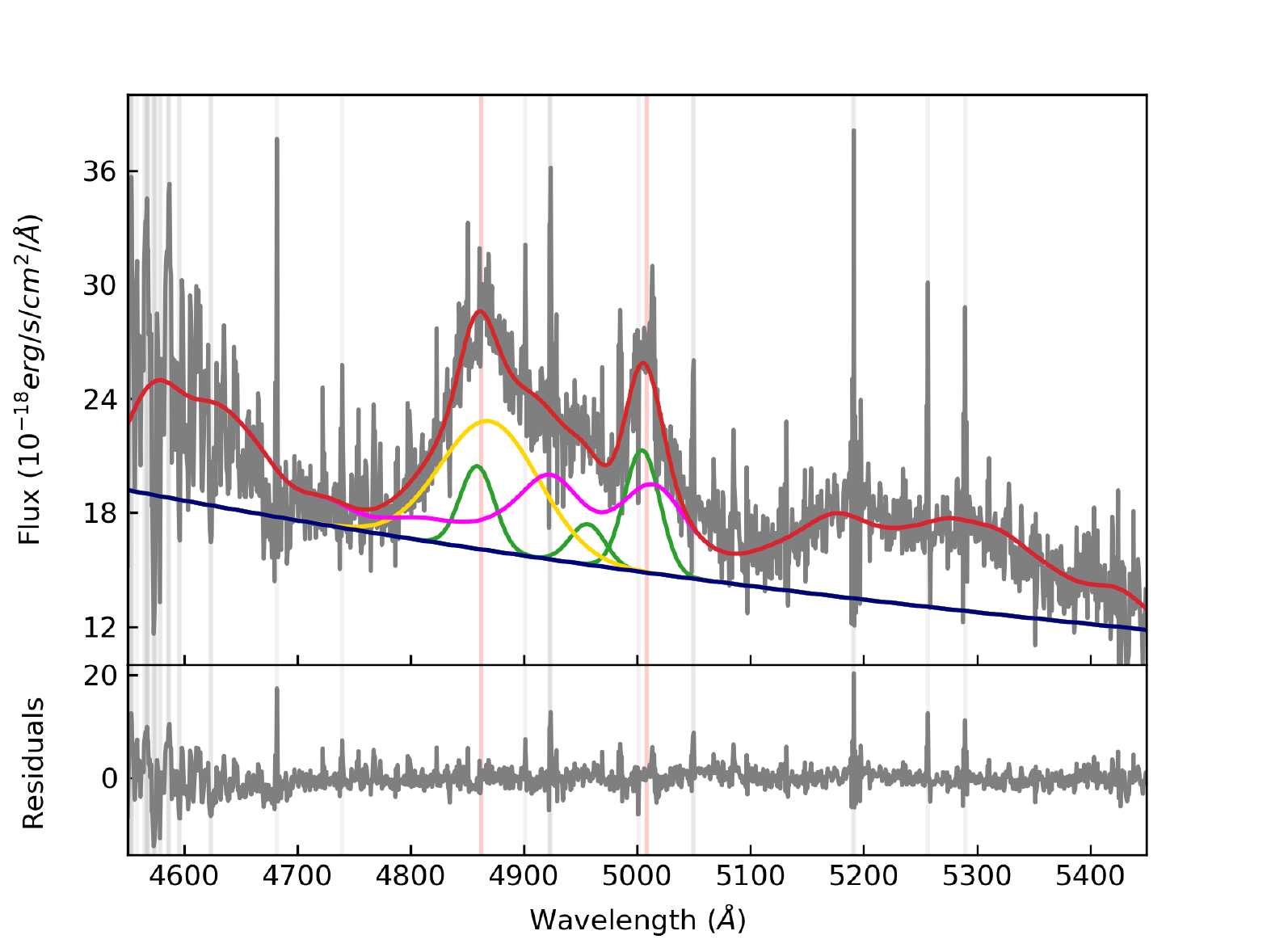}   
      \includegraphics[width=0.4\textwidth]{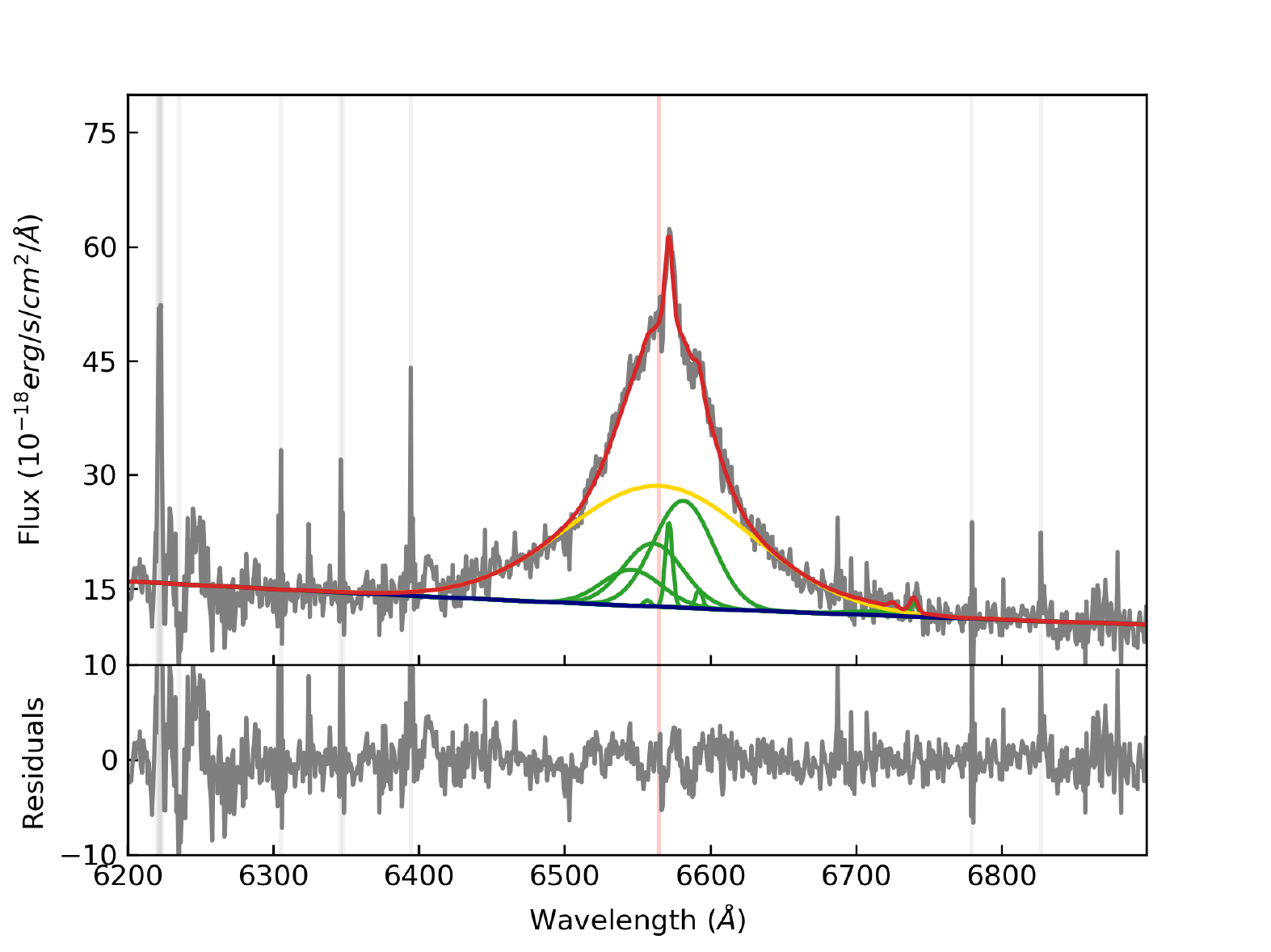}
\caption{cid\_346. The modeling is the same as in Fig. \ref{fig:app}}
   \end{figure*}

    \begin{figure*}
 \center
    \includegraphics[width=0.4\textwidth]{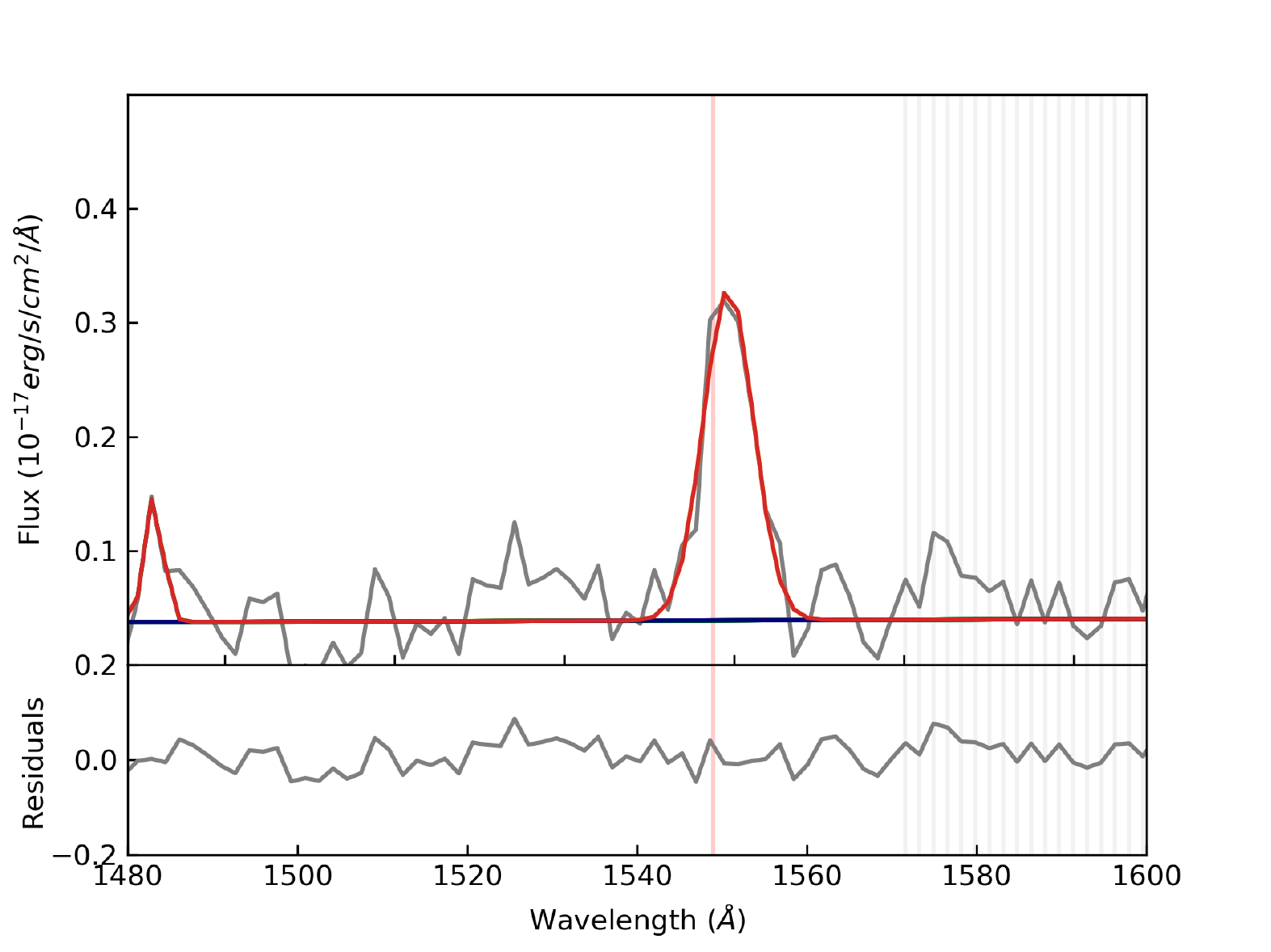}
      \includegraphics[width=0.4\textwidth]{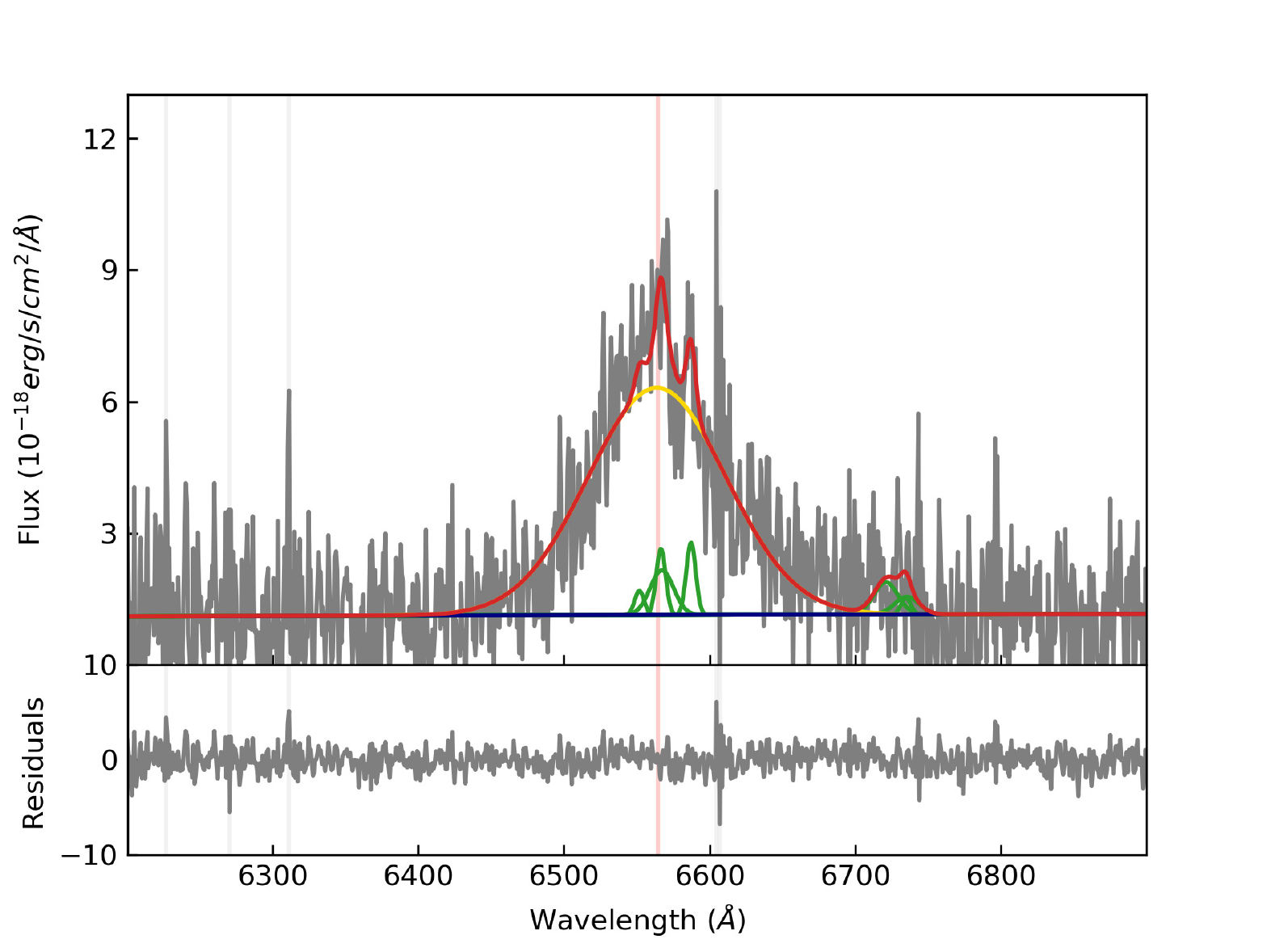}
\caption{cid\_1205. The modeling is the same as in Fig. \ref{fig:app}}
   \end{figure*}
   
    \begin{figure*}
 \center
    \includegraphics[width=0.4\textwidth]{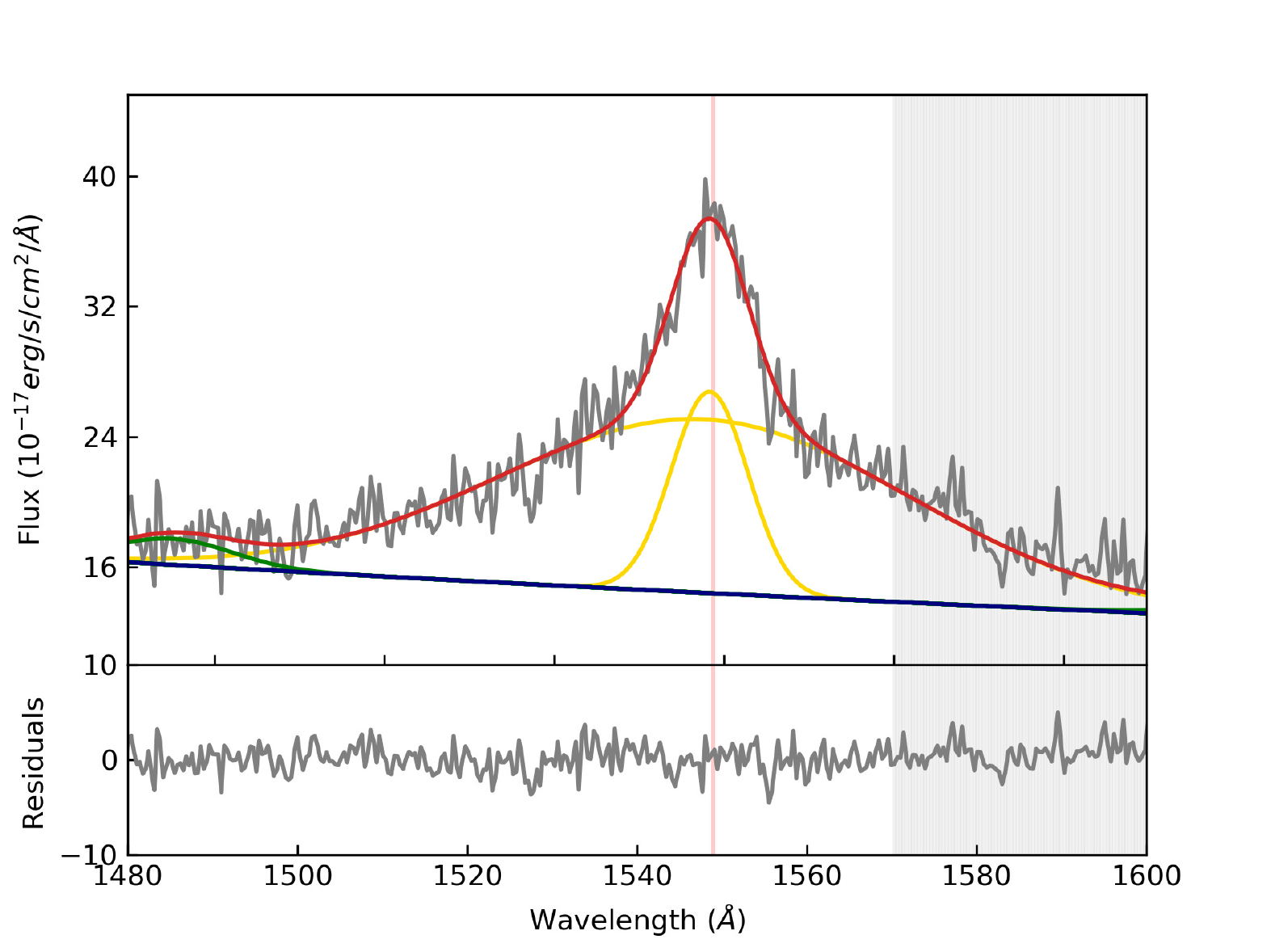}
   \includegraphics[width=0.4\textwidth]{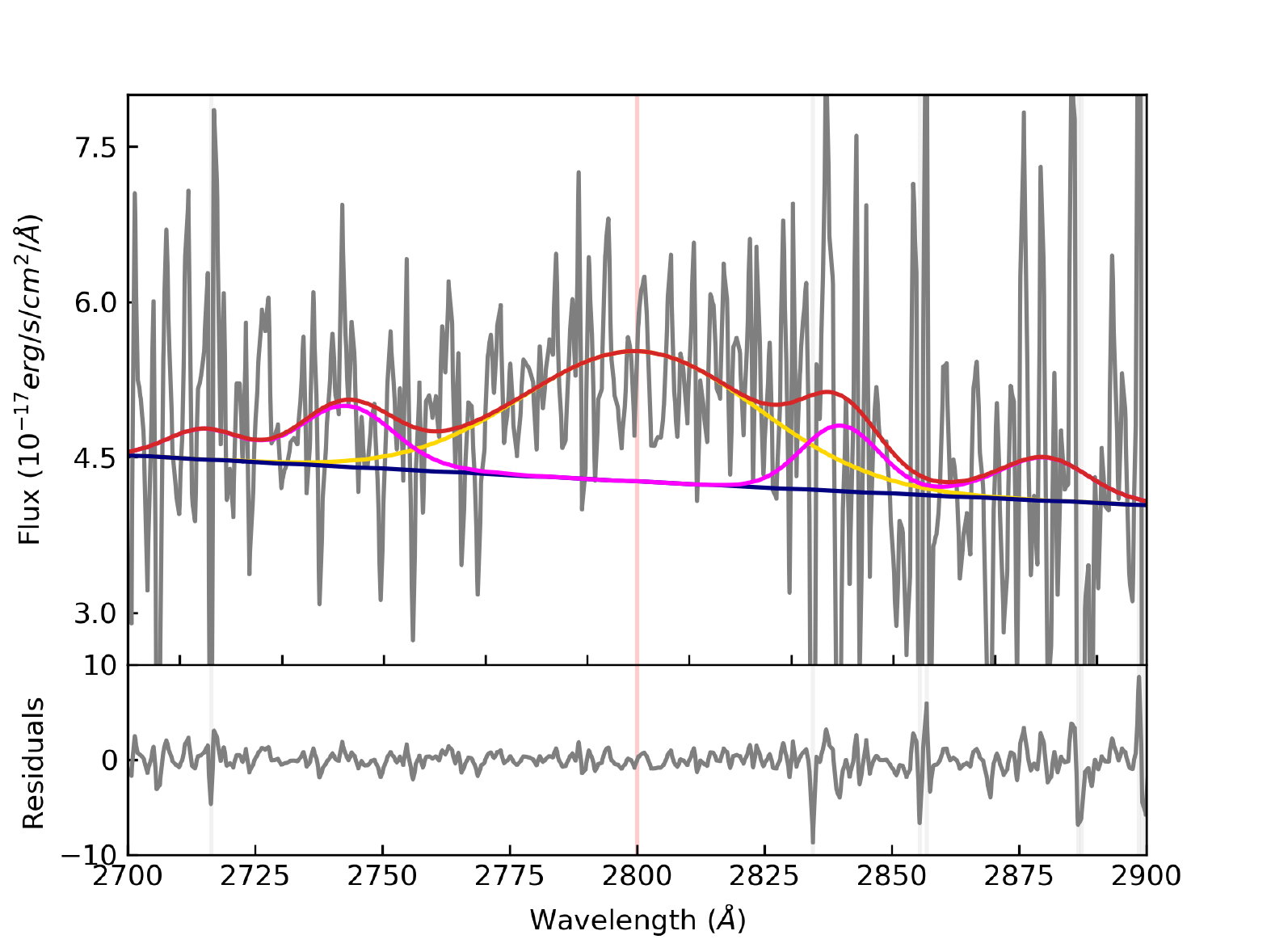}
   \includegraphics[width=0.4\textwidth]{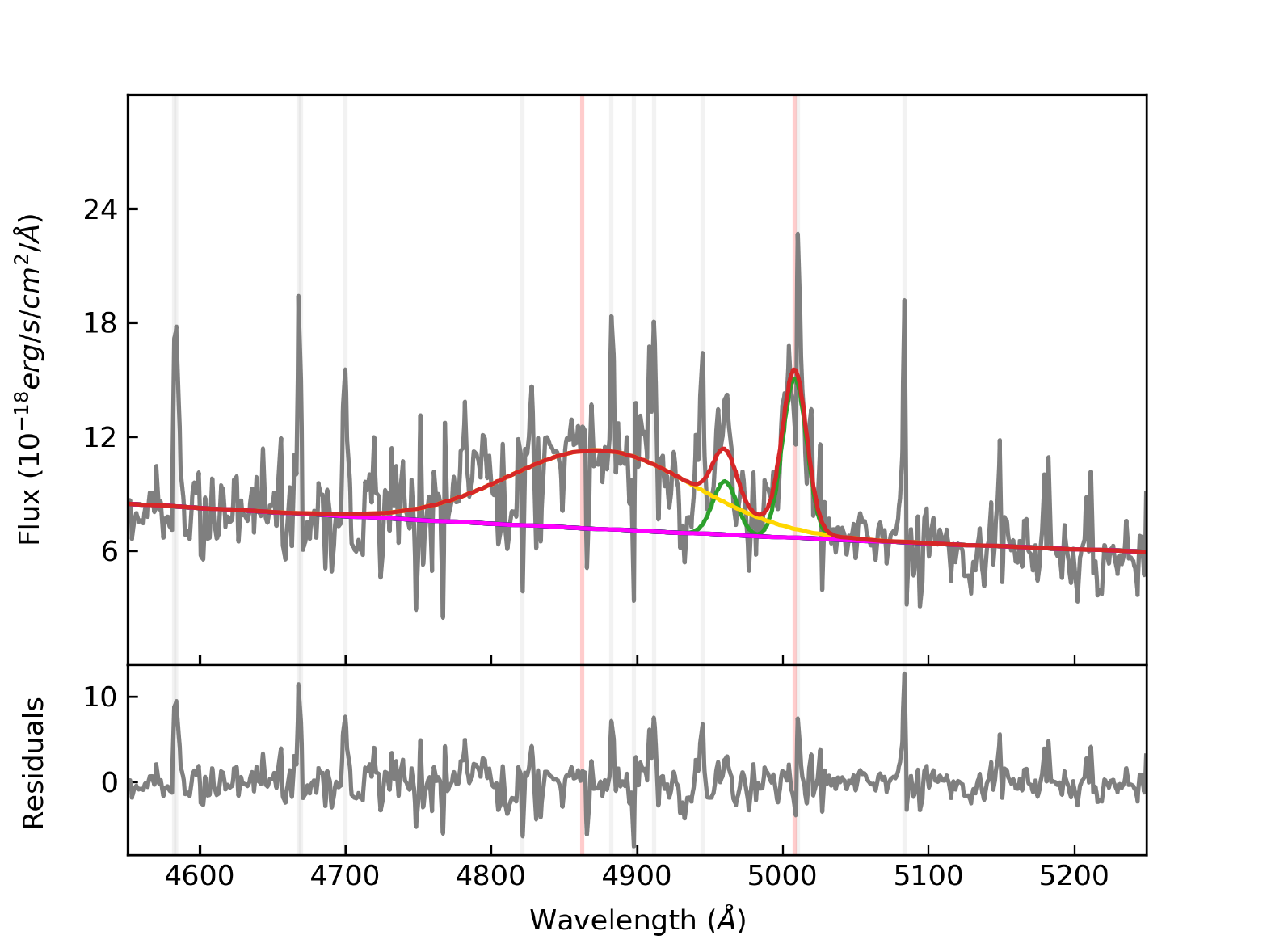}
         \includegraphics[width=0.4\textwidth]{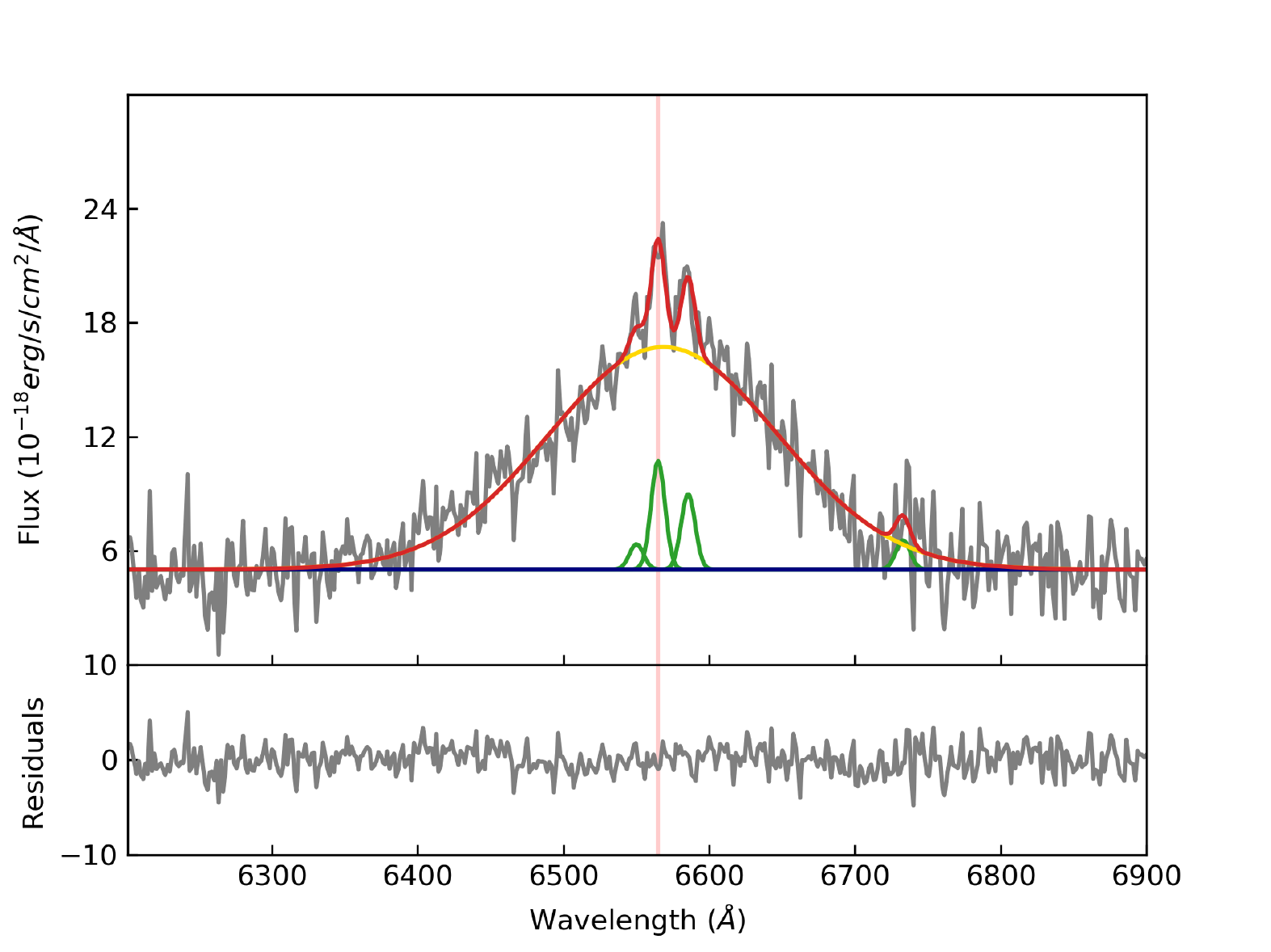}
\caption{cid\_467. The modeling is the same as in Fig. \ref{fig:app}}
   \end{figure*}

   \begin{figure*}
   \center
   \includegraphics[width=0.4\textwidth]{./figure/plot_UV/Spectrum_SDSS_1333_CIV_paper-eps-converted-to.pdf}
   \includegraphics[width=0.4\textwidth]{./figure/plot_UV/Spectrum_MgII_1_gauss_mod_SDSS_1333_paper-eps-converted-to.pdf}
   \includegraphics[width=0.4\textwidth]{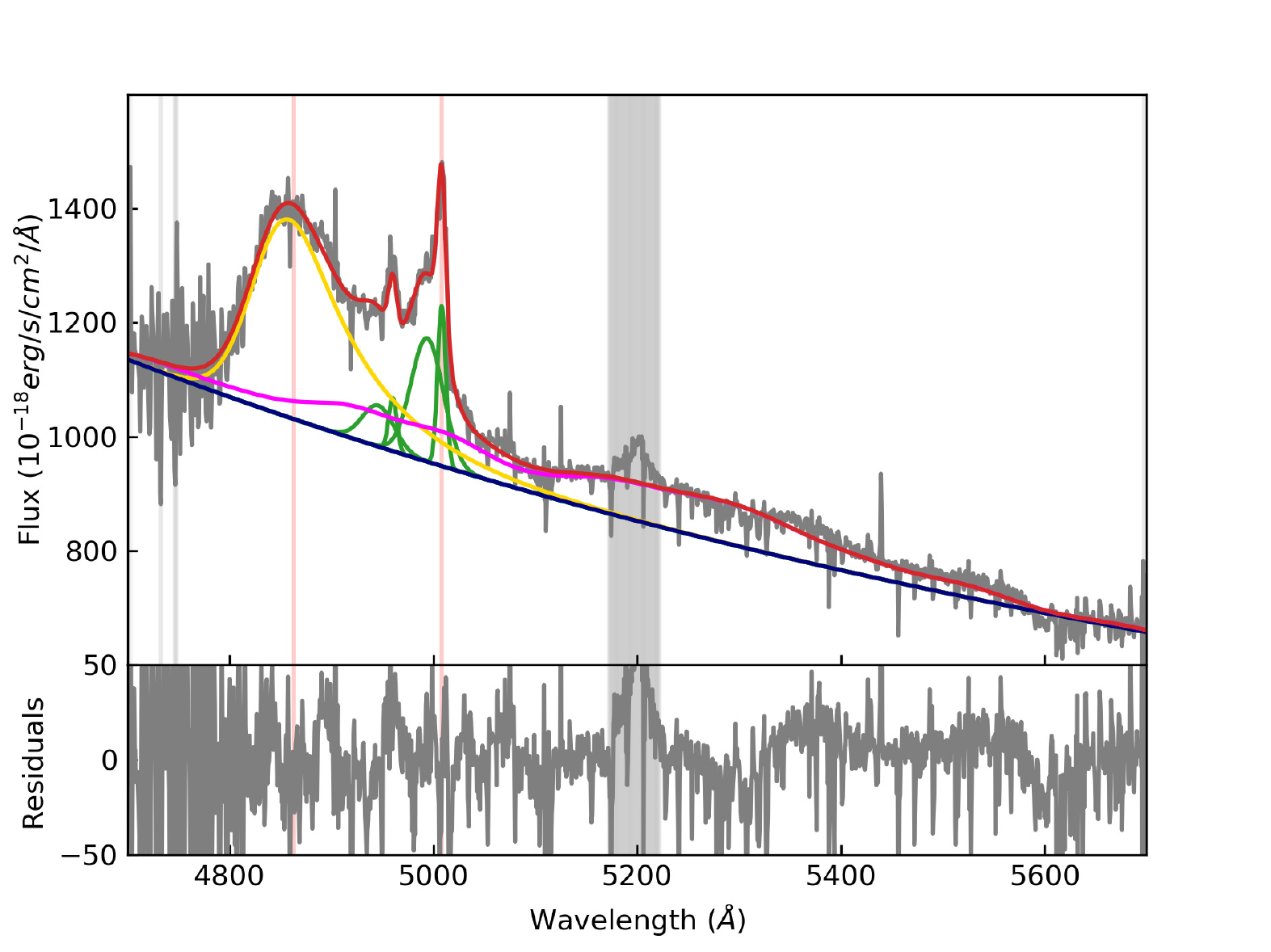}      \includegraphics[width=0.4\textwidth]{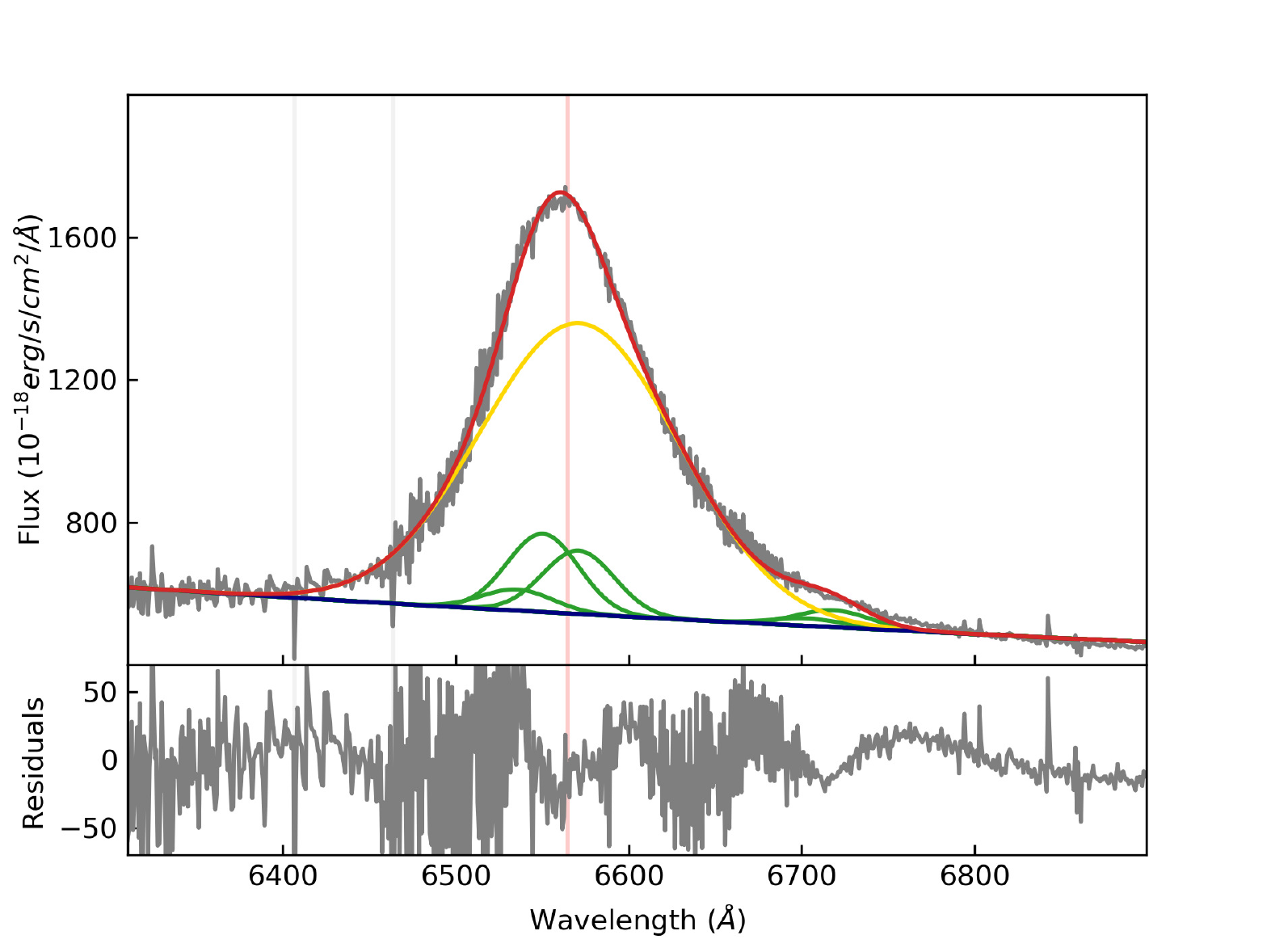}
\caption{J1333+1649. The modeling is the same as in Fig. \ref{fig:app}}
   \end{figure*}

    \begin{figure*}
   \center
   \includegraphics[width=0.4\textwidth]{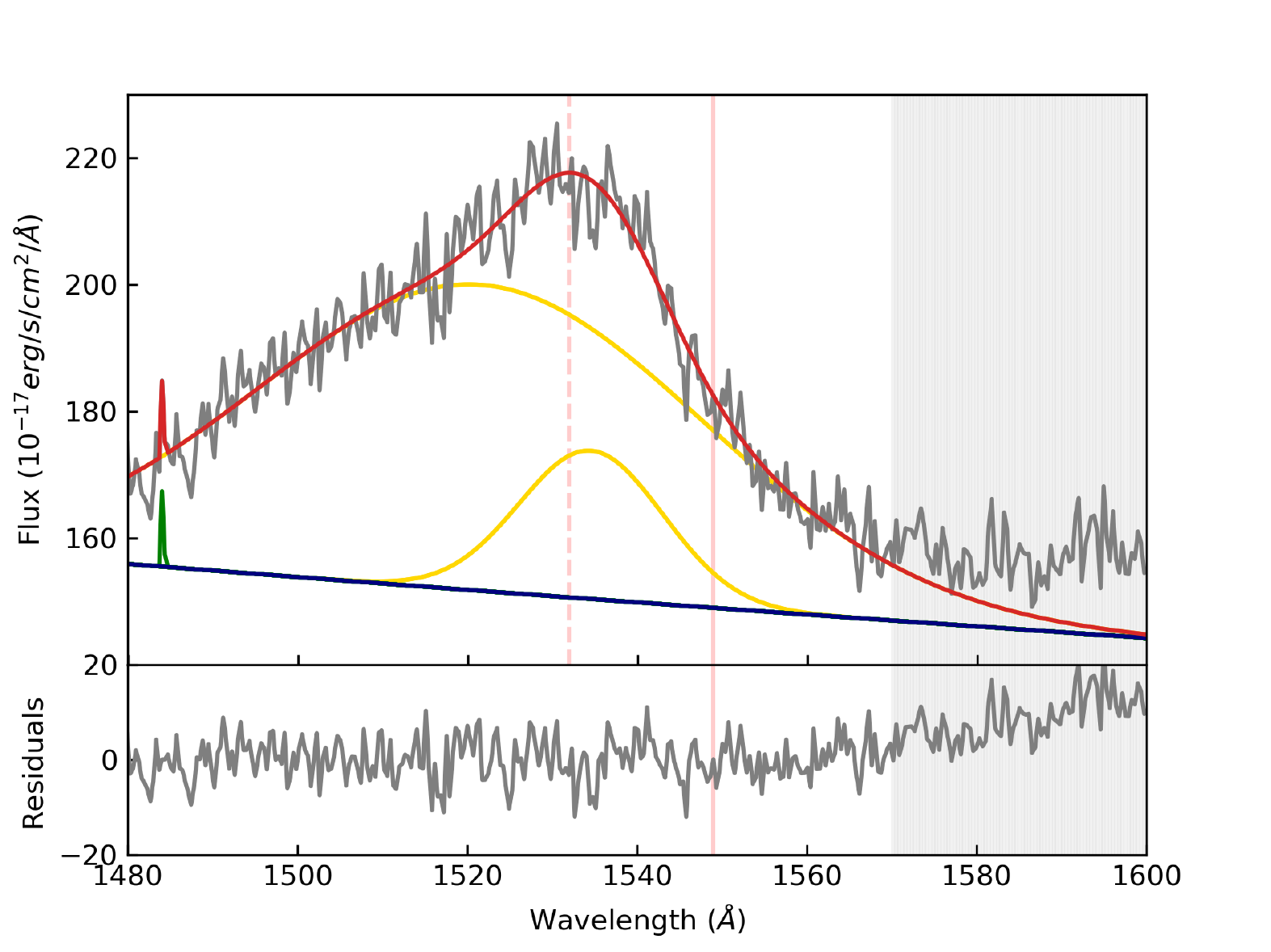}
   \includegraphics[width=0.4\textwidth]{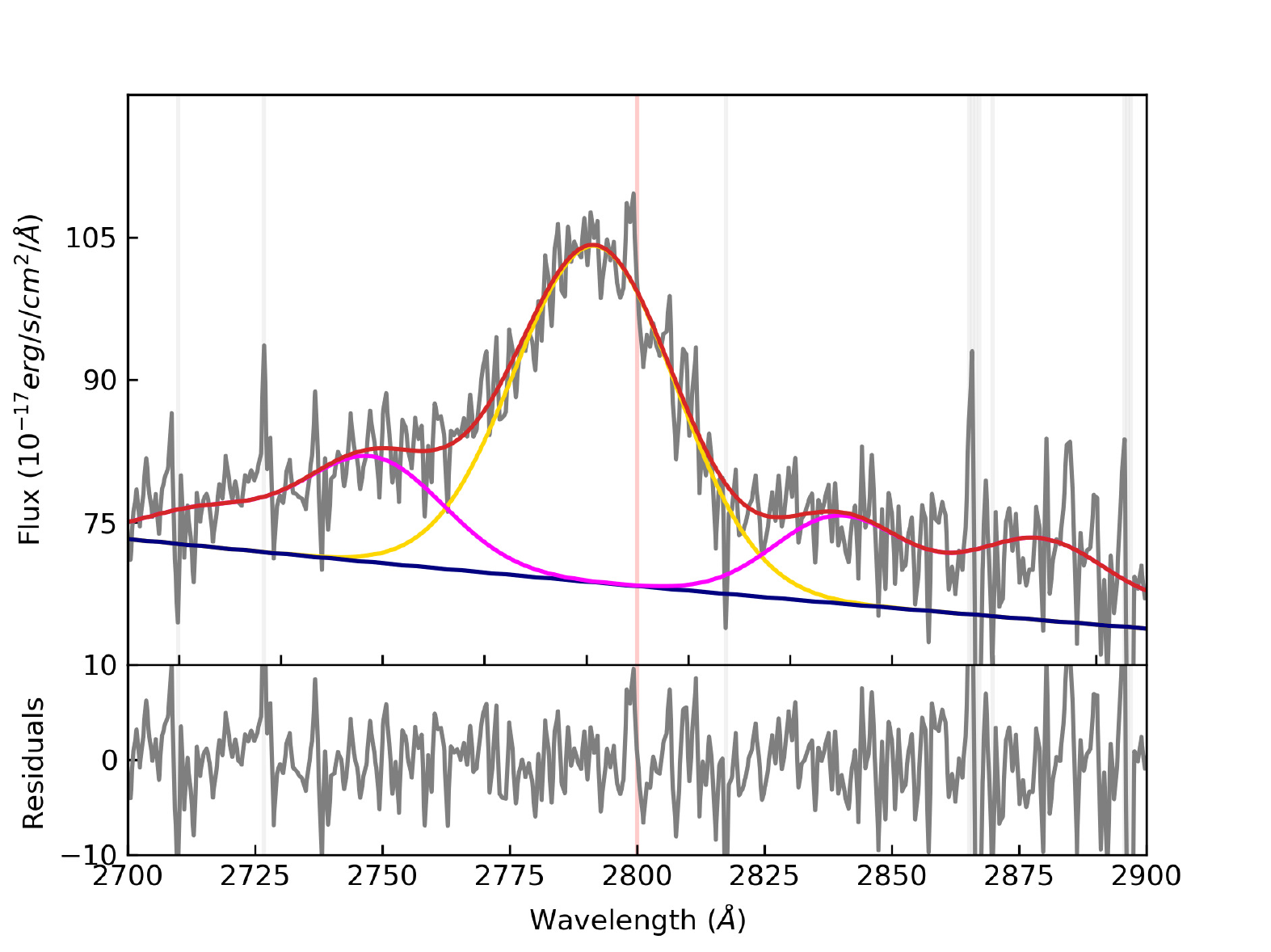}
   \includegraphics[width=0.4\textwidth]{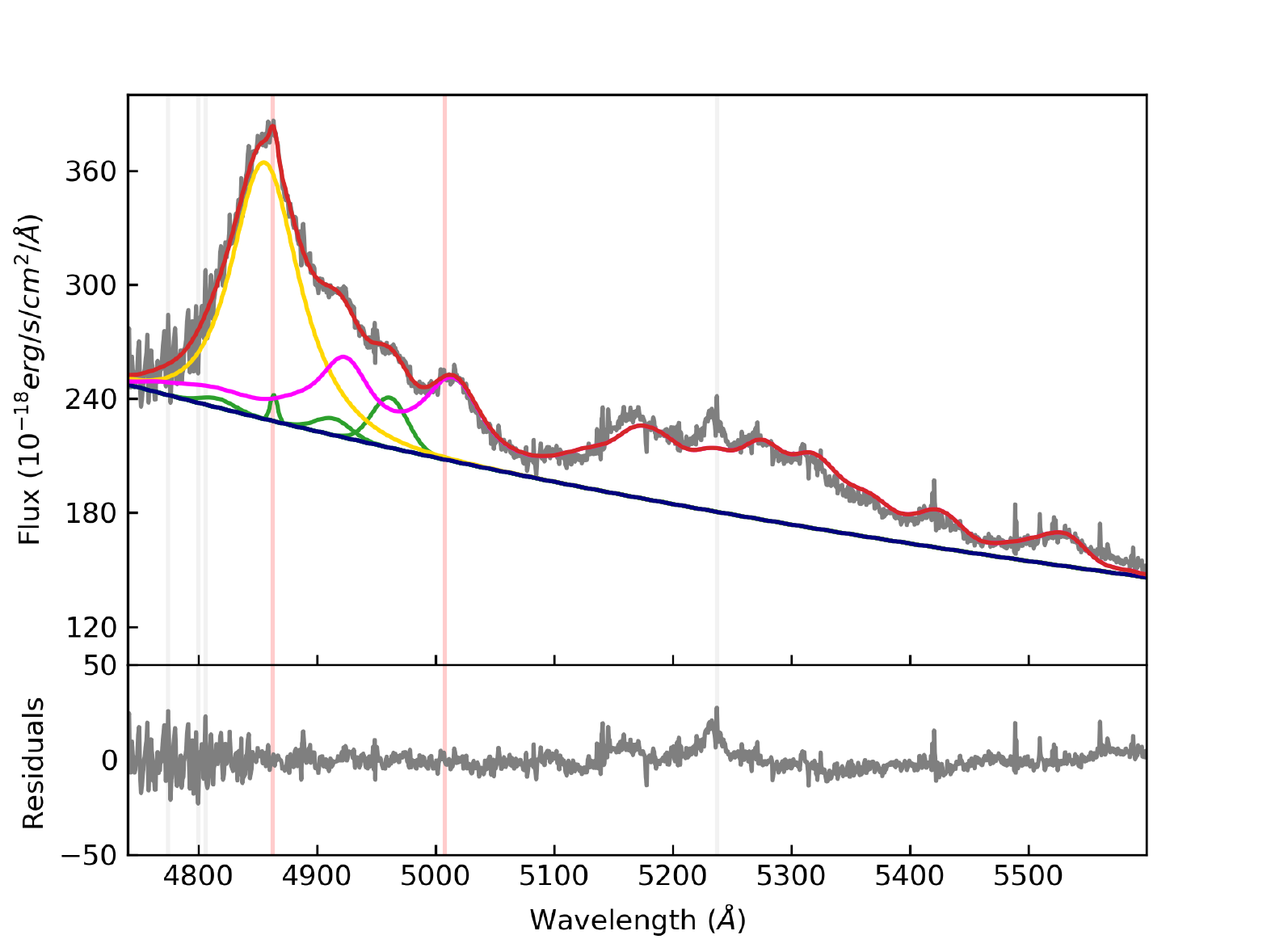}      \includegraphics[width=0.4\textwidth]{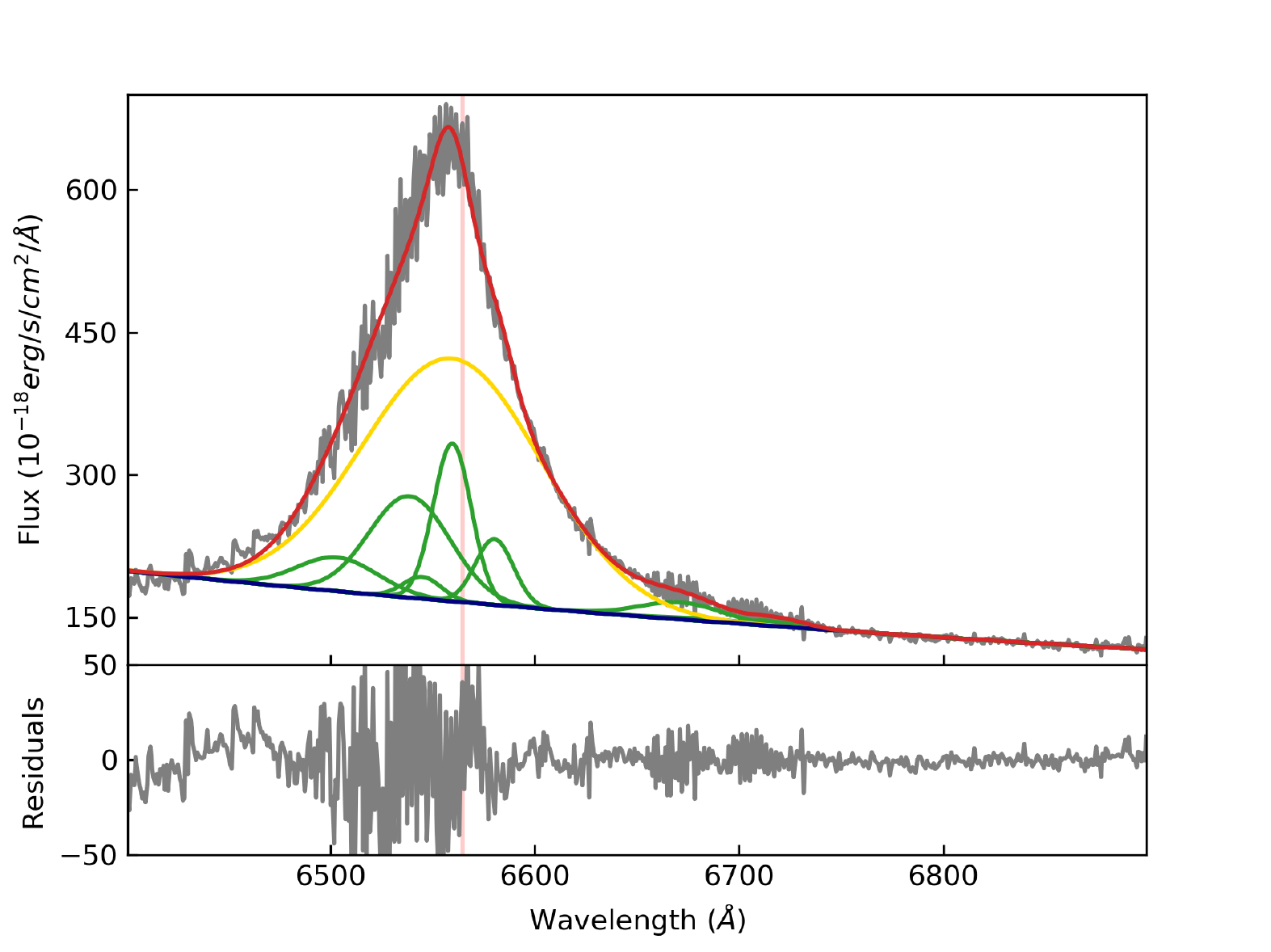}
\caption{J1441+0454. The modeling is the same as in Fig. \ref{fig:app}}
   \end{figure*}

    \begin{figure*}
 \center
    \includegraphics[width=0.4\textwidth]{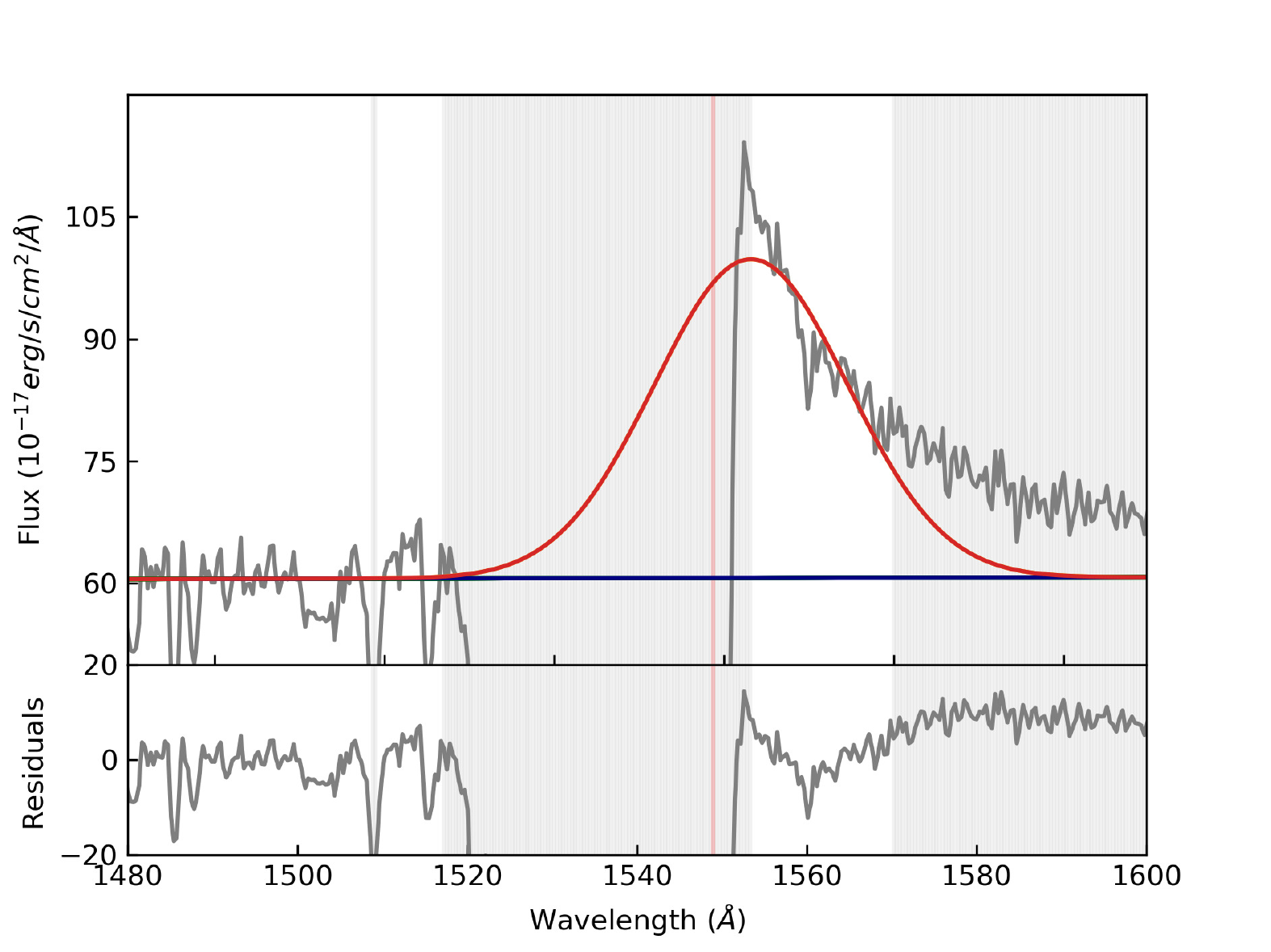}
   \includegraphics[width=0.4\textwidth]{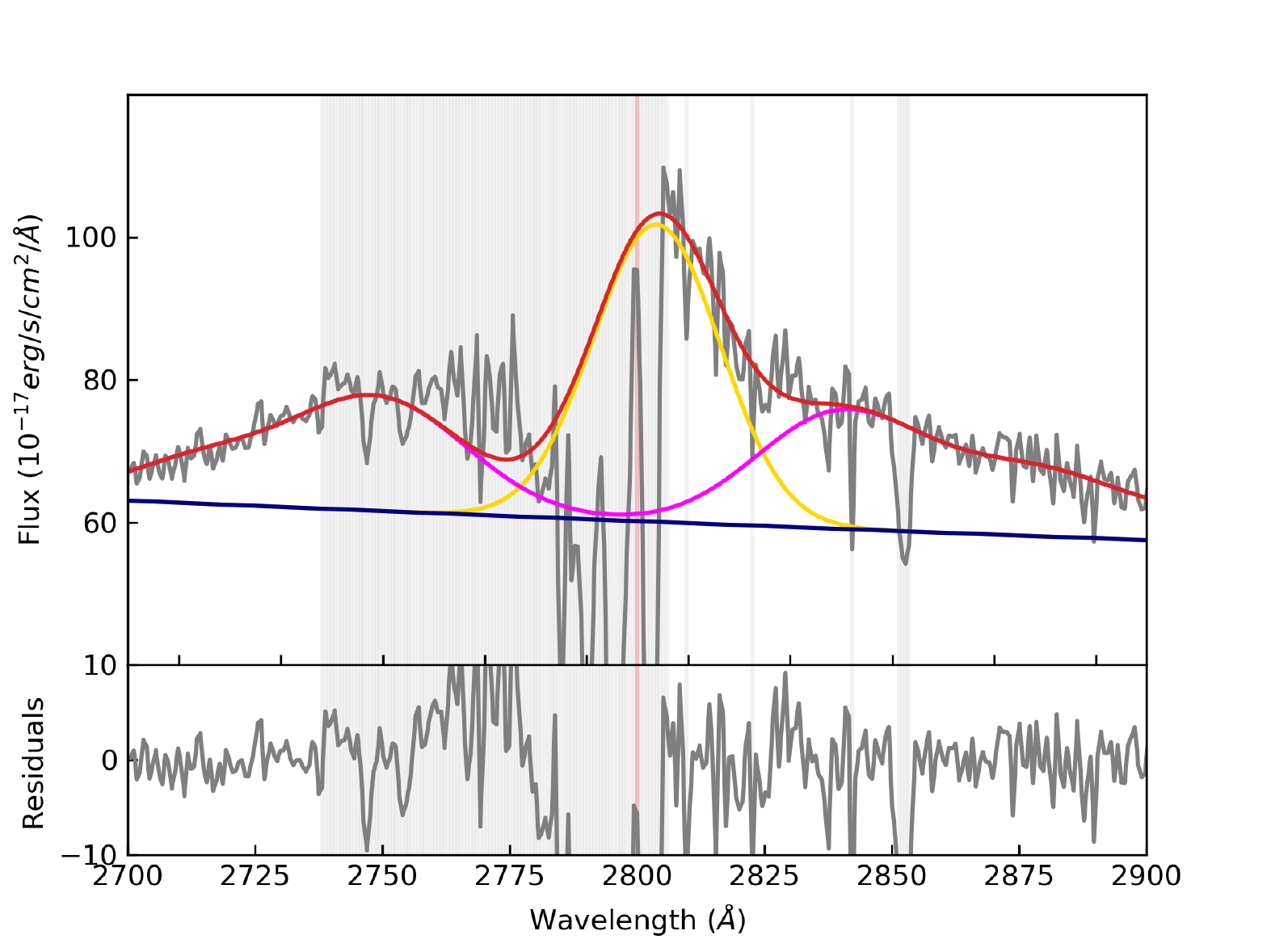}
   \includegraphics[width=0.4\textwidth]{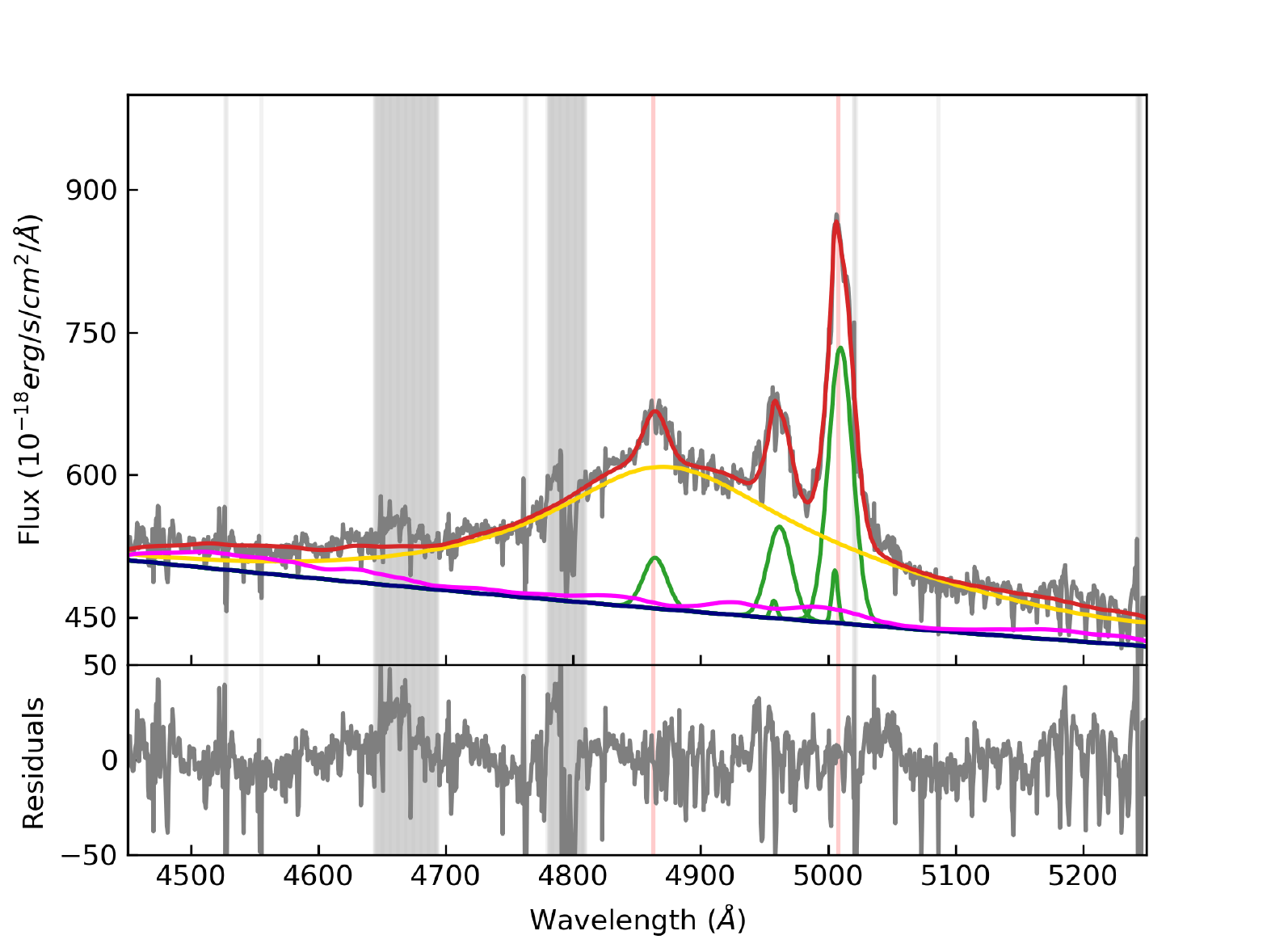}     
    \includegraphics[width=0.4\textwidth]{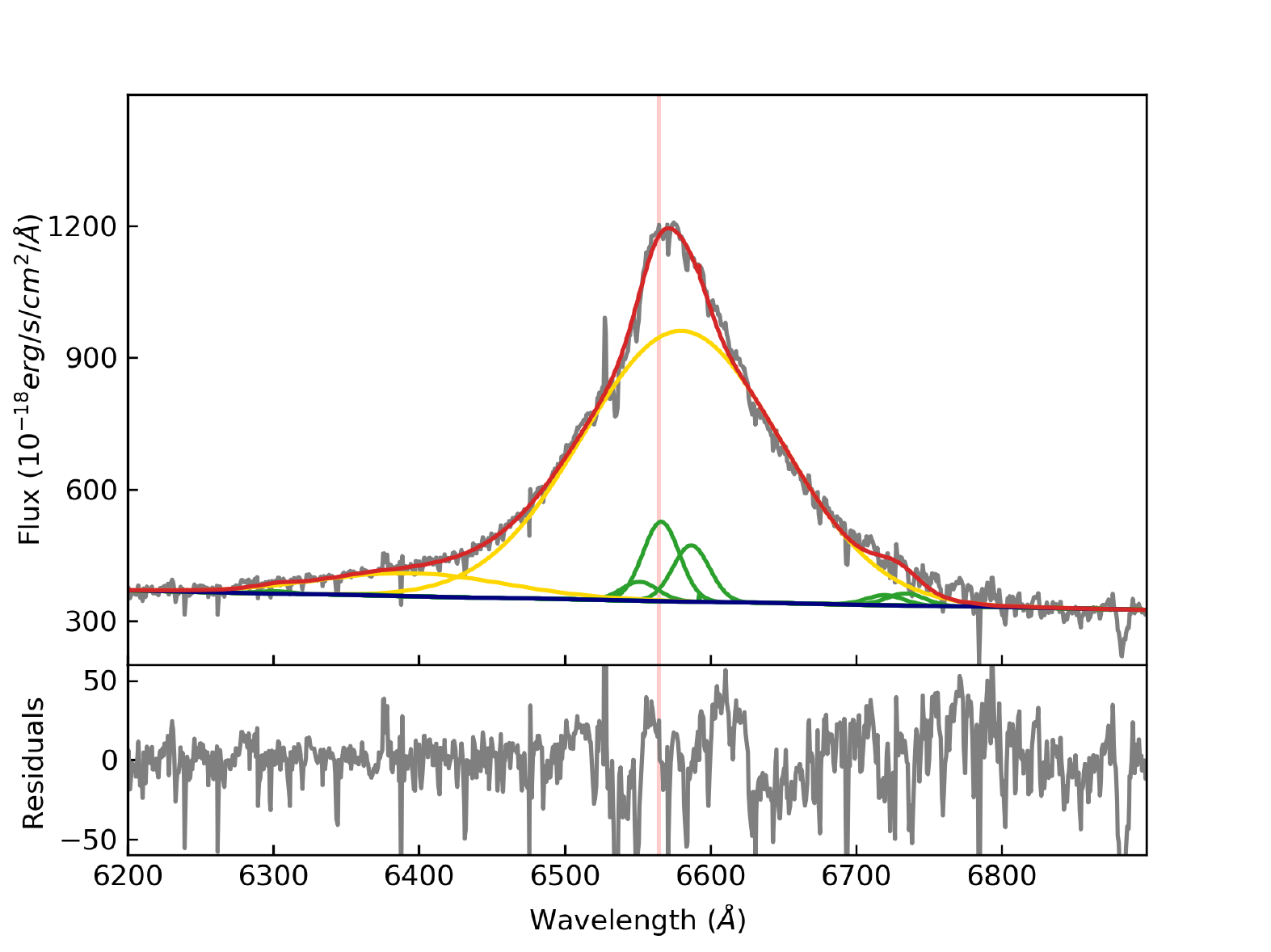}
\caption{J1549+1245. The modeling is the same as in Fig. \ref{fig:app}. We added one Gaussian component around 6380 \AA\ to reduce the $\chi^2$ of our fit (e.g. \citealt{Carniani2016})}
   \end{figure*}

 \begin{figure*}
 \center
    \includegraphics[width=0.4\textwidth]{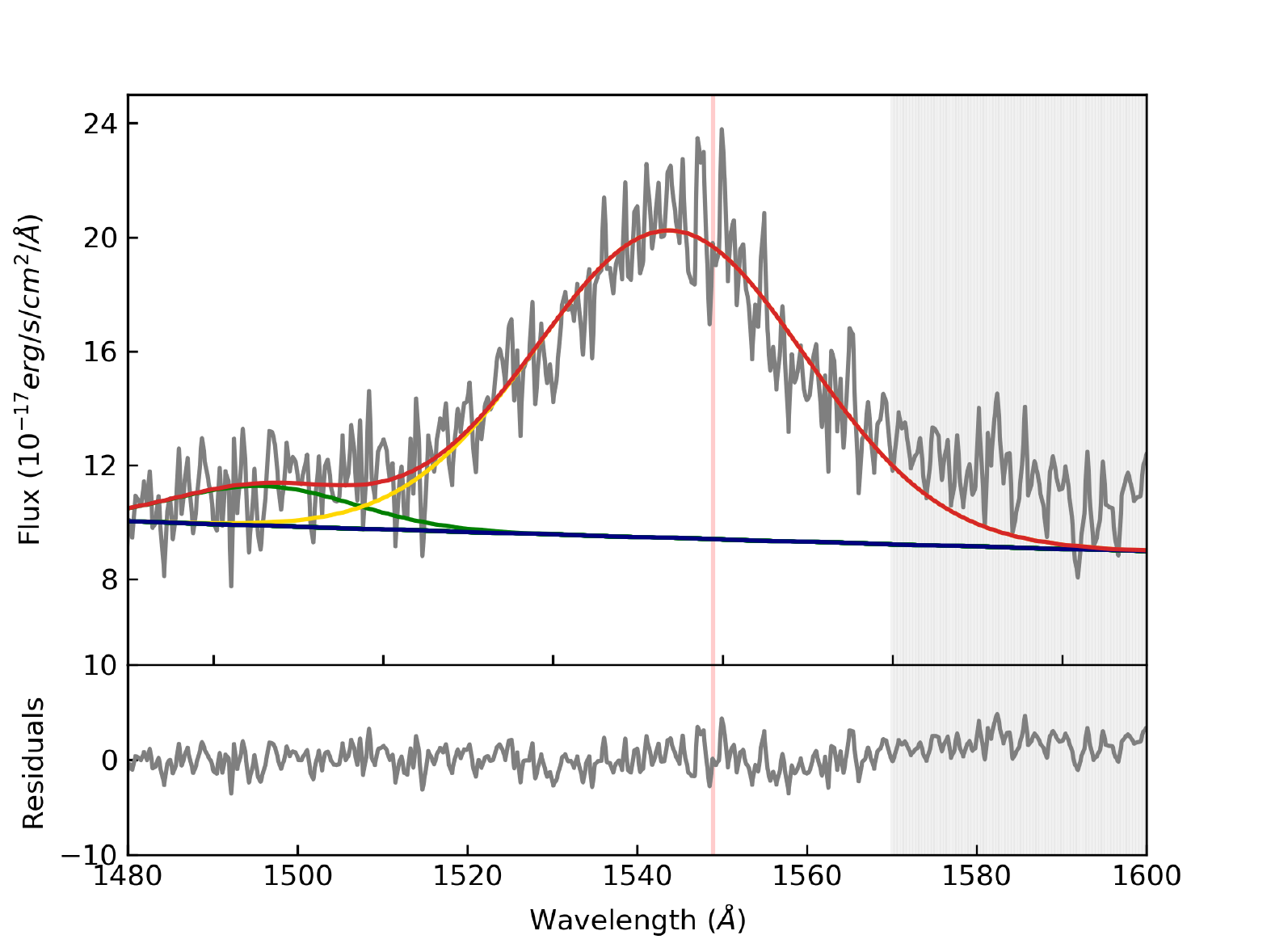}
   \includegraphics[width=0.4\textwidth]{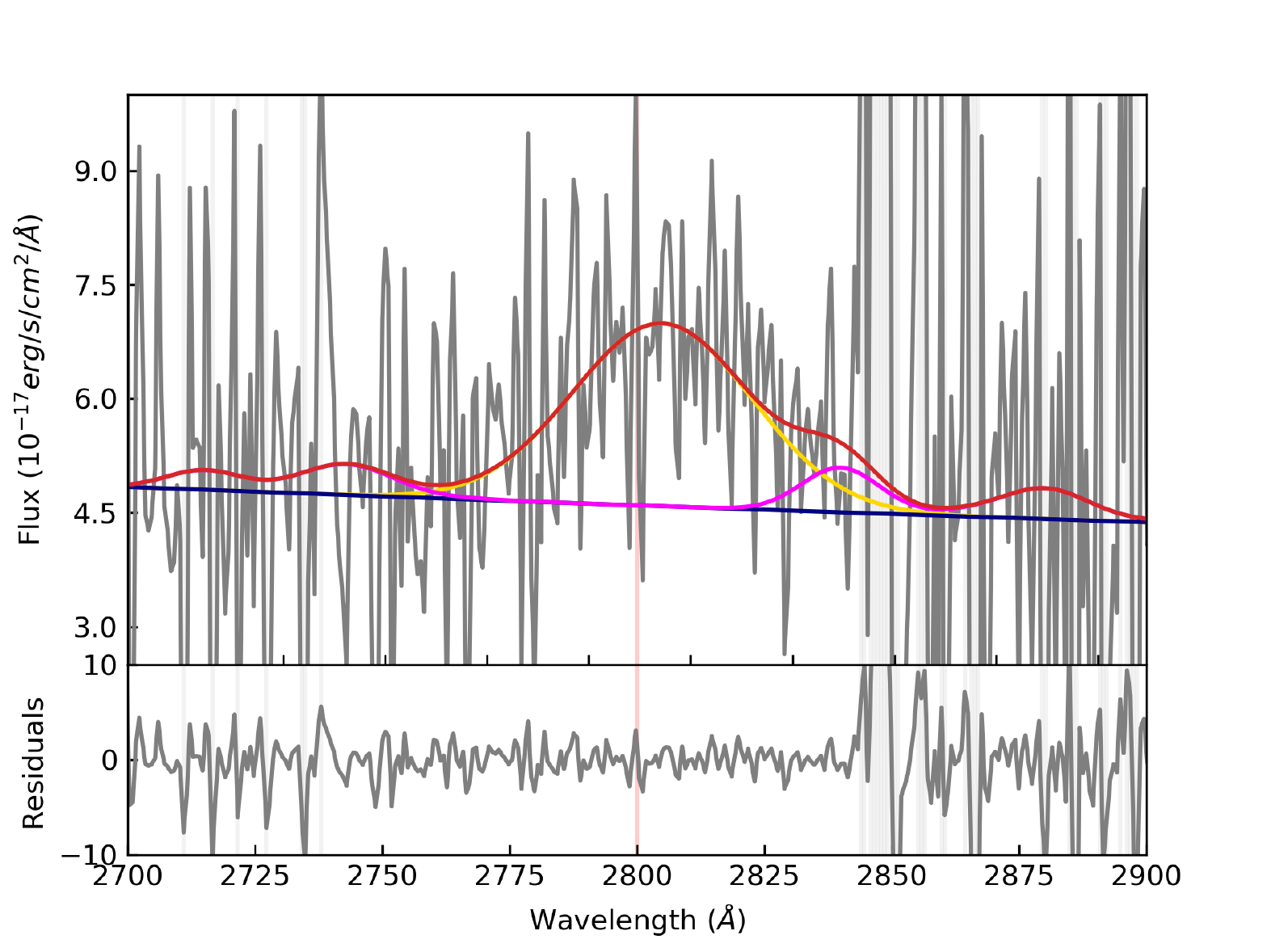}
   \includegraphics[width=0.4\textwidth]{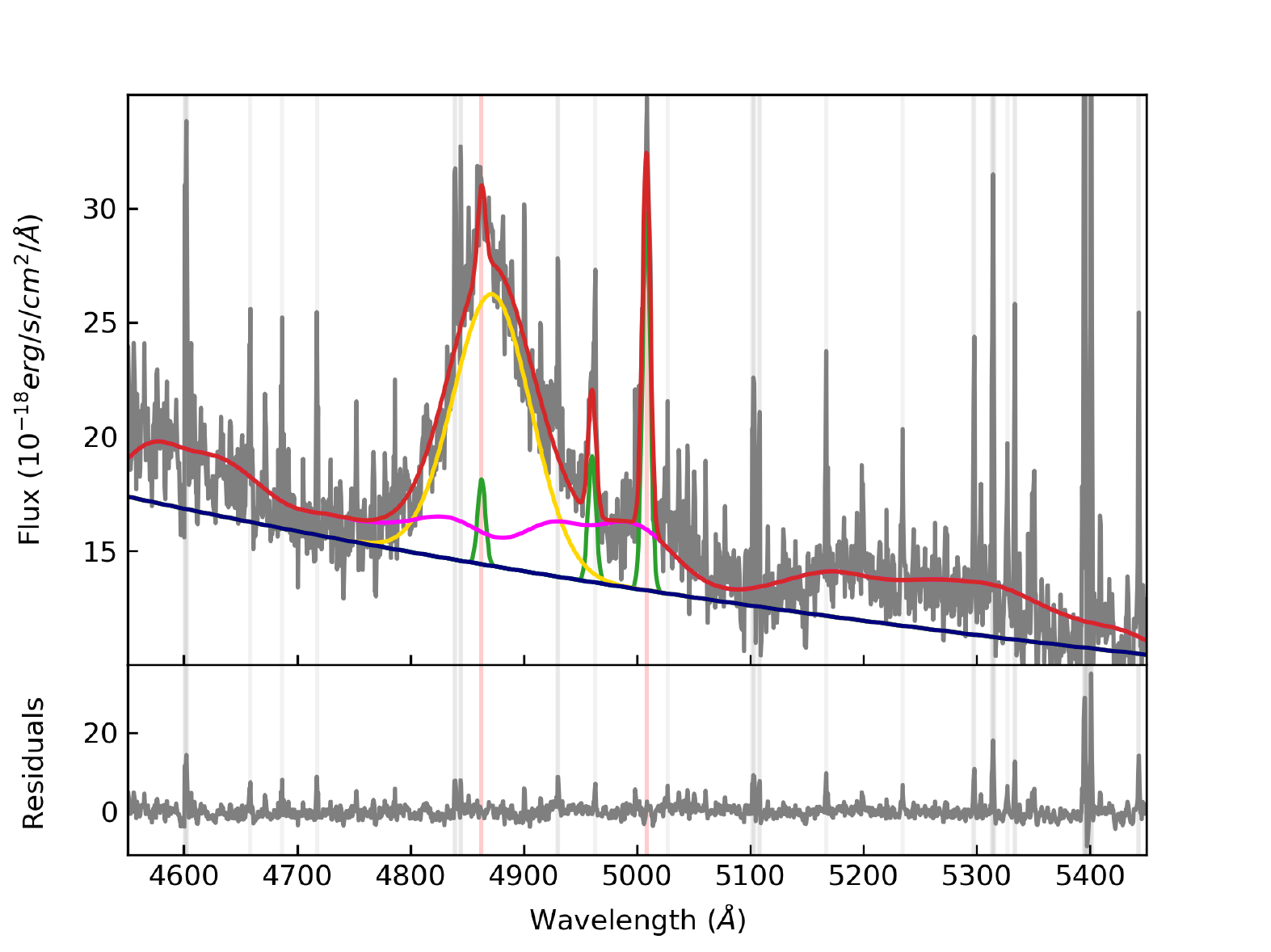}      \includegraphics[width=0.4\textwidth]{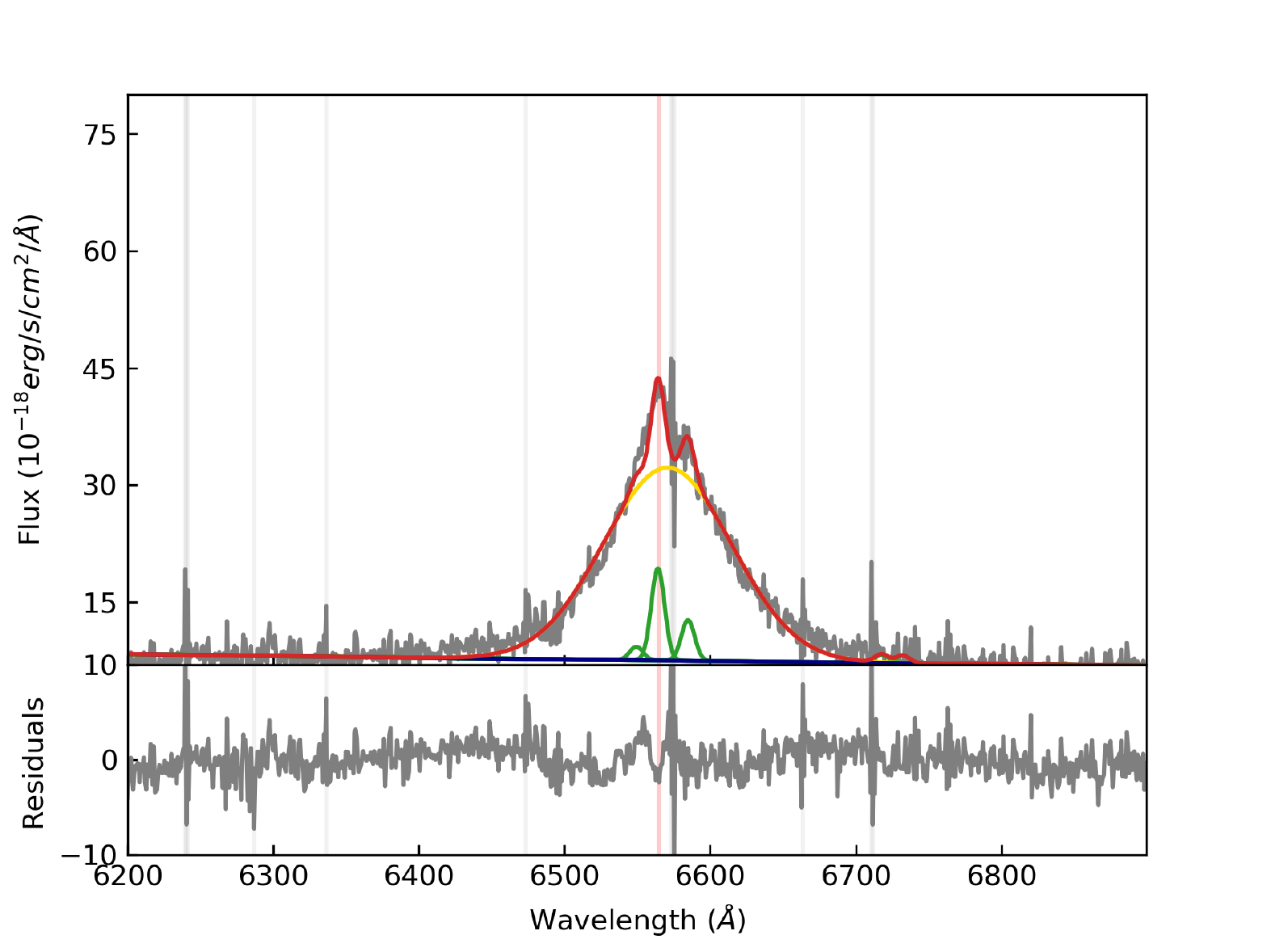}
\caption{S82X1905. The modeling is the same as in Fig. \ref{fig:app}}
   \end{figure*}

 \begin{figure*}
 \center
    \includegraphics[width=0.4\textwidth]{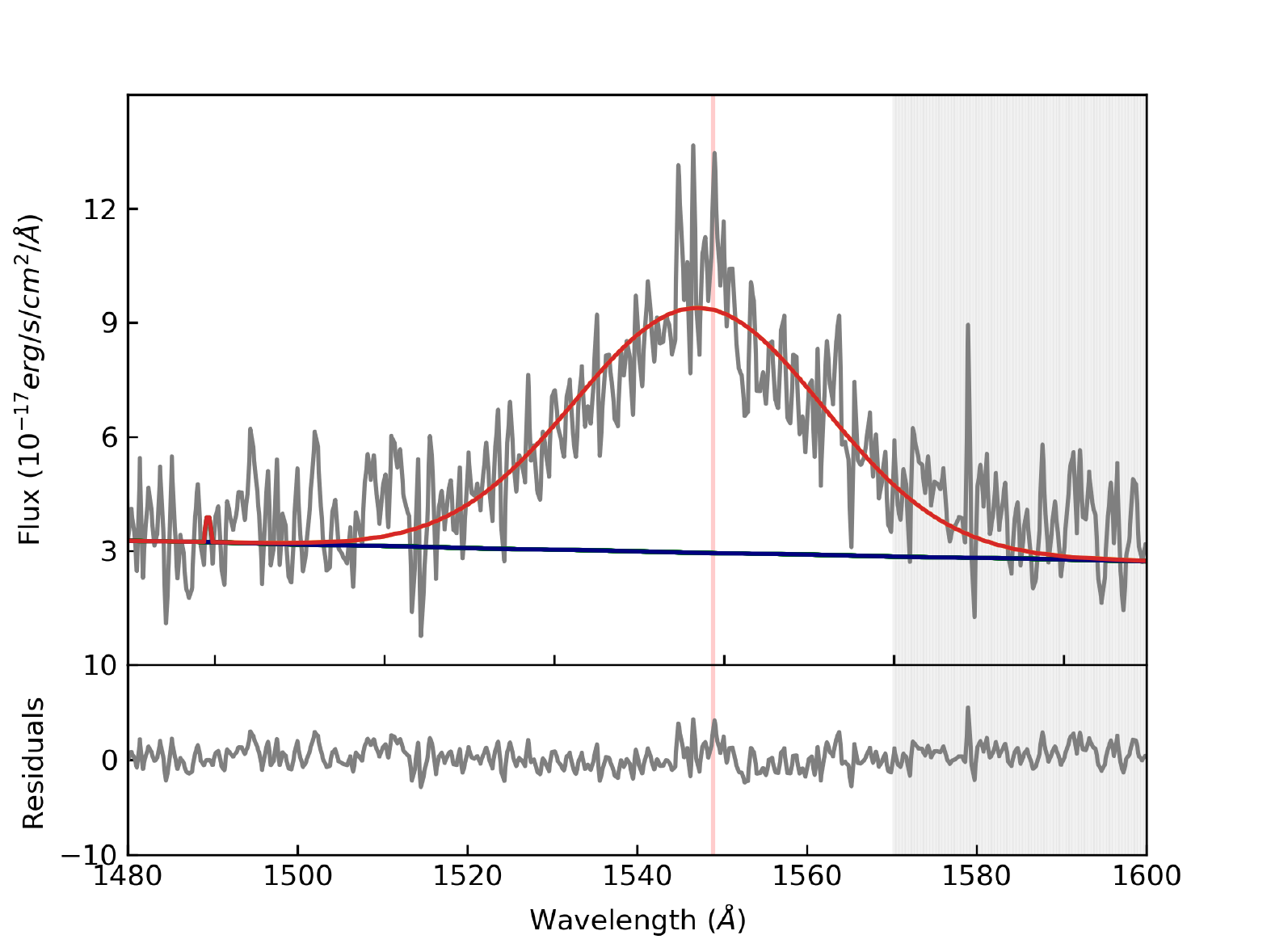}
   \includegraphics[width=0.4\textwidth]{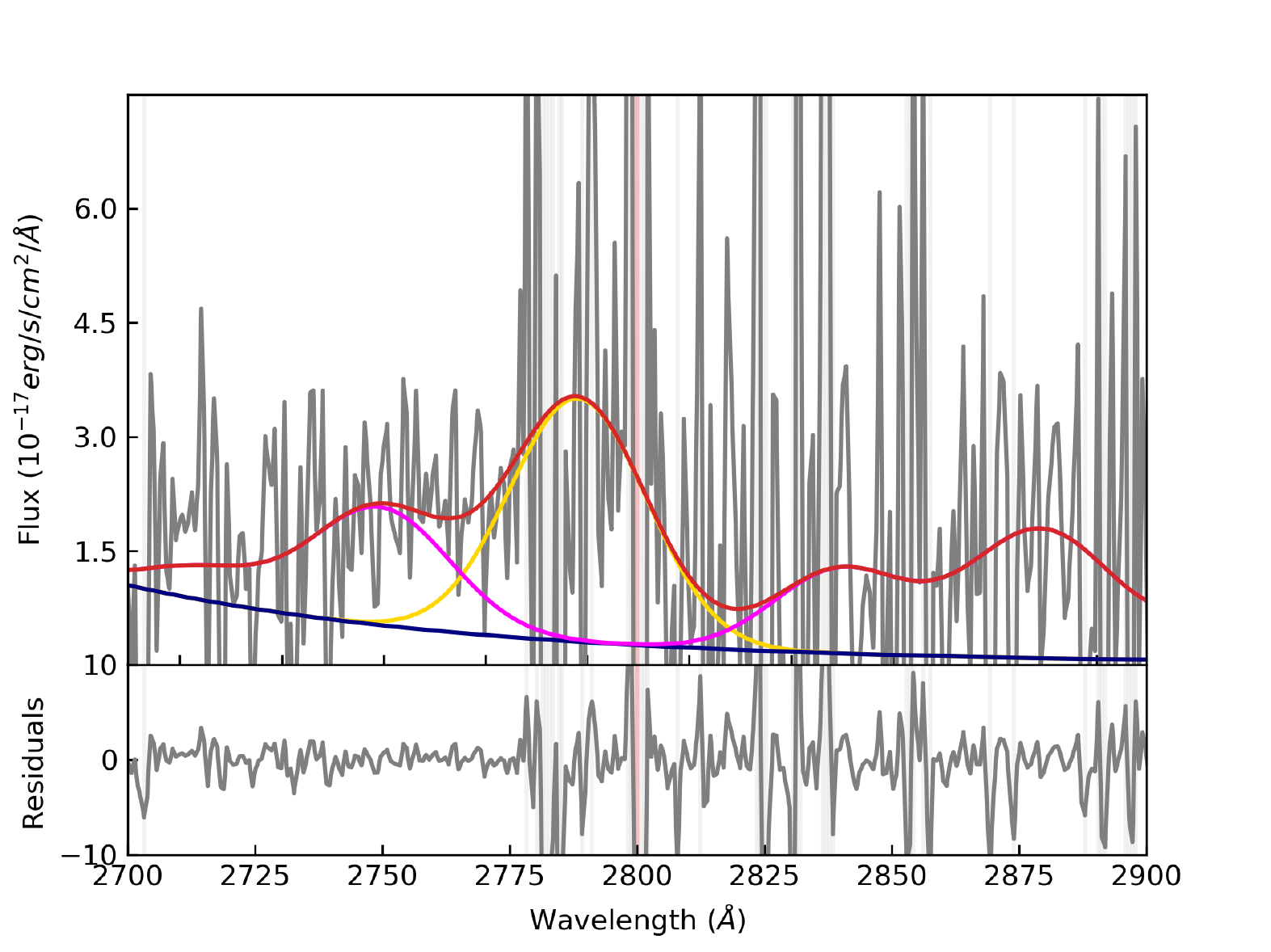}
   \includegraphics[width=0.4\textwidth]{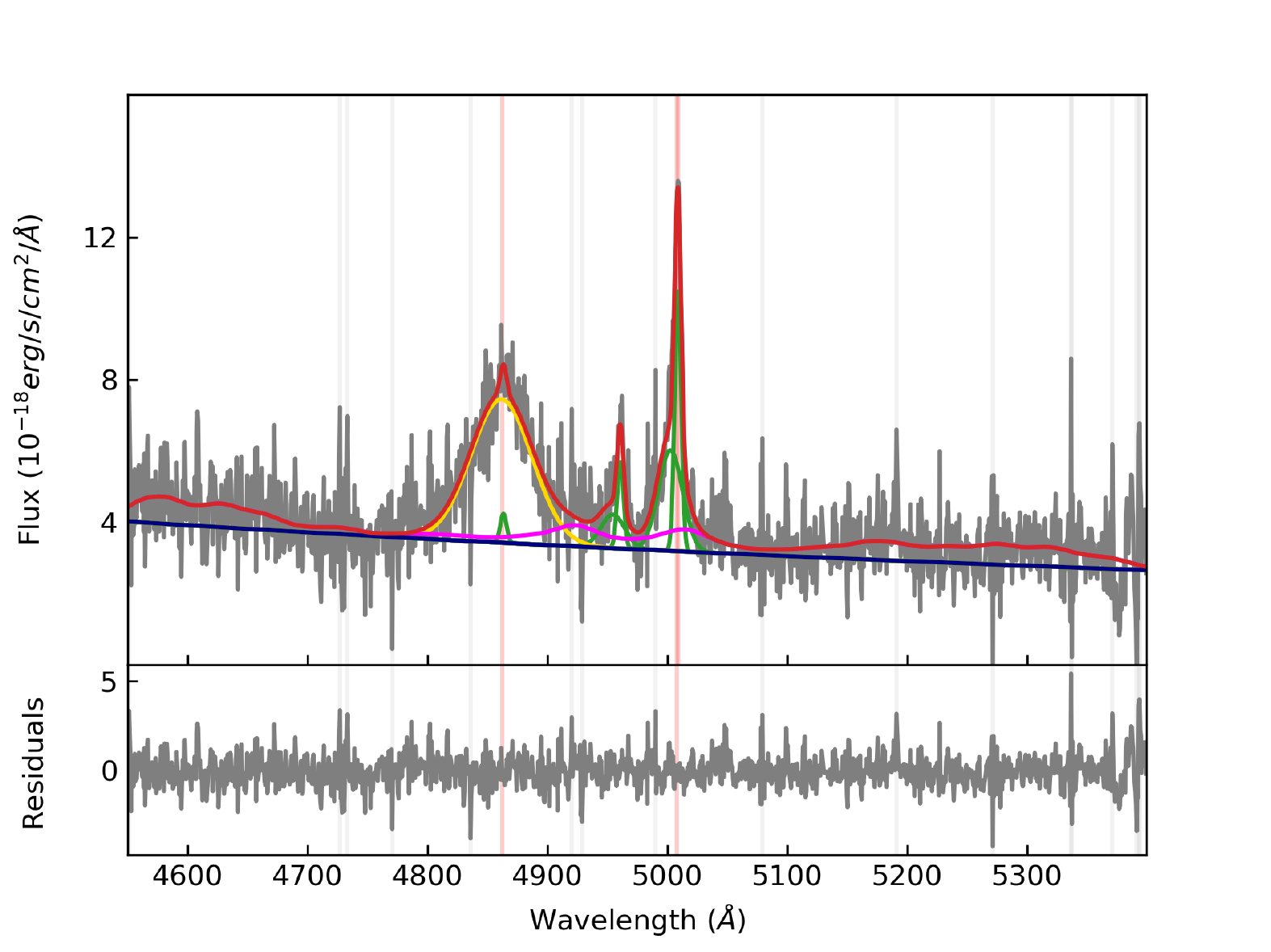}      \includegraphics[width=0.4\textwidth]{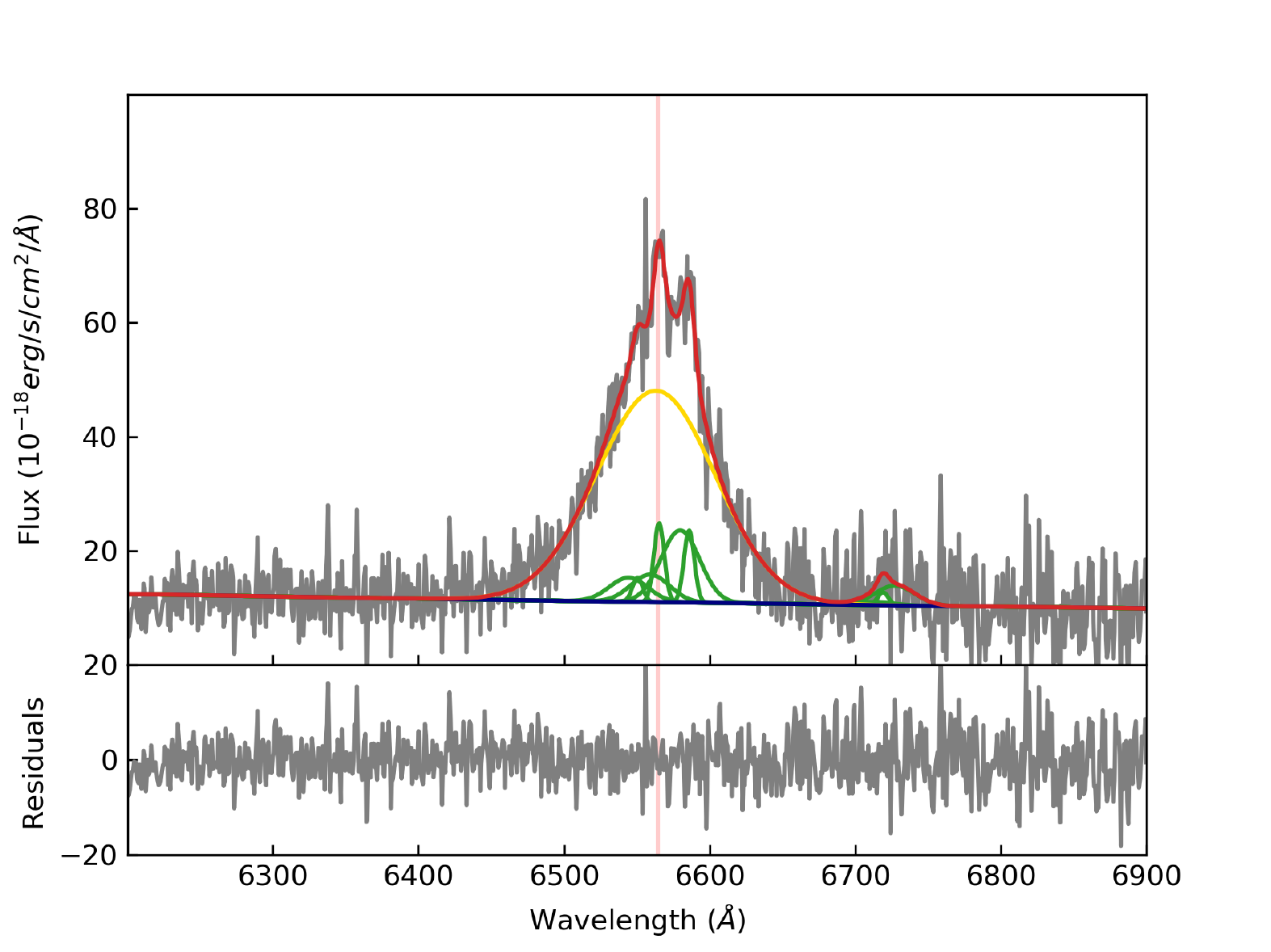}
\caption{S82X1940. The modeling is the same as in Fig. \ref{fig:app}}
   \end{figure*}

 \begin{figure*}
 \center
    \includegraphics[width=0.4\textwidth]{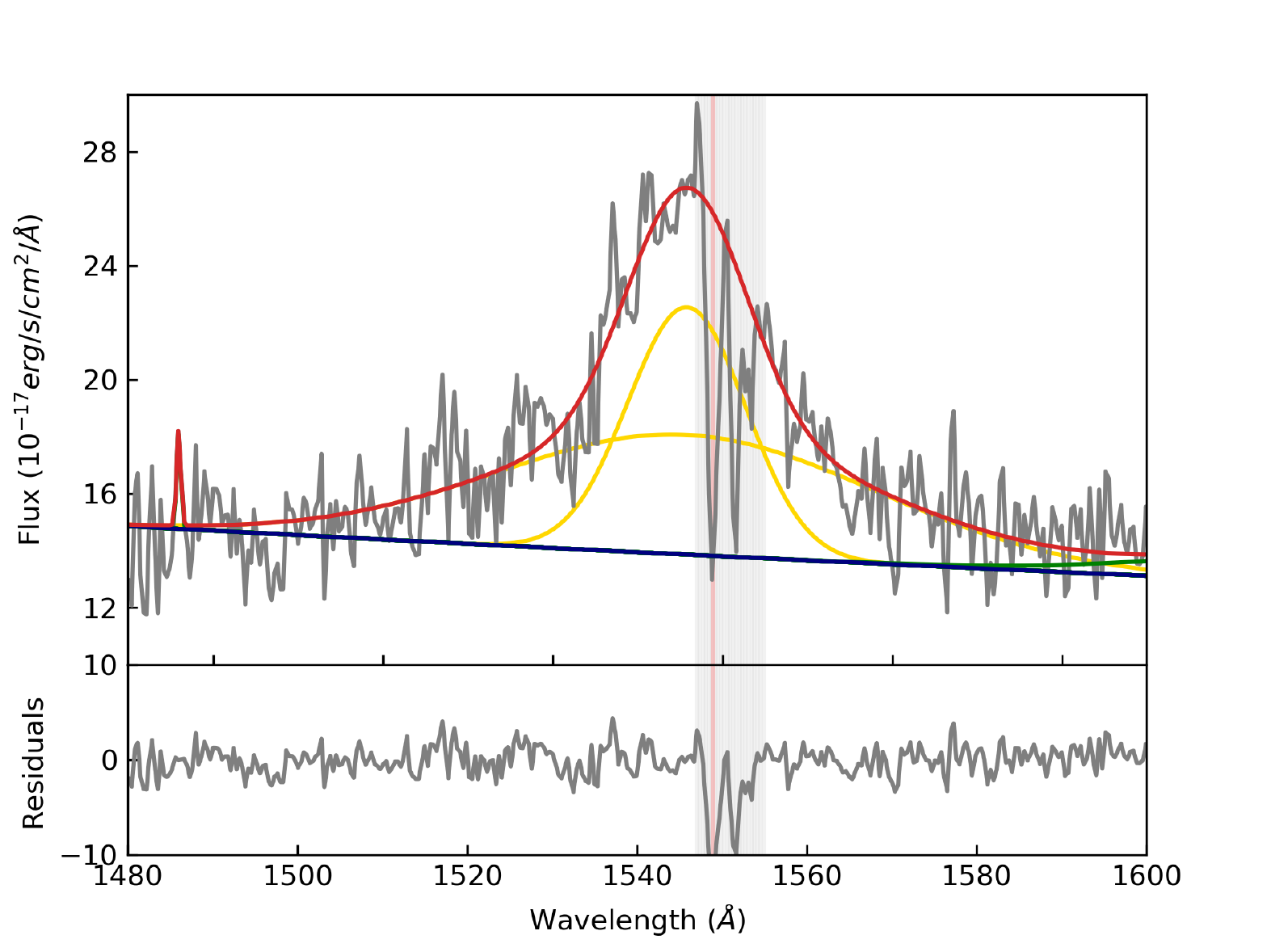}
   \includegraphics[width=0.4\textwidth]{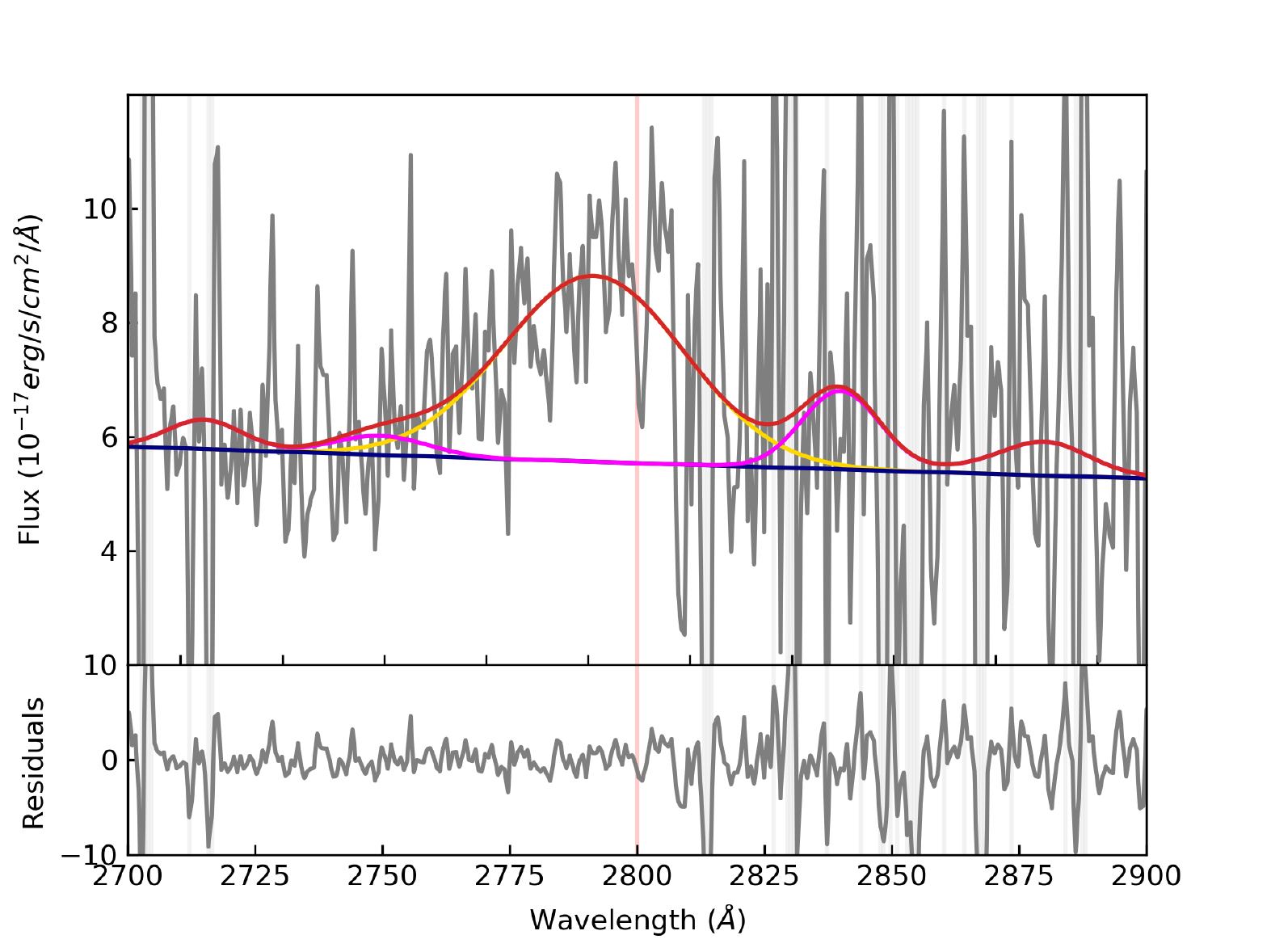}
   \includegraphics[width=0.4\textwidth]{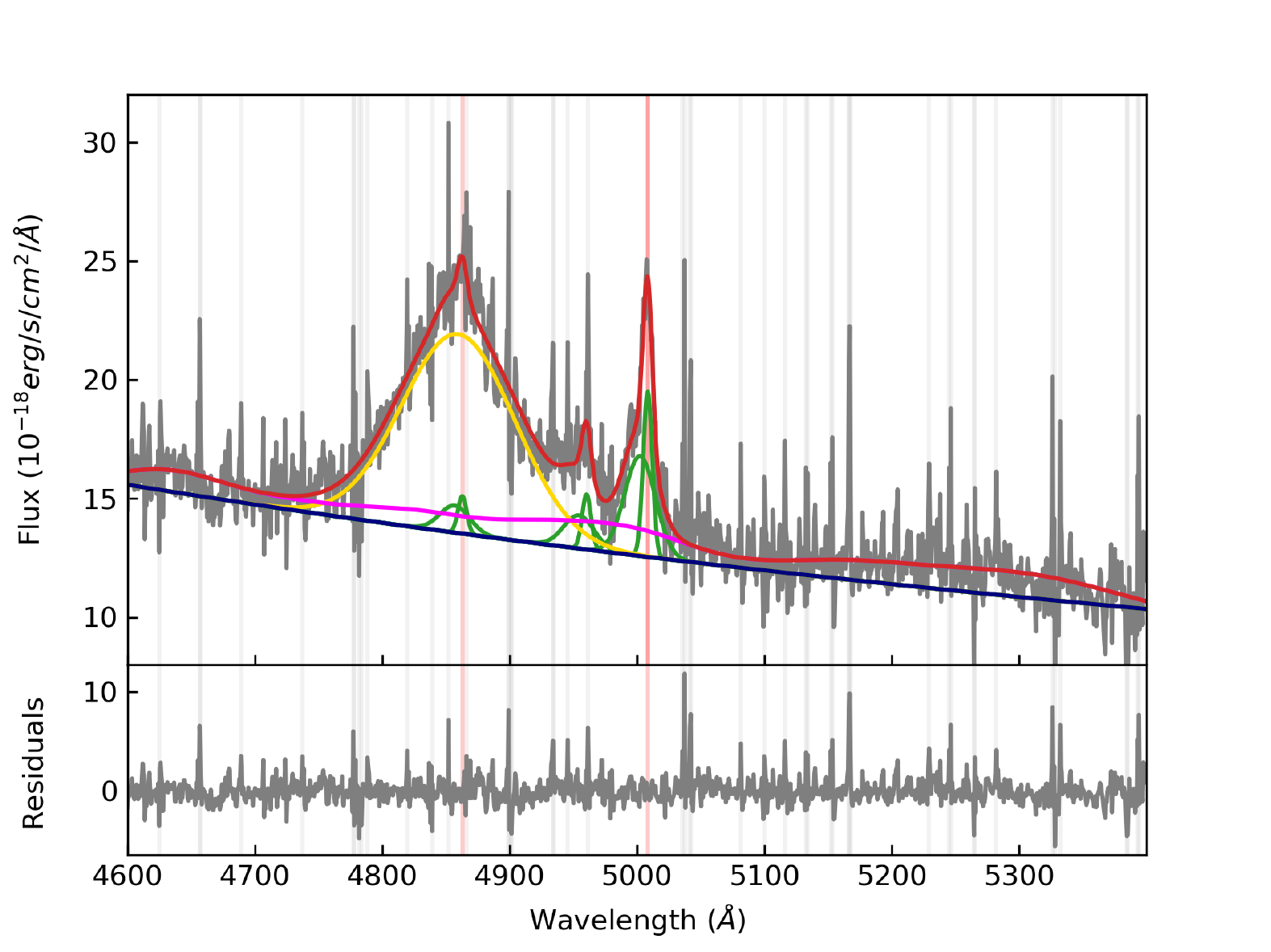}      \includegraphics[width=0.4\textwidth]{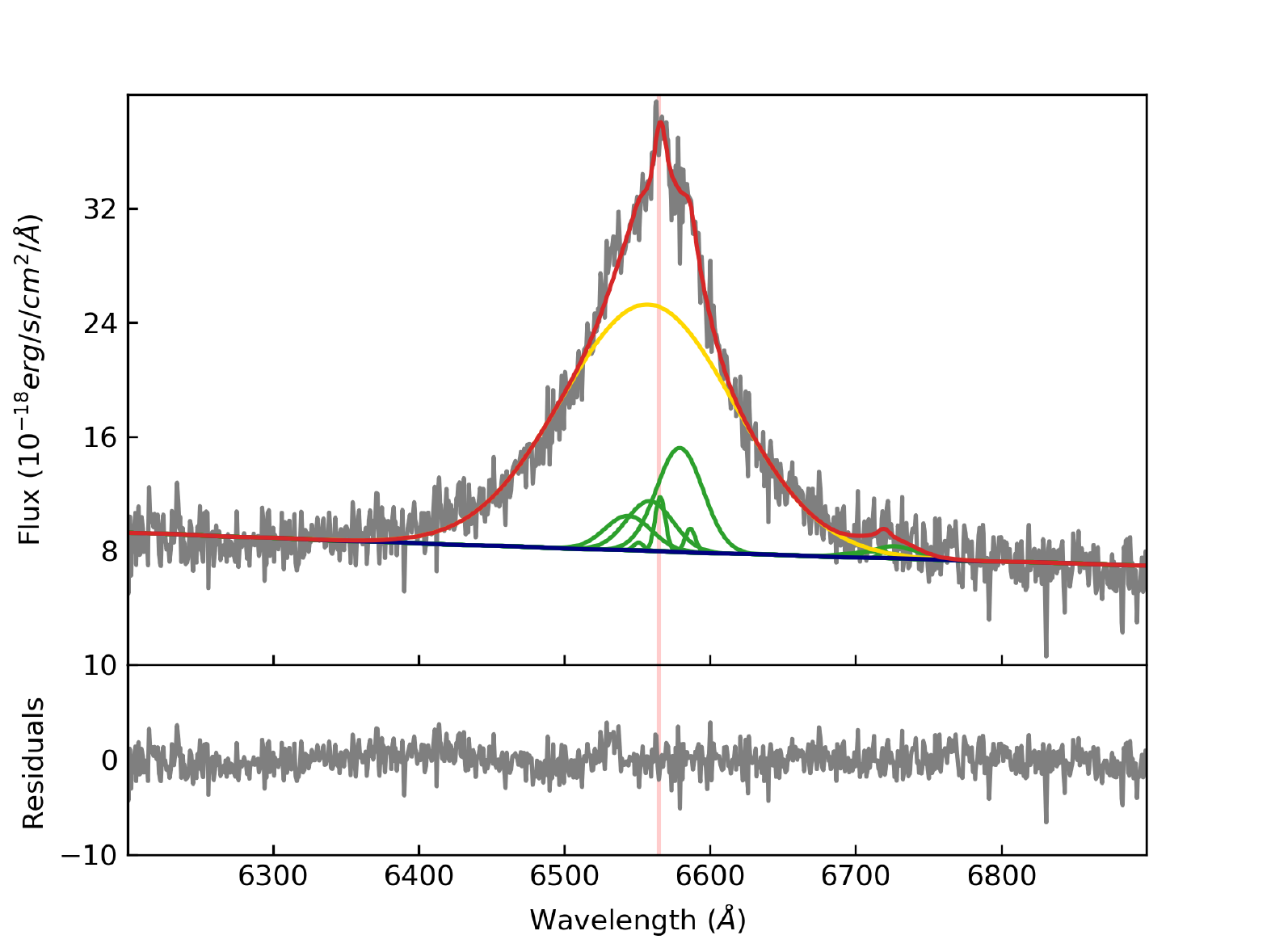}
\caption{S82X2058. The modeling is the same as in Fig. \ref{fig:app}}
   \end{figure*}

\end{appendix}
\end{document}